\RequirePackage{fix-cm}
\documentclass[smallextended]{svjour3}       
\smartqed  
\usepackage{graphicx}
\usepackage{mathptmx}      

\usepackage{latexsym}
\usepackage{amsfonts}
\usepackage{amsmath}
\usepackage{amssymb}
\usepackage{booktabs}
\usepackage{mathrsfs}
\usepackage{times}
\usepackage{lineno}
\usepackage{yhmath}
\usepackage{natbib}
\usepackage{aas_macros}
\usepackage{subfig}
\usepackage{pdflscape}
\usepackage[colorlinks=true,citecolor=blue,urlcolor=blue,breaklinks]{hyperref}

\newcommand{\f}[2]{\frac{#1}{#2}}
\newcommand{\pderiv}[2]{\frac{\partial #1}{\partial #2}}
\newcommand{\pd}[2]{\partial_{#2}{#1}}
\newcommand{\deriv}[2]{\frac{d #1}{d #2}}
\newcommand{\lo}{\textnormal{\tiny\textsc{L}}}
\newcommand{\hi}{\textnormal{\tiny\textsc{H}}}

\newcommand{\dr}{\Delta r}

\newcommand{\de}{\Delta \varepsilon}
\newcommand{\dx}{\Delta x}
\newcommand{\dt}{\Delta t}
\newcommand{\dV}{\Delta V}

\newcommand{\bK}{\mathbf{K}}
\newcommand{\tV}{\tilde{V}}
\newcommand{\tK}{\tilde{K}}
\newcommand{\IN}{\mbox{\tiny{\sc In}}}
\newcommand{\OUT}{\mbox{\tiny{\sc Out}}}
\newcommand{\ABEM}{\mbox{\tiny{\sc AbEm}}}
\newcommand{\SCAT}{\mbox{\tiny{\sc Scat}}}
\newcommand{\ISO}{\mbox{\tiny{\sc Iso}}}
\newcommand{\PAIR}{\mbox{\tiny{\sc Pair}}}
\newcommand{\nSpecies}{N_{\mbox{\tiny{\sc Sp}}}}
\newcommand{\LEFT}{\mbox{\tiny{\sc L}}}
\newcommand{\RIGHT}{\mbox{\tiny{\sc R}}}
\newcommand{\TOT}{\mbox{\tiny{\sc Tot}}}
\newcommand{\ME}{\mbox{\tiny{\sc Me}}}

\def\InvisibleSpace{}

\def\Mvariable{}
\def\Mfunction{}
\def\dispSFNumberedEquationmath{}

\def\multsp{}
\def\IndentingNewLine{}

\def\ScriptCapitalL{{\cal L}}

\def\ScriptCapitalT{{\cal T}}

\def\RawWedge{"705E}
\def\_{"7016}
\def\overvar#1#2{\mathaccent #2 #1}

\newcommand{\dd}{\mathrm{d}}

\newcommand{\npo}{{n+1}}

\newcommand{\lhalf}{(1/2)}

\newcommand{\lthreehalf}{(3/2)}
\newcommand{\kn}{{R_\epsilon}}
\newcommand{\varv}{{\upsilon}}

\def\ldbrack{\lbrack\!\!\!\left\lbrack}
\def\rdbrack{\rbrack\!\!\!\right\rbrack}

\newcommand{\ds}{\displaystyle}

\newcommand{\p}{\partial}
\newcommand{\ve}{\varepsilon}
\newcommand{\vk}{\varkappa}

\begin{document}

\title{Physical, numerical, and computational challenges of modeling neutrino transport in core-collapse supernovae
}

\titlerunning{Challenges of modeling neutrino transport in core-collapse supernovae}        

\author{Anthony Mezzacappa
\and
Eirik Endeve
\and
O.~E.~Bronson Messer
\and Stephen W.\ Bruenn
}


\institute{
A. Mezzacappa \at
              Department of Physics and Astronomy, University of Tennessee, Knoxville, TN 37996 \\
              \email{mezz@utk.edu} \\
              \\
E. Endeve \at
              Computer Science and Mathematics Division, Oak Ridge National Laboratory, Oak Ridge, TN 37831 \\
              Department of Physics and Astronomy, University of Tennessee, Knoxville, TN 37996 \\
              \email{endevee@ornl.gov} \\
              \\
O.~E.~B. Messer \at 
		National Center for Computational Sciences, Oak Ridge National Laboratory, Oak Ridge, TN 37831 \\
		Physics Division, Oak Ridge National Laboratory, Oak Ridge, TN 37831 \\
		Department of Physics and Astronomy, University of Tennessee, Knoxville, TN 37996 \\
              \email{bronson@ornl.gov} \\
              \\
S.~W. Bruenn \at
		Department of Physics \\
		Florida Atlantic University \\
		Boca Raton, FL 33431 \\
		\email{bruenn@fau.edu} \\
}

\date{Received: date / Accepted: date}

\maketitle

\begin{abstract}
The proposal that core collapse supernovae are neutrino driven is still the subject of active investigation
more than fifty years after the seminal paper by Colgate and White. The modern version of this paradigm,
which we owe to Wilson, proposes that the supernova shock wave is powered by neutrino heating, mediated
by the absorption of electron-flavor neutrinos and antineutrinos emanating from the proto-neutron star surface,
or neutrinosphere. Neutrino weak interactions with the stellar core fluid, the theory of which is still evolving, are 
flavor and energy dependent. The associated neutrino mean free paths extend over many orders of magnitude and are never always small relative 
to the stellar core radius. Thus, neutrinos are never always fluid like. Instead, a kinetic description of them in 
terms of distribution functions that determine the number density of neutrinos in the six-dimensional phase 
space of position, direction, and energy, for both neutrinos and antineutrinos of each flavor, or in terms of 
angular moments of these neutrino distributions that instead provide neutrino number densities in the 
four-dimensional phase-space subspace of position and energy, is needed. In turn, the computational 
challenge is twofold: (i) to map the kinetic equations governing the evolution of these distributions or 
moments onto discrete representations that are stable, accurate, and, perhaps most important, respect 
physical laws such as conservation of lepton number and energy and the Fermi--Dirac nature of neutrinos 
and (ii) to develop efficient, supercomputer-architecture-aware solution methods for the resultant nonlinear 
algebraic equations. In this review, we present the current state of the art in attempts to meet this challenge.
\end{abstract}


\setcounter{tocdepth}{3}
\tableofcontents


\section{Preface}

At this stage in the development of the theory of core-collapse supernovae two possible explosion mechanisms are most often discussed: neutrino-driven and magneto-rotationally-driven. The state of the theory is not sufficiently well developed to determine whether or not there is a clear break between these two cases or whether they represent limiting cases of a continuum. Nonetheless, in any scenario, the physics discussed here is relevant. It either dominates, leading to a neutrino-driven explosion, or sets the stage for a magneto-rotationally-driven supernova. That is, core-collapse supernova theorists have no choice but to first master and, more important, implement realistic models of neutrino transport in core-collapse supernova environments. What is meant by ``realistic'' will hopefully become clear as we progress through this review, but what will also hopefully become clear: Challenges to achieving realism will be faced on multiple fronts: physical, numerical, and computational.

When charged to write this review, we were asked not to provide an encyclopedic review of past work in the field but, rather, to present the current issues and challenges faced by the core-collapse supernova modeling community, particularly as they pertain to what is arguably the most difficult aspect to model: neutrino transport. Thus, with this charge in mind, we have written our review with an emphasis on the future, on what modelers must and will face to develop realistic models of these most important events.

\section{Setting the stage}
\label{sec:SettingTheStage}

The idea that core-collapse supernovae could be neutrino driven was first proposed more than fifty years ago by \citet{CoWh66} in their seminal numerical study. This work set neutrinos front and center in core-collapse supernova theory, which has remained the case ever since. The Colgate and White studies were followed by the early studies of \citet{Wilson1971} that cast doubt on the efficacy of their proposal. But the development of the electroweak theory, which predicted the existence of weak neutral currents, would change all that. Given weak neutral currents, Freedman recognized that it would be possible for neutrinos to scatter off of the nucleons in a nucleus \emph{collectively}. The cross sections for such scattering would be proportional to the nuclear neutron number, $N$, and would consequently be large. Shortly thereafter, \citet{Wilson1974}, using the new weak interaction cross sections for this process, demonstrated that the Colgate and White proposal was in fact viable. The recognition of this intertwined relationship between core-collapse supernova physics and neutrino weak interaction physics drives continued research to this day. Nearly forty years of further study in the context of the assumption of spherical symmetry was set in motion by this early and foundational work, which traversed a range of descriptions of neutrino transport in stellar cores, a range of sophistication of the treatment of the microphysics input included in the models, which includes the neutrino weak interaction physics and the equations of state describing a stellar core's nuclear, leptonic, and photonic degrees of freedom.

Neutrino mediation of core-collapse supernova dynamics in its modern instantiation is through charged-current absorption of electron neutrinos and antineutrinos on neutrons and protons, respectively. The nucleons become available as the stalled supernova shock wave dissociates the nuclei in the infalling stellar core material as the material passes through it. The neutrino absorption heats the material, depositing energy behind the shock. The shock loses energy initially to dissociation and neutrino losses. When sufficient energy is deposited by neutrino heating, the shock again becomes dynamical, propagates outward in radius, and reverses the infall of material passing through it, to disrupt the star in a core-collapse supernova \citep{Wilson1985,BeWi85}. This modern instantiation of neutrinos' role in the supernova mechanism relies on the developments surrounding the large neutrino--nucleus scattering cross sections discussed earlier. \citet{Arnett1977} was the first to show that such cross sections led to the trapping of the electron neutrinos produced during stellar core collapse through electron capture on nuclei and protons. He demonstrated that, despite their nature as weakly interacting particles, the densities in the stellar core rise sufficiently rapidly to render the electron neutrino mean free paths smaller than the size of the stellar core. Neutrino trapping gives rise to a trapped degenerate sea of electron neutrinos in the inner stellar core that emerge after stellar core bounce and the launch of the supernova shock wave from the proto-neutron star on diffusive time scales.

\begin{figure}[htbp]
\includegraphics[width=\textwidth]{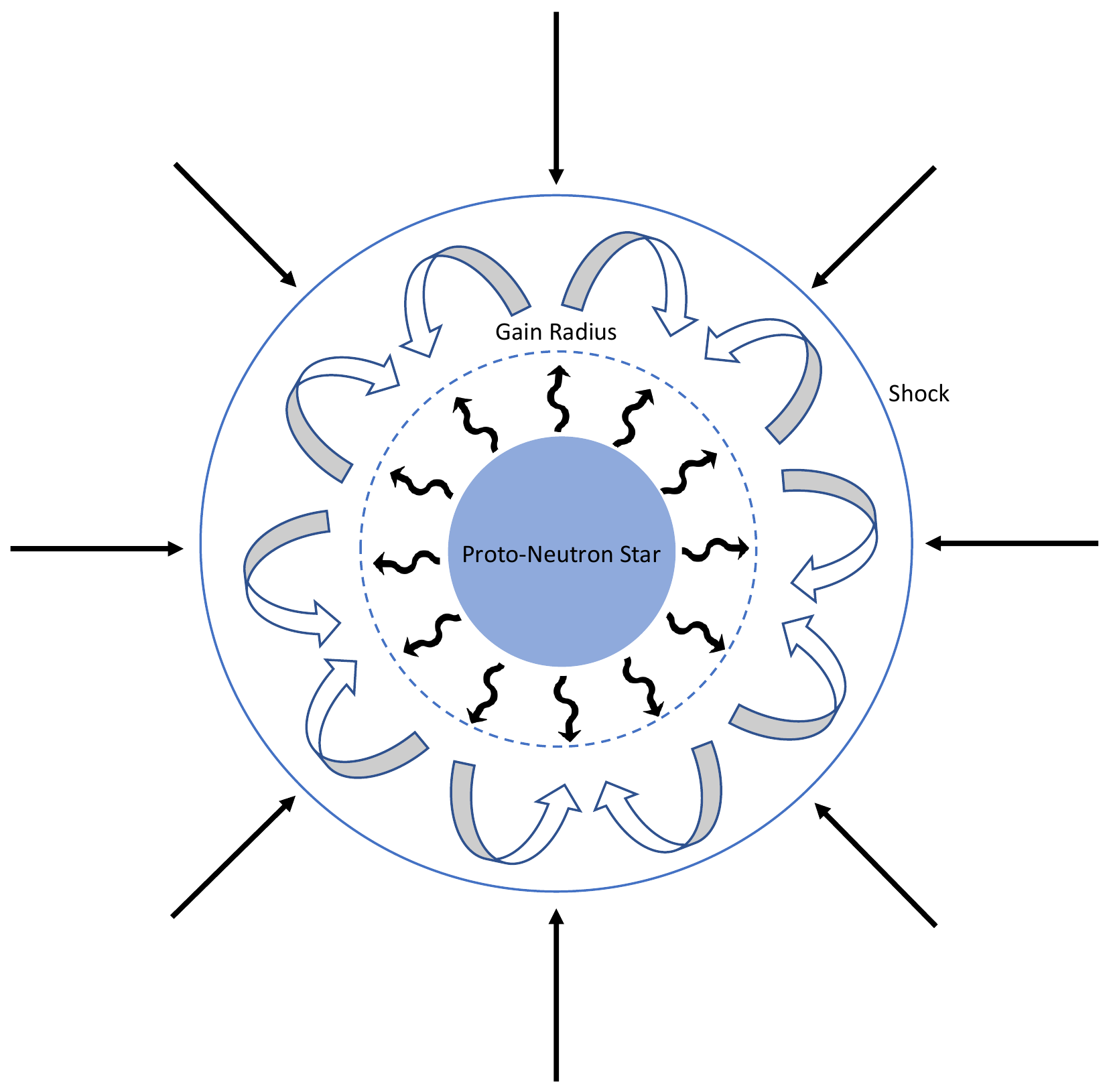}
\caption{Schematic showing the characteristic structure after stellar core bounce and the stall of the supernova shock wave seen in all core-collapse supernova models. All three flavors of neutrinos, together with their antineutrino partners, emanate from the proto-neutron star. Here, a single surface characterizes the proto-neutron star surface and the ``neutrinosphere,'' the surface of last scattering for the neutrinos. In reality, there are multiple surfaces, although they are very close together. The neutrino interaction cross sections are flavor and energy dependent. Consequently, there is a neutrinosphere for each neutrino flavor and energy ``group'' in core-collapse supernova models. Between the proto-neutron star surface and the stalled shock wave is the so called gain radius, separating the region of net neutrino cooling (below the gain radius) from net neutrino heating (above the gain radius). Neutrino heating is mediated by charged-current absorption of electron neutrinos (antineutrinos) on neutrons (protons) below the shock, liberated by shock dissociation of nuclei as they pass through it. Cooling is mediated by the inverse weak interactions. Neutrino heating in the ``gain region'' between the gain radius and the shock is central to the neutrino-driven core-collapse supernova mechanism. Given this neutrino heating, the gain region becomes convectively unstable. Neutrino-driven turbulent convection in this region assists neutrino heating to generate a supernova. The goal is to reverse the infall of the material ahead of the shock and for the shock itself to propagate outward. The neutrino heating in the gain region is sensitive to the neutrino luminosities, spectra, and angular distributions there, all of which depend on the transport of neutrinos through the semitransparent neutrinospheric region, where the neutrinos are neither diffusive nor free streaming.}
\label{fig:PNSGainShockProfile}
\end{figure}

The proto-neutron star comprises the inner cold unshocked core and a hot shocked mantle of material above it that is not ejected by the shock. Electron degeneracy is lifted in the hot mantle, leading to a significant population of electron--positron pairs, which in turn leads to the production of neutrinos and antineutrinos of all three flavors via electron--positron annihilation. The densities in the mantle are sufficiently high that neutrinospheres for all three flavors of neutrinos and antineutrinos exist, all lying within kilometers of each other, as a function of flavor and energy, in the density cliff that defines the proto-neutron star surface. The post-bounce stratification of the core, setting the stage for neutrino shock revival is shown in Fig.~\ref{fig:PNSGainShockProfile}. Neutrinos of all three flavors emerge from their respective neutrinospheres at the proto-neutron star surface. Between the proto-neutron star surface and the shock, neutrino heating and cooling take place through charged-current electron neutrino and antineutrino absorption on and emission by nucleons, respectively. The different radial dependencies of neutrino heating and cooling lead to net heating above the ``gain radius'' and net cooling below it. The region between the gain radius and the shock, where net neutrino heating takes place, is known as the gain region.

The energy deposition rate per gram of material in the gain region can be expressed in terms of the electron neutrino and antineutrino luminosities, squared rms energies, and inverse flux factor as
\begin{equation}
\dot{\epsilon}=\frac{X_n}{\lambda_{0}^{a}}\frac{L_{\nu_e}}{4\pi r^2} \left\langle E^{2}_{\nu_e} \right\rangle \left\langle \frac{1}{\mathcal{F}_{\nu_e}} \right\rangle
+\frac{X_p}{\bar{\lambda}_{0}^{a}}\frac{L_{\bar{\nu}_e}}{4\pi r^2} \left\langle E^{2}_{\bar{\nu}_e} \right\rangle \left\langle \frac{1}{\mathcal{F}_{\bar{\nu}_e}} \right\rangle,
\label{eq:heatingrate}
\end{equation}
where $\epsilon$ is the internal energy of the stellar core fluid per gram, $X_{n,p}$ are the neutron and proton mass fractions, respectively, $L_{\nu_e,\bar{\nu}_e}$ are the electron neutrino and antineutrino luminosities, respectively, $\mathcal{F}_{\nu_e,\bar{\nu}_e}$ are the inverse flux factors for the electron neutrinos and antineutrinos, respectively, and $\lambda_{0}^{a}, \bar{\lambda}_{0}^{a}$ are constants related to the weak interaction coupling constants. Thus, knowledge of the neutrino luminosities, spectra, and angular distributions are needed to compute the neutrino heating rates. This requires knowledge of the neutrino distribution functions, $f_{\nu_e,\bar{\nu}_e}(r,\theta,\phi,E,\theta_{p},\phi_{p},t)$, from which these quantities can be calculated. The neutrino distribution functions are determined by solving their respective Boltzmann kinetic equations, which will be discussed later. Thus, the core-collapse supernova problem is a phase space problem, in the end involving 6 dimensions plus time. The common parlance, dividing core-collapse supernova models between ``1D'' (spherical symmetry), ``2D'' (axisymmetry), or ``3D'' models is quite misleading. In reality, the dimensionality is 3D for spherical symmetry, involving 1 spatial dimension (radius) and 2 momentum-space dimensions (neutrino energy and a single direction cosine), 5D for axisymmetry, involving 2 spatial dimensions (radius and $\theta$) and 3 momentum-space dimensions (neutrino energy and 2 direction cosines), and 6D, involving 3 spatial dimensions (radius, $\theta$, and $\phi$) and 3 momentum-space dimensions (neutrino energy and 2 direction cosines).

The central densities of the proto-neutron star reach values between $10^{14}$ and $10^{15}\mathrm{\ g\ cm}^{-3}$. Its mass, which is $O(1)\,M_\odot$, is initially contained within a radius $O(100)$ km. Such conditions are not Newtonian. Detailed comparisons made in the context of spherically symmetric models of core-collapse supernovae \citep{BrDeMe01} between Newtonian and general relativistic models revealed the dramatic differences in the overall ``compactification'' of the postbounce core configuration defined by the neutrinosphere, gain, and shock radii, as well as the dramatic difference between the magnitudes of the infall velocities through the gain region. Moreover, neutrino luminosities and rms energies were increased in the general relativistic case due to the higher core temperatures. These studies made obvious the fact the core-collapse supernova environment is a general relativistic environment.

Models that assume spherical symmetry reached the needed level of sophistication only fairly recently, with fully general relativistic models that included Boltzmann neutrino transport, an extensive set of neutrino weak interactions, and, at the time, an industry-standard equation of state \citep{LiMeTh01,LeMeMe12a}. 
The outcomes of these models were quite discouraging.
In all cases, the shock radius reaches a maximum and then recedes with time until the simulations are terminated. Explosion does not occur, and the end outcome of each simulation would be the formation of a stellar-mass black hole.

With the exception of the lowest-mass massive stars \citep{KiJaHi06} it became clear the Colgate and White proposal was doomed to fail without the aid of additional physics. Specifically, the assumption of spherical symmetry had to be eliminated. In retrospect, it is now obvious why: Neutrino emission by the proto-neutron star, driving the explosion above, is fueled by the accretion of stellar core material onto it. Explosion in spherical symmetry would cut off such accretion entirely once initiated, cutting off the fuel that drives the neutrino emission that drives the explosion. Unless accretion and explosion can occur simultaneously, we are presented with a Goldilocks problem: Enough energy has to be deposited behind the shock before explosion occurs. But for sufficiently energetic explosions, an explosion cannot occur too soon. And given that the accretion rates decrease with time, due to stellar core density profiles, an explosion also cannot occur too late.

\begin{figure}[htb]
\includegraphics[width=\textwidth]{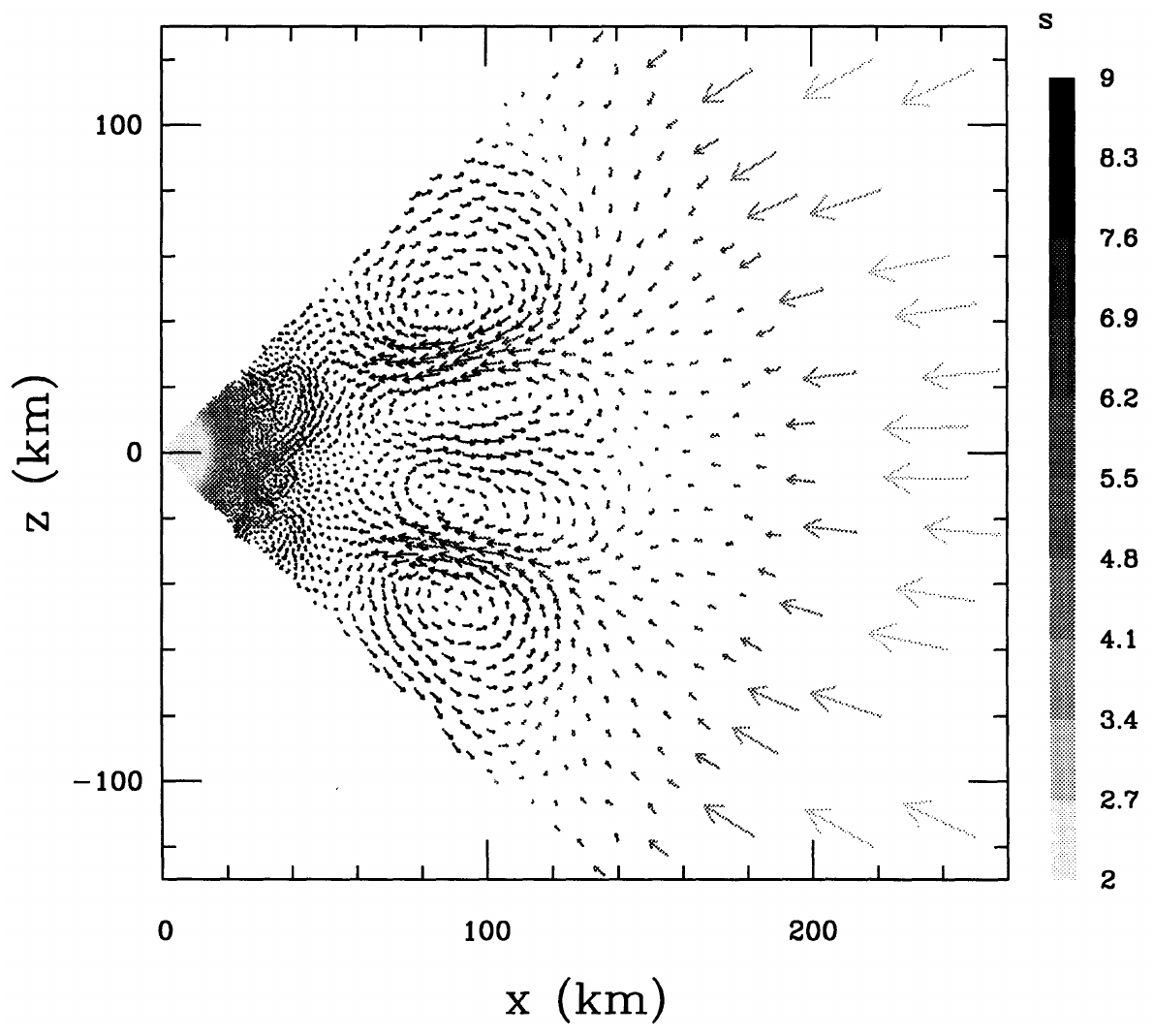}
\caption{Snapshot of neutrino-driven convection at 25 ms after bounce in the two-dimensional core-collapse supernova model of \citet{HeBeHi94} initiated from a $25\,M_\odot$ progenitor.}
\label{fig:NDConvection}
\end{figure}

The first two-dimensional core-collapse supernova simulations by \citet{HeBeCo92,HeBeHi94} demonstrated that accretion and explosion naturally coexist in the postshock flow. Heating by the proto-neutron star from below generates convection in the gain region. Such ``neutrino-driven'' convection allows continued accretion while some of the material is heated, expands, and moves outward. Lower-entropy, accreting fingers are evident in Fig.~\ref{fig:NDConvection}, as well as higher-entropy rising plumes. The Herant et~al.\ studies opened the next, much-needed chapter in core-collapse supernova theory. As with spherically symmetric modeling, axisymmetric modeling continues to this day. (See \citealt{Mueller2020} for a focused and comprehensive review on convection and other fluid instabilities in core collapse supernova environments that are integral to the supernova explosion mechanism.)

The core-collapse supernova modeling community has not yet produced general relativistic axisymmetric models with Boltzmann neutrino transport and with industry-standard weak interaction physics and equations of state, but significant progress has been made. The first simulations to evolve both the neutrino spectra and their angular distributions were performed by \citet{OtBuDe08}. Included were the spatial advection terms on the left-hand side of the Boltzmann equation (corresponding to neutrino transport in each of the spatial dimensions) and the collision term on the right-hand side of the equation (corresponding to neutrino sources and sinks due to emission, absorption, and scattering) with a subset of the weak interactions considered complete today. The simulations were purely Newtonian. Neglected were all relativistic effects in the Boltzmann kinetic equations, describing special relativistic Doppler shift of neutrino energies, general relativistic blue and red shift of neutrino energies, angular aberration of neutrino propagation, etc. Outcomes from their multi-angle, multi-frequency approach were compared with outcomes from a similar simulation performed with multigroup flux-limited diffusion. Notable differences were obtained between the two transport approaches in the results obtained for neutrino radiation field quantities entering the expression for neutrino heating, Eq.~\eqref{eq:heatingrate}---specifically, the inverse flux factors and rms energies --, which translated into notable differences in neutrino heating, which were up to a factor of 3 for rapidly rotating cores.
More recent studies assuming axisymmetry by \citet{NaIwFu18} implemented special relativistic Boltzmann neutrino transport with a subset of the neutrino weak interactions regarded as essential in today's leading multi-physics models, coupled to Newtonian hydrodynamics and gravity. In light of their Boltzmann implementation, these authors were able to make assessments regarding the fundamental assumption at the heart of the most commonly used closure prescription---the so-called M1 closure---currently in use in most multi-dimensional supernova studies deploying multidimensional neutrino transport in a moments approach we will discuss shortly. Nagakura et~al.\ find that the assumption that the neutrino radiation field is not in fact axisymmetric about the outward radial direction, reflected in non-negligible off-diagonal components of the Eddington tensor---specifically, $k^{r\theta}$. The authors emphasize how such components play a non-negligible role in the evolution of the neutrino fluxes, increasing the neutrino luminosities by $\sim$10\%. The neutrino heating rate, Eq.~\eqref{eq:heatingrate}, is then increased commensurately. Experience has shown that corrections at this level in any or all of the quantities entering the neutrino heating rate are noteworthy and warrant continued exploration, perhaps for all models, but especially in light of marginal cases of explosion for some, perhaps many, progenitors.

\begin{figure}[htbp]
\includegraphics[width=\textwidth]{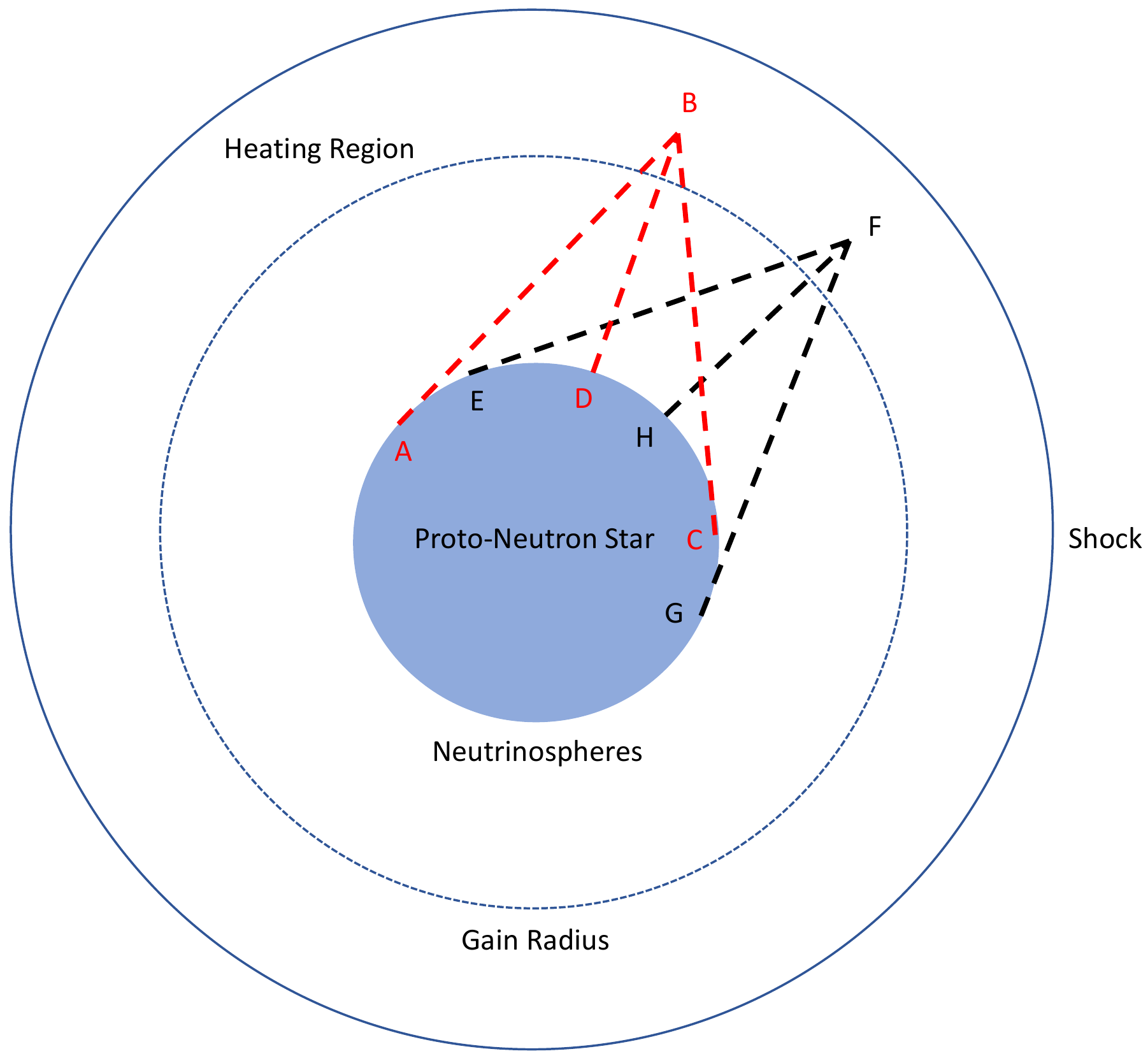}
\caption{Schematic showing the key characteristics of the ray-by-ray neutrino transport approximation. Along each radial ray (e.g., along segments DB or HF), a complete solution to the spherically symmetric neutrino transport equations is obtained assuming spherical conditions given by the conditions along each ray. This approximation afforded the ability to implement sophisticated transport solvers that had been developed in the context of models of core-collapse supernovae assuming spherical symmetry, at the expense of ignoring \emph{net} lateral transport that would occur in multiple spatial dimensions. In spherical symmetry, neutrinos can propagate along the segment AB, which is clearly not a purely radial segment. Therefore, there \emph{is} lateral transport. However, in spherical symmetry, every neutrino propagating along AB is matched by a neutrino propagating along CB, and the net flux at point B is purely radial. The lateral fluxes cancel exactly. Focusing on neutrino heating at point B, the ray-by-ray approach assumes that the thermodynamic conditions across the proto-neutron star surface (i..e., the neutrinosphere) between points A and C are uniform and given by the thermodynamic conditions at point D. Given a temporary hot spot at point D on the surface, the neutrino heating at point B would be overestimated. Moreover, were point H significantly cooler, relatively speaking, at the same instant, heating at point F would be underestimated because the hot spot at point D would be ignored even though it is within the cone of neutrino trajectories contributing to the neutrino heating at F. Thus, the ray-by-ray approximation may lead to larger angular variations in the neutrino radiation field than would be present were three-dimensional transport used---particularly if the hot spots on the proto-neutron star surface persist.}
\label{fig:RbR}
\end{figure}

Not unexpectedly, given the physical complexity and the computational cost, no simulations have been performed to date that deploy three-dimensional general relativistic Boltzmann neutrino transport in general relativistic core-collapse supernova models---i.e., including general relativistic hydrodynamics and gravity. This is a long-term goal and, as made clear by what we have learned in the context of studies in spherical and axisymmetry, a needed goal. Nonetheless, three-dimensional core-collapse supernova modeling of increasing sophistication is ongoing. The first three-dimensional core-collapse supernova models were performed by \citet{FrWa04} using gray (neutrino angle- and energy-integrated) radiation hydrodynamics. The first spectral (neutrino-angle integrated) three-dimensional models were performed by \citet{HaMuWo13}. The current stable of spectral three-dimensional models fall under two categories. Both implement spectral (but not multi-angle) neutrino transport in a one- or two-moment approach. In one category, the so-called ``ray-by-ray'' approximation is used. In the other, the neutrino transport is three dimensional. (A clarifying remark: The simulations by Hanke et al. used a Boltzmann solver in the context of their ray-by-ray approach. As such, some angular dependence was kept. However, three-dimensional models require two angles to describe a neutrino's propagation direction, and in the ray-by-ray approach the angular dependence in one of the angles is approximate in the sense that it is computed assuming spherical symmetry.)

The earliest three-dimensional models---e.g., those of Hanke et~al.\---implemented ray-by-ray transport. In the ray-by-ray approach, the three-dimensional neutrino transport problem is broken up into $N=N_{\theta}\times N_{\phi}$ spherically symmetric problems, where $N_{\theta,\phi}$ are then number of $\theta$, $\phi$ zones used in the simulation. The ray-by-ray approximation follows lateral neutrino transport under the assumption of spherical symmetry, meaning there is lateral transport of individual neutrinos, but the net lateral flux is zero. (For example, neutrinos can propagate along the segment between A and B in Fig.~\ref{fig:RbR}, but an equal number of neutrinos must propagate along the path between C and B, such that the net flux at point B is purely radial.) Moreover, as illustrated by Fig.~\ref{fig:RbR}, neutrino heating at a point in the gain region may be over- or under-estimated. Consider the point B in the heating region. The backward cone emanating from point B subtends a portion of the neutrinosphere, between points A and C, that is the source of the neutrinos that heat the material at point B. The ray-by-ray approximation, which assumes spherical symmetry for each ray, assumes that the thermodynamic conditions across the neutrinosphere between points A and C are the same as those at point D. If point D is a hot spot, the ray-by-ray approximation will compute the heating at point B assuming the neutrinosphere between points A and C is hot. For neutrino heating at point F, and assuming that point H is not a hot spot, the ray-by-ray approximation will assume that conditions at point H are mimicked across the portion of the neutrinosphere between points E and G, regardless of the fact that point D is hot and within that portion of the surface. Thus, the neutrino heating at point B will be overestimated, whereas the neutrino heating at point F will be underestimated. Whether or not the ray-by-ray approximation leads to significant over- or under-estimations of the neutrino heating over the course of the shock reheating epoch will of course depend on whether or not such variations in the thermodynamic conditions across the neutrinosphere persist, which requires a comparison taking into consideration the time dependence of such thermodynamic conditions.
Comparisons between ray-by-ray and non-ray-by-ray approaches in the context of axisymmetric core-collapse supernova models found notable differences in, among other outcomes, the time to explosion \citep{SkBuDo16}. However, more recent comparisons in the context of three-dimensional models found no significant differences between the two approaches \citep{GlJuJa19}. Of course, without three-dimensional transport implementations, it would be difficult to assess the efficacy of using the ray-by-ray approach, or other approximations. In the end, such approximations must be removed, if only just to check them. The ray-by-ray approach of the Oak Ridge group is based on one-moment closure through flux-limited diffusion \citep{BrBlHi20}. They follow the evolution for the lowest angular moment of the neutrino distribution: the number density. The Max Planck group's ray-by-ray implementation is based on two-moment closure \citep{RaJa00}. They solve an approximate Boltzmann equation for the purposes of computing the variable Eddington factor needed to close the system of equations describing the evolution of the first two moments of the neutrino distribution (in spherical symmetry, there is only one first moment, corresponding to the radial number flux, together with the zeroth moment, the neutrino number density). 

For both two- and three-dimensional core-collapse supernova models that attempt to include general relativity at some level of approximation, if not exactly, Newtonian or general relativistic hydrodynamics, and two- or three-dimensional neutrino transport are all based on the solution of the neutrino moments equations describing the evolution of the lowest angular moments of the neutrino distribution function. For example, in terms of the neutrino distribution function, the number moments (spectral number density, spectral number flux) are defined as
\begin{equation}
\mathcal{N}(r,\theta,\phi,E,t)\equiv\int_{0}^{2\pi}d\phi_p\int_{-1}^{+1}d\mu f(r,\theta,\phi,\mu,\phi_p,E,t),
\label{eq:zerothmoment}
\end{equation}
\begin{equation}
\mathcal{F}^{i}(r,\theta,\phi,E,t)\equiv\int_{0}^{2\pi}d\phi_p\int_{-1}^{+1}d\mu n^{i}f(r,\theta,\phi,\mu,\phi_p,E,t),
\label{eq:firstmoment}
\end{equation}
where $\mu\equiv\cos\theta_p$ is the neutrino direction cosine defined by $\theta_p$, one of the angles of propagation defined in terms of the outward pointing radial vector defining the neutrino's position at time $t$. In three dimensions, two angles are needed to uniquely define a neutrino propagation direction. The angle $\phi_p$ provides the second. $n^i$ is the neutrino direction cosine in the $i^{\rm th}$ direction, whose components are given as functions of $\mu$ and $\phi_p$. $E$ is the neutrino energy. $E,\theta_p,\phi_p$ can be viewed as spherical momentum space coordinates. Above, $\mathcal{N}$ and $\mathcal{F}^i$ are the number density and number fluxes, respectively. In three dimensions, there is of course a number flux for each of the three spatial dimensions, delineated by the superscript $i$. Integration of the neutrino Boltzmann equation over the angles $\theta_p$ and $\phi_p$, weighted by $1$, $n^i$, $n^{i}n^j$, ... defines an infinite set of evolution equations for the infinite number of angular moments of the distribution function, which is obviously impossible to solve. In a moments approach to neutrino transport, the infinite set of equations is rendered finite by truncation, after the equation for the zeroth moment in the case of one-moment closure (e.g., flux-limited diffusion) or after the equations for the first moments in the case of two-moment closure (e.g., M1 closure). In the latter case, closure can be ``prescribed'' (e.g., M1 closure) or computed (e.g., through a variable Eddington tensor approach). We will discuss these approaches in greater detail later in our review. It is important to understand the essence of the approximations being made in moments approaches to neutrino transport in core-collapse supernova models. One does not integrate out all of the angular information contained within the neutrino distribution function. Some angular information remains. The higher the closure is made in the order of moment equations, the more angular information is kept. For example, two-moment closure keeps the fundamental angular dependencies. The ratio of the number flux in any of the three dimensions to the number density, at any spacetime point, is a measure of how forward peaked the neutrino angular distribution is in that dimension at that point. Thus, a moments approach retains much of the information of the neutrino radiation field contained within the neutrino distribution functions, while providing a sophisticated modeling path forward that is achievable on present leadership-class computing systems. Direct Boltzmann solutions for the neutrino radiation field will have to wait until sustained exascale computing platforms become available over the next decade.

%
Three-dimensional models that include an approximation to general relativistic gravity in the form of an ``effective potential,'' Newtonian hydrodynamics, 
ray-by-ray one- or two-moment neutrino transport with some corrections for special relativity ($O(v/c)$) or general relativity (gravitational redshift of neutrino energies), 
and a state-of-the-art set of neutrino weak interactions have been performed by the Max Planck and Oak Ridge groups \citep{HaMuWo13,LeBrHi15,MeJaMa15,MeJaBo15,SuJaMe18}. 
Three-dimensional models that include general relativistic hydrodynamics and gravity, and three-dimensional, general relativistic, $O(1)$ or fully relativistic (special and general) 
two-moment neutrino transport with an extensive set of neutrino weak interactions have been performed by \citet{RoOtHa16} and \citet{KuTaKo16}, respectively. Three-dimensional 
models that couple Newtonian hydrodynamics and approximate general relativistic gravity, as above, to three-dimensional two-moment neutrino transport with corrections for 
special and general relativity, as above, and an extensive set of neutrino weak interactions were performed by \citet{OcCo18,VaBuRa19,BuRaVa19}.

%
It is clear the core collapse supernova modeling state of the art in three dimensions is evolving, with some models classifiable as more complete macrophysically 
--- i.e., that implement three-dimensional, general relativistic gravity, hydrodynamics, and neutrino transport --- and some models classifiable 
as more complete microphysically --- i.e., that include state-of-the-art microphysics.

Note that the computational cost associated with the solution of the neutrino moment or Boltzmann transport equations is dominated by the computations associated with the ``collision term''---i.e., with the neutrino interactions with the stellar core fluid. It is also clear---and of course at this point should come as no surprise---the above history of the development of core-collapse supernova theory over the last fifty-plus years centers on the development of neutrino transport theory and its implementation in this context.

Neutrino mass, albeit small in relation to the neutrino energies attained in core collapse supernovae, leads to neutrino flavor transformations. There is growing, though still inconclusive, evidence that such transformations may play a role in neutrino shock reheating (e..g., see \cite{TaHuRa17,AbDuSu19,AzYaMo19}).  The existence of so called ``fast'' flavor transformations, which can exist even in the baryon-laden environment below the supernova shock wave, was first brought to the attention of the supernova modeling community by Sawyer (\cite{Sawyer05}). Prior to this work, it was assumed that quantum mechanical coherence among the neutrinos in the region beneath the shock would de-cohere due to neutrino--matter collisions, thereby rendering such effects unimportant to neutrino shock reheating. However, fast modes operate on scales much shorter than a neutrino mean free path and, in fact, are not wiped out by collisions and beg to be considered. As in the classical case, the story boils down to capturing the neutrino angular distributions for all three flavors of neutrinos, as a function of space and time during the evolution of the supernova. The neutrinospheres for the three neutrino flavors are distinguished first and foremost by their interactions with the stellar fluid, with electron neutrinos and antineutrinos interacting through both charged and neutral currents and the muon and tau neutrinos interacting only through neutral currents. Moreover, the preponderance of neutrons over protons reduces the opacity of the stellar fluid to electron antineutrinos, and a hierarchy sets in, with the muon and tau neutrinospheres at the highest densities, followed by the neutrinosphere associated with the electron antineutrinos, followed in turn by the neutrinosphere associated with the electron neutrinos, at the lowest densities, relatively speaking. Given the layering of the neutrinospheres, at a given time during neutrino shock revival, the neutrino angular distributions at a given spatial location in the cavity between the neutrinospheres and the shock will differ by flavor. It is the differences between the angular distributions of each flavor that sets the stage for fast flavor transformation.

Thus, the need, in the classical case, for a Boltzmann description of the neutrino radiation field is multifold: (1) Moments approaches are approximations, whose efficacy cannot be known {\em a priori} and must be checked against the exact (classical) result. Examples of this will be discussed here. (2) The development of closure prescriptions for moment models is rife with difficulty, partially because of nonlinearities introduced by the closure procedure. For example, a numerical method for two-moment, multifrequency, general relativistic neutrino transport that respects Fermi--Dirac statistics does not yet exist and will be difficult to develop.  Furthermore, the development of nonlinear moment models beyond the two-moment approximation, to capture more kinetic effects, will be even more challenging. (3) Boltzmann and low-order moments approaches can be used together to accelerate convergence of the solution to the Boltzmann equation, potentially becoming competitive, in terms of speed and memory use, with nonlinear, high-order moments approaches. (4) The exploration of fast flavor transformations on the core collapse supernova mechanism will require precise knowledge of the neutrino angular distributions for all three flavors across spacetime of a supernova model. Such information can be obtained only through a solution of the classical Boltzmann kinetic equations for each neutrino flavor in association with simulation of the coherent quantum effects -- i.e., through a solution of the multi-angle, multi-frequency neutrino {\em quantum} kinetics equations for all neutrino flavors.

While the justification for deploying Boltzmann kinetics in the classical case can be made, it is through a combination of Boltzmann and moments approaches that progress will be made in both the near and the long term. We are attempting to address myriad science questions, and past experience already tells us that the answer to these questions will vary with characteristics of the massive progenitors in which core collapse supernovae occur. How do massive stars explode? Which explode and which do not? Among those that explode, what elements do they produce? How do they contribute to galactic chemical evolution? And the list goes on. At present, there is no foreseeable time at which all of these questions will be addressable with Boltzmann methods, let alone quantum kinetics. An uncountable number of models will ultimately be required to understand the death of the diverse population of stars we are presented with in nature, as well as the death of any one of them. Our understanding of stellar death will not come from a single ``hero'' simulation, but from many simulations. Thus, it is in the application of both Boltzmann (classically) and moments approaches and, through this, the development of ever more realistic moments approaches that we will be able to advance our knowledge of one of the most important phenomena in the Universe. This is already clear from the modeling history to date. We have come a long way since Colgate and White's seminal work through precisely the hybrid approach discussed here. Hence, this review will focus on both approaches, as well as point to potentially efficacious hybrid approaches that could be developed and deployed in the future.

\section{Design specifications}
There have been many lessons learned during the 54 years that have passed since the first numerical simulations of core-collapse supernovae were performed by Colgate and White. These lessons can now be used to construct a list of design specifications for models of neutrino transport that will be used in future core-collapse supernova models:
\begin{enumerate}
\item Ultimately, definitive simulations of core-collapse supernovae in the classical limit will require a Boltzmann kinetic description of neutrino transport for all three flavors of neutrinos and their antineutrino partners.
\item In the event sufficient evidence points to the need to consider in greater detail the impact of neutrino quantum kinetics on the supernova explosion mechanism, a quantum kinetics description of neutrino transport would be required. A classical Boltzmann description would be the natural, and required, starting point for the development of a such a quantum kinetics treatment.
\item The simulations must be general relativistic. They must include special and general relativistic effects such as Doppler and red/blue shifts of neutrino energy, respectively, and angular aberration in both cases, due to fluid motion and spacetime curvature.
\item These simulations must include all of the neutrino weak interactions that have been to date demonstrated to be important, and the description of the interactions must be state of the art.
\item The quality of core-collapse supernova simulations will ultimately be gauged by, among other things, the degree to which lepton number and energy are conserved.
More specifically, the discretizations of the integro-partial differential Boltzmann equations must conserve lepton number and energy \emph{simultaneously}.
\item The discretizations of the Boltzmann equations---in particular, the collision terms---must accommodate both small- and large-energy scattering.
\item The numerical methods must also accommodate realistic equations of state for the nuclear, leptonic, and photonic components. In cases where the neutrino opacities depend on the nuclear force model, the neutrino opacities and the equation of state must be consistent.
\item In the interim when moments approaches to neutrino transport must be used until Boltzmann approaches become feasible, all of the above design specifications still hold.
\item For moments models, the closures used must respect the Fermi--Dirac statistics of neutrinos, reflecting the fact that the neutrino distribution functions are bounded.
\end{enumerate}
\section{The equations of neutrino radiation hydrodynamics}

In core-collapse supernova models, the stellar fluid is modeled as a perfect fluid, augmented by an equation for the electron density in order to accommodate a nuclear equation of state.  
(For brevity of presentation, we will not include effects due to electromagnetic fields.)  
The relevant equations are then
\begin{align}
  \nabla_{\nu}J_{\mbox{\tiny B}}^{\nu} &= 0, \label{eq:BaryonMassConservation} \\
  \nabla_{\nu}T_{\mbox{\tiny fluid}}^{\mu\nu} &= - G^{\mu}(f_{\nu_{e}},f_{\bar{\nu}_{e}},\ldots), \label{eq:fluidFourMomentumConservation} \\
  \nabla_{\nu}J_{e}^{\nu} &= - m_{\mbox{\tiny B}}\,L(f_{\nu_{e}},f_{\bar{\nu}_{e}},\ldots), \label{eq:ElectronNumberConservation}
\end{align}
where the baryon rest-mass density current is
\begin{equation}
  J_{\mbox{\tiny B}}^{\nu} = \rho\,u^{\nu},
\end{equation}
where $\rho=m_{\mbox{\tiny B}}\,n_{\mbox{\tiny B}}$ is the baryon rest-mass density, $m_{\mbox{\tiny B}}$ the average baryon (rest) mass, $n_{\mbox{\tiny B}}$ the baryon density, and $u^{\nu}$ is the fluid four-velocity.  
The fluid energy-momentum tensor is
\begin{equation}
  T_{\mbox{\tiny fluid}}^{\mu\nu} = \rho\,h\,u^{\mu}\,u^{\nu} + p\,g^{\mu\nu},
\end{equation}
where $h=1+(e+p)/\rho$ is the specific enthalpy, $e$ the internal energy density, and $p$ the pressure.  
The electron density current is given by
\begin{equation}
  J_{e}^{\nu} = \rho\,Y_{e}\,u^{\nu},
\end{equation}
where $Y_{e}$ is the electron fraction.  
The electron density (technically electron minus positron density) is $n_{e}=\rho\,Y_{e}/m_{\mbox{\tiny B}}$.  
To close the system given by Eqs.~\eqref{eq:BaryonMassConservation}--\eqref{eq:ElectronNumberConservation}, the pressure $p$ is given by an equation of state (EOS); e.g., $p=p(\rho,e,Y_{e})$.  

The source terms on the right-hand sides of Eqs.~\eqref{eq:fluidFourMomentumConservation} and \eqref{eq:ElectronNumberConservation}, $-G^{\mu}$ and $-L$, describe four-momentum and lepton exchange between the fluid and neutrinos.  
These terms depend on the neutrino distribution functions (or moments of the neutrino distribution functions), as already noted in Sect.~\ref{sec:SettingTheStage} , as well as on thermodynamic properties of the stellar fluid.  
This nonlinear coupling is the key to the supernova mechanism, and associated observables, and is the topic of the present review.  

\subsection{The need for a kinetic description of neutrinos}
\label{sec:needForKineticDescription}

\begin{figure}[h]
	\captionsetup[subfigure]{justification=centering}
	{	\includegraphics[width=0.5\linewidth]{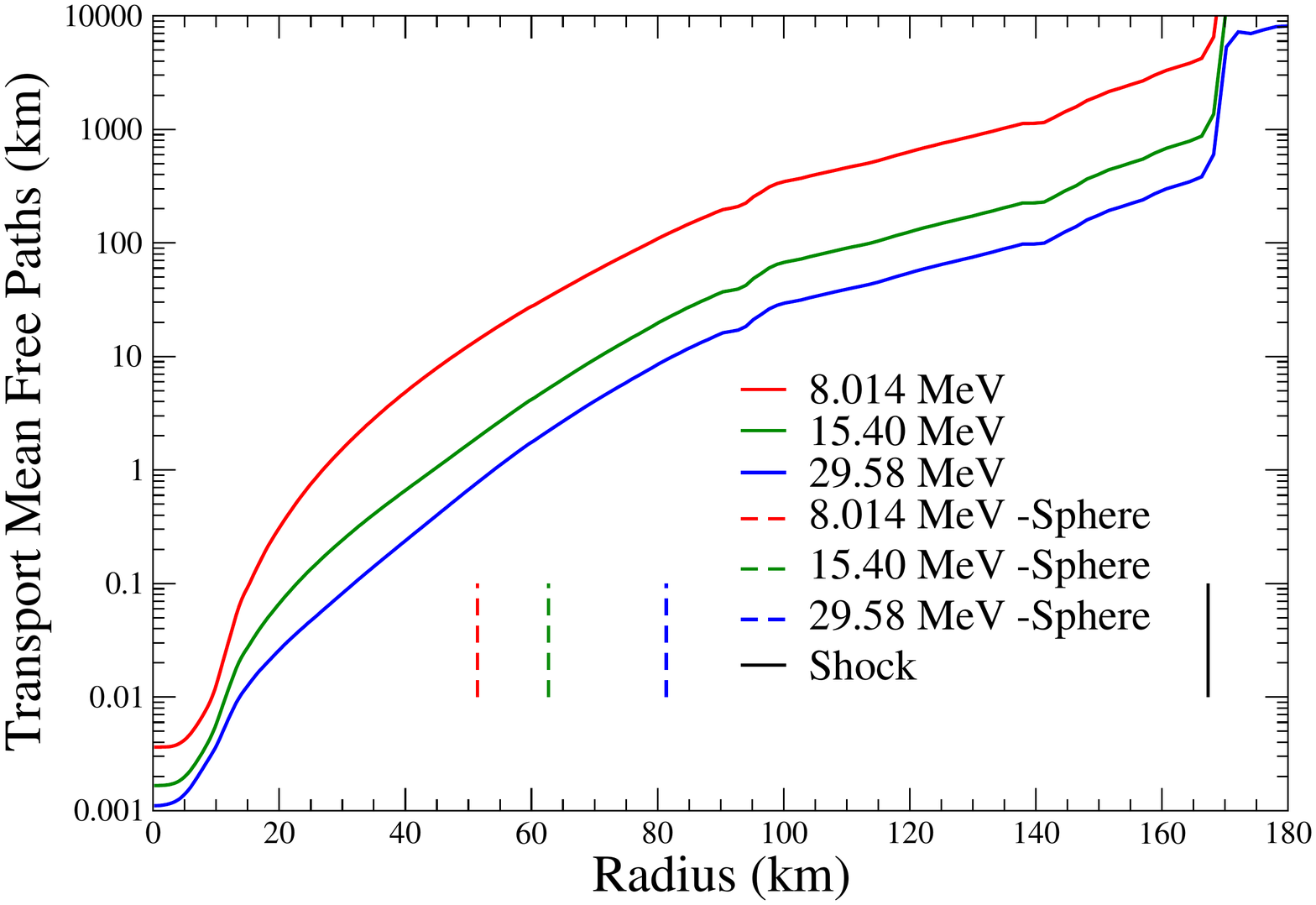}
			\label{fig:tmfp_nue_mag}}~
	{	\includegraphics[width=0.5\linewidth]{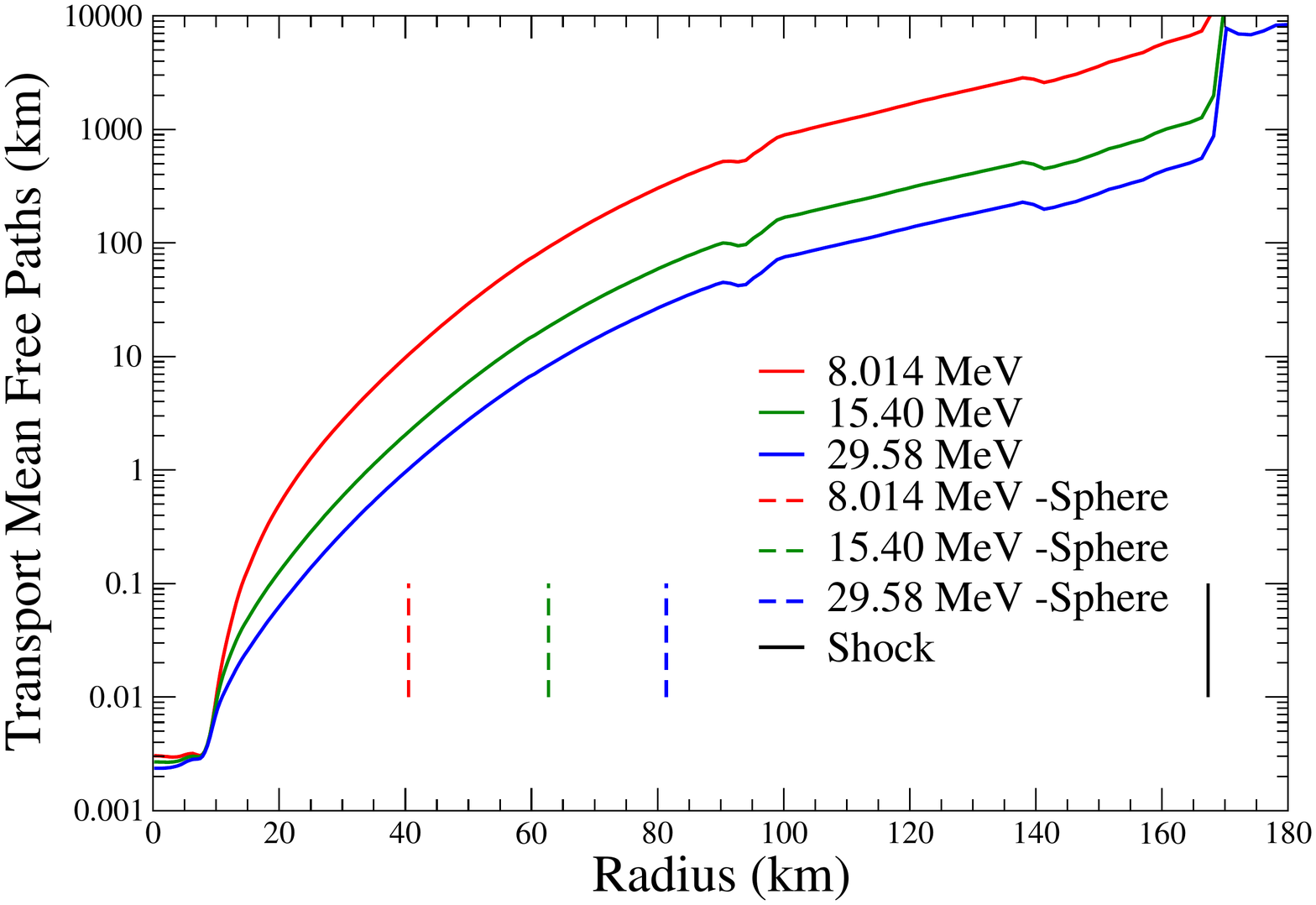}
		\label{fig:tmfp_nuebar_mag}}\\
	{	\includegraphics[width=0.5\linewidth]{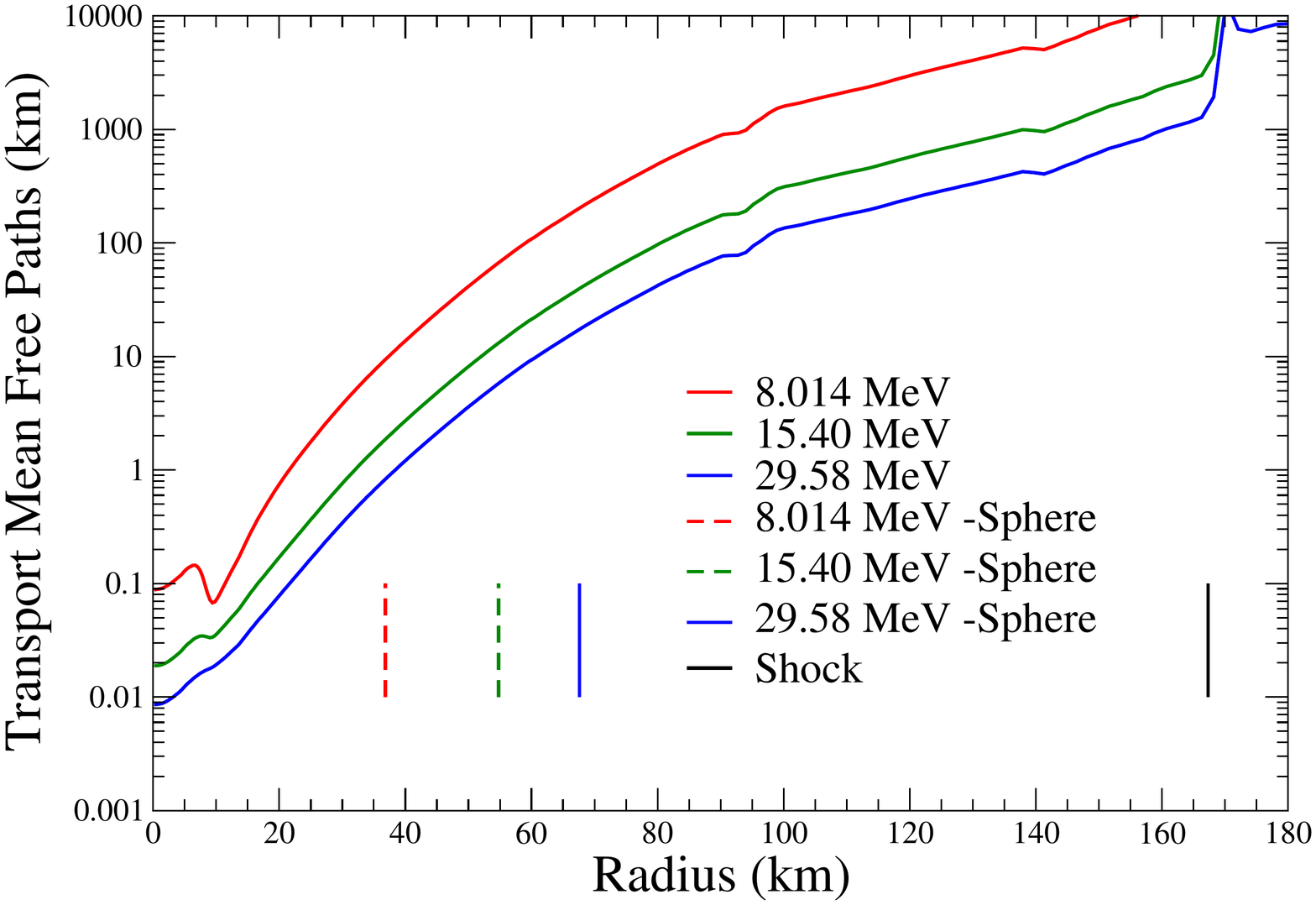}
		\label{fig:tmfp_nux_mag}}~
	{	\includegraphics[width=0.5\linewidth]{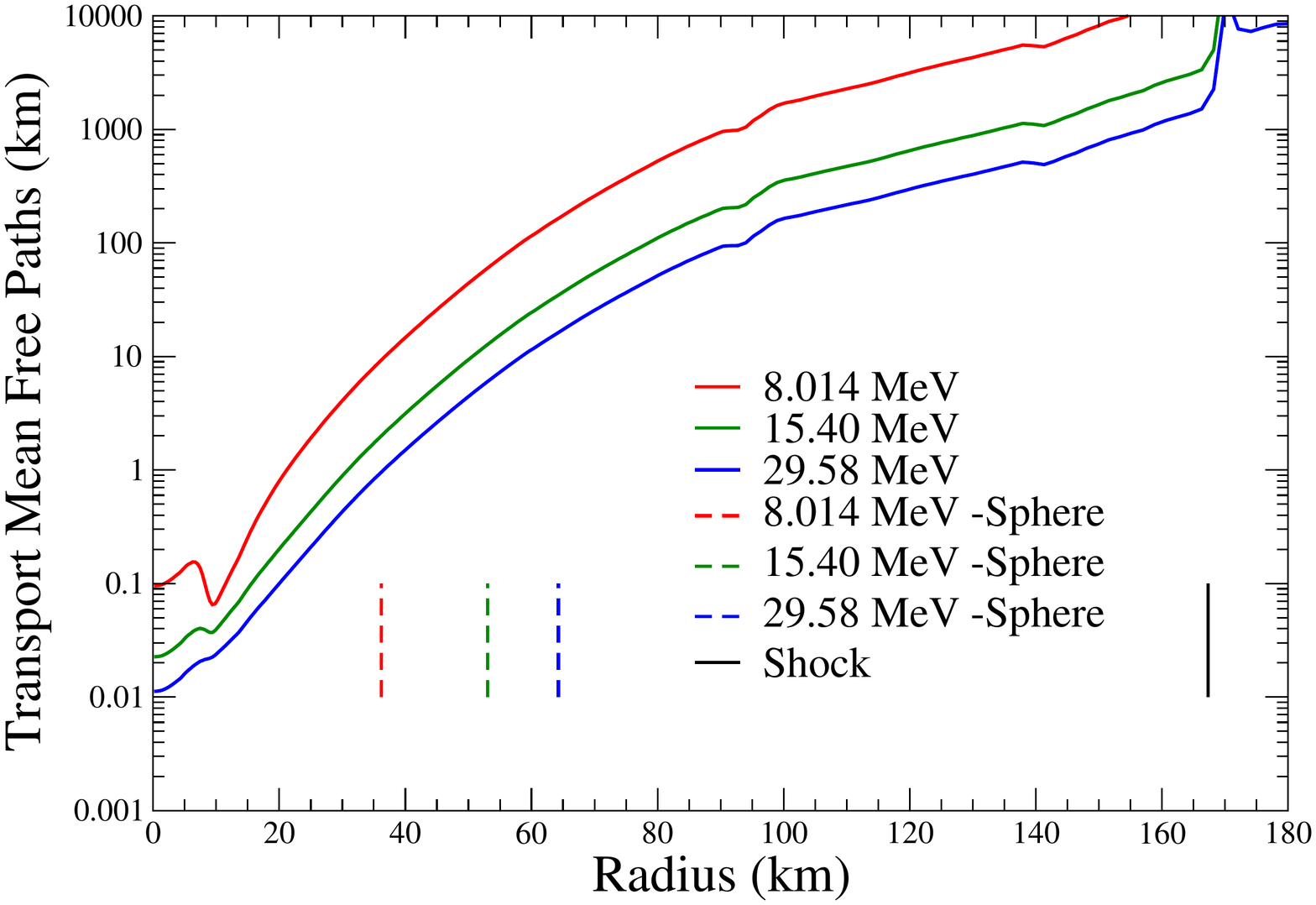}
		\label{fig:tmfp_nuxbar_mag}}~
		
	\caption{Plots of the neutrino and antineutrino mean free paths at 100 ms after bounce, during the neutrino shock reheating epoch, for all three flavors of neutrinos at select energies. The upper left and right panels show plots of the electron-neutrino and anti-neutrino mean free paths, respectively. The lower left and right panels show plots of the heavy-flavor ($\mu$ and $\tau$) neutrinos and antineutrinos, respectively. The data used to generate the plots are taken from a supernova model beginning with a $12\,M_\odot$ progenitor and evolved with the \textsc{Chimera} supernova code. To set the correct physical scale against which the mean free paths can be compared, we indicate the location of the various neutrinospheres and the shock wave. All four plots demonstrate that, as we move out in radius to lower densities, all of the mean free paths plotted vary from being much less than to much greater than the neutrinosphere radii---i.e., to the characteristic spatial scale of the proto-neutron star. Consequently, the neutrinos will not behave in a fluid-like manner everywhere, and a kinetic rather than a fluid description of them is necessary.}
	\label{fig:tmfp}
\end{figure}

Figure \ref{fig:tmfp} shows the magnitude of the neutrino transport mean free paths for the electron neutrino, electron antineutrino, and heavy-flavor neutrinos (muon and tau neutrinos and their antineutrinos). The mean free paths are given at a time of 100 ms after bounce, during the critical shock reheating epoch, in the context of a \textsc{Chimera} supernova simulation of a $12\,M_\odot$ star. They are given as a function of radius, for select neutrino energies. Also shown are the neutrinospheres for the select energies, as well as the radius of the stalled shock wave. For all neutrino flavors and energies, the mean free paths exceed the respective neutrinosphere radii, as well as the shock radius, at some radius as we move outward. That is, the neutrino mean free paths exceed the scale of the proto-neutron star, as well as the shock radius scale, before we reach the shock radius. Under these circumstances, the neutrinos are not well described as components of the proto-neutron star fluid everywhere within it, and therefore, they are certainly not well described as a fluid in the critical heating layer between the proto-neutron star and the shock. A kinetic description of the neutrinos is required. Such a description, based on the Boltzmann kinetic equations, would supply the neutrino distributions functions, $f(r,\theta,\phi,\mu,\phi_{p},E,t)$, for each species of neutrino and antineutrino, where $\mu$ is a the direction cosine taken with respect to the outward radial direction, $\phi_{p}$ is the corresponding second angle describing the neutrino propagation direction in these momentum-space spherical polar coordinates, and $E=|p|$ is the neutrino energy. Deep in the proto-neutron star, neutrinos and the proto-neutron star fluid are in weak-interaction equilibrium. The distribution functions are then given by their equilibrium counterparts and the neutrinos are well described as an additional component of the fluid. Of course, the neutrinos fall out of weak equilibrium as the neutrinospheres are approached, and beyond them stream freely. Thus a fluid description of them would be limited to only a small portion of the simulation domain and would be of equally limited utility. The nature of the weak interactions demands the greater computational challenge and the higher computational cost of a kinetic description of neutrino transport in the proto-neutron star and above it in the cavity between it and the shock.
\subsection{The choice of phase-space coordinates}

The expansion from the four dimensions of spacetime to the seven dimensions of relativistic phase space brings with it additional choices. Now, in addition to making what will hopefully be optimal choices for spacetime coordinates, we will also need to consider optimal choices for momentum-space coordinates. And this is not without some give and take. Simplification in some respects afforded by one choice is always accompanied by complexification in other respects.

There is, however, an overarching consideration that guides the typical choice made by most modelers: Neutrino--matter interactions are most naturally and, consequently, most easily described in the frame of reference of the inertial observer instantaneously comoving with the fluid. (The fluid is accelerating, but the instantaneously comoving observer is not.) In this frame, the matter is instantaneously at rest, and the neutrino four-momentum components that enter the expressions for the neutrino weak interaction rates are the components measured by the comoving observer. However, while the description of neutrino--matter interactions are simplified in this picture, the choice to use four-momenta measured by instantaneously comoving observers introduces additional terms on the left-hand side of the Boltzmann equation that correspond to relativistic angular aberration and Doppler shift due to the fact that two spatially-adjacent instantaneously-comoving observers do not necessarily have the same velocity---in general, they will measure different neutrino angles of propagation and energies. In the context of Newtonian gravity, this would certainly add considerable complexity to the left-hand side of the Boltzmann equation. But in the general relativistic case, such momentum-space advection terms that involve derivatives with respect to the neutrino angles of propagation (or their direction cosines) and the neutrino energy are already there in light of general relativistic angular aberration and frequency shift in curved spacetime. While the character of the physical effects---special versus general relativistic---is different and, as such, presents different numerical challenges, the relative additional complexity of adding terms corresponding to special relativistic effects---e.g., relativistic Doppler shift and angular aberration---to the left-hand side of the Boltzmann equation versus the significant simplification of the collision term when comoving-frame neutrino four-momenta are used has led most modelers to choose comoving frame neutrino four momenta as phase-space coordinates. With regard, then, to the difficulties associated with the terms/effects added to the advection of neutrinos in phase-space, as we will see in this review, very different numerical approaches have been taken to describe them.

In what follows, we will adopt the following notation: We will designate the neutrino four-momentum components measured by an inertial observer instantaneously comoving with the fluid as $p^{\hat{\mu}}$. Neutrino four-momentum components measured by an Eulerian observer will be designated as $p^{\bar{\mu}}$. Finally, the neutrino four-momentum components in the coordinate basis will be designated as $p^{\mu}$.

\subsection{The general relativistic Boltzmann equation}
\label{sec:GeneralRelativisticBoltzmannEquation}
		
In light of the need to conserve simultaneously both energy and lepton number, we wish to begin with a version of the Boltzmann equation that is \emph{manifestly} conservative across all phase-space dimensions. As we will show, this is not true of the standard formulation of the general relativistic Boltzmann equation. In this section, we outline the derivation of both as presented by \cite{CaMe03} to illustrate the differences and, of course, to arrive at a form of the Boltzmann equation that is better suited to numerical application. Before we begin, we emphasize the following: While spacetime is endowed with a natural metric, $g_{{\mu}{\nu}}$, which is determined by Einstein's equations given the stress--energy content of spacetime, phase space is not. Consequently, the development of general relativistic neutrino radiation hydrodynamics \emph{requires} the full machinery of the metric-free language of the differential and integral calculus of forms. That is, the derivation we present below is not a matter of taste. Treatments of non-relativistic kinetic theory typically assume that phase space is endowed with a Euclidean metric. This can serve as a bookkeeping device at best, and it is important to interpret the theory accordingly.

The one-particle phase space for particles of arbitrary mass is an eight-dimensional space, which we label $M$, of spacetime position $x$ and four-momentum $p$. If we specify a mass for the particle, $m$, which satisfies
\begin{equation}
m^2=-g_{\mu\nu}p^{\mu}p^{\nu},
\end{equation}
we confine ourselves to a hypersurface of $M$, which we write as $M_m$, which is the phase space for particles of mass $m$. The flow in $M_m$ defined by the particle trajectories 
$(x,p)$ is generated by the Liouville operator
\begin{equation}
\dispSFNumberedEquationmath{{L_m}={p^{\overvar{\mu }{\RawWedge }}}{{{{\ScriptCapitalL }^{\mu }}\InvisibleSpace }_{\overvar{\mu }{\RawWedge }}}\frac{\partial
}{\partial {x^{\mu }}}
-{{{{\Gamma
}^{\overvar{i}{\RawWedge }}}\InvisibleSpace }_{\overvar{\nu }{\RawWedge }\overvar{\rho }{\RawWedge }}}{p^{\overvar{\nu }{\RawWedge }}}{p^{\overvar{\rho
}{\RawWedge }}}
\frac{\partial }{\partial {p^{\overvar{i}{\RawWedge }}}}}.
\label{liouville}
\end{equation}
$\cal L^{\hat{\mu}}_{\mu}$ is the composite transformation that takes us, first, from the coordinate basis to the orthonormal frame of the
Eulerian observer at rest with respect to the ``laboratory'' and, second, via a Lorentz transformation, from the Eulerian frame to the frame 
of reference comoving with the stellar core fluid:
\begin{equation}
{\cal L}^{\hat{\mu}}_{\mu} = \Lambda^{\hat{\mu}}_{\bar{\mu}}e^{\bar{\mu}}_{\mu}.
\label{eq:Ltrans}
\end{equation}
$\cal L^{\mu}_{\hat{\mu}}$ is the inverse transformation. $\Gamma^{\hat{\mu}}_{\hat{\mu}\hat{\nu}}$ are the Ricci Rotation Coefficients 
and are given by
\begin{equation}
{\Gamma^{\hat{\mu}}}_{\hat{\nu}\hat{\rho}}
=  {\cal L^{\hat{\mu}}}_{{\mu}} {\cal L^\nu}_{\hat{\nu}} {\cal L^\rho}_{\hat{\rho}} \,{\Gamma^\mu}_{\nu\rho} 
+ {\cal L^{\hat{\mu}}}_{{\mu}} {\cal L^\rho}_{\hat{\rho}} 
\frac{\partial {\cal L^\mu}_{\hat{\nu}}}{\partial x^\rho},
\label{eq:ConnectionComoving}
\end{equation}
where $\Gamma^{\mu}_{\nu\rho}$ are the Levi-Civita connection coefficients corresponding to the spacetime metric $g_{\mu\nu}$.

For a given type of particle of mass $m$, the distribution function, $f$, gives the density of such particles in phase space. An equation for the distribution function, the Boltzmann equation, is derived by considering a closed six-dimensional hypersurface $\partial D$ bounding a region $D$ in $M_m$. The net number of particles flowing through the boundary of $D$ is given by the generalized Stokes' Theorem
\begin{equation}
\dispSFNumberedEquationmath{\Mfunction{N}[\partial D]=\int_{\partial D} f\multsp \omega =\int_D d(f\multsp \omega ),}\label{stokes}
\end{equation}
where the infinitesimal surface element $\omega$ normal to the flow across $D$ is given by
\begin{equation}
\omega={L_m}\cdot \Omega
\end{equation}
and $\Omega$ is an infinitesimal volume element in $M_m$. The product rule gives
\begin{equation}
\dispSFNumberedEquationmath{d(f\multsp \omega )=\Mvariable{df}\wedge \omega =\Mvariable{df}\wedge ({L_m}\cdot \Omega ),}\label{almostLiouville}
\end{equation}
where we have used the fact that $d\omega=0$ (an expression of the general relativistic 
Liouville's Theorem that tells us that the phase-space flow is incompressible). But $f$, $L_m$, and 
$\Omega$ obey the identity
\begin{equation}
\dispSFNumberedEquationmath{\Mvariable{df}\wedge (L_m\cdot \Omega )=L_m[f]\Omega .}
\end{equation}
Then
\begin{equation}
\dispSFNumberedEquationmath{\Mfunction{N}[\partial D]=\int_D {L_m}[f]\Omega .}
\label{almostBoltzmann}
\end{equation}
Finally, the number of particles crossing the boundary $\partial D$ of $D$ in $M_m$ is given by the change in the number of particles in $D$ due to
emission, absorption, and scattering. Defining the ``collision term,'' $\mathcal{C}[f]$, as the spacetime density of such events, we have
\begin{equation}
\dispSFNumberedEquationmath{\Mfunction{N}[\partial D]=\int_D \mathcal{C}[f]\Omega ,}
\label{almostBoltzmann}
\end{equation}
and
\begin{equation}
\dispSFNumberedEquationmath{{L_m}[f]=\mathcal{C}[f].}
\end{equation}
Substituting for $L_m$ using Eq.~\eqref{liouville}, we arrive at the Boltzmann equation in ``standard'' form:
\begin{equation}
\dispSFNumberedEquationmath{{p^{\overvar{\mu }{\RawWedge }}}{{{{\ScriptCapitalL }^{\mu }}\InvisibleSpace }_{\overvar{\mu }{\RawWedge }}}\frac{\partial
f}{\partial {x^{\mu }}}
-{{{{\Gamma }^{\overvar{j}{\RawWedge }}}\InvisibleSpace }_{\overvar{\nu }{\RawWedge }\overvar{\rho }{\RawWedge }}}{p^{\overvar{\nu }{\RawWedge
}}}{p^{\overvar{\rho }{\RawWedge }}}
\frac{\partial {u^{\overvar{i}{\RawWedge }}}}{\partial {p^{\overvar{j}{\RawWedge }}}}\frac{\partial f}{\partial
{u^{\overvar{i}{\RawWedge }}}}=\mathcal{C}[f].}\label{fullBoltzmann}
\end{equation}
Note that to obtain the Boltzmann equation, we had to consider integration on our phase-space manifold $M_m$ 
on which there is no natural metric. This \emph{necessitates} the use of the language of differential forms. 

If we integrate over momentum space, we obtain the balance equation for particle number
\begin{equation}
\dispSFNumberedEquationmath{\frac{1}{{\sqrt{-g}}}\frac{\partial }{\partial {x^{\mu }}}\big({\sqrt{-g}}{N^{\mu }}\big)=\int \mathcal{C}[f]{{\pi
}_m},}\label{numberConservation}
\end{equation}
where
\begin{equation}
\dispSFNumberedEquationmath{{N^{\mu }(x)}=\int {{{{f \ScriptCapitalL }^{\mu }}\InvisibleSpace }_{\overvar{\mu
}{\RawWedge }}}{p^{\overvar{\mu }{\RawWedge }}}\multsp \multsp {{\pi }_m}=\int f{p^{\mu }}\multsp \multsp {{\pi }_m}} \label{numberVector}
\end{equation}
is the particle 4-current density and
\begin{equation}
\dispSFNumberedEquationmath{{{\pi }_m}=\frac{1}{E({\bf p})}\Bigg|\det \big[\frac{\partial {\bf p}}{\partial {\bf u}}\big]\Bigg|{{\Mvariable{du}}^{\overvar{1}{\RawWedge
}\overvar{2}{\RawWedge }\overvar{3}{\RawWedge }}}}\label{momentumElement}
\end{equation}
is the invariant momentum-space 3-volume expressed in terms of the spherical momentum-space coordinates: 
$u^{\hat{i}}=(E=||p||/c,\mu\equiv\cos\theta_{p},\phi_{p})$. 
But in light of the fact that the Boltzmann equation is not expressed in manifestly conservative form it is 
not obvious how we arrive at Eq.~\eqref{numberConservation} by integrating over momentum space.
We desire to reexpress the Boltzmann equation in terms of spacetime and momentum-space divergences
so that it is manifestly conservative with respect to an integration over a spacetime region, a momentum-space
region, or both---i.e., a phase-space region.

Of course, the generalized Stokes' Theorem, Eq.~\eqref{stokes}, is an expression of manifest conservation, 
equating the change in a quantity within a volume of phase space in terms of a surface term involving its flux on 
the volume's boundary.
The key insight by \citet{CaMe03} was to recognize that the total exterior derivative $d(f\omega)$ in Eq.~\eqref{stokes} can instead be
expressed as
\begin{equation}
\dispSFNumberedEquationmath{d(f\multsp \omega )=\mathcal{N}[f]\Omega ,}
\label{conservativeNumberOperator}
\end{equation}
where 
\begin{eqnarray}
\mathcal{N}[f]&\equiv&\frac{1}{{\sqrt{-g}}}\frac{\partial }{\partial {x^{\mu }}}\big({\sqrt{-g}}{{{{\ScriptCapitalL
}^{\mu }}\InvisibleSpace }_{\overvar{\mu }{\RawWedge }}}
{p^{\overvar{\mu }{\RawWedge }}}f\multsp \big)
-\nonumber  \\
& & E({\bf p})\Bigg|\det \big[\frac{\partial {\bf p}}{\partial {\bf u}}\big]{{\Bigg|}^{-1}}  \frac{\partial }{\partial {u^{\overvar{i}{\RawWedge }}}}\Bigg(\frac{1}{E({\bf p})}\Bigg|\det \big[\frac{\partial {\bf p}}{\partial
{\bf u}}\big]\Bigg|
{{{{\Gamma
}^{\overvar{j}{\RawWedge }}}\InvisibleSpace }_{\overvar{\nu }{\RawWedge }\overvar{\rho }{\RawWedge }}}{p^{\overvar{\nu }{\RawWedge }}}{p^{\overvar{\rho
}{\RawWedge }}}
\frac{\partial {u^{\overvar{i}{\RawWedge }}}}{\partial {p^{\overvar{j}{\RawWedge }}}}f\Bigg).\label{conservativeNumberOperator2}
\end{eqnarray}
Substituting Eq.~\eqref{conservativeNumberOperator} in Eq.~\eqref{stokes} and using Eq.~\eqref{almostBoltzmann}, we arrive at
\begin{eqnarray}
\label{consBE}
& &\frac{1}{{\sqrt{-g}}}\frac{\partial }{\partial {x^{\mu }}}\big({\sqrt{-g}}{{{{\ScriptCapitalL
}^{\mu }}\InvisibleSpace }_{\overvar{\mu }{\RawWedge }}}
{p^{\overvar{\mu }{\RawWedge }}}f\multsp \big)
-\nonumber  \\
& & E({\bf p})\Bigg|\det \big[\frac{\partial {\bf p}}{\partial {\bf u}}\big]{{\Bigg|}^{-1}}  \frac{\partial }{\partial {u^{\overvar{i}{\RawWedge }}}}\Bigg(\frac{1}{E({\bf p})}\Bigg|\det \big[\frac{\partial {\bf p}}{\partial
{\bf u}}\big]\Bigg|
{{{{\Gamma
}^{\overvar{j}{\RawWedge }}}\InvisibleSpace }_{\overvar{\nu }{\RawWedge }\overvar{\rho }{\RawWedge }}}{p^{\overvar{\nu }{\RawWedge }}}{p^{\overvar{\rho
}{\RawWedge }}}
\frac{\partial {u^{\overvar{i}{\RawWedge }}}}{\partial {p^{\overvar{j}{\RawWedge }}}}f\Bigg)\label{conservativeNumberOperator2} \\ \nonumber
& & = \mathcal{C}[f],
\end{eqnarray}
which is the manifestly conservative formulation of the Boltzmann equation. It is now obvious that upon integration over momentum space, for example, the momentum derivative terms on the left-hand side of the Boltzmann equation in Eq.~\eqref{consBE} will give rise only to surface terms. The counterpart equation for 4-momentum conservation can be derived in the same way \citep{CaMe03} and is given by
\begin{eqnarray}
& & \frac{1}{{\sqrt{-g}}}\frac{\partial }{\partial {x^{\nu }}}\big({\sqrt{-g}}{{\ScriptCapitalT }^{\Mvariable{\mu \nu }}}\big) - \nonumber \\
& & {E({\bf p})}\Bigg|\det \big[\frac{\partial {\bf p}}{\partial {\bf u}}\big]{{\Bigg|}^{-1}} \frac{\partial }{\partial {u^{\overvar{i}{\RawWedge }}}}\Bigg(\frac{1}{{E({\bf p})}}\Bigg|\det \big[\frac{\partial {\bf p}}{\partial
{\bf u}}\big]\Bigg|
{{{{\Gamma }^{\overvar{j}{\RawWedge }}}\InvisibleSpace
}_{\overvar{\nu }{\RawWedge }\overvar{\rho }{\RawWedge }}}{p^{\overvar{\rho }{\RawWedge }}}
\frac{\partial {u^{\overvar{i}{\RawWedge }}}}{\partial
{p^{\overvar{j}{\RawWedge }}}}{{{{\ScriptCapitalL }^{\overvar{\nu }{\RawWedge }}}\InvisibleSpace }_{\nu }}{{\ScriptCapitalT }^{\Mvariable{\mu \nu
}}}\Bigg)\IndentingNewLine \nonumber\\ 
&  & 
= -
{{{{\Gamma }^{\mu }}\InvisibleSpace
}_{\Mvariable{\nu \rho }}}{{\ScriptCapitalT }^{\Mvariable{\nu \rho }}}
+
{{{{\ScriptCapitalL }^{\mu }}\InvisibleSpace }_{\overvar{\mu }{\RawWedge
}}}{p^{\overvar{\mu }{\RawWedge }}}\mathbb{C}[f],\label{eq:fourMomentumConservativeBoltzmann}
\end{eqnarray}
where
\begin{equation}
\dispSFNumberedEquationmath{{{\ScriptCapitalT }^{\Mvariable{\mu \nu }}}\equiv {{{{\ScriptCapitalL }^{\mu }}\InvisibleSpace }_{\overvar{\mu }{\RawWedge
}}}{{{{\ScriptCapitalL }^{\nu }}\InvisibleSpace }_{\overvar{\nu }{\RawWedge }}}{p^{\overvar{\mu }{\RawWedge }}}{p^{\overvar{\nu }{\RawWedge }}}f}
\end{equation}
is the specific particle stress-energy tensor.

As an illustrative example, we specialize Eq.~\eqref{consBE} to the case of spherical symmetry, Lagrangian coordinates, and $\mathcal{O}(v/c)$ transport, as 
in \citet{MeBr93a,MeBr93b,MeBr93c}. As shown by Cardall and Mezzacappa, Eq.~\eqref{consBE} reduces to
\begin{eqnarray}
& & {\partial \over \partial t}\left(f\over\rho\right)
+ 
 {\partial\over\partial m}\left(4\pi r^2\rho\mu\, {f\over\rho}\right)
+
{1\over\epsilon^2}{\partial\over\partial\epsilon}
\left(\epsilon^3\left[\mu^2\left({3 v\over r} +
{\partial\ln\rho\over\partial t}\right)-{v\over r}\right]{f\over\rho}\right)
\nonumber \\
& & + {\partial \over\partial\mu}\left(
(1-\mu^2)\left[{1\over r}+\mu\left({3 v\over r}+
{\partial\ln\rho\over\partial t}\right)
\right]{f\over\rho}\right)
={1\over\rho\, \epsilon}\,\mathcal{C}[f],
\label{numberConservativeBoltzmann}
\end{eqnarray}
in agreement with the conservative formulation of the Boltzmann equation used in \citet{MeBr93a,MeBr93b,MeBr93c}. 
In spherical symmetry and to $\mathcal{O}(v/c)$ one can arrive at a manifestly conservative form of the 
Boltzmann equation through trial and error. However, in three dimensions and with full general relativity, 
such trial and error approaches are doomed to failure. A manifestly conservative starting point becomes 
paramount.

\subsection{The 3+1 formulation of general relativity}

The fundamental building blocks of the ``3+1'' formulation of general relativity are the spacelike hypersurfaces
corresponding to surfaces of constant $\tau$, where $\tau$ is some scalar function of the spacetime coordinates $x^\mu$:
$\tau=\tau(x^0,x^1,x^2,x^3)$. It is natural to choose $\tau$ to be $x^0=t$. The spacelike hypersurfaces, $\Sigma_t$,
are threaded by a timelike congruence of constant-spatial-coordinate curves. The points of constant 
$x^i(t)$ between two hypersurfaces separated by $dt$ are connected by the four-vector $t$. At each point of the hypersurface
$\Sigma_t$, there is a unit timelike normal four-vector $n$ satisfying $n_\mu n^\mu = -1$. $n$ corresponds to the four-velocity
of the observer at rest with respect to the hypersurface. This is the generalization of the definition of the Eulerian observer 
familiar from non-relativistic formulations. The four-vector $\beta$, known as the ``shift'' vector, describes how the spatial 
coordinates move within each hypersurface. The proper time between two hypersurfaces $\Sigma_t$ and $\Sigma_{t+dt}$ 
is given by $\alpha dt$. $\alpha$ is known as the ``lapse'' function. Given such a foliation of spacetime and such a 
coordinatization, the squared spacetime line element becomes
\begin{equation}
  ds^{2} = - (\alpha^{2}-\beta_{i}\beta^{i})dt^{2} + 2\beta_{i}dx^{i}dt+\gamma_{ij}dx^{i}dx^{j},
  \label{3+1metric}
\end{equation}
where $\gamma_{ij}$ is the metric on the hypersurface $\Sigma_t$. 
From Eq.~\eqref{3+1metric}, the spacetime metric can be read off as
\begin{equation}
  g_{\mu\nu} = 
  \left(
  \begin{array}{cc}
    -\alpha^{2}+\beta_{i}\beta^{i} & \beta_{i} \\
    \beta_{i} & \gamma_{ij}
  \end{array}
  \right),
\end{equation}
whose determinant $g$ can be computed directly to find $\sqrt{-g}=\alpha\,\sqrt{\gamma}$, where $\gamma$ is the determinant of the spatial metric.  

In addition to the intrinsic geometry---specifically, the intrinsic curvature---of each spacelike hypersurface, which is determined by the metric $\gamma_{ij}$, we describe how such a hypersurface is embedded in the four-dimensional spacetime by its extrinsic curvature, $\mathsf{K}_{ij}$, which is related to the three-metric by
\begin{equation}
\partial_{t}\gamma_{ij}=-2\alpha \mathsf{K}_{ij}+D_{i}\beta_{j}+D_{j}\beta_{i}.
\label{eq:extrinsiccurvature}
\end{equation}
Here $D_{i}$ corresponds to the covariant derivative on $\Sigma_{t}$ corresponding to the Levi--Civita connection 
associated with $\gamma_{ij}$. 
We can regard the coordinates of this formulation as the metric components, $\gamma_{ij}$, and the components
of the extrinsic curvature, $\mathsf{K}_{ij}$, as the velocities. The dynamics is supplied by the Einstein equations, which 
provide the following evolution equations for the six independent components of $\mathsf{K}_{ij}$:
\begin{eqnarray}
\label{eq:extcurvevolequation}
\partial_{t}\mathsf{K}_{ij} & = & -D_{i}D_{j}\alpha +\beta^{k}\partial_{k}\mathsf{K}_{ij}+\mathsf{K}_{ik}\partial_{j}\beta^{k}+\mathsf{K}_{kj}\partial_{i}\beta^{k} \\ \nonumber
& + & \alpha \left( ^{(3)}R_{ij}+\mathsf{K}\mathsf{K}_{ij}-2\mathsf{K}_{ik}\mathsf{K}^{k}_{j} \right) +4\pi\alpha [\gamma_{ij} (S-E) - 2S_{ij}], \nonumber
\end{eqnarray}
where $\mathsf{K}$ is the trace of the extrinsic curvature tensor, and $^{(3)}R_{ij}$ is the Ricci curvature tensor for the spacelike
hypersurface. The source terms in Eq.~\eqref{eq:extcurvevolequation} are given in terms of the stress--energy tensor, 
$T_{\alpha\beta}$, by
\begin{eqnarray}
\label{eq:sourceterms}
S_{\mu\nu} & = & \gamma^{\alpha}_{\mu}\gamma^{\beta}_{\nu}T_{\alpha\beta}, \\
S_{\mu} & = & -\gamma^{\alpha}_{\mu}n^{\beta}T_{\alpha\beta}, \\
S & = & S^{\mu}_{\mu}, \\
E & = & n^{\alpha}n^{\beta}T_{\alpha\beta},
\end{eqnarray}
\noindent 
where 
\begin{equation}
  n^{\mu} = \f{1}{\alpha}(1,-\beta^{i}) \quad\text{with}\quad
  n_{\mu} = (-\alpha,0)
\end{equation}
and
\begin{equation}
  \gamma^{\alpha}_{\hspace{4pt}\mu}=\delta^{\alpha}_{\hspace{4pt}\mu}+n^{\alpha}\,n_{\mu}
\end{equation}
provide timelike and spacelike projections, respectively. 
While not drawn here, there is a corresponding spacelike hypersurface to which the fluid four-velocity
\begin{equation}
  u^{\mu} = W\,(\,n^{\mu}+v^{\mu}\,)
  \label{eq:fluidFourVelocityEulerian}
\end{equation}
is the unit timelike normal, which defines the timelike basis element of the orthonormal frame of reference of the inertial observer instantaneously comoving with the fluid and at rest with respect to the hypersurface.  
This is our generalized Lagrangian observer in this formalism.  
The projection into the slice defined by the normal $u^{\mu}$ is given by
\begin{equation}
  h^{\alpha}_{\hspace{4pt}\mu}=\delta^{\alpha}_{\hspace{4pt}\mu}+u^{\alpha}\,u_{\mu}.  
  \label{eq:projectorLagrangian}
\end{equation}
Here, $W=-n_{\mu}u^{\mu}$ is the Lorentz factor and $v^{\mu}=(\gamma^{\mu}_{\hspace{4pt}\nu}u^{\nu})/W$ the fluid three-velocity.  

\subsection{3+1 general relativistic hydrodynamics}
\label{sec:hydrodynamics3p1}

The 3+1 slicing of spacetime allows us to formulate the radiation-hydrodynamics equations in a form suitable for numerical solution.  
Here we briefly summarize the 3+1 form of the hydrodynamics equations given by Eqs.~\eqref{eq:BaryonMassConservation}--\eqref{eq:ElectronNumberConservation} (see, e.g., \citealt{Anile89,ReZa13} for details).  
The mass conservation equation (cf.\ Eq.~\eqref{eq:BaryonMassConservation}) becomes
\begin{equation}
  \f{1}{\alpha\sqrt{\gamma}}
  \big[\,
    \pd{}{t}\big(\,\sqrt{\gamma}\,D\,\big) + \pd{}{i}\big(\,\sqrt{\gamma}\,D\,\big[\,\alpha\,v^{i}-\beta^{i}\,\big]\,\big)
  \,\big]
  =0,
  \label{eq:BaryonMassConservation3p1}
\end{equation}
where $D=W\,\rho$, while the electron number conservation equation (cf.\ Eq.~\eqref{eq:ElectronNumberConservation}) becomes
\begin{equation}
  \f{1}{\alpha\sqrt{\gamma}}
  \big[\,
    \pd{}{t}\big(\,\sqrt{\gamma}\,D\,Y_{e}\,\big) + \pd{}{i}\big(\,\sqrt{\gamma}\,D\,Y_{e}\,\big[\,\alpha\,v^{i}-\beta^{i}\,\big]\,\big)
  \,\big]
  =-m_{\mbox{\tiny B}}\,L.  
  \label{eq:ElectronNumberConservation3p1}
\end{equation}
Conservative forms of the energy and momentum equations are derived by decomposing Eq.~\eqref{eq:fluidFourMomentumConservation} into components relative to the spatial hypersurface.  
The energy equation becomes
\begin{align}
  &\f{1}{\alpha\sqrt{\gamma}}
  \big[\,
    \pd{}{t}\big(\,\sqrt{\gamma}\,\tau_{\mbox{\tiny fluid}}\,\big)
    +\pd{}{i}\big(\,\sqrt{\gamma}\,\big[\,\alpha\,(S^{i}-D\,v^{i})-\tau_{\mbox{\tiny fluid}}\,\beta^{i}\,\big]\,\big)
  \,\big] \nonumber \\
  &\hspace{12pt}
  =\f{1}{\alpha}\,\big[\,\alpha\,S^{ik}\,\mathsf{K}_{ik}-S^{i}\pd{\alpha}{i}\,\big]+n_{\mu}\,G^{\mu},
  \label{eq:fluidEnergyEquation3p1}
\end{align}
where $\tau_{\mbox{\tiny fluid}}=E-D$, $E=\rho\,h\,W^{2}-p$, $S^{\mu}=\rho\,h\,W^{2}\,v^{\mu}$, and $S^{\mu\nu}=\rho\,h\,W^{2}\,v^{\mu}\,v^{\nu}+p\,\gamma^{\mu\nu}$, while the momentum equation is given by
\begin{align}
  &\f{1}{\alpha\sqrt{\gamma}}
  \big[\,
    \pd{}{t}\big(\,\sqrt{\gamma}\,S_{j}\,\big)
    +\pd{}{i}\big(\,\sqrt{\gamma}\,\big[\,\alpha\,S^{i}_{\hspace{2pt}j}-\beta^{i}\,S_{j}\,\big]\,\big)
  \,\big] \nonumber \\
  &\hspace{12pt}
  =\f{1}{\alpha}\,\big[\,S_{i}\,\pd{\beta^{i}}{j}+\f{1}{2}\,\alpha\,S^{ik}\pd{\gamma_{ik}}{j}-E\,\pd{\alpha}{j}\,\big] - \gamma_{j\mu}\,G^{\mu}.
  \label{eq:fluidMomentumEquation3p1}
\end{align}
The source terms modeling lepton and four-momentum exchange due to neutrino--matter interactions ($-L$, $-n_{\mu}\,G^{\mu}$, and $\gamma_{j\mu}\,G^{\mu}$, respectively) will be discussed in detail in Sect.~\ref{sec:interactions}.  

\subsection{The 3+1 general relativistic Boltzmann equation}

The general relativistic Boltzmann equation in both conservative form and using the spacetime coordinates associated with the 3+1 decomposition of spacetime was derived by \citet{CaEnMe13b}. Essential to the derivation is 
the recognition that the composite transformation $L^{\mu}_{\hat{\mu}}$ can be viewed as the coordinate basis components 
($\mu$) of the element of the tetrad  of four-vectors ($\hat{\mu}$) corresponding to the frame carried by the observer instantaneously 
comoving with the fluid. The Eulerian decomposition of $L^{\mu}_{\hat{\mu}}$ into timelike and spacelike components is
\begin{equation}
L^{\mu}_{\hat{\mu}}={\cal L}_{\hat{\mu}}n^{\mu}+l^{\mu}_{\hat{\mu}},
\label{eq:tetraddecomposition}
\end{equation}
where ${\cal L}_{\hat{\mu}}$ is the coefficient of the timelike component of the tetrad element (four-vector) 
designated by $\hat{\mu}$, and $l^{\mu}_{\hat{\mu}}$ is the spacelike component of this tetrad element. Explicit expressions
for ${\cal L}_{\hat{\mu}}$ and $l^{\mu}_{\hat{\mu}}$ can be found in \citep{CaEnMe13b}. The Ricci Rotation Coefficients can 
be expressed as
\begin{equation}
\Gamma^{\hat{\rho}}_{\hat{\nu}\hat{\mu}}= L^{\hat{\rho}}_{\nu} L^{\mu}_{\hat{\mu}}\nabla_{\mu} L^{\nu}_{\hat{\nu}}.
\label{eq:riccirotcoeffdecomp}
\end{equation}
Using the decomposition (\ref{eq:tetraddecomposition}), we are left with three terms to evaluate:
\begin{equation}
L^{\hat{\rho}}_{\nu} L^{\mu}_{\hat{\mu}}
\left(
{\cal L}_{\hat{\nu}}\nabla_{\mu}n^{\nu}
+n^{\nu}\nabla_{\mu}{\cal L}_{\hat{\nu}}
+\nabla_{\mu}l^{\nu}_{\hat{\nu}}
\right).
\label{eq:threeterms}
\end{equation}
The results can be found in \citet{CaEnMe13b}. With the decomposition of the momentum-space
transformation matrix $P^{\tilde{i}}_{\hat{i}}$ into elements parallel and perpendicular to the three-momentum
$p^{\hat{i}}$,
\begin{equation}
P^{\tilde{i}}_{\hat{i}} =  \frac{Q^{\tilde{i}}p_{\hat{i}}}{p} + U^{\tilde{i}}_{\hat{i}},
\label{eq:momentumdecomp}
\end{equation}
with
\begin{eqnarray}
\label{eq:momentumdecomp2}
Q^{\tilde{i}} & = & \frac{P^{\tilde{i}}_{\hat{i}} p^{\hat{i}}}{p}, \\
p & = & \sqrt{p^{\hat{i}}p_{\hat{i}}}, \\
U^{\tilde{i}}_{\hat{i}} & = & P^{\tilde{i}}_{\hat{j}} k^{\hat{j}}_{\hat{i}}, \\
k^{\hat{j}}_{\hat{i}} & = & \delta^{\hat{j}}_{\hat{i}} + \frac{p^{\hat{j}}p_{\hat{i}}}{p^2}.
\end{eqnarray}
The 3+1 general relativistic Boltzmann equation can now be written as
\begin{equation}
S_N + M_N = C[f],
\label{eq:ConservativeCovariant}
\end{equation}
where the spacetime divergence is
\begin{equation}
S_N = \frac{\left( -p_{\hat 0} \right)}{\alpha\sqrt{\gamma}}\left[ \frac{\partial \left(D_N \right)}{\partial t} + \frac{\partial \left( F_N \right)^i }{\partial x^i} \right],
\label{eq:Spacetime_N_31}
\end{equation}
with 
\begin{eqnarray}
D_N &=& \frac{\sqrt{\gamma}}{\left( -p_{\hat 0} \right)} \, \mathcal{L}_{\hat\mu} \, p^{\hat\mu} f, 
\label{eq:Density_N} \\
\left( F_N \right)^i &=& \frac{\sqrt{\gamma}}{\left( -p_{\hat 0} \right)} \left( \alpha\, {\ell^i}_{\hat\mu} - \beta^i \mathcal{L}_{\hat\mu} \right) p^{\hat\mu} f.
\label{eq:Flux_N}
\end{eqnarray}
$D_N$ and $\left( F_N \right)^i $  are, respectively, the conserved number density and number flux.
The momentum-space divergence, $M_N$, can be expressed as 
\begin{eqnarray}
M_N &=& \frac{1}{\alpha\sqrt{\gamma}} \frac{\left( -p_{\hat 0} \right)}{\sqrt{\lambda}} \frac{\partial}{\partial p^{\tilde\imath}} \left\{
\sqrt{\lambda} \, \frac{Q^{\tilde\imath} \left(-p_{\hat 0}\right)}{p}\! \left[ \left( R_N \right)^{\hat 0} + \left( O_N \right)^{\hat 0} \right] \right. \nonumber \\
&& \left. + \sqrt{\lambda} \, {U^{\tilde \imath}}_{\hat \imath} \left[ \left( R_N \right)^{\hat \imath} + \left( O_N \right)^{\hat \imath} \right] \right\},
\label{eq:Momentum_N_31}
\end{eqnarray}
where
\begin{eqnarray}
\left( R_N \right)^{\hat\rho} &=& \frac{\alpha\sqrt{\gamma}}{\left( -p_{\hat 0} \right)}\, p^{\hat\nu} p^{\hat\mu} f \nonumber \\
& & \times \left[  \mathcal{L}^{\hat\rho}\, {\ell^j}_{\hat\nu} \left( \frac{ \mathcal{L}_{\hat\mu}  }{\alpha} \frac{\partial \alpha}{\partial x^j} 
-  {\ell^k}_{\hat\mu} \, \mathsf{K}_{jk} \right) 
 \right. \nonumber \\
& & \left. - {\ell^{\hat\rho j}} \! \left( \! \frac{\mathcal{L}_{\hat\nu} \mathcal{L}_{\hat\mu}  }{\alpha} \frac{\partial \alpha}{\partial x^j} 
\!-\! \frac{\ell_{k\hat\nu} \, \mathcal{L}_{\hat\mu}}{\alpha} \frac{\partial \beta^k}{\partial x^j} 
\!-\!  \frac{{\ell^k}_{\hat\nu} \,{\ell^i}_{\hat\mu}}{2} \frac{\partial \gamma_{ki} }{\partial x^j} \!\right) \!\right]\!  
\label{eq:Redshift_N}
\end{eqnarray}
describes momentum shifts (i.e., redshift and angular aberration in momentum-space spherical coordinates) due to gravity as embodied in the spacetime geometry,  
\begin{eqnarray}
\left( O_N \right)^{\hat\rho} &=& \frac{\sqrt{\gamma}}{\left( -p_{\hat 0} \right)} \, p^{\hat\nu} p^{\hat\mu} f \nonumber \\
& & \times \left\{ \mathcal{L}^{\hat\rho} \left[  \mathcal{L}_{\hat\mu} \frac{\partial \mathcal{L}_{\hat\nu}}{\partial t} + \left( \alpha\, {\ell^j}_{\hat\mu} - \beta^j \mathcal{L}_{\hat\mu}\right) \frac{\partial \mathcal{L}_{\hat\nu}}{\partial x^j} \right] \right. \nonumber \\
& & \left. 
- \ell^{\hat\rho k} \left[ \mathcal{L}_{\hat\mu} \frac{\partial \ell_{k \hat\nu}}{\partial t}  +  \left(\alpha\, {\ell^j}_{\hat\mu} - \beta^j \mathcal{L}_{\hat\mu}\right) \frac{\partial \ell_{k \hat\nu}}{\partial x^j}
\right] \right\} 
\label{eq:Observer_N}
\end{eqnarray}
are `observer corrections' due to the acceleration of the fluid and, consequently, changing comoving observers with different velocities (and partially entangled with the geometry as well), and 

\begin{equation}
\sqrt{\lambda}=\Bigg|\det \big[\frac{\partial {\bf p}}{\partial {\bf u}}\big]\Bigg| .
\end{equation}

\subsection{Multi-frequency moment kinetics and the closure problem}
\label{sec:MomentKineticsAndClosure}

Because of the prohibitively high computational cost associated with solving the Boltzmann equation with sufficient phase-space resolution, most (all in three spatial dimensions) supernova models to date employ a moments approach to neutrino transport.  
In the moments approach, one solves for a finite number of moments of the distribution function (instead of the distribution function), and the hierarchy of moment equations is closed by a closure procedure, relating higher-order moments to the evolved lower-order moments.  

The basic idea of the moments approach can be illustrated by considering the Boltzmann equation in one spatial dimension
\begin{equation}
  \pd{f}{t}+\mu\pd{f}{x} = \chi\,(f_{0}-f) + \sigma\,(\langle f \rangle-f),
  \label{eq:boltzmannSimple}
\end{equation}
where, for simplicity, we let the distribution function depend on spatial position, $x$, momentum-space angle cosine, $\mu$, and time, $t$. $\chi$ is the absorption opacity, $f_{0}$ is the isotropic equilibrium distribution, and $\sigma$ is the scattering opacity due to isotropic and isoenergetic scattering.  
A finite number ($N+1$) of angular moments of the distribution function can be formed as weighted integrals over angle:
\begin{equation}
  m^{(k)}(x,t)=\langle\,f,\,\mu^{k}\,\rangle\equiv\f{1}{2}\int_{-1}^{1}f(\mu,x,t)\,\mu^{k}\,d\mu,\quad k=0,1,\ldots,N.
  \label{eq:momentsSimple}
\end{equation}
Thus, in a truncated moments approach the distribution function is approximated by the moments vector
\begin{equation}
  \mathbf{m}_{N}=\big(\,m^{(0)},m^{(1)},\ldots,m^{(N)}\,\big)^{T}
\end{equation}
so that 
\begin{equation}
  f(\mu,x,t) \approx \sum_{k=0}^{N}c^{(k)}\,m^{(k)}(x,t)\,\mu^{k},
  \label{eq:expansionSimple}
\end{equation}
where $c^{(k)}$ are normalization constants.  
Similarly, by taking moments of the Boltzmann equation in Eq.~\eqref{eq:boltzmannSimple}, the hierarchy of moment equations is given by
\begin{align}
  \pd{m^{(0)}}{t}+\pd{m^{(1)}}{x} &=\chi\,(f_{0} - m^{(0)}), \label{eq:zerothMomentEquationSimple} \\
  \pd{m^{(k)}}{t}+\pd{m^{(k+1)}}{x} &= \chi\,(\langle f_{0},\mu^{k} \rangle-m^{(k)}) + \sigma\,(m_{0}^{(k)}-m^{(k)}), \quad\text{for}\quad k>0, \label{eq:higherMomentEquationSimple}
\end{align}
where on the right-hand sides we have defined
\begin{equation}
  \langle f_{0},\mu^{k} \rangle=f_{0}\,\f{[1+(-1)^{k}]}{2\,(k+1)}
  \quad\mbox{and}\quad
  m_{0}^{(k)}=m^{(0)}\,\f{[1+(-1)^{k}]}{2\,(k+1)}.  
\end{equation}
When considering the expansion in Eq.~\eqref{eq:expansionSimple}, the moments approach is simply an approximation to the angular dependence of the distribution function in terms of the monomial basis $\{\mu_{k}\}_{k=0}^{N}$.  
The power of the moments approach becomes evident when collisions are moderate to strong.  
In this case, collisions tend to drive the zeroth moment $m^{(0)}$ towards the isotropic distribution $f_{0}$, the higher-order odd moments decay exponentially to zero ($m^{(k)}\to0$; $k$ odd), and the higher-order even moments tend to $m^{(k)}\to m^{(0)}/(k+1)$ ($k$ even).  
Thus, the angular dependence of the distribution is captured well by only a few moments.  
In the absence of collisions, more moments are typically needed to capture the angular shape of the distribution function.  
Note in particular that in Eq.~\eqref{eq:higherMomentEquationSimple}, the equation for the $k$-th moment contains the $k+1$-th moment.  
Thus, in a truncated moment model based on $N+1$ moments, $\mathbf{m}_{N}$, the equation for $m^{(N)}$ contains the moment $m^{(N+1)}$, which must be related to the lower order moments by a closure procedure---i.e., $m^{(N+1)}:=g(\mathbf{m}_{N})$---in order to form a closed system of equations.  
This is referred to as the closure problem.  
Typically, the closure function $g$ is a nonlinear function of $\mathbf{m}_{N}$, which can make the construction of numerical methods for moment models more difficult.  
There are several challenges associated with the construction of closures for moment hierarchies \citep[see, e.g.,][]{Le96}, one being the construction of closures that preserve the hyperbolic character of the system of moment equations; see, e.g., \citet{PoIbMi00}, for a discussion of this topic in the context of two-moment models.  
In the remainder of this section, we will discuss relativistic two-moment models ($N=1$ in the simpler formalism above).  
In the multi-dimensional setting, the two-moment model evolves four unknowns (e.g., the energy density and three components of the momentum density), and, in the relativistic setting considered here, second and third moments appear in the equations for the first moments.  

Conservative, 3+1 general relativistic, multi-frequency (or multi-energy) two-moment formalisms have been developed by \citet{ShKiSe11,CaEnMe13a}.  
The formalism of \citet{ShKiSe11} is based on the formalism of \citet{Th81}, while the formalism of \citet{CaEnMe13a} starts out with the conservative formulation of kinetic theory from \citet{CaMe03} discussed in Sect.~\ref{sec:GeneralRelativisticBoltzmannEquation}.  
Both approaches, of course, lead to the same result, which we summarize here.  

Covariant expressions for the first few moments of the distribution function $f$ are given by
\begin{align}
  N^{\mu}(x,t)
  &= \int_{V_{p}} f(p,x,t)\,p^{\mu}\,\pi_{m}, \label{eq:numberMoments} \\
  T^{\mu\nu}(x,t)
  &=\int_{V_{p}} f(p,x,t)\,p^{\mu}\,p^{\nu}\,\pi_{m}, \label{eq:stressEnergyMoments} \\
  Q^{\mu\nu\rho}(x,t)
  &=\int_{V_{p}} f(p,x,t)\,p^{\mu}\,p^{\nu}\,p^{\rho}\,\pi_{m},  \label{eq:heatFluxMoments}
\end{align}
where $N^{\mu}$ is the four-current density, $T^{\mu\nu}$ the stress-energy tensor, and the rank three tensor of moments $Q^{\mu\nu\rho}$ is sometimes referred to as the tensor of fluxes or heat flux tensor.  
When expressed in terms of comoving frame spherical-polar momentum coordinates $(\varepsilon,\vartheta,\varphi)$, the invariant momentum-space 3-volume in Eq.~\eqref{momentumElement} is
\begin{equation}
  \pi_{m} = \varepsilon\,\sin\vartheta\,d\vartheta\,d\varphi\,d\varepsilon.  
\end{equation}
Higher-order moments can be constructed similarly in a straightforward way, but we will limit the discussion to moment models involving the moments in Eqs.~\eqref{eq:numberMoments}-\eqref{eq:heatFluxMoments}.  
Note that the moments defined above depend only on position $x$ and time $t$.  
However, because neutrino heating and cooling rates are sensitive to the neutrino energy (cf.\ Sect.~\ref{sec:SettingTheStage}), supernova models based on moment descriptions for neutrino transport retain the energy dimension and solve for \emph{angular moments}, or \emph{spectral moments}, defined by
\begin{align}
  \mathcal{N}^{\mu}(\varepsilon,x,t)
  &=\f{1}{4\pi}\int_{\mathbb{S}^{2}}f\,p^{\mu}\,\f{d\omega}{\varepsilon}, \label{eq:numberAngularMoments} \\
  \mathcal{T}^{\mu\nu}(\varepsilon,x,t)
  &=\f{1}{4\pi}\int_{\mathbb{S}^{2}}f\,p^{\mu}\,p^{\nu}\,\f{d\omega}{\varepsilon}, \label{eq:stressEnergyAngularMoments} \\
  \mathcal{Q}^{\mu\nu\rho}(\varepsilon,x,t)
  &=\f{1}{4\pi}\int_{\mathbb{S}^{2}}f\,p^{\mu}\,p^{\nu}\,p^{\rho}\,\f{d\omega}{\varepsilon}, \label{eq:heatFluxAngularMoments}
\end{align}
where $d\omega=\sin\vartheta d\vartheta d\varphi$ and the integrals extend over the sphere
\begin{equation}
  \mathbb{S}^{2} = \big\{\,\omega\in(\vartheta,\varphi)~|~\vartheta\in[0,\pi],\,\varphi\in[0,2\pi)\,\big\},
\end{equation}
where $\vartheta$ and $\varphi$ are momentum-space angular coordinates.  
The angular moments defined in Eqs.~\eqref{eq:numberAngularMoments}-\eqref{eq:heatFluxAngularMoments} depend on the neutrino energy, $\varepsilon$, position, $x$, and time, $t$.  
They are related to the moments in Eqs.~\eqref{eq:numberMoments}-\eqref{eq:heatFluxMoments} by the integral over energy
\begin{equation}
  \big\{\,N^{\mu},\,T^{\mu\nu},\,Q^{\mu\nu\rho}\,\big\}(x,t)
  =\int_{0}^{\infty}\big\{\,\mathcal{N}^{\mu},\,\mathcal{T}^{\mu\nu},\,\mathcal{Q}^{\mu\nu\rho}\,\big\}(\varepsilon,x,t)\,dV_{\varepsilon},
\end{equation}
where the infinitesimal energy-space shell-volume element is $dV_{\varepsilon}=4\pi\varepsilon^{2}d\varepsilon$.  
In forming the angular moments we have used the freedom in choosing distinct spacetime and momentum space coordinates: 
$x$ and $t$ are spacetime coordinates in a global coordinate basis, while $\{\varepsilon,\vartheta,\varphi\}$ are momentum coordinates in 
a comoving basis.  

Moment equations governing the evolution of the angular moments are derived from the general relativistic Boltzmann equation discussed in Sect.~\ref{sec:GeneralRelativisticBoltzmannEquation}.  
Since current supernova modelers employing angular moment models use either a flux-limited diffusion (one-moment) or a two-moment approach, we will limit the discussion to these approaches.  
In this context, we will need evolution equations for the spectral neutrino number density, energy density, and three-momentum density.  
The evolution equation for the neutrino number density is obtained by multiplying Eq.~\eqref{consBE} by $1/(4\pi\varepsilon)$ and integrating over $\mathbb{S}^{2}$:
\begin{equation}
  \nabla_{\nu}\mathcal{N}^{\nu}
  -\f{1}{\varepsilon^{2}}\pderiv{}{\varepsilon}\big(\,\varepsilon^{2}\,\mathcal{T}^{\mu\nu}\,\nabla_{\mu}u_{\nu}\,\big)
  =\f{1}{4\pi}\int_{\mathbb{S}^{2}}\mathcal{C}[f]\,\f{d\omega}{\varepsilon},
  \label{eq:spectralNumberEquation}
\end{equation}
where $u_{\nu}$ is the four-velocity of the observer measuring neutrino energy $\varepsilon$ (i.e., the comoving observer).  
Note that the left-hand side of Eq.~\eqref{eq:spectralNumberEquation} is in divergence form, and the use of spherical momentum-space coordinates is apparent from the form of the second term.  
Integrating over energy ($dV_{\varepsilon}$) gives rise to the balance equation
\begin{equation}
  \nabla_{\nu}N^{\nu} = \int_{V_{p}}\mathcal{C}[f]\,\pi_{m},
  \label{eq:numberEquation}
\end{equation}
where the left-hand side is in conservative form.  
The right-hand side gives rise to lepton exchange sources and sinks due to neutrino--matter interactions (e.g., emission and absorption).  
In a similar manner, conservative evolution equations for the neutrino four-momentum are obtained by multiplying the four-momentum conservative Boltzmann equation in Eq.~\eqref{eq:fourMomentumConservativeBoltzmann} by $1/(4\pi\varepsilon)$ and integrating over $\mathbb{S}^{2}$:
\begin{equation}
  \nabla_{\nu}\mathcal{T}^{\mu\nu}
  -\f{1}{\varepsilon^{2}}\pderiv{}{\varepsilon}\big(\,\varepsilon^{2}\,\mathcal{Q}^{\mu\nu\rho}\,\nabla_{\nu}u_{\rho}\,\big)
  =\f{1}{4\pi}\int_{\mathbb{S}^{2}}\mathcal{C}[f]\,p^{\mu}\,\f{d\omega}{\varepsilon}.  
  \label{eq:spectralFourMomentumEquation}
\end{equation}
Again, integrating this equation over energy results in the balance equation
\begin{equation}
  \nabla_{\nu}T^{\mu\nu} = \int_{V_{p}}\mathcal{C}[f]\,p^{\mu}\,\pi_{m},
  \label{eq:fourMomentumEquation}
\end{equation}
where the left-hand side is in conservative form, and the right-hand side gives rise to four-momentum exchange with the fluid.  

Eq.~\eqref{eq:spectralFourMomentumEquation} forms a basis for the two-moment model for neutrino transport.  
Since neutrinos exchange lepton number and four-momentum with the fluid, Eq.~\eqref{eq:spectralNumberEquation} needs to be considered, as well.  
However, these equations are not independent.  
Due to the relations (obvious from the definitions in Eqs.~\eqref{eq:numberAngularMoments}-\eqref{eq:heatFluxAngularMoments})
\begin{equation}
  \mathcal{N}^{\nu} = -\f{u_{\mu}}{\varepsilon}\,\mathcal{T}^{\mu\nu}
  \quad\text{and}\quad
  \mathcal{T}^{\nu\rho} = -\f{u_{\mu}}{\varepsilon}\,\mathcal{Q}^{\mu\nu\rho},
\end{equation}
Eqs.~\eqref{eq:spectralNumberEquation} and \eqref{eq:spectralFourMomentumEquation} are related in a similar way: Eq~\eqref{eq:spectralNumberEquation} can be obtained from Eq.~\eqref{eq:spectralFourMomentumEquation} by a contraction with $-u_{\mu}/\varepsilon$.  
In a numerical implementation targeting both lepton and four-momentum exchange between neutrinos and the stellar fluid, such consistency is desirable since the numerical method then preserves a critical structure of the moment system.  
In the following, we provide versions of the two-moment model in the 3+1 framework of general relativity.  
Before we delve into the details, we briefly discuss two useful decompositions of the angular moments.  

\subsubsection{Lagrangian decompositions}
\label{sec:LagrangianDecompositions}

With comoving frame four-momentum coordinates, Lagrangian decompositions of tensors is a natural way to express the angular moments in Eqs.~\eqref{eq:numberAngularMoments}-\eqref{eq:heatFluxAngularMoments} in terms of elementary moments of the distribution function.  
This is achieved with the Lagrangian decomposition of the particle four-momentum
\begin{equation}
  p^{\mu} = \varepsilon\,(\,u^{\mu}+\ell^{\mu}\,),
  \label{eq:fourMomentumLagrangianDecomposition}
\end{equation}
where $u^{\mu}$ is the four-velocity of the Lagrangian observer, and $\ell^{\mu}$ is a unit four-vector orthogonal to $u^{\mu}$; i.e., $\ell_{\mu}\ell^{\mu}=1$ and $u_{\mu}\ell^{\mu}=0$.  
Then, $\varepsilon=-u_{\mu}p^{\mu}$ is the neutrino energy measured by the Lagrangian observer.  
In terms of the composite transformation of the neutrino four-momentum, $p^{\mu}=\mathcal{L}^{\mu}_{\hspace{6pt}\hat{\mu}}p^{\hat{\mu}}=\varepsilon\big(\mathcal{L}^{\mu}_{\hspace{6pt}\hat{0}}+\mathcal{L}^{\mu}_{\hspace{6pt}\hat{\imath}}\,\ell^{\hat{\imath}}\big)$, a comparison with Eq.~\eqref{eq:fourMomentumLagrangianDecomposition} implies that $\mathcal{L}^{\mu}_{\hspace{6pt}\hat{0}}=u^{\mu}$ and $\ell^{\mu}=\mathcal{L}^{\mu}_{\hspace{6pt}\hat{\imath}}\,\ell^{\hat{\imath}}$, where
\begin{equation}
  \ell^{\hat{\imath}} = \big\{\,\cos\vartheta,\,\sin\vartheta\cos\varphi,\,\sin\vartheta\sin\varphi\,\big\}
\end{equation}
are components of the spatial unit vector in the orthonormal comoving frame.  
(See Section~\eqref{sec:GeneralRelativisticBoltzmannEquation} for the definition of $\mathcal{L}^{\mu}_{\hspace{6pt}\hat{\mu}}$.)
Inserting Eq.~\eqref{eq:fourMomentumLagrangianDecomposition} into Eq.~\eqref{eq:numberAngularMoments} results in the Lagrangian decomposition of the spectral neutrino four-current density
\begin{equation}
  \mathcal{N}^{\mu} 
  = \mathcal{D}\,u^{\mu} + \mathcal{I}^{\mu},
  \label{eq:numberCurrentLagrangianDecomposition}
\end{equation}
where the angular moments
\begin{equation}
  \big\{\mathcal{D},\mathcal{I}^{\mu}\big\}(\varepsilon,x,t) = \f{1}{4\pi}\int_{\mathbb{S}^{2}}f(\omega,\varepsilon,x,t)\,\big\{\,1,\,\ell^{\mu}\,\big\}\,d\omega
  \label{eq:numberMomentsLagrangian}
\end{equation}
are the comoving spectral number density and number flux, respectively.  
Using the fluid four-velocity $u^{\mu}$ and the projector in Eq.~\eqref{eq:projectorLagrangian}, these components are obtained from $\mathcal{D}=-u_{\mu}\mathcal{N}^{\mu}$ and $\mathcal{I}^{\mu}=h^{\mu}_{\hspace{6pt}\nu}\mathcal{I}^{\nu}$.  
The moments in Eq.~\eqref{eq:numberMomentsLagrangian} are the most elementary in the moment hierarchy, and for the two-moment model, these are used in the closure procedure to determine the higher-order moments in terms of $\mathcal{D}$ and $\mathcal{I}^{\mu}$.  
Note that for an isotropic distribution function $f=f_{0}$ (where $f_{0}$ is independent of $\omega$), $\mathcal{D}=f_{0}$ and $\mathcal{I}^{\mu}=0$.  

In a similar way, using Eq.~\eqref{eq:fourMomentumLagrangianDecomposition} in Eq.~\eqref{eq:stressEnergyAngularMoments}, the Lagrangian decomposition of the stress-energy tensor is given by
\begin{equation}
  \mathcal{T}^{\mu\nu} 
  = \mathcal{J}\,u^{\mu}\,u^{\nu} + \mathcal{H}^{\mu}\,u^{\nu} + u^{\mu}\,\mathcal{H}^{\nu} + \mathcal{K}^{\mu\nu},
  \label{eq:stressEnergyLagrangianDecomposition}
\end{equation}
where
\begin{equation}
  \big\{\,\mathcal{J},\,\mathcal{H}^{\mu},\,\mathcal{K}^{\mu\nu}\,\big\}(\varepsilon,x,t)
  = \f{\varepsilon}{4\pi}\int_{\mathbb{S}^{2}}f(\omega,\varepsilon,x,t)\,\big\{1,\,\ell^{\mu},\,\ell^{\mu}\ell^{\nu}\,\big\}\,d\omega,
  \label{eq:energyMomentsLagrangian}
\end{equation}
and $\mathcal{H}^{\mu}$ and $\mathcal{K}^{\mu\nu}$ are orthogonal to $u_{\mu}$ (spacelike in the comoving frame); i.e., $u_{\mu}\mathcal{H}^{\mu}=u_{\mu}\mathcal{K}^{\mu\nu}=u_{\nu}\mathcal{K}^{\mu\nu}=0$.  
In Eq.~\eqref{eq:energyMomentsLagrangian}, $\mathcal{J}$, $\mathcal{H}^{\mu}$, and $\mathcal{K}^{\mu\nu}$ are respectively the spectral energy density, momentum density, and stress measured by a Lagrangian observer.  
The four-velocity $u_{\mu}$ and the associated orthogonal projector $h_{\mu\nu}$ are used to extract components of the Lagrangian decompositions of $\mathcal{T}^{\mu\nu}$:
\begin{equation}
  \mathcal{J}
  = u_{\mu}\,u_{\nu}\,\mathcal{T}^{\mu\nu},
  \quad
  \mathcal{H}^{\mu}
  =-u_{\nu}\,h^{\mu}_{\hspace{6pt}\rho}\,\mathcal{T}^{\nu\rho},
  \quad\text{and}\quad
  \mathcal{K}^{\mu\nu}
  =h^{\mu}_{\hspace{6pt}\rho}\,h^{\nu}_{\hspace{6pt}\sigma}\,\mathcal{T}^{\rho\sigma}.
  \label{eq:stressEnergyLagrangianExtractions}
\end{equation}
Note that the Lagrangian energy density and momentum density are related to the number density and flux by a factor $\varepsilon$; i.e.,
\begin{equation}
  \big\{\,\mathcal{J},\,\mathcal{H}^{\mu}\,\big\} = \varepsilon\,\big\{\,\mathcal{D},\,\mathcal{I}^{\mu}\,\big\}.  
\end{equation}

Finally, a Lagrangian decomposition of the rank-three tensor in Eq.~\eqref{eq:heatFluxAngularMoments} gives
\begin{align}
  \mathcal{Q}^{\mu\nu\rho} 
  &= \varepsilon\,\big(\,
    \mathcal{J}\,u^{\mu}\,u^{\nu}\,u^{\rho} + \mathcal{H}^{\mu}\,u^{\nu}\,u^{\rho} + \mathcal{H}^{\nu}\,u^{\mu}\,u^{\rho} + \mathcal{H}^{\rho}\,u^{\mu}\,u^{\nu} \nonumber \\
    &\hspace{32pt}
    + \mathcal{K}^{\mu\nu}\,u^{\rho} +\mathcal{K}^{\mu\rho}\,u^{\nu} +\mathcal{K}^{\nu\rho}\,u^{\mu} + \mathcal{L}^{\mu\nu\rho}
  \,\big),
  \label{eq:heatFluxLagrangianDecomposition}
\end{align}
where the spectral rank-three tensor measured by a Lagrangian observer,
\begin{equation}
  \mathcal{L}^{\mu\nu\rho}(\varepsilon,x,t)
  = \f{\varepsilon}{4\pi}\int_{\mathbb{S}^{2}}f(\omega,\varepsilon,x,t)\,\ell^{\mu}\ell^{\nu}\ell^{\rho}\,d\omega,
  \label{eq:heatFluxMomentsLagrangian}
\end{equation}
is orthogonal to $u_{\mu}$---i.e., $u_{\mu}\mathcal{L}^{\mu\nu\rho}=u_{\nu}\mathcal{L}^{\mu\nu\rho}=u_{\rho}\mathcal{L}^{\mu\nu\rho}=0$---and is obtained from $\mathcal{Q}^{\mu\nu\rho}$ using the orthogonal projector:
\begin{equation}
  \mathcal{L}^{\mu\nu\rho} 
  = \f{1}{\varepsilon}\,h^{\mu}_{\hspace{4pt}\sigma}\,h^{\nu}_{\hspace{4pt}\kappa}\,h^{\rho}_{\hspace{4pt}\lambda}\,\mathcal{Q}^{\sigma\kappa\lambda}.  
\end{equation}

\subsubsection{Eulerian decompositions}
\label{sec:EulerianDecompositions}

Eulerian projections of tensors are particularly useful when deriving evolution equations in the context of moment models for neutrino transport, as it is the Eulerian number density, energy density, and three-momentum density that are governed by conservation laws.  
In a manner similar to the Lagrangian decomposition in Eq.~\eqref{eq:numberCurrentLagrangianDecomposition}, the Eulerian decomposition of the spectral number current density is
\begin{equation}
  \mathcal{N}^{\mu} = \mathcal{N}\,n^{\mu} + \mathcal{G}^{\mu},
  \label{eq:numberCurrentEulerianDecomposition}
\end{equation}
where $n_{\mu}\,\mathcal{G}^{\mu}=0$.  
The four-velocity $n_{\mu}$ and the projector $\gamma_{\mu\nu}=g_{\mu\nu}+n_{\mu}\,n_{\nu}$ can be used to extract the Eulerian components
\begin{equation}
  \mathcal{N} = -n_{\mu}\,\mathcal{N}^{\mu}
  \quad\text{and}\quad
  \mathcal{G}^{\mu} = \gamma^{\mu}_{\hspace{6pt}\nu}\,\mathcal{N}^{\nu},
  \label{eq:numberCurrentEulerianExtractions}
\end{equation}
where $\mathcal{N}$ and $\mathcal{G}^{\mu}$ are the spectral number density and number flux density measured by an Eulerian observer, respectively.  
Note that $\mathcal{N}$ and $\mathcal{G}^{\mu}$ are still considered functions of $\varepsilon$, the neutrino energy measured by a Lagrangian observer.  
Thus, the definition in Eq.~\eqref{eq:numberCurrentEulerianDecomposition} should merely be viewed as a decomposition of $\mathcal{N}^{\mu}$ in a different basis than in Eq.~\eqref{eq:numberCurrentLagrangianDecomposition}, not as moments of the distribution with respect to Eulerian momentum coordinates.  
Inserting the Lagrangian decomposition in Eq.~\eqref{eq:numberCurrentLagrangianDecomposition} into the expressions in Eq.~\eqref{eq:numberCurrentEulerianExtractions}, the Eulerian number density and number flux density are expressed in terms of the Lagrangian number density and number flux density as
\begin{align}
  \mathcal{N}
  &= W\,\mathcal{D} + v_{\mu}\,\mathcal{I}^{\mu}, \label{eq:eulerianNumberInTermsOfLagrangianMoments} \\
  \mathcal{G}^{\mu}
  &=\big[\,\delta^{\mu}_{\hspace{6pt}\nu}-n^{\mu}v_{\nu}\,\big]\mathcal{I}^{\nu} + W\,\mathcal{D}\,v^{\mu}. \label{eq:eulerianNumberFluxInTermsOfLagrangianMoments}
\end{align}

Similarly, the Eulerian decomposition of the stress-energy tensor is
\begin{equation}
  \mathcal{T}^{\mu\nu} = \mathcal{E}\,n^{\mu}\,n^{\nu} + \mathcal{F}^{\mu}\,n^{\nu} + n^{\mu}\,\mathcal{F}^{\nu} + \mathcal{S}^{\mu\nu},
  \label{eq:stressEnergyEulerianDecomposition}
\end{equation}
where $\mathcal{E}$, $\mathcal{F}^{\mu}$, and $\mathcal{S}^{\mu\nu}$ are respectively the spectral energy density, momentum density, and stress measured by an Eulerian observer.  
The Eulerian momentum density and stress are spacelike (i.e., $n_{\mu}\mathcal{F}^{\mu}=n_{\mu}\mathcal{S}^{\mu\nu}=n_{\nu}\mathcal{S}^{\mu\nu}=0$), and the components of the Eulerian decomposition of $\mathcal{T}^{\mu\nu}$ are extracted using $n_{\mu}$ and the associated orthogonal projector $\gamma_{\mu\nu}$:
\begin{equation}
  \mathcal{E}
  = n_{\mu}\,n_{\nu}\,\mathcal{T}^{\mu\nu},
  \quad
  \mathcal{F}^{\mu}
  =-n_{\nu}\,\gamma^{\mu}_{\hspace{6pt}\rho}\,\mathcal{T}^{\nu\rho},
  \quad
  \mathcal{S}^{\mu\nu}
  =\gamma^{\mu}_{\hspace{6pt}\rho}\,\gamma^{\nu}_{\hspace{6pt}\sigma}\,\mathcal{T}^{\rho\sigma}.
  \label{eq:stressEnergyEulerianExtractions}
\end{equation}
Inserting the Lagrangian decomposition in Eq.~\eqref{eq:stressEnergyLagrangianDecomposition} into the expressions in Eq.~\eqref{eq:stressEnergyEulerianExtractions}, the Eulerian energy density, momentum density, and stress are expressed in terms of the corresponding Lagrangian quantities as \citep[cf.\ Equations~(B8)--(B10) in][]{CaEnMe13a}
\begin{align}
  \mathcal{E}
  &=W^{2}\mathcal{J} + 2\,W\,v_{\mu}\,\mathcal{H}^{\mu} + v_{\mu}\,v_{\nu}\,\mathcal{K}^{\mu\nu}, 
  \label{eq:eulerianEnergyInTermsOfLagrangianMoments} \\
  \mathcal{F}^{\mu}
  &=W\,v^{\mu}\,\big(\,W\mathcal{J} + v_{\nu}\,\mathcal{H}^{\nu}\,\big) + \big[\,\delta^{\mu}_{\hspace{6pt}\rho}-n^{\mu}\,v_{\rho}\,\big]\,\big(\,W\mathcal{H}^{\rho}+v_{\nu}\mathcal{K}^{\nu\rho}\,\big), 
  \label{eq:eulerianMomentumInTermsOfLagrangianMoments} \\
  \mathcal{S}^{\mu\nu}
  &=W^{2}\mathcal{J}v^{\mu}v^{\nu} + Wv^{\nu}\big[\,\delta^{\mu}_{\hspace{6pt}\rho}-n^{\mu}v_{\rho}\,\big]\,\mathcal{H}^{\rho}
  +Wv^{\mu}\big[\,\delta^{\nu}_{\hspace{6pt}\sigma}-n^{\nu}v_{\sigma}\,\big]\mathcal{H}^{\sigma} \nonumber \\
  &\hspace{24pt}
  +\big[\,\delta^{\mu}_{\hspace{6pt}\rho}-n^{\mu}v_{\rho}\,\big]\big[\,\delta^{\nu}_{\hspace{6pt}\sigma}-n^{\nu}v_{\sigma}\,\big]\mathcal{K}^{\rho\sigma}.  
  \label{eq:eulerianStressInTermsOfLagrangianMoments}
\end{align}

Finally, and similar to Eqs.~\eqref{eq:numberCurrentEulerianDecomposition} and \eqref{eq:stressEnergyEulerianDecomposition}, the Eulerian decomposition of the rank three tensor in Eq.~\eqref{eq:heatFluxAngularMoments} is given by
\begin{align}
  \mathcal{Q}^{\mu\nu\rho}
  &=\varepsilon\,\big(\,
    \mathcal{X}\,n^{\mu}\,n^{\nu}\,n^{\rho} + \mathcal{Y}^{\mu}\,n^{\nu}\,n^{\rho} + \mathcal{Y}^{\nu}\,n^{\mu}\,n^{\rho} + \mathcal{Y}^{\rho}\,n^{\mu}\,n^{\nu} \nonumber \\
    &\hspace{32pt}
    + \mathcal{Z}^{\mu\nu}\,n^{\rho} +\mathcal{Z}^{\mu\rho}\,n^{\nu} +\mathcal{Z}^{\nu\rho}\,n^{\mu} + \mathcal{W}^{\mu\nu\rho}
  \,\big),
  \label{eq:heatFluxEulerianDecomposition}
\end{align}
where the Eulerian components are obtained from
\begin{align}
  \mathcal{X}
  &= - \f{1}{\varepsilon}\,n_{\mu}\,n_{\nu}\,n_{\rho}\,\mathcal{Q}^{\mu\nu\rho}, \\
  \mathcal{Y}^{\mu}
  &=\f{1}{\varepsilon}\,\gamma^{\mu}_{\hspace{4pt}\sigma}\,n_{\nu}\,n_{\rho}\,\mathcal{Q}^{\sigma\nu\rho}, \\
  \mathcal{Z}^{\mu\nu}
  &=-\f{1}{\varepsilon}\,\gamma^{\mu}_{\hspace{4pt}\sigma}\,\gamma^{\nu}_{\hspace{4pt}\kappa}\,n_{\rho}\,\mathcal{Q}^{\sigma\kappa\rho}, \\
  \mathcal{W}^{\mu\nu\rho}
  &=\f{1}{\varepsilon}\,\gamma^{\mu}_{\hspace{4pt}\sigma}\,\gamma^{\nu}_{\hspace{4pt}\kappa}\,\gamma^{\rho}_{\hspace{4pt}\lambda}\,\mathcal{Q}^{\sigma\kappa\lambda}.  
\end{align}
These components can be expressed in terms of the Lagrangian moments by inserting the Lagrangian decomposition in Eq.~\eqref{eq:heatFluxLagrangianDecomposition}.  
We will not repeat these tedious expressions here, but see Eqs.~(B15), (B16), (B17), and (B18) in \citet{CaEnMe13a} for expressions relating respectively $\mathcal{Z}$, $\mathcal{Y}_{\mu}$, $\mathcal{Z}_{\mu\nu}$, and $\mathcal{W}_{\mu\nu\rho}$ in terms of the Lagrangian moments $\mathcal{J}$, $\mathcal{H}^{\mu}$, $\mathcal{K}^{\mu\nu}$, and $\mathcal{L}^{\mu\nu\rho}$ (note the difference of the factor of $\varepsilon$ between our definition of $\mathcal{Q}^{\mu\nu\rho}$ and the corresponding variable in \citet{CaEnMe13a}).  

While components of Lagrangian decompositions are more closely related to the distribution function, Eulerian decompositions appear to be more natural to use in the 3+1 approach, and powerful in simplifying terms appearing in the moment equations, especially for the energy derivative terms in Eqs.~\eqref{eq:spectralNumberEquation} and \eqref{eq:spectralFourMomentumEquation}, which contain contractions with the covariant derivative of the fluid four-velocity.  
As elaborated on in \citet{CaEnMe13a}, Eulerian decompositions of $\mathcal{T}^{\mu\nu}$ and $\mathcal{Q}^{\mu\nu\rho}$, in combination with the Eulerian decomposition of $u^{\mu}$, in Eq.~\eqref{eq:fluidFourVelocityEulerian} result in surprisingly simple expressions, without explicit reference to connection coefficients (cf.\ Eq.~\eqref{eq:ConnectionComoving}).  
Moreover, as emphasized by \citet{CaEnMe13a}, consistent use of Eulerian decompositions in spacetime and momentum-space divergences in the moment equations turns out to simplify the elucidation of the relationship between the equations for four-momentum and number conservation in the 3+1 case.  

\subsubsection{Two-moment kinetics}
\label{sec:TwoMoment}

In this section we review two-moment models in the 3+1 formulation of general relativity, which can serve as a basis for the development of numerical methods and their implementation in codes to model neutrino transport in core-collapse supernovae.  
We present three versions, all based on Eq.~\eqref{eq:spectralFourMomentumEquation}, but using different projections.  
The projection of Eq.~\eqref{eq:spectralFourMomentumEquation} orthogonal and tangential to the spacelike slice of the Eulerian observer (using $n_{\mu}$ and $\gamma_{\mu\nu}$) gives rise to the \emph{Eulerian} two-moment model, while the projection of Eq.~\eqref{eq:spectralFourMomentumEquation} orthogonal and tangential to the spacelike slice of the Lagrangian observer (using $u_{\mu}$ and $h_{\mu\nu}$) gives rise to the \emph{Lagrangian} two-moment model.  
We also present a \emph{number conservative} two-moment model, which is closely related to the Lagrangian two-moment model, but uses projections based on $u_{\mu}/\varepsilon$ and $h_{\mu\nu}/\varepsilon$.  
This results in one of the evolved equations being Eq.~\eqref{eq:spectralNumberEquation}, which is neutrino number conservative.  
Analytically, all these formulations are equivalent, but they could have different numerical properties.  

\paragraph{Eulerian two-moment model}

The Eulerian two-moment model evolves the spectral energy density and momentum density measured by an Eulerian observer ($\mathcal{E}$ and $\mathcal{F}_{j}$, respectively).  
The energy equation is obtained as the projection of Eq.~\eqref{eq:spectralFourMomentumEquation} onto the four-velocity of the Eulerian observer (i.e., contracting $-n_{\mu}$ with Eq.~\eqref{eq:spectralFourMomentumEquation}).  
The result is:
\begin{align}
  &\f{1}{\alpha\sqrt{\gamma}}
  \big[\,\pd{}{t}\big(\,\sqrt{\gamma}\,\mathcal{E}\,\big)+\pd{}{i}\big(\,\sqrt{\gamma}\,\big[\,\alpha\,\mathcal{F}^{i}-\beta^{i}\,\mathcal{E}\,\big]\,\big)\,\big]
  -\f{1}{\varepsilon^{2}}\pderiv{}{\varepsilon}\big(\,\varepsilon^{2}\,(-n_{\mu})\,\mathcal{Q}^{\mu\nu\rho}\,\nabla_{\nu}u_{\rho}\,\big) \nonumber \\
  &\hspace{0pt}
  =\f{1}{\alpha}\,\big[\,\alpha\,\mathcal{S}^{ij}\,\mathsf{K}_{ij}-\mathcal{F}^{i}\,\pd{\alpha}{i}\,\big]
  +\f{W}{4\pi}\int_{\mathbb{S}^{2}}\mathcal{C}(f)\,d\omega+\f{v^{j}}{4\pi}\int_{\mathbb{S}^{2}}\mathcal{C}(f)\,\ell_{j}\,d\omega,
  \label{eq:spectralEulerianEnergyEquation_3p1}
\end{align}
where the sources on the right-hand side are due to spacetime geometry and energy exchange between neutrinos and the fluid.  
The left-hand side is in divergence form, where the divergence operates on the spacetime-plus-energy phase-space.  
In expressing the terms inside the energy derivative (last term on the left-hand side), we make use of the Eulerian decomposition in Eq.~\eqref{eq:heatFluxEulerianDecomposition} to write
\begin{align}
  &-\f{n_{\mu}}{\varepsilon}\,\mathcal{Q}^{\mu\nu\rho}\,\nabla_{\nu}u_{\rho} \nonumber \\
  &=
  \big(\,
    \mathcal{X}\,n^{\nu}\,n^{\rho} + \mathcal{Y}^{\nu}\,n^{\rho} + n^{\nu}\,\mathcal{Y}^{\rho} + \mathcal{Z}^{\nu\rho}
  \,\big)\,\nabla_{\nu}u_{\rho} \nonumber \\
  &=
  \f{W}{\alpha}\,
  \Big\{\,
    \big(\,\mathcal{Y}^{i} - \mathcal{X}\,v^{i}\,\big)\,\pd{\alpha}{i}
    +\mathcal{Y}_{k}\,v^{i}\,\pd{\beta^{k}}{i}
    +\alpha\,\mathcal{Z}^{ki}\,\big(\,\f{1}{2}\,v^{m}\,\pd{\gamma_{ki}}{m} - \mathsf{K}_{ki}\,\big)
  \,\Big\} \nonumber \\
  &\hspace{12pt}
  +\f{1}{\alpha}\,
  \Big\{\,
    \mathcal{Y}_{k}\,\pd{}{t}\big(Wv^{k}\big)
    - \mathcal{X}\,\pd{W}{t}
    - \big(\,\alpha\,\mathcal{Y}^{i} - \mathcal{X}\,\beta^{i}\,\big)\,\pd{W}{i} \nonumber \\
    &\hspace{64pt}
    + \big(\,\alpha\,\mathcal{Z}_{k}^{\hspace{4pt}i} - \mathcal{Y}_{k}\,\beta^{i}\,\big)\,\pd{}{i}\big(Wv^{k}\big)
  \,\Big\}, 
  \label{eq:observerCorrectionsEulerianEnergyEquation_3p1}
\end{align}
which account for changes in the spectral energy density due to gravitational energy shifts and the fact that adjacent 
comoving observer velocities in spacetime are generally different.

The momentum equation is obtained as the projection of Eq.~\eqref{eq:spectralFourMomentumEquation} into the slice with normal given by $n^{\mu}$ (i.e., contracting $\gamma_{j\mu}$ with Eq.~\eqref{eq:spectralFourMomentumEquation}), which results in
\begin{align}
  &\f{1}{\alpha\sqrt{\gamma}}
  \big[\pd{}{t}\big(\,\sqrt{\gamma}\mathcal{F}_{j}\,\big)+\pd{}{i}\big(\,\sqrt{\gamma}\big[\,\alpha\mathcal{S}^{i}_{\hspace{4pt}j}-\beta^{i}\mathcal{F}_{j}\,\big]\,\big)\big]
  -\f{1}{\varepsilon^{2}}\pderiv{}{\varepsilon}\big(\,\varepsilon^{2}\,\gamma_{j\mu}\,\mathcal{Q}^{\mu\nu\rho}\,\nabla_{\nu}u_{\rho}\,\big) \label{eq:spectralEulerianMomentumEquation_3p1} \\
  &\hspace{0pt}
  =\f{1}{\alpha}\,\big[\,\mathcal{F}_{i}\,\pd{\beta^{i}}{j}+\f{1}{2}\,\alpha\,\mathcal{S}^{ik}\,\pd{\gamma_{ik}}{j}-\mathcal{E}\,\pd{\alpha}{j}\,\big]
  +\f{1}{4\pi}\int_{\mathbb{S}^{2}}\mathcal{C}(f)\,\ell_{j}\,d\omega+\f{Wv_{j}}{4\pi}\int_{\mathbb{S}^{2}}\mathcal{C}(f)\,d\omega, \nonumber
\end{align}
where the right-hand side gives rise to changes in the spectral momentum density due to spacetime geometry and neutrino--matter interactions.  
Again, using the Eulerian decomposition in Eq.~\eqref{eq:heatFluxEulerianDecomposition}, the terms inside the energy derivative can be written as
\begin{align}
  &\f{\gamma_{j\mu}}{\varepsilon}\,\mathcal{Q}^{\mu\nu\rho}\,\nabla_{\nu}u_{\rho} \nonumber \\
  &=
  \big(\,
    \mathcal{Y}_{j}\,n^{\nu}\,n^{\rho} + \mathcal{Z}_{j}^{\hspace{4pt}\nu}\,n^{\rho} + \mathcal{Z}_{j}^{\hspace{4pt}\rho}\,n^{\nu} + \mathcal{W}_{j}^{\hspace{4pt}\nu\rho}
  \,\big)\,\nabla_{\nu}u_{\rho} \nonumber \\
  &=
  \f{W}{\alpha}\,
  \Big\{\,
    \big(\,\mathcal{Z}_{j}^{\hspace{2pt}i} - \mathcal{Y}_{j}\,v^{i}\,\big)\,\pd{\alpha}{i}
    +\mathcal{Z}_{jk}\,v^{i}\,\pd{\beta^{k}}{i}
    +\alpha\,\mathcal{W}_{j}^{\hspace{2pt}ki}\,\big(\,\f{1}{2}\,v^{m}\,\pd{\gamma_{ki}}{m} - \mathsf{K}_{ki}\,\big)
  \,\Big\} \nonumber \\
  &\hspace{12pt}
  +\f{1}{\alpha}\,
  \Big\{\,
    \mathcal{Z}_{jk}\,\pd{}{t}\big(Wv^{k}\big)
    - \mathcal{Y}_{j}\,\pd{W}{t}
    - \big(\,\alpha\,\mathcal{Z}_{j}^{\hspace{2pt}i} - \mathcal{Y}_{j}\,\beta^{i}\,\big)\,\pd{W}{i} \nonumber \\
    &\hspace{64pt}
    + \big(\,\alpha\,\mathcal{W}_{jk}^{\hspace{4pt}i} - \mathcal{Z}_{jk}\,\beta^{i}\,\big)\,\pd{}{i}\big(Wv^{k}\big)
  \,\Big\},
  \label{eq:observerCorrectionsEulerianMomentumEquation_3p1}
\end{align}
which account for changes in the spectral momentum density due to gravitational comoving observer effects.  

An obvious advantage of the Eulerian two-moment model given by Eqs.~\eqref{eq:spectralEulerianEnergyEquation_3p1} and \eqref{eq:spectralEulerianMomentumEquation_3p1} is the conservative form.  
Integrating these equations over energy space (using $dV_{\varepsilon}=4\pi\varepsilon^{2}d\varepsilon$) results in the radiation energy equation
\begin{align}
  &\f{1}{\alpha\sqrt{\gamma}}
  \big[\,\pd{}{t}\big(\,\sqrt{\gamma}\,E\,\big)+\pd{}{i}\big(\,\sqrt{\gamma}\,\big[\,\alpha\,F^{i}-\beta^{i}\,E\,\big]\,\big)\,\big] \label{eq:EulerianEnergyEquation_3p1} \\
  &=\f{1}{\alpha}\,\big[\,\alpha\,S^{ij}\,\mathsf{K}_{ij}-F^{i}\,\pd{\alpha}{i}\,\big]
  +W\int_{V_{p}}\mathcal{C}[f]\,\varepsilon\,\pi_{m}
  +v^{j}\int_{V_{p}}\mathcal{C}[f]\,\varepsilon\,\ell_{j}\,\pi_{m} \nonumber
\end{align}
and radiation momentum equation
\begin{align}
  &\f{1}{\alpha\sqrt{\gamma}}
  \big[\,\pd{}{t}\big(\,\sqrt{\gamma}F_{j}\,\big)+\pd{}{i}\big(\,\sqrt{\gamma}\big[\,\alpha S^{i}_{\hspace{4pt}j}-\beta^{i}F_{j}\,\big]\,\big)\,\big] \label{eq:EulerianMomentumEquation_3p1} \\
  &=\f{1}{\alpha}\,\big[\,F_{i}\,\pd{\beta^{i}}{j}+\f{1}{2}\,\alpha\,S^{ik}\,\pd{\gamma_{ik}}{j}-E\,\pd{\alpha}{j}\,\big]
  +\int_{V_{p}}\mathcal{C}[f]\,\varepsilon\,\ell_{j}\,\pi_{m}
  +Wv_{j}\int_{V_{p}}\mathcal{C}[f]\,\varepsilon\,\pi_{m}, \nonumber
\end{align}
where the energy-integrated Eulerian moments are given by
\begin{equation}
  \big\{\,E,\,F^{\mu},\,S^{\mu\nu}\,\big\} = \int_{0}^{\infty}\big\{\,\mathcal{E},\,\mathcal{F}^{\mu},\,\mathcal{S}^{\mu\nu}\,\big\}\,dV_{\varepsilon}.  
\end{equation}
Eqs.~\eqref{eq:EulerianEnergyEquation_3p1} and \eqref{eq:EulerianMomentumEquation_3p1} are conservation laws for radiation energy and momentum in the sense that in the case of Cartesian coordinates in flat spacetime, with no neutrino--matter interactions, the right-hand sides vanish, and the equations express exact conservation of radiation energy and momentum.  

The Eulerian two-moment model presented here is the basis for several codes developed to model neutrino transport in core-collapse supernovae \citep{OCon15,KuTaKo16,RoOtHa16}: the \textsc{GR1D} code \citep{OCon15} solves the equations in spherical symmetry; the \textsc{Zelmani} code \citep{RoOtHa16} solves the equations in three spatial dimensions, but does not include velocity dependent terms (i.e., $v^{i}=0$ in the transport equations); and \citet{KuTaKo16} solve the full system in three spatial dimensions.  

\paragraph{Lagrangian two-moment model}

The Lagrangian two-moment model is an alternative to the Eulerian two-moment model discussed above, where the spectral energy density and momentum density measured by the Lagrangian observer with four-velocity $u_{\mu}$ are evolved ($\mathcal{J}$ and $\mathcal{H}_{j}$, respectively).  
The energy equation is obtained as the projection of Eq.~\eqref{eq:spectralFourMomentumEquation} along the four-velocity of the Lagrangian observer (i.e., contracting $-u_{\mu}$ with Eq.~\eqref{eq:spectralFourMomentumEquation}), which results in
\begin{align}
  &\f{1}{\alpha\sqrt{\gamma}}
  \big[\,\pd{}{t}\big(\,\sqrt{\gamma}\,\big[\,W\mathcal{J}+v^{i}\mathcal{H}_{i}\,\big]\,\big)
  +\pd{}{i}\big(\,\sqrt{\gamma}\,\big[\,\alpha\,\mathcal{H}^{i}+\big(\,\alpha\,v^{i}-\beta^{i}\,\big)\,W\mathcal{J}\,\big]\,\big)\,\big] \nonumber \\
  &\hspace{6pt}
  -\f{1}{\varepsilon^{2}}\pderiv{}{\varepsilon}\big(\,\varepsilon^{3}\,\mathcal{T}^{\mu\nu}\,\nabla_{\mu}u_{\nu}\,\big)
  =-\mathcal{T}^{\mu\nu}\nabla_{\mu}u_{\nu} + \f{1}{4\pi}\int_{\mathbb{S}^{2}}\mathcal{C}(f)\,d\omega,
  \label{eq:spectralLagrangianEnergyEquation_3p1}
\end{align}
where the contraction of the stress-energy equation with the covariant derivative of the Lagrangian observer's four-velocity is given in $3+1$ form as
\begin{align}
  &\mathcal{T}^{\mu\nu}\,\nabla_{\mu}u_{\nu} \nonumber \\
  &=
  \big(\,
    \mathcal{E}\,n^{\mu}\,n^{\nu} + \mathcal{F}^{\mu}\,n^{\nu} + n^{\mu}\,\mathcal{F}^{\nu} + \mathcal{S}^{\mu\nu}
  \,\big)\,\nabla_{\mu}u_{\nu} \nonumber \\
  &=
  \f{W}{\alpha}\,
  \Big\{\,
    \big(\,\mathcal{F}^{i} - \mathcal{E}\,v^{i}\,\big)\,\pd{\alpha}{i}
    +\mathcal{F}_{k}\,v^{i}\,\pd{\beta^{k}}{i}
    +\alpha\,\mathcal{S}^{ki}\,\big(\,\f{1}{2}\,v^{m}\,\pd{\gamma_{ki}}{m} - \mathsf{K}_{ki}\,\big)
  \,\Big\} \nonumber \\
  &\hspace{12pt}
  +\f{1}{\alpha}\,
  \Big\{\,
    \mathcal{F}_{k}\,\pd{}{t}\big(Wv^{k}\big)
    - \mathcal{E}\,\pd{W}{t}
    - \big(\,\alpha\,\mathcal{F}^{i} - \mathcal{E}\,\beta^{i}\,\big)\,\pd{W}{i} \nonumber \\
    &\hspace{64pt}
    + \big(\,\alpha\,\mathcal{S}_{k}^{\hspace{4pt}i} - \mathcal{F}_{k}\,\beta^{i}\,\big)\,\pd{}{i}\big(Wv^{k}\big)
  \,\Big\},
  \label{eq:observerCorrectionsLagrangianEnergyEquation_3p1}
\end{align}
which accounts for changes to the spectral energy density from gravitational effects and from the fact that adjacent comoving observers in spacetime have different velocities.
In Eq.~\eqref{eq:observerCorrectionsLagrangianEnergyEquation_3p1}, we made use of the Eulerian decomposition of the stress-energy tensor, which, as discussed at the end of Sect.~\ref{sec:EulerianDecompositions}, is more convenient than using the Lagrangian decomposition, since it keeps the number of terms in the expression to a minimum and simplifies book-keeping.  
The components of the Eulerian decomposition are related to the Lagrangian components by Eqs.~\eqref{eq:eulerianEnergyInTermsOfLagrangianMoments}-\eqref{eq:eulerianStressInTermsOfLagrangianMoments}.  

The Lagrangian momentum equation is obtained by projecting Eq.~\eqref{eq:spectralFourMomentumEquation} tangential to the slice with $u^{\mu}$ as the normal (i.e., contracting $h_{j\mu}$ with Eq.~\eqref{eq:spectralFourMomentumEquation}), which gives
\begin{align}
  &\f{1}{\alpha\sqrt{\gamma}}
  \big[\,\pd{}{t}\big(\,\sqrt{\gamma}\,\big[\,W\mathcal{H}_{j}+v^{i}\mathcal{K}_{ij}\,\big]\,\big)
  +\pd{}{i}\big(\,\sqrt{\gamma}\,\big[\,\alpha\,\mathcal{K}^{i}_{\hspace{4pt}j}+\big(\,\alpha\,v^{i}-\beta^{i}\,\big)\,W\mathcal{H}_{j}\,\big]\,\big)\,\big]
  \label{eq:spectralLagrangianMomentumEquation_3p1} \\
  &\hspace{6pt}
  -\f{1}{\varepsilon^{2}}\pderiv{}{\varepsilon}\big(\,\varepsilon^{2}\,h_{j\mu}\,\mathcal{Q}^{\mu\nu\rho}\,\nabla_{\nu}u_{\rho}\,\big)
  =\mathcal{T}^{\mu\nu}\,\big(\,\nabla_{\nu}h_{j\mu} + \Gamma^{\rho}_{\hspace{4pt}j\nu}h_{\rho\mu}\,\big) 
  + \f{1}{4\pi}\int_{\mathbb{S}^{2}}\mathcal{C}(f)\,\ell_{j}\,d\omega, \nonumber
\end{align}
where the ``geometry'' source on the right-hand side can be written as
\begin{align}
  &\mathcal{T}^{\mu\nu}\,\big(\,\nabla_{\nu}h_{j\mu} + \Gamma^{\rho}_{\hspace{4pt}j\nu}h_{\rho\mu}\,\big) \nonumber \\
  &\hspace{12pt}
  =\f{1}{2}\,\mathcal{T}^{\mu\nu}\,\pd{g_{\mu\nu}}{j}
  +Wv_{j}\,\mathcal{T}^{\mu\nu}\nabla_{\mu}u_{\nu}
  +u_{\mu}\mathcal{T}^{\mu\nu}\,\pd{}{\nu}\big(\,Wv_{j}\,\big).
  \label{eq:spectralLagrangianMomentumEquationGeometrySource_3p1}
\end{align}
Again, using the Eulerian decomposition of $\mathcal{T}^{\mu\nu}$, the first term on the right-hand side of Eq.~\eqref{eq:spectralLagrangianMomentumEquationGeometrySource_3p1} can be written as
\begin{equation}
  \f{1}{2}\,\mathcal{T}^{\mu\nu}\,\pd{g_{\mu\nu}}{j}
  =\f{1}{\alpha}\,\big[\,\mathcal{F}_{i}\,\pd{\beta^{i}}{j}+\f{1}{2}\,\alpha\,\mathcal{S}^{ik}\,\pd{\gamma_{ik}}{j}-\mathcal{E}\,\pd{\alpha}{j}\,\big],
\end{equation}
which also appears on the right-hand side of Eq.~\eqref{eq:spectralEulerianMomentumEquation_3p1}.  
Similarly, the third term on the right-hand side of Eq.~\eqref{eq:spectralLagrangianMomentumEquationGeometrySource_3p1} can be written as
\begin{align}
  u_{\mu}\mathcal{T}^{\mu\nu}\,\pd{}{\nu}\big(\,Wv_{j}\,\big)
  &=-\f{W}{\alpha}\,\Big\{\,\mathcal{E}-v^{k}\,\mathcal{F}_{k}\,\Big\}\,\pd{}{t}\big(Wv_{j}\big) \\
  &\hspace{12pt}
  -\f{W}{\alpha}\,
  \Big\{\,
    \big(\,\alpha\,\mathcal{F}^{i}-\beta^{i}\,\mathcal{E}\,\big)
    -v^{k}\,\big(\,\alpha\mathcal{S}^{i}_{\hspace{4pt}k}-\beta^{i}\,\mathcal{F}_{k}\,\big)
  \,\Big\}\,\pd{}{i}\big(Wv_{j}\big), \nonumber
\end{align}
while the second term on the right-hand side of Eq.~\eqref{eq:spectralLagrangianMomentumEquationGeometrySource_3p1} contains the expression in Eq.~\eqref{eq:observerCorrectionsLagrangianEnergyEquation_3p1}.  
Finally, the expression inside the energy derivative term on the left-hand side of Eq.~\eqref{eq:spectralLagrangianMomentumEquation_3p1} can be written as
\begin{align}
  &\f{h_{j\mu}}{\varepsilon}\,\mathcal{Q}^{\mu\nu\rho}\,\nabla_{\nu}u_{\rho} \nonumber \\
  &=\big(\,\f{1}{\varepsilon}\,\mathcal{Q}_{j}^{\hspace{4pt}\nu\rho}-Wv_{j}\,\mathcal{T}^{\nu\rho}\,\big)\,\nabla_{\nu}u_{\rho} \nonumber \\
  &=
  \Big\{\,
    \big(\mathcal{Y}_{j}-Wv_{j}\,\mathcal{E}\big)\,n^{\nu}\,n^{\rho} 
    +\big(\mathcal{Z}_{j}^{\hspace{4pt}\nu}-Wv_{j}\,\mathcal{F}^{\nu}\big)\,n^{\rho} \nonumber \\
  &\hspace{24pt}
  +\big(\mathcal{Z}_{j}^{\hspace{4pt}\rho}-Wv_{j}\mathcal{F}^{\rho}\big)\,n^{\nu}
  +\big(\mathcal{W}_{j}^{\hspace{4pt}\nu\rho}-Wv_{j}\mathcal{S}^{\nu\rho}\big)\,\Big\}\,\nabla_{\nu}u_{\rho},
  \label{eq:observerCorrectionsLagrangianMomentumEquation_3p1}
\end{align}
which is a contraction of Eulerian decompositions of rank two tensors, with components $\big(\mathcal{Y}_{j}-Wv_{j}\,\mathcal{E}\big)$, $\big(\mathcal{Z}_{j}^{\hspace{4pt}\nu}-Wv_{j}\,\mathcal{F}^{\nu}\big)$, and $\big(\mathcal{W}_{j}^{\hspace{4pt}\nu\rho}-Wv_{j}\mathcal{S}^{\nu\rho}\big)$, contracted with the covariant derivative of the fluid four-velocity, and can be written in a form similar to Eq.~\eqref{eq:observerCorrectionsEulerianMomentumEquation_3p1}.  

The Lagrangian two-moment model presented here (Eqs.~\eqref{eq:spectralLagrangianEnergyEquation_3p1} and \eqref{eq:spectralLagrangianMomentumEquation_3p1}) is the basis for several codes used to model neutrino transport in core-collapse supernovae: \citet{MuJaDi10} used it in conjunction with the conformal flatness approximation to GR (CFA) and ray-by-ray neutrino transport transport; and \citet{JuObJa15} and \citet{SkDoBu19} used this model in its $\mathcal{O}(v/c)$ limit to develop multi-dimensional neutrino transport codes.  

\paragraph{Number conservative two-moment model}

The number conservative model is yet another formulation of two-moment transport, which evolves the spectral number density as measured by the Eulerian observer (with four-velocity $n_{\mu}$) and the spectral number flux.  
The equation for the number density is obtained (1) directly from Eq.~\eqref{eq:spectralNumberEquation}, (2) by contraction of Eq.~\eqref{eq:spectralFourMomentumEquation} with $-u_{\mu}/\varepsilon$, or (3) by dividing Eq.~\eqref{eq:spectralLagrangianEnergyEquation_3p1} by $\varepsilon$.  
In $3+1$ form it is given by
\begin{align}
  &\f{1}{\alpha\sqrt{\gamma}}
  \big[\,\pd{}{t}\big(\,\sqrt{\gamma}\,\big[\,W\mathcal{D}+v^{i}\mathcal{I}_{i}\,\big]\,\big)
  +\pd{}{i}\big(\,\sqrt{\gamma}\,\big[\,\alpha\,\mathcal{I}^{i}+\big(\,\alpha\,v^{i}-\beta^{i}\,\big)\,W\mathcal{D}\,\big]\,\big)\,\big] \nonumber \\
  &\hspace{12pt}
  -\f{1}{\varepsilon^{2}}\pderiv{}{\varepsilon}\big(\,\varepsilon^{2}\,\mathcal{T}^{\mu\nu}\,\nabla_{\mu}u_{\nu}\,\big)
  =\f{1}{4\pi}\int_{\mathbb{S}^{2}}\mathcal{C}(f)\,\f{d\omega}{\varepsilon},
  \label{eq:spectralNumberEquation_3p1}
\end{align}
where the expression inside the energy derivative (last term on the left-hand side) is given by Eq.~\eqref{eq:observerCorrectionsLagrangianEnergyEquation_3p1}.  
Eq.~\eqref{eq:spectralNumberEquation_3p1} is conservative in the sense that an integration over energy space gives the balance equation Eq.~\eqref{eq:numberEquation}, which in $3+1$ form is given by
\begin{equation}
  \f{1}{\alpha\sqrt{\gamma}}
  \big[\,\pd{}{t}\big(\,\sqrt{\gamma}\,N\,\big)
  +\pd{}{i}\big(\,\sqrt{\gamma}\,\big[\,\alpha\,G^{i}-\beta^{i}\,N\,\big]\,\big)\,\big]
  =\int_{V_{p}}\mathcal{C}[f]\,\pi_{m},
  \label{eq:numberEquation_3p1}
\end{equation}
expressing exact particle conservation in the absence of particle-converting neutrino--matter interactions (e.g., emission and absorption).  

The equation for the number flux density is obtained by contraction of $h_{j\mu}/\varepsilon$ with Eq.~\eqref{eq:spectralFourMomentumEquation} (or by dividing Eq.~\eqref{eq:spectralLagrangianMomentumEquation_3p1} by $\varepsilon$):
\begin{align}
  &\f{1}{\alpha\sqrt{\gamma}}
  \big[\,\pd{}{t}\big(\,\sqrt{\gamma}\,\big[\,W\mathcal{I}_{j}+v^{i}\widehat{\mathcal{K}}_{ij}\,\big]\,\big)
  +\pd{}{i}\big(\,\sqrt{\gamma}\,\big[\,\alpha\,\widehat{\mathcal{K}}^{i}_{\hspace{4pt}j}+\big(\,\alpha\,v^{i}-\beta^{i}\,\big)\,W\mathcal{I}_{j}\,\big]\,\big)\,\big] \nonumber \\
  &\hspace{6pt}
  -\f{1}{\varepsilon^{2}}\pderiv{}{\varepsilon}\Big(\,\varepsilon^{2}\,h_{j\mu}\,\widehat{\mathcal{Q}}^{\mu\nu\rho}\,\nabla_{\nu}u_{\rho}\,\Big)
  =\f{1}{2}\,\widehat{\mathcal{T}}^{\mu\nu}\,\pd{g_{\mu\nu}}{j}
  +\f{1}{\varepsilon}\,\widehat{\mathcal{Q}}_{j}^{\hspace{4pt}\mu\nu}\,\nabla_{\nu}u_{\mu}
  -\mathcal{N}^{\nu}\,\pd{}{\nu}u_{j} \nonumber \\
  &\hspace{150pt}
  +\f{1}{4\pi}\int_{\mathbb{S}^{2}}\mathcal{C}(f)\,\ell_{j}\,\f{d\omega}{\varepsilon}.  
  \label{eq:spectralNumberFluxEquation_3p1}
\end{align}
Here, we use the ``hat'' to denote previously-defined moments divided by $\varepsilon$; e.g.,
\begin{equation}
  \big\{\,\widehat{\mathcal{T}}^{\mu\nu},\,\widehat{\mathcal{Q}}^{\mu\nu\rho}\,\big\} = \f{1}{\varepsilon}\big\{\,\mathcal{T}^{\mu\nu},\,\mathcal{Q}^{\mu\nu\rho}\,\big\}.  
\end{equation}
The expression in the energy derivative in Eq.~\eqref{eq:spectralNumberFluxEquation_3p1} is given by Eq.~\eqref{eq:observerCorrectionsLagrangianMomentumEquation_3p1}.  
The first term on the right-hand side of Eq.~\eqref{eq:spectralNumberFluxEquation_3p1} can be written as (cf.\ Eq.~\eqref{eq:spectralLagrangianMomentumEquation_3p1})
\begin{equation}
  \f{1}{2}\,\widehat{\mathcal{T}}^{\mu\nu}\,\pd{g_{\mu\nu}}{j}
  =\f{1}{\alpha}\,\Big\{\,\widehat{\mathcal{F}}_{i}\,\pd{\beta^{i}}{j}+\f{1}{2}\,\alpha\,\widehat{\mathcal{S}}^{ik}\,\pd{\gamma_{ik}}{j}-\widehat{\mathcal{E}}\,\pd{\alpha}{j}\,\Big\},
\end{equation}
while the third term on the right-hand side of Eq.~\eqref{eq:spectralNumberFluxEquation_3p1} can be written as
\begin{equation}
  \mathcal{N}^{\nu}\pd{}{\nu}u_{j}
  =\f{1}{\alpha}\,\Big\{\,\mathcal{N}\,\pd{}{t}\,\big(Wv_{j}\big)+\big(\alpha\,\mathcal{G}^{i}-\beta^{i}\mathcal{N}\big)\,\pd{}{i}\big(Wv_{j}\big)\,\Big\},
\end{equation}
where $\mathcal{N}$ and $\mathcal{G}^{i}$ are written in terms of Lagrangian moments in Eqs.~\eqref{eq:eulerianNumberInTermsOfLagrangianMoments} and \eqref{eq:eulerianNumberFluxInTermsOfLagrangianMoments}.  
The second term on the right-hand side of Eq.~\eqref{eq:spectralNumberFluxEquation_3p1} can be written as
\begin{align}
  \f{1}{\varepsilon}\,\widehat{\mathcal{Q}}_{j}^{\hspace{4pt}\mu\nu}\,\nabla_{\nu}u_{\mu}
  &=
  \Big\{\,
    \widehat{\mathcal{Y}}_{j}\,n^{\mu}\,n^{\nu}
    +\widehat{\mathcal{Z}}_{j}^{\hspace{4pt}\mu}\,n^{\nu}
    +n^{\mu}\,\widehat{\mathcal{Z}}_{j}^{\hspace{4pt}\nu}
    +\widehat{\mathcal{W}}_{j}^{\hspace{4pt}\mu\nu}
  \,\Big\}\,\nabla_{\nu}u_{\mu},
\end{align}
which is in the same form as Eq.~\eqref{eq:observerCorrectionsLagrangianEnergyEquation_3p1}, but where $\widehat{\mathcal{Y}}_{j}$, $\widehat{\mathcal{Z}}_{j}^{\hspace{4pt}\mu}$, and $\widehat{\mathcal{W}}_{j}^{\hspace{4pt}\mu\nu}$, replace $\mathcal{E}$, $\mathcal{F}^{\mu}$, and $\mathcal{S}^{\mu\nu}$, respectively.  

This number conservative two-moment model was presented in spherical symmetry, assuming the conformal flatness approximation (CFA) to general relativity, by \citet{MuJaDi10}, and was also presented in the $\mathcal{O}(v/c)$ limit by \citet{JuObJa15}, but it was not explicitly used in the numerical techniques developed by either of these authors.  
The model presented here is the 3+1 general relativistic version of that model, without approximation.  
It should also be mentioned that \citet{RaJa02} developed a two-moment, variable Eddington factor method based on solving both the Lagrangian two-moment model and the number conservative two-moment model simultaneously, in spherical symmetry and in the $\mathcal{O}(v/c)$ limit, treating the radiation energy density, momentum density, number density and number flux density as independent variables.  
However, resulting from inconsistency between the energy and number equations, in this approach the mean energy in an energy group, $\mathcal{J}/\mathcal{D}$, is not constrained to the group boundaries, and can even move outside the group \citep{MuJaDi10}.

\subsubsection{The closure problem}
\label{sec:closure}

The two-moment models discussed above are not closed.  
The rank-two tensor $\mathcal{K}^{\mu\nu}$ defined in Eq.~\eqref{eq:energyMomentsLagrangian} and the rank-three tensor $\mathcal{L}^{\mu\nu\rho}$ defined in Eq.~\eqref{eq:heatFluxMomentsLagrangian} are present in various terms in the two-moment model: components of $\mathcal{K}^{\mu\nu}$ are present in spacetime derivative terms, while components of $\mathcal{K}^{\mu\nu}$ and $\mathcal{L}^{\mu\nu\rho}$ are present in energy derivative terms and source terms.  
These tensor components must be expressed in terms of the evolved moments to close the system of equations.  
For the Eulerian and the Lagrangian two-moment models, the evolved quantities are ultimately the energy density and momentum density measured by a comoving observer; $\mathcal{J}$ and $\mathcal{H}_{j}$, respectively.  
(For the number conservative two-moment model, the evolved quantities are the number density and number flux density measured by a comoving observer; $\mathcal{D}$ and $\mathcal{I}_{j}$, respectively.)

Following \citet{Le84,AnPeSa92}, the general symmetric, rank-two tensor $\mathcal{K}^{\mu\nu}$, depending on $\mathcal{J}$ and $\mathcal{H}^{\mu}$, that is orthogonal to the fluid four-velocity $u_{\mu}$ and that satisfies the trace condition $\mathcal{K}^{\mu}_{\hspace{6pt}\mu}=\mathcal{J}$ takes the form
\begin{equation}
  \mathcal{K}^{\mu\nu}
  =\f{1}{2}\,\Big[\,\big(\,1-\mathfrak{k}\,\big)\,h^{\mu\nu}+\big(\,3\,\mathfrak{k}-1\,\big)\,\widehat{\mathsf{h}}^{\mu}\,\widehat{\mathsf{h}}^{\nu}\,\Big]\,\mathcal{J},
  \label{eq:radiationStressTensor}
\end{equation}
where $\mathfrak{k}(\mathcal{J},\mathfrak{h})$ is the Eddington factor, $\mathfrak{h}=\mathcal{H}/\mathcal{J}$ is the flux factor, $\mathcal{H}=\sqrt{\mathcal{H}_{\mu}\mathcal{H}^{\mu}}$, and $\widehat{\mathsf{h}}^{\mu}=\mathcal{H}^{\mu}/\mathcal{H}$ is a unit four-vector parallel to $\mathcal{H}^{\mu}$.  
It is straightforward to show that the Eddington factor can be written as
\begin{equation}
  \mathfrak{k}=\f{\widehat{\mathsf{h}}_{\mu}\widehat{\mathsf{h}}_{\nu}\,\mathcal{K}^{\mu\nu}}{\mathcal{J}}
  =\f{\f{1}{4\pi}\int_{\mathbb{S}^{2}}f(\omega)\,(\widehat{\mathsf{h}}_{\mu}\ell^{\mu})^{2}\,d\omega}{\f{1}{4\pi}\int_{\mathbb{S}^{2}}f(\omega)\,d\omega}
  =\f{\f{1}{2}\int_{-1}^{1}\mathfrak{f}(\mu)\,\mu^{2}\,d\mu}{\f{1}{2}\int_{-1}^{1}\mathfrak{f}(\mu)\,d\mu},
  \label{eq:eddingtonFactor}
\end{equation}
where we have defined
\begin{equation}
  \mathfrak{f}(\mu)=\f{1}{2\pi}\int_{0}^{2\pi}f(\mu,\varphi)\,d\varphi.  
\end{equation}
In the last step in Eq.~\eqref{eq:eddingtonFactor} we have aligned the momentum-space coordinate system in the comoving frame so that $\widehat{\mathsf{h}}_{\mu}\ell^{\mu}=\widehat{\mathsf{h}}_{\hat{\mu}}\ell^{\hat{\mu}}=\cos\vartheta=\mu$. (Note, this is not the same $\mu$ that will be defined later, in Sect.~\ref{sec:PhaseSpaceCoordinates}. The angle here is defined in terms of the direction specified by $\widehat{\mathsf{h}}_{\hat{\mu}}$, whereas in Sect.~\ref{sec:PhaseSpaceCoordinates} it will be defined in terms of $\hat{r}$.)
The two-moment closure for $\mathcal{K}^{\mu\nu}$ requires the Eddington factor to be specified in terms of $\mathcal{J}$ and $\mathfrak{h}$ (or equivalently $\mathcal{D}$ and $\mathfrak{h}$).  
We will discuss some specific approaches further below.  

In a similar way, we can construct the third-order moment, $\mathcal{L}^{\mu\nu\rho}$, depending on $\mathcal{J}$ and $\mathcal{H}^{\mu}$, as the symmetric rank-three tensor that is orthogonal to $u_{\mu}$ and that satisfies the trace conditions $\mathcal{L}^{\mu\nu}_{\hspace{12pt}\nu}=\mathcal{H}^{\mu}$. From \citep[e.g.,][]{Pen92,CaEnMe13a,JuObJa15},
\begin{equation}
  \mathcal{L}^{\mu\nu\rho}
  =\f{1}{2}\,
  \Big[\,
    \big(\,\mathfrak{h}-\mathfrak{q}\,\big)\,
    \Big(\,\widehat{\mathsf{h}}^{\mu}\,h^{\nu\rho}+\widehat{\mathsf{h}}^{\nu}\,h^{\mu\rho}+\widehat{\mathsf{h}}^{\rho}\,h^{\mu\nu}\,\Big)
    +\big(\,5\,\mathfrak{q}-3\,\mathfrak{h}\,\big)\,\widehat{\mathsf{h}}^{\mu}\,\widehat{\mathsf{h}}^{\nu}\,\widehat{\mathsf{h}}^{\rho}
  \,\Big]\,\mathcal{J},
  \label{eq:radiationHeatFluxTensor}
\end{equation}
where we have defined the ``heat flux'' factor $\mathfrak{q}(\mathcal{J},\mathfrak{h})$:  
\begin{equation}
  \mathfrak{q} = \f{\widehat{h}_{\mu}\,\widehat{h}_{\nu}\,\widehat{h}_{\rho}\,\mathcal{L}^{\mu\nu\rho}}{\mathcal{J}}
  =\f{\f{1}{4\pi}\int_{\mathbb{S}^{2}}f(\omega)\,(\widehat{\mathsf{h}}_{\mu}\ell^{\mu})^{3}\,d\omega}{\f{1}{4\pi}\int_{\mathbb{S}^{2}}f(\omega)\,d\omega}
  =\f{\f{1}{2}\int_{-1}^{1}\mathfrak{f}(\mu)\,\mu^{3}\,d\mu}{\f{1}{2}\int_{-1}^{1}\mathfrak{f}(\mu)\,d\mu}.
  \label{eq:heatFluxFactor}
\end{equation}
The two-moment closure for $\mathcal{L}^{\mu\nu\rho}$ requires that we specify the heat flux factor in terms of $\mathcal{J}$ and $\mathfrak{h}$ (or $\mathcal{D}$ and $\mathfrak{h}$).  

To complete the specification of the two-moment closure, the Eddington and heat flux factors must be specified in terms of the zeroth and first moments.  
To this end, several approaches have been proposed for the Eddington factor, including maximum entropy closure \citep[e.g.,][]{Mine78}, Kershaw-type closure \citep[e.g.,][]{Kers76}, and closures derived from fits to results obtained with higher-fidelity models \citep[e.g.,][]{Janka91}.  
In the context of spherically symmetric proto-neutron star models, \citet{MuAbUr17} carried out a comprehensive comparison of results obtained with two-moment neutrino transport, using analytic Eddington factors, to results obtained with Monte Carlo transport calculations.  
\citet{MuAbUr17} included Eddington factors from \citet{WiCoCo75,Kers76,Le84,Mine78,CeBl94,Janka91,Janka92}, and found no closure to perform consistently better than the others in the test cases considered.  
Because the maximum entropy closures of \citet{Mine78} and \citet{CeBl94} gave practically identical results and never yielded the worst results, and given the simplicity of the closure by \citet{Mine78} relative to the closure by \citet{CeBl94}, \citet{MuAbUr17} concluded that the \citet{Mine78} closure is the most attractive choice for neutrino transport around proto-neutron stars.  
The closures provided by \citet{Mine78} and \citet{Le84} are probably the most widely used in core-collapse supernova simulations employing two-moment neutrino transport.  
Recently, \citet{JuObJa15}, comparing the closures of \citet{Mine78,CeBl94,Le84} in the context of a simulation of collapse and post-bounce evolution of a 13~$M_{\odot}$ star in spherical symmetry, showed that the differences in shock radii, neutrino luminosities, and mean energies are practically indistinguishable.  
This may be because the closures are very similar for the values of $\mathcal{J}$ and $\mathfrak{h}$ encountered.  
\citet{ChEnHa19} considered Eddington factors by \citet{Mine78,CeBl94,LaBa11,BaLa17} and found that, under certain conditions, results obtained with closures based on Fermi--Dirac statistics can differ significantly from results obtained with the \citet{Mine78} closure, which is based on Boltzmann statistics.  

\begin{figure}[htb]
\centering
\includegraphics[width=0.6\textwidth]{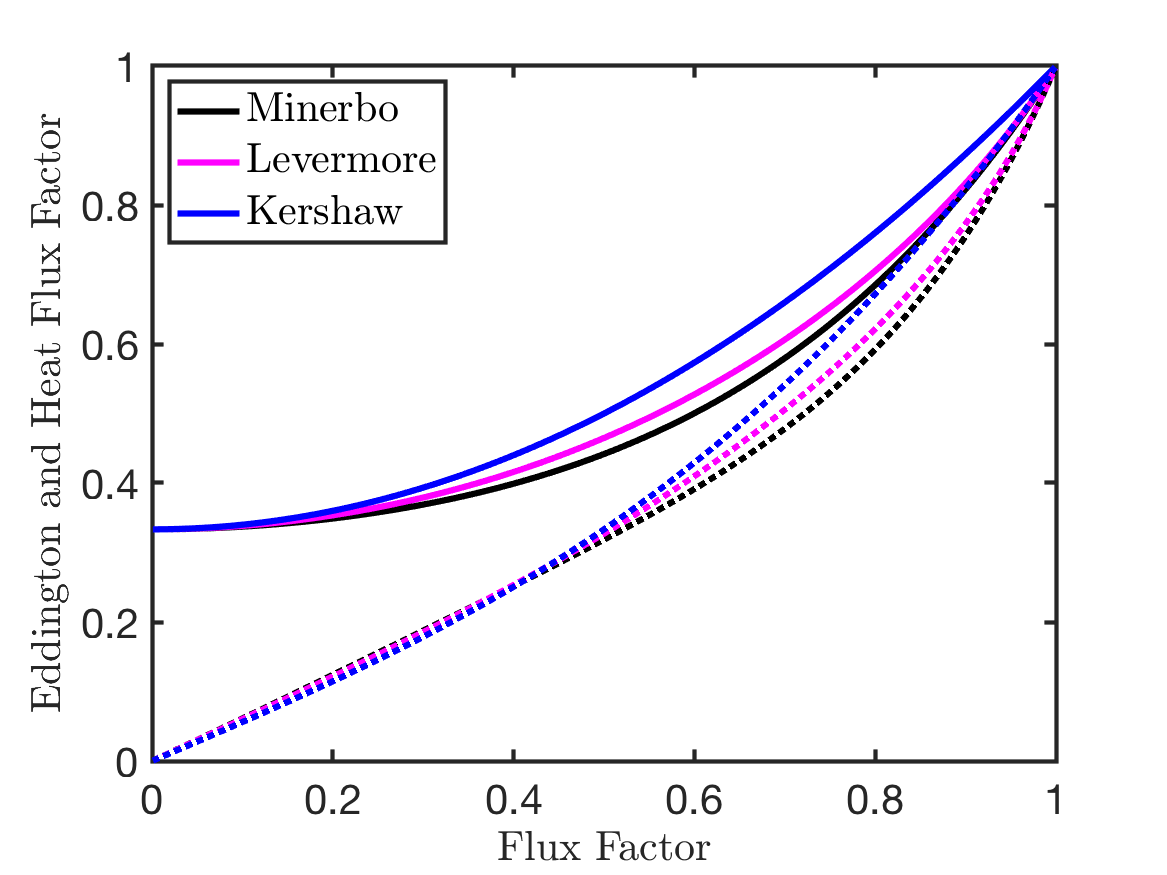}
\caption{Plot of Eddington factors $\mathfrak{k}$ (solid lines) and heat flux factors $\mathfrak{q}$ (dotted lines) versus flux factor $\mathfrak{h}$ for the closures due to \citet{Mine78} (black), \citet{Le84} (magenta), and \citet{Kers76} (blue).}
\label{fig:eddingtonFactors}
\end{figure}

We discuss the closures due to \citet{Mine78}, \citet{Le84}, and \citet{Kers76} in further detail and give explicit expressions for Eddingon and heat flux factors, which are also plotted in Figure~\ref{fig:eddingtonFactors} (see figure caption for details).  

\paragraph{Maximum entropy closure}

The maximum entropy approach to specifying the Eddington and heat flux factors comes from statistical mechanics, and has been used extensively in moment models for radiation transport \citep[e.g.][]{Mine78,CeBl94}.  
In this approach, the ``most probable'' values of $\mathfrak{k}$ and $\mathfrak{q}$ are determined by finding the distribution function $\mathfrak{f}_{\ME}$ that maximizes the entropy functional $s[\mathfrak{f}_{\ME}]$, subject to the constraints that $\mathfrak{f}_{\ME}$ reproduces the known moments (e.g., $\mathcal{D}$ and $\mathfrak{h}$).  
The unknowns can then be computed from Eqs.~\eqref{eq:eddingtonFactor} and \eqref{eq:heatFluxFactor} by setting $\mathfrak{f}=\mathfrak{f}_{\ME}$.  
For the two-moment model, the maximum entropy distribution is obtained by extremizing 
\begin{equation}
  S = \int_{-1}^{1}s[\mathfrak{f}_{\ME}]\,d\mu + \alpha_{0}\int_{-1}^{1}\mathfrak{f}_{\ME}\,d\mu + \alpha_{1}\int_{-1}^{1}\mathfrak{f}_{\ME}\,\mu\,d\mu
  \label{eq:maximumEntropyObjectiveFunction}
\end{equation}
with respect to $\mathfrak{f}_{\ME}$, where the Lagrange multipliers $\alpha_{0}$ and $\alpha_{1}$ are introduced for the constraints.  
A particularly simple closure is obtained by considering the case of Boltzmann statistics, where $s[\mathfrak{f}_{\ME}]=\mathfrak{f}_{\ME}\ln\mathfrak{f}_{\ME}-\mathfrak{f}_{\ME}$.  
This case was considered in detail by \citet{Mine78}, and is the low-occupancy limit ($\mathcal{D}\ll1$) of the more appropriate case (for neutrino transport) of Fermi--Dirac statistics considered by \citet{CeBl94}.  
For the case of Boltzmann statistics, the maximum entropy distribution is easily found to be given by
\begin{equation}
  \mathfrak{f}_{\ME}(\mu) = \exp\big(\,\alpha_{0}+\alpha_{1}\,\mu\,\big),
  \label{eq:maximumEntropyDistribution}
\end{equation}
where $\alpha_{0}$ and $\alpha_{1}$ are found from the known moments.  
Direct integration of Eq.~\eqref{eq:maximumEntropyDistribution} gives \citep{Mine78}
\begin{align}
  \mathcal{D}
  = \f{1}{2}\int_{-1}^{1}\mathfrak{f}_{\ME}(\mu)\,d\mu
  &=e^{\alpha_{0}}\,\sinh(\alpha_{1})/\alpha_{1}, \label{eq:numberDensityME} \\
  \mathcal{I}
  = \f{1}{2}\int_{-1}^{1}\mathfrak{f}_{\ME}(\mu)\,\mu\,d\mu
  &=e^{\alpha_{0}}\,\big(\,\alpha_{1}\,\cosh(\alpha_{1})-\sinh(\alpha_{1})\,\big)/\alpha_{1}^{2}, \label{eq:numberFluxME}
\end{align}
which can be solved for $\alpha_{0}$ and $\alpha_{1}$.  
In particular, the flux factor is given by the Langevin function $L(\alpha_{1})$,
\begin{equation}
  \mathfrak{h}
  =\mathcal{I}/\mathcal{D}
  =\coth(\alpha_{1})-1/\alpha_{1}
  \equiv L(\alpha_{1}),
\end{equation}
and is independent of $\alpha_{0}$.  
Thus, $\alpha_{1}(\mathfrak{h})=L^{-1}(\mathfrak{h})$.  
(The inversion of the Langevin function must be done numerically.)
Once $\alpha_{1}$ is obtained, $\alpha_{0}$ can be obtained directly from either Eq.~\eqref{eq:numberDensityME} or Eq.~\eqref{eq:numberFluxME}, which completes the specification of $\mathfrak{f}_{\ME}$.  
Then the Eddington factor and heat flux factor can be computed by direct integration
\begin{equation}
  \mathfrak{k}(\mathfrak{h}) = 1-2\,\mathfrak{h}/\alpha_{1}(\mathfrak{h})
  \quad\mbox{and}\quad
  \mathfrak{q}(\mathfrak{h}) = \coth\big(\alpha_{1}(\mathfrak{h})\big) - 3\,\mathfrak{k}(\mathfrak{h})/\alpha_{1}(\mathfrak{h}),
  \label{eq:eddingtonAndHeatFluxFactorsME}
\end{equation}
which closes the two-moment model under the simplifying assumption of Boltzmann statistics, which is a reasonable approximation for neutrinos only when the occupation density is low; i.e., when $\mathcal{D}\ll1$.  
This closure is referred to as the Minerbo closure, and is a commonly used closure in simulations employing spectral two-moment neutrino transport \citep[e.g.,][]{KuTaKo16,JuBoJa18,OcCo18}.  
In practice, to avoid inverting the Langevin function for $\alpha_{1}$, the Eddington and heat flux factors can be approximated as polynomials in the flux factor.  
This leads to algebraic expressions, which are computationally more efficient.  
The algebraic form of the Eddingon factor, which approximates the one in Eq.~\eqref{eq:eddingtonAndHeatFluxFactorsME} to better than one percent, is given by \citep{CeBl94}
\begin{equation}
  \mathfrak{k}_{\mbox{\tiny Alg}}(\mathfrak{h})
  =\f{1}{3} + \f{2}{15}\,\big(\,3\,\mathfrak{h}^{2} - \mathfrak{h}^{3} + 3\,\mathfrak{h}^{4}\,\big).  
  \label{eq:eddingtonFactorMinerbo}
\end{equation}
Similarly, the algebraic form of the heat flux factor, which approximates the one in Eq.~\eqref{eq:eddingtonAndHeatFluxFactorsME} to within a few percent, is given by \citep{JuObJa15}
\begin{equation}
  \mathfrak{q}_{\mbox{\tiny Alg}}(\mathfrak{h})
  =\mathfrak{h}\,\big(\,45 + 10\,\mathfrak{h} - 12\,\mathfrak{h}^{2} - 12\,\mathfrak{h}^{3} + 38\,\mathfrak{h}^{4} - 12\,\mathfrak{h}^{5} + 18\,\mathfrak{h}^{6}\,\big) / 75.  
\end{equation}
In Figure~\ref{fig:eddingtonFactors}, the Eddington and heat flux factors $\mathfrak{k}_{\mbox{\tiny Alg}}$ and $\mathfrak{q}_{\mbox{\tiny Alg}}$ are plotted versus the flux factor $\mathfrak{h}$ (denoted Minerbo in the legend, using solid and dotted black lines, respectively).  

Another two-moment closure based on the maximum entropy principle is the so-called M1 closure \cite[e.g.,][]{Le84,DuFu99}.  
The M1 closure is thus based on the same principle as the Minerbo closure, but a different entropy functional is considered; namely the entropy functional for Bose--Einstein statistics $s[\mathfrak{f}_{\ME}]=(1+\mathfrak{f}_{\ME})\ln(1+\mathfrak{f}_{\ME})-\mathfrak{f}_{\ME}\,\ln\mathfrak{f}_{\ME}$.
For the M1 closure the Eddington factor is given by
\begin{equation}
  \mathfrak{k}_{\mbox{\tiny M1}}(\mathfrak{h}) 
  = \f{3+4\,\mathfrak{h}^{2}}{5+2\sqrt{4-3\,\mathfrak{h}^{2}}}.
  \label{eq:eddingtonFactorM1}
\end{equation}
It should be noted that \citet{Le84} derived this result without the maximum entropy principle.  
More recently, \citet{VaAuDu11} proposed a numerical method for multi-group radiation hydrodynamics in the $\mathcal{O}(v/c)$ limit, and provided an expression for the heat flux factor in the M1 model:
\begin{equation}
  \mathfrak{q}_{\mbox{\tiny M1}}(\mathfrak{h}) = 3\,\varphi_{1}(\mathfrak{h})\,\mathfrak{h} + \varphi_{2}(\mathfrak{h})\,\mathfrak{h}^{3},
  \label{eq:heatfluxFactorM1}
\end{equation}
where
\begin{align}
  \varphi_{1}(\mathfrak{h})
  &=\f{(\mathfrak{h}-2+a)(\mathfrak{h}+2-a)}{4\mathfrak{h}(a-2)^{5}}
  \Big[\,
    12\ln\Big(\f{\mathfrak{h}-2+a}{\mathfrak{h}+2-a}\Big)\big(\mathfrak{h}^{4}+2a\mathfrak{h}^{2}-7\mathfrak{h}^{2}-4a+8\big) \nonumber \\
    &\hspace{108pt}
    +48\mathfrak{h}^{3}-9a\mathfrak{h}^{3}-80\mathfrak{h}+40a\mathfrak{h}
  \,\Big], \label{eq:phi1M1} \\
  \varphi_{2}(\mathfrak{h})
  &=\f{1}{\mathfrak{h}^{3}(a-2)^{5}}
  \Big[\,
    60\ln\Big(\f{\mathfrak{h}-2+a}{\mathfrak{h}+2-a}\Big)\big(-\mathfrak{h}^{6}+15\mathfrak{h}^{4}-3a\mathfrak{h}^{4}+15a\mathfrak{h}^{2}-42\mathfrak{h}^{2}-16a+32\big) \nonumber \\
    &\hspace{50pt}
    +54a\mathfrak{h}^{5}-465\mathfrak{h}^{5}-674a\mathfrak{h}^{3}+2140\mathfrak{h}^{3}+1056a\mathfrak{h}-2112\mathfrak{h}
  \,\Big], \label{eq:phi1M2}
\end{align}
and $a=\sqrt{4-3\mathfrak{h}^{2}}$.  
The M1 closure is another commonly used closure in simulations employing spectral two-moment neutrino transport \citep[e.g.,][]{SkDoBu19}.  
In Figure~\ref{fig:eddingtonFactors}, the Eddington and heat flux factors $\mathfrak{k}_{\mbox{\tiny M1}}$ and $\mathfrak{q}_{\mbox{\tiny M1}}$ are plotted versus the flux factor $\mathfrak{h}$ (denoted ``Levermore'' in the legend, using solid and dotted magenta lines, respectively).  
When plotting the heat flux factor, we found $\varphi_{1}$ and $\varphi_{2}$ to exhibit oscillatory behavior as $\mathfrak{h}\to0$.  
To avoid these oscillations in $\mathfrak{q}_{\mbox{\tiny M1}}$, we used Taylor expansions of $\varphi_{1}$ (around $\mathfrak{h}=0.1$) and $\varphi_{2}$ (around $\mathfrak{h}=0.2$) to plot $\mathfrak{q}_{\mbox{\tiny M1}}$ for smaller values of $\mathfrak{h}$.  

The low occupancy assumption used for the Minerbo closure does not hold everywhere in a supernova simulation, but may be a reasonable approximation in the neutrino heating region.  
The M1 closure based on Bose--Einstein statistics is also not a good approximation when the phase space occupation is high.  
In this case, a more realistic treatment for neutrinos must consider the entropy functional for Fermi--Dirac statistics, where $s[\mathfrak{f}_{\ME}]=\mathfrak{f}_{\ME}\,\ln\mathfrak{f}_{\ME}+(1-\mathfrak{f}_{\ME})\ln(1-\mathfrak{f}_{\ME})$, and follow the procedure as outlined above, as was done by \citet{CeBl94}, and more recently in further detail by \citet{LaBa11}.  
For the maximum entropy closure derived by \citet{CeBl94}, the Eddington factor is
\begin{equation}
  \mathfrak{k}_{\mbox{\tiny CB}}(\mathcal{D},\mathfrak{h}) = \f{1}{3} + \f{2\,(1-\mathcal{D})\,(1-2\mathcal{D})}{3}\,\Theta\Big(\f{\mathfrak{h}}{1-\mathcal{D}}\Big),
  \label{eq:eddingtonFactorCB}
\end{equation}
where $\Theta(x)=x^{2}(3-x+3x^{2})/5$.  
To account for Fermi--Dirac statistics, the Eddington factor in Eq.~\eqref{eq:eddingtonFactorCB} depends on both the number density $\mathcal{D}$ and the flux factor $\mathfrak{h}$.  
In the low-occupancy limit when, $\mathcal{D}\ll1$, this Eddington factor reduces to the Eddington factor due to Minerbo in Eq.~\eqref{eq:eddingtonFactorMinerbo}.  
\citet{CeBl94} did not provide an expression for the heat flux factor.  

It should be noted that the term ``M1 closure,'' used here to refer to the closure in Eqs.~\eqref{eq:eddingtonFactorM1} and \eqref{eq:heatfluxFactorM1}, derives from the more general term ``M$N$ closure,'' which is used in transport theory to refer to maximum entropy closures applied to $N$-moment hierarchies.  
As such, all the closures discussed in this section are M1 closures, but they differ in the entropy functional that is maximized.  

\paragraph{Kershaw closure}

A different approach to the closure problem was proposed by \citet{Kers76}.  
The key idea behind the Kershaw closure is to consider the bounds on the moments generated by the underlying distribution function.  
For a nonnegative distribution function ($\mathfrak{f}\ge0$), the set generated by the normalized moments $\{\,1,\,\mathfrak{h},\,\mathfrak{k},\,\mathfrak{q}\,\}$ is convex and bounded, which in turn allows one to construct any sequence of moments in this set by a convex combination of moment vectors on the boundary of this domain.  
The moments constructed by this procedure are then ``good'' in the sense that they can be obtained from a nonnegative distribution function.  

For the two-moment model, the Kershaw closure procedure can be used to specify $\mathfrak{k}$ and $\mathfrak{q}$ in terms of $\mathfrak{h}$.  
For $\mathfrak{f}\ge0$, it is straightforward to show that $-1\le\mathfrak{h}\le1$, while the bounds on the Eddington factor are given by
\begin{equation}
  \mathfrak{h}^{2}\equiv\mathfrak{k}_{\lo}(\mathfrak{h})\le\mathfrak{k}\le\mathfrak{k}_{\hi}(\mathfrak{h})\equiv1.  
\end{equation}
For $\zeta\in[0,1]$, the Eddington factor can be written as the convex combination
\begin{equation}
  \mathfrak{k}_{\mbox{\tiny{\sc K}}}(\mathfrak{h}) = \zeta\,\mathfrak{k}_{\lo}(\mathfrak{h})+(1-\zeta)\,\mathfrak{k}_{\hi}(\mathfrak{h}).  
\end{equation}
Demanding that this expression be correct in the limit when $\mathfrak{h}=0$, i.e., $\mathfrak{k}(0)=1/3$, gives $\zeta=2/3$, so that
\begin{equation}
  \mathfrak{k}_{\mbox{\tiny{\sc K}}}(\mathfrak{h}) = \f{1}{3} + \f{2}{3}\,\mathfrak{h}^{2}.  
  \label{eq:eddingtonFactorKershaw}
\end{equation}
Similarly, for the heat flux factor, it can be shown that the following bounds hold \citep[e.g.,][]{Schn16}:
\begin{align}
  -\mathfrak{k}+\f{(\mathfrak{h}+\mathfrak{k})^{2}}{1+\mathfrak{h}}
  \equiv\mathfrak{q}_{\lo}(\mathfrak{h},\mathfrak{k})\le\mathfrak{q}\le\mathfrak{q}_{\hi}(\mathfrak{h},\mathfrak{k})\equiv
  \mathfrak{k}-\f{(\mathfrak{h}-\mathfrak{k})^{2}}{1-\mathfrak{h}}.
\end{align}
Constructing the heat flux factor from a convex combination of these bounds, and using $\mathfrak{k}_{\mbox{\tiny{\sc K}}}(\mathfrak{h})$, gives
\begin{equation}
  \mathfrak{q}_{\mbox{\tiny{\sc K}}}(\mathfrak{h})
  =\zeta\,\mathfrak{q}_{\lo}(\mathfrak{h},\mathfrak{k}_{\mbox{\tiny{\sc K}}}(\mathfrak{h}))
  +(1-\zeta)\,\mathfrak{q}_{\hi}(\mathfrak{h},\mathfrak{k}_{\mbox{\tiny{\sc K}}}(\mathfrak{h})).
\end{equation}
Demanding that $\mathfrak{q}_{\mbox{\tiny{\sc K}}}(0)=0$ (isotropic limit) gives $\zeta=1/2$, so that
\begin{equation}
  \mathfrak{q}_{\mbox{\tiny{\sc K}}}(\mathfrak{h})
  =\f{\mathfrak{h}\,\big(\,\mathfrak{h}^{2}+\mathfrak{k}_{\mbox{\tiny{\sc K}}}(\mathfrak{h})^{2}-2\,\mathfrak{k}_{\mbox{\tiny{\sc K}}}(\mathfrak{h})\,\big)}{(\mathfrak{h}^{2}-1)}.  
  \label{eq:heatFluxFactorKershaw}
\end{equation}
In Fig.~\ref{fig:eddingtonFactors}, the Eddington and heat flux factors $\mathfrak{k}_{\mbox{\tiny{\sc K}}}$ and $\mathfrak{q}_{\mbox{\tiny{\sc K}}}$ are plotted versus the flux factor $\mathfrak{h}$ (denoted ``Kershaw'' in the legend; solid and dotted blue lines, respectively).  
The Kershaw closure considered here only assumes $\mathfrak{f}\ge0$, which holds for Bose--Einstein and Boltzmann statistics.  
Kershaw-type closures for Fermi--Dirac statistics, which is appropriate for neutrinos where $\mathfrak{f}\in[0,1]$, was recently considered by \citet{BaLa17}.  

\subsubsection{One-moment kinetics}
\label{sec:oneMomentKinetics}

One-moment models (commonly referred to as flux-limited diffusion models \citep{LePo81}) are among the earliest transport models adopted for neutrino transport in core-collapse supernova simulations \citep{Bruenn1975}, and are still in use today \citep[e.g.,][]{BrBlHi20,RaJuJa19}.  
Essentially, one-moment models evolve only the zeroth moment of the distribution function, while higher-order moments are specified through a closure procedure.  
Specifically, the radiation flux is specified in terms of the zeroth moment in a way that it is correct in both the diffusion and streaming regimes.  
In order to be correct in the streaming regime, a flux limiter is applied to transition the model from parabolic (diffusion) to hyperbolic (streaming).  
Here we consider the 3+1 general relativistic formulation presented by \citet{RaJuJa19}, which was derived using the formalisms from \citet{ShKiSe11,EnCaMe12c,CaEnMe13a}.  
We start from a slightly different perspective, since we already have presented the main evolution equation in Eq.~\eqref{eq:spectralLagrangianEnergyEquation_3p1}.  
\citet{RaJuJa19} define their angular moments with an additional factor of $\varepsilon^{2}$ relative to our definitions in Eq.~\eqref{eq:energyMomentsLagrangian} and absorb $\sqrt{\gamma}$ into the variables; hence, we make the following definitions
\begin{equation}
  \big\{\,\hat{\mathcal{J}},\,\hat{\mathcal{H}}^{\mu},\,\hat{\mathcal{T}}^{\mu\nu},\ldots\big\}
  =\sqrt{\gamma}\,\varepsilon^{2}\,\big\{\,\mathcal{J},\,\mathcal{H}^{\mu},\,\mathcal{T}^{\mu\nu},\ldots\big\};
\end{equation}
i.e., similar definitions hold for other moments appearing in the equations.  
(They also do not normalize their moments by the factor of $4\pi$, but that should not cause confusion in the presentation here.)
We can then write Eq.~\eqref{eq:spectralLagrangianEnergyEquation_3p1} as
\begin{align}
  &\f{1}{\alpha}
  \big[\,\pd{}{t}\big(\,\big[\,W\hat{\mathcal{J}}+v^{i}\hat{\mathcal{H}}_{i}\,\big]\,\big)
  +\pd{}{i}\big(\,\big[\,\alpha\,\hat{\mathcal{H}}^{i}+\big(\,\alpha\,v^{i}-\beta^{i}\,\big)\,W\hat{\mathcal{J}}\,\big]\,\big)\,\big] \nonumber \\
  &\hspace{6pt}
  +\hat{\mathcal{R}}_{\varepsilon} - \pd{}{\varepsilon}\big(\,\varepsilon\,\hat{\mathcal{R}}_{\varepsilon}\,\big)
  =\f{1}{4\pi}\int_{\mathbb{S}^{2}}\hat{\mathcal{C}}(f)\,d\omega,
  \label{eq:spectralLagrangianEnergyEquationFLD_3p1}
\end{align}
where we have defined
\begin{align}
  \hat{\mathcal{R}}_{\varepsilon} 
  &= \hat{\mathcal{T}}^{\mu\nu}\nabla_{\mu}u_{\nu} \nonumber \\
  &=W\,\Big[\,\hat{\mathcal{F}}_{k}\,\pd{v^{k}}{\tau} + \hat{\mathcal{S}}_{k}^{\hspace{4pt}i}\pd{v^{k}}{i} 
  + \big(\hat{\mathcal{F}}^{i}-\hat{\mathcal{E}}v^{i}\big)\,\pd{\ln\alpha}{i} + \alpha^{-1}\hat{\mathcal{F}}_{k}v^{i}\pd{\beta^{k}}{i} \label{eq:observerCorrectionsLagrangianEnergyEquationFLD_3p1} \\
  &\hspace{24pt}
  +\hat{\mathcal{S}}^{ik}\big(\,\f{1}{2}v^{m}\pd{\gamma_{ik}}{m}-\mathsf{K}_{ik}\,\big)\,\Big]
  -\big(\,\hat{\mathcal{E}}-v^{k}\hat{\mathcal{F}}_{k}\,\big)\,\pd{W}{\tau} - \big(\,\hat{\mathcal{F}}^{i}-\hat{\mathcal{S}}_{k}^{\hspace{4pt}i}v^{k}\,\big)\,\pd{W}{i}, \nonumber
\end{align}
where in the second step, we used Eq.~\eqref{eq:observerCorrectionsLagrangianEnergyEquation_3p1}, re-expressed in the form given by \citet{RaJuJa19} (cf.\ their Eq.~(A14)), and we have defined $\pd{}{\tau}=n^{\mu}\pd{}{\mu}$.  

\citet{RaJuJa19} solve for moments defined in an orthonormal comoving frame, and write
\begin{align}
  \hat{\mathcal{H}}^{i}
  &=L^{i}_{\hspace{2pt}\hat{\mu}}\hat{\mathcal{H}}^{\hat{\mu}} = e^{i}_{\hspace{2pt}\bar{\mu}}\,\Lambda^{\bar{\mu}}_{\hspace{4pt}\hat{\mu}}\hat{\mathcal{H}}^{\hat{\mu}} \nonumber \\
  &=e^{i}_{\hspace{2pt}\hat{i}}\,\hat{\mathcal{H}}^{\hat{i}} + W\,\Big(\,\f{W}{W+1}\,v^{i}-\f{\beta^{i}}{\alpha}\,\Big)\,\bar{v}_{\hat{i}}\,\hat{\mathcal{H}}^{\hat{i}},
  \label{eq:coordinateBasisInTermsOfComovingH}
\end{align}
where
\begin{equation}
  \hat{\mathcal{H}}^{\hat{i}}(\varepsilon)=\sqrt{\gamma}\,\f{\varepsilon^{3}}{4\pi}\int_{\mathbb{S}}f(\omega,\varepsilon)\,\ell^{\hat{i}}(\omega)\,d\omega.
\end{equation}
(Similar expressions can be made for higher-order moments; see \citet{EnCaMe12c,RaJuJa19}.)
In Eq.~\eqref{eq:coordinateBasisInTermsOfComovingH}, we remind the reader that $\Lambda^{\bar{\mu}}_{\hspace{4pt}\hat{\mu}}$ is the Lorentz transformation between the orthonormal comoving frame basis and the orthonormal laboratory frame basis, while $e^{i}_{\hspace{2pt}\bar{\mu}}$ is a transformation between the orthonormal laboratory frame basis and the coordinate basis.  
We have made the choice $e^{\mu}_{\hspace{2pt}\bar{0}}=n^{\mu}$, and $\bar{v}_{\hat{i}}$ are three-velocity components in the orthonormal laboratory frame basis ($\bar{v}_{\bar{i}}=\bar{v}^{\bar{i}}=\bar{v}_{\hat{i}}=\bar{v}_{\hat{i}}$), so that $v^{i}=e^{i}_{\hspace{2pt}\hat{i}}\bar{v}^{\hat{i}}$, where the notation $e^{i}_{\hspace{2pt}\hat{i}}=e^{i}_{\hspace{2pt}\bar{i}}\delta^{\bar{i}}_{\hspace{2pt}\hat{i}}$ is used.  

To close the one-moment (MGFLD) model, \citet{RaJuJa19} replace the momentum density by the gradient of the energy density:
\begin{eqnarray}
  \mathcal{H}^{\hat i}\longrightarrow - D \frac{e^{k \hat i}}{\alpha^3} \partial_k (\alpha^3 \mathcal{J})~,
	\label{eq:tr_fld_flux}
\end{eqnarray}
where $D$ is the diffusion coefficient, which they express in terms of the flux-limiter $\lambda\in[0,1/3]$ and
the total opacity $\kappa_t$ as
\begin{eqnarray}
	D \equiv \frac{\lambda}{\kappa_\mathrm{t}}.
\end{eqnarray}
For Levermore--Pomraning and Wilson flux-limiting,
\begin{eqnarray}\label{eq:limiterLPW}
    \lambda_\mathrm{LP} & \equiv &  \frac{2+R}{6+3R+R^2}~, \nonumber \\
    \lambda_\mathrm{Wilson} & \equiv &  \frac{1}{3+R}~,
\end{eqnarray}
respectively, where \citet{RaJuJa19} define the generalized Knudsen number 
as
\begin{eqnarray}\label{eq:knudsen}
	R & \equiv & \frac{|e^{k \hat i}\partial_{k} (\alpha^3\mathcal{J})|}{\kappa_\mathrm{t}\alpha^3\mathcal{J}}.
\end{eqnarray}
Thus, when the opacity is high, $R\to0$ and $\lambda\to1/3$.  
On the other hand, when the opacity is low, $\lambda\to1/R$ and
\begin{equation}
  \mathcal{H}^{\hat i}\to-\f{e^{k \hat i}\partial_{k} (\alpha^3\mathcal{J})}{|e^{k \hat i}\partial_{k} (\alpha^3\mathcal{J})|}\,\mathcal{J}.  
\end{equation}
The Eddington tensor is related to the neutrino radiation stress tensor:
\begin{eqnarray}\label{eq:edd_tensor1}
	\chi^{\hat i \hat j} & = &  \frac{\mathcal{K}^{\hat i \hat j}}{\mathcal{J}}~.
\end{eqnarray}
In the MGFLD approximation, the Eddington tensor, which appears in the expression for $\hat{\mathcal{R}}_{\varepsilon}$, takes a form analogous to Eq.~\eqref{eq:radiationStressTensor}:
\begin{eqnarray}\label{eq:edd_tensor_fld}
    \chi^{\hat i \hat j} = \frac{1}{2} [(1-\chi)\delta^{\hat i \hat j} 
    + (3\chi-1)h^{\hat i}h^{\hat j}]~.
\end{eqnarray}
In Eq.~\eqref{eq:edd_tensor_fld}, $h^{\hat{i}}$ is a unit vector in the direction of the neutrino flux, $\mathcal{H}^{\hat {i}}$, and
$\chi$ is the Eddington factor, which is given by
\begin{eqnarray}\label{eq:edd_factor_fld}
    \chi = \lambda + (\lambda R)^2~.
\end{eqnarray}

\section{Neutrino interactions}
\label{sec:interactions}

The phenomenon of core-collapse supernovae is a magnificent juxtaposition of the macroscopic physics of neutrino radiation hydrodynamics and the microscopic physics of neutrino weak interactions and the nuclear equation of state. In particular, the weak interactions between the neutrinos and the matter are what make neutrinos important to this phenomenon. Thus, any review of neutrino transport in core-collapse supernovae must include a discussion of such interactions. In the history of core-collapse supernova modeling, there have been many important examples of studies that have demonstrated the impact of additional weak interaction physics and/or improved treatments of such physics in supernova models. Here we select a subset of these studies, each selected to investigate one of the dimensions of this component of supernova modeling: (1) The impact of the addition of new weak-interaction channels. (2) The impact of improved treatments of channels that have already been included in the models. (3) The interplay between different weak-interaction channels and the impact of adding/changing more than one weak-interaction channel at a time in a model. (4) The uncertainties in the weak-interaction rates currently used in core-collapse supernova models and their ramifications for core-collapse supernova modeling.

\subsection{An intertwined history}
		
Looking back at the history of the development of the theory of weak interactions and of core-collapse supernovae, especially during the time frame after the discovery and publication of the electroweak theory, it becomes obvious that (1) the first period of what can be called modern core-collapse supernova theory, after the publication of the seminal work of Colgate and White, was greatly influenced and greatly accelerated by the new electroweak theory, for more than a decade, and (2) the interplay between advancing descriptions of neutrino weak interactions and core-collapse supernovae continued well beyond this period, even to this day. 

A year after the publication of the Colgate and White work, the electroweak theory was published \citep{Wein67,Salam1968}. It was specifically the advent of weak neutral currents that would turn out to be a game changer for core-collapse supernova theory. Seven years after the publication of the electroweak theory, \citet{Freedman1974} showed that, owing to weak neutral currents, neutrinos could scatter coherently off the nucleons in a nucleus, introducing an $A^2$ dependence in the cross section, where $A$ is the atomic number. During stellar core collapse, the core is neutronized through the emission and escape of electron neutrinos. As a result, the core nuclei become large---i.e., have large $A$---given that the nuclear size is a competition between Coulomb repulsion and surface tension, the former favoring smaller nuclei, the latter favoring larger nuclei, and the latter winning out. In turn, coherent nuclear scattering cross sections become large. Following Freedman's discovery and publication, Tubbs and Schramm provided an electroweak-theory-based set of cross sections for problems of astrophysical interest \citep{TuSc75}. Subsequently, these were implemented in the pioneering work of \citet{Arnett1977}, wherein he showed that coherent nuclear scattering led to the trapping of electron neutrinos during stellar core collapse and to the development of a trapped Fermi sea of them in the core. This provided the foundation for the discovery five years later by Wilson that the stalled core-collapse supernova shock wave could be revived by charged-current mediated electron neutrino and antineutrino absorption on the shock-liberated nucleons behind it \citep{Wilson1985,BeWi85}, which marked the beginning of contemporary core-collapse supernova theory, which has largely operated within the framework of the delayed-shock or, equivalently, the neutrino-reheating mechanism. The fifteen years between 1966 and 1982 saw the fundamental and significant advance from the first models of core-collapse supernovae to the establishment of the framework within which all core-collapse supernova modelers operate today. The developments in core-collapse supernova theory during these first fifteen years were very tightly intertwined with the development of weak interaction physics. 
While this period was certainly unique in this regard, additional milestones, owing to further development in the theory of neutrino interactions in the environments of interest here, occurred since. 

In 1985, Bruenn published a landmark paper on the physics of stellar core collapse \citep{Bruenn1985}. Bruenn included the following electron neutrino emissivities and opacities in his models, which have come to be known as the ``Bruenn 85'' opacity set. Subsequent to Bruenn's publication and prior to the publications discussed below, this set was frozen in as the canonical neutrino opacity set. It is still used today in code tests and comparisons. Bruenn included electron capture on (free) protons and nuclei and the inverse interactions of electron neutrino absorption, as well as scattering on (free) nucleons and electrons and coherent scattering on nuclei in his models. For electron antineutrino and heavy-flavor neutrino production, electron--positron pair annihilation served as the dominant source after core bounce and shock formation. 

\begin{figure}[htb]
\centering
\includegraphics[width=0.8\textwidth]{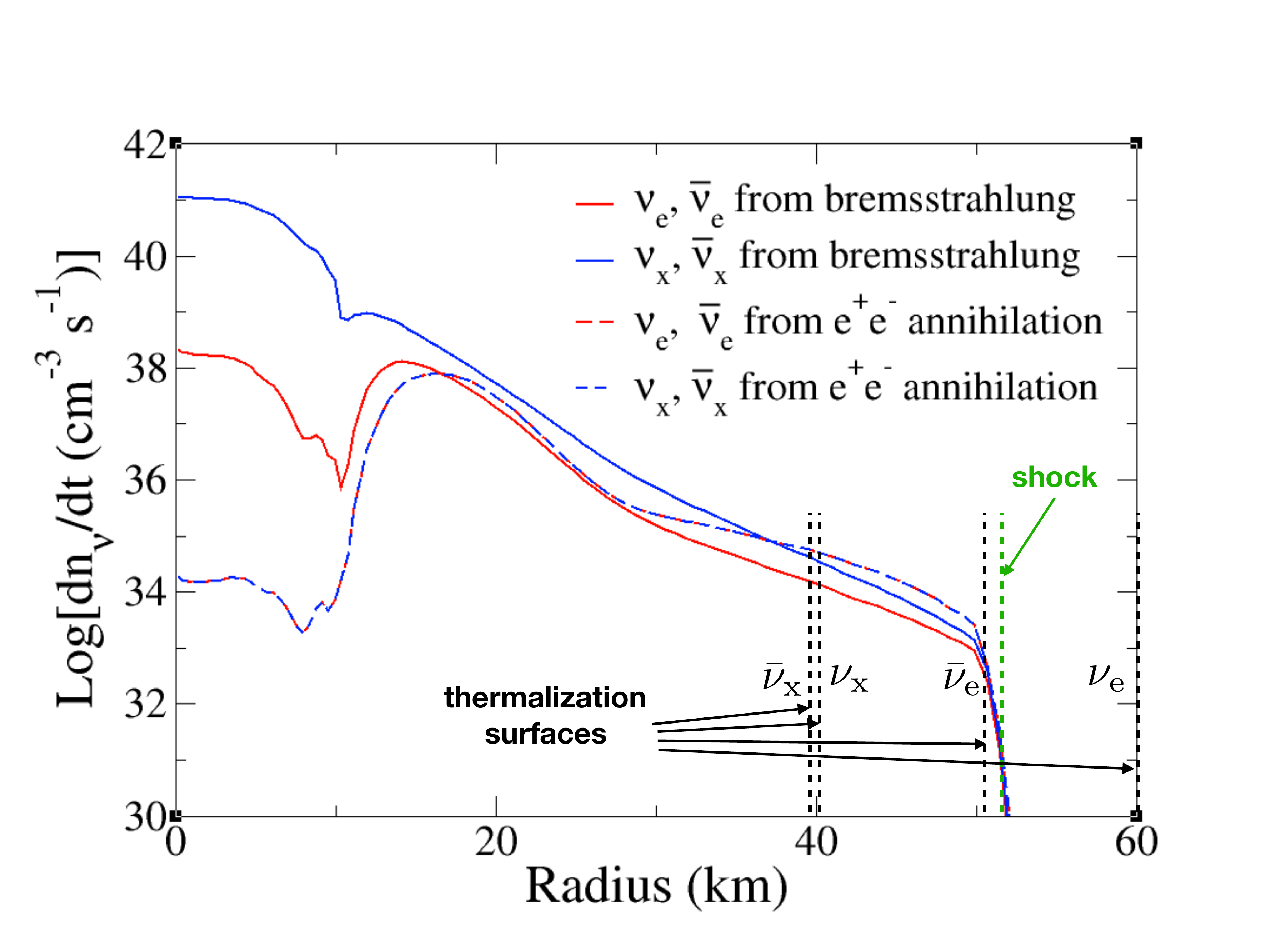}
\caption{The neutrino number production rate due to neutrino--antineutrino pair production via electron--positron annihilation and nucleon--nucleon bremsstrahlung are plotted. Also shown are the thermalization surfaces for the different neutrino and antineutrino flavors, as well as the shock location at this time after bounce: 4 ms. The data used to generate the plot are taken from a \textsc{Chimera} model using an $18\,M_\odot$ progenitor. At the high densities present at radii below $\sim$10 km in the core, pair production from bremsstrahlung dominates. On the other hand, between the heavy-flavor thermalization surfaces and the shock, production by electron--positron pair annihilation is consistently larger.}
\label{fig:brem4ms}
\end{figure}

\begin{figure}[htb]
\centering
\includegraphics[width=0.8\textwidth]{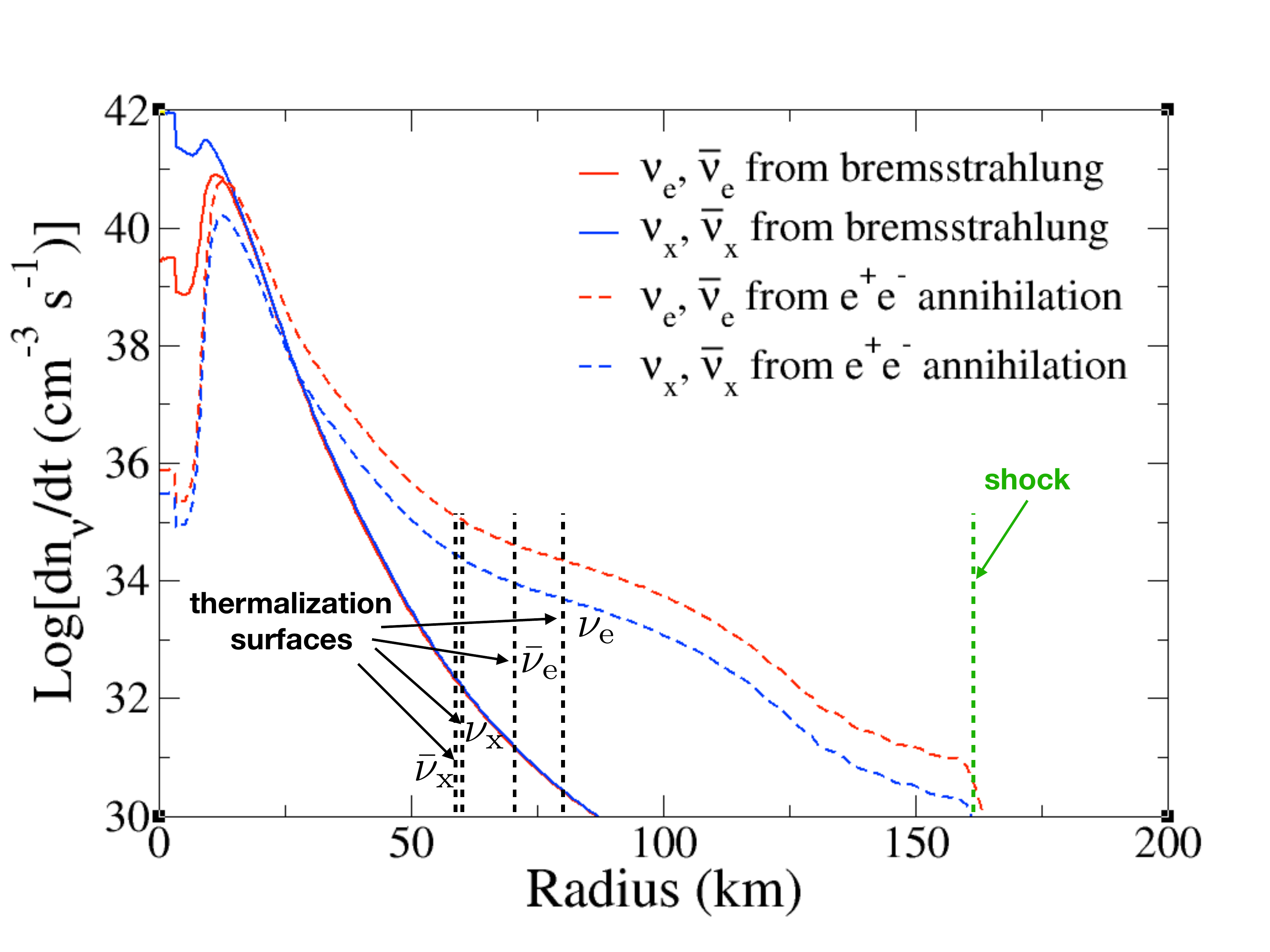}
\caption{The same as in Fig.~\ref{fig:brem4ms} but at a time of 100 ms after bounce, during the critical shock reheating epoch. At this time, at radii above $\sim$25 km, which is well below the thermalization spheres where the neutrino spectra set in, neutrino pair production is dominated by electron--positron pair annihilation. At high densities, below $\sim$10 km, production due to bremsstrahlung continues to dominate.}
\label{fig:brem100ms}
\end{figure}

\citet{HaRa98} computed the production of neutrino--antineutrino pairs from nucleon--nucleon bremsstrahlung. Prior to the recognition that such brems\-strahlung could lead to, and perhaps dominate, neutrino pair production, pair production occurred only through electron--positron pair annihilation. Thus, particularly for the muon and tau neutrino flavors, which have only pair production as sources, brems\-strahlung production introduced a fundamental change. Figures~\ref{fig:brem4ms} and \ref{fig:brem100ms} show the relative importance of nucleon--nucleon bremsstrahlung for the production of electron neutrino--antineutrino pairs of all three flavors, relative to the production by electron--positron annihilation. The results shown are for two times after core bounce, at 4 and 100 ms, in a core-collapse supernova model performed with the \textsc{Chimera} code, initiated from an $18\,M_\odot$ progenitor.

\begin{figure}[htb]
\includegraphics[width=\textwidth]{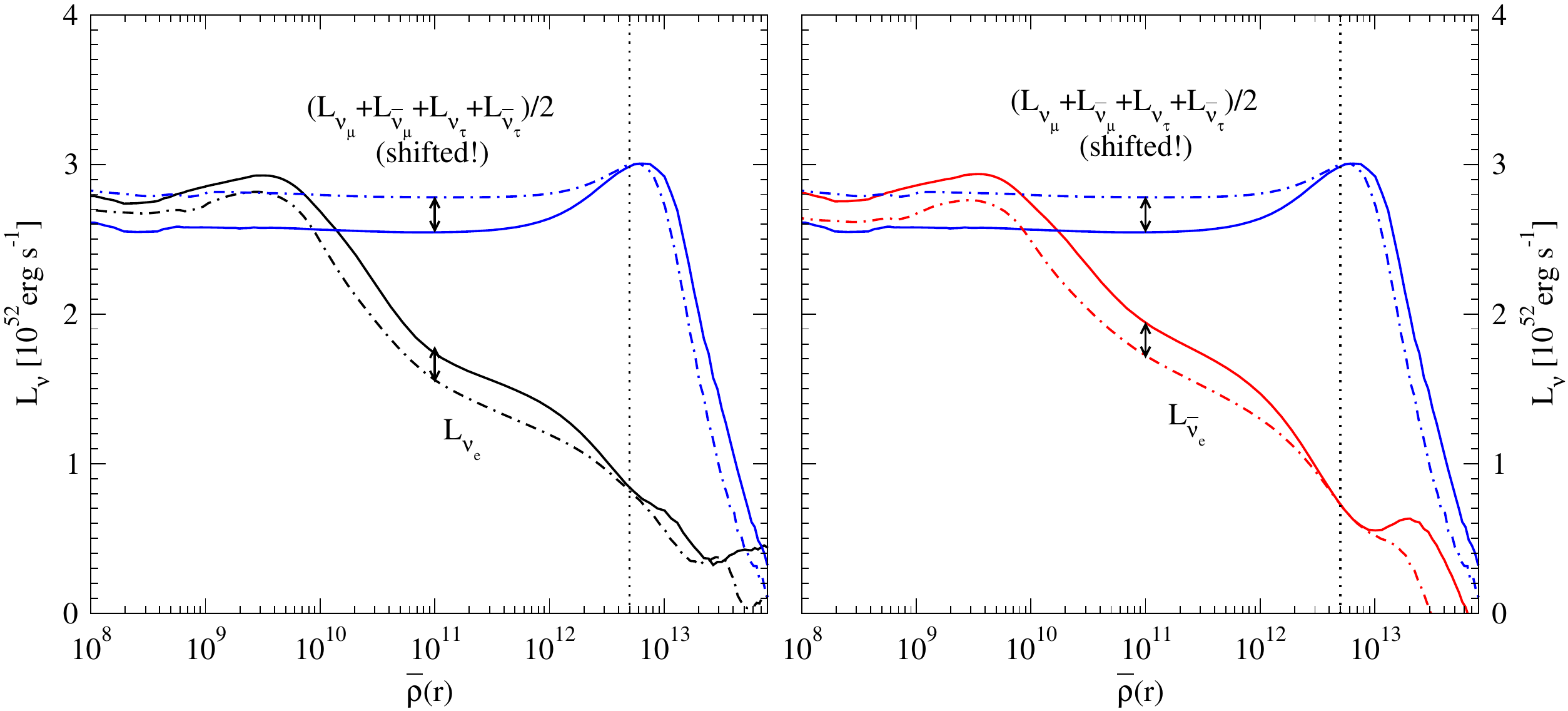}
\caption{Plotted are the neutrino and antineutrino luminosities for all three flavors of neutrinos, as a function of density, at 400 ms after bounce in the general relativistic model of \citet{MuJaMa12} initiated from a $15\,M_\odot$ progenitor. Solid lines show data from the model that includes neutrino--nucleon small-energy scattering. Evident in the plots is the $\sim$20\% increase in both the electron neutrino and antineutrino luminosities, at a density of $10^{11}\mathrm{\ g\ cm}^{-3}$, due to the heating of the electron neutrinospheres resulting from the scattering of higher-energy heavy flavor neutrinos, emanating from deeper regions, on nucleons in the neutrinospheric region.}
\label{fig:MuJaMa12Fig17}
\end{figure}

In the same year, \citet{BuSa98} and \citet{RePrLa98} took on the long-term challenge to understand neutrino interactions in dense, \emph{interacting}, nuclear matter, taking into account nucleon recoil, degeneracy, relativity, thermal motions, and correlations. In particular, these authors computed new differential scattering rates (and new charged-current absorption and emission rates), which were no longer iso-energetic, as had been assumed before (e.g., in the Bruenn 85 opacity set), but resulted in small energy transfer between the neutrinos and the nucleons. Per scattering, the amount of energy transferred would be of little consequence, but taken over all of the scattering events in the dense environment in the vicinity of the neutrinospheres, such small-energy scattering has a notable impact. \citet{MuJaMa12} were the first to demonstrate this. In particular, they showed that small-energy scattering of heavy flavor neutrinos by nucleons at the electron neutrino- and antineutrino-spheres led to heating of these neutrinospheres and, consequently, an increase in the electron neutrino flavor luminosities. Their results are shown in Fig.~\ref{fig:MuJaMa12Fig17}. This in turn impacted neutrino shock reheating. In the absence of small-energy scattering on nucleons, shock revival was delayed by 50--100 ms relative to their baseline model.

In 2003, yet another source of heavy-flavor neutrino pair production was introduced. \citet{BuJaKe03} examined the production of heavy-flavor neutrino pairs through the annihilation of electron-neutrino pairs. They too found that heavy-flavor pair production by electron-flavor pair annihilation dominated the production of such pairs through electron--positron pair annihilation. Moreover, they found that the inclusion of this mode of heavy-flavor production in their model boosted the heavy-flavor luminosities during the first $\sim 150$ ms after bounce and decreased the electron-flavor luminosities after $\sim 200$ ms. They found too that their shock was weaker and reached a smaller peak radius when electron-flavor pair annihilation was included. While the differences were not ``dramatic,'' they concluded they were also not ``negligible.''

And once again in the same year, progress was made on a different front. 
The rates for electron capture on nuclei in the Bruenn 85 opacity set are based on the Independent Particle Model for the nucleons in the nucleus. That is, the IPM assumes that the nucleons are noninteracting. Under this assumption, the final neutron states are filled for nuclei with $N>40$, which is true of the stellar core nuclei, and electron capture is blocked, relying in turn solely on capture on protons. This assumption was finally removed, as nuclear structure models developed. In 2003, rates for electron capture on nuclei using a ``hybrid'' model, wherein thermal excitation and nucleon--nucleon correlations were both accounted for, were recomputed \citep{LaMaSa03}. Owing to the improved description, capture in nuclei was no longer blocked and in fact dominates capture on protons during core collapse, resulting in a more neutronized/deleptonized core, a smaller inner core, and a deeper shock formation mass \citep{HiMeMe03}.

\begin{figure}[htbp]
\includegraphics[width=\textwidth]{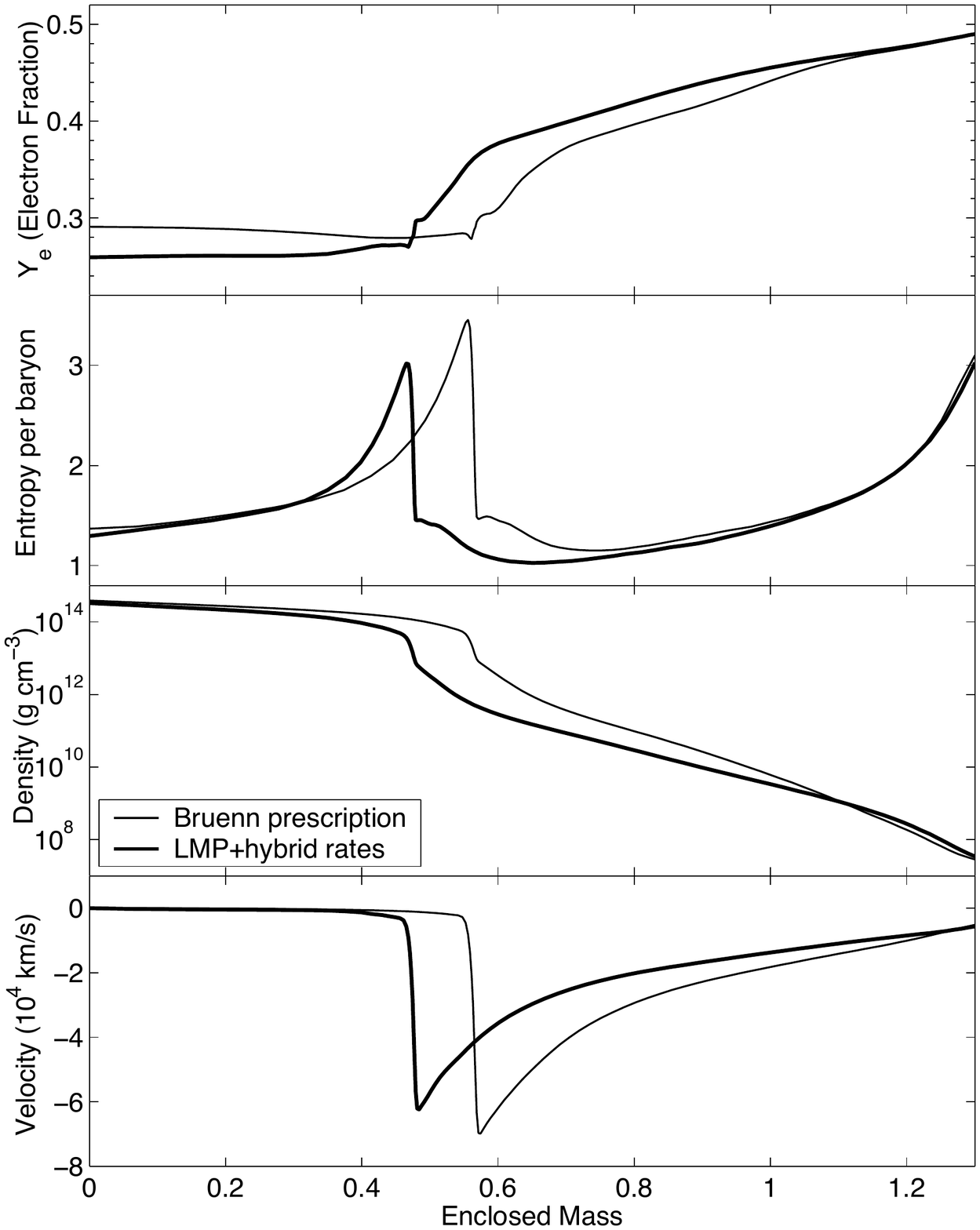}
\caption{From \citet{HiMeMe03}, plots of the density, entropy per baryon, electron fraction, and fluid velocity at core bounce using data from two models: one implementing the ``Bruenn 85'' electron capture rates \citep{Bruenn1985}, based on the Independent Particle Model of nucleons, and one implementing the rates of \citet{LaMaSa03}, based on the ``Hybrid'' model, which includes correlations between the nucleons in RPA and finite-temperature effects. The former (latter) data correspond to the thin (thick) black lines in the plots. Given the inclusion of the hybrid model electron capture rates, electron capture is unblocked and proceeds, leading to increased electron capture in the core in this model and, consequently, to a significant (inward) change in the location of the bounce shock in mass.}
\label{fig:bounce_prl}
\end{figure}

The importance of the above additions and modifications to the neutrino opacities in core collapse supernova models were reinforced in the 
context of later two-dimensional models developed by other groups \citep{BuVaDo18,JuBoJa18,KoTaFi18}.

\begin{figure}[htb]
\includegraphics[width=\textwidth]{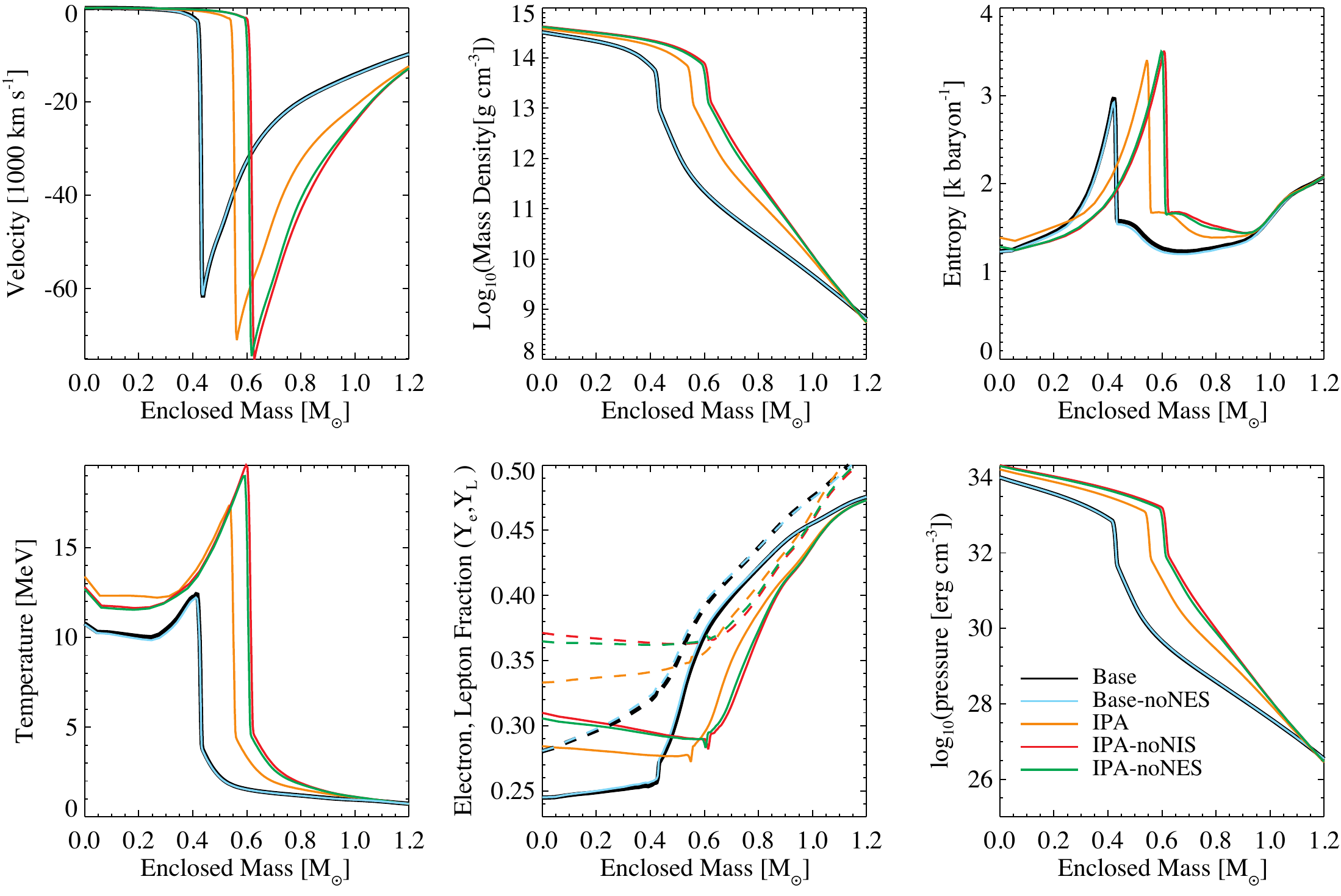}
\caption{Plots of velocity, density, entropy, temperature, electron and lepton fraction, and pressure at core bounce across five models with different input physics \citep{LeMeMe12b}. The model ``Base'' includes all weak interactions and uses the modern, hybrid-model electron capture rates. Model Base-NoNES includes the same weak interaction physics, with one exception: neutrino-electron scattering (NES) is not included. Similarly, model ``IPA'' includes all weak interaction channels, as does model Base, but uses the Independent Particle Approximation (IPA) rates for nuclear electron capture. And model ``IPA-NoNES'' includes the same weak interaction physics except neutrino--electron scattering. Comparing models Base and Base-NoNES, no significant changes result when NES is excluded. On the other hand, comparing models IPA and IPA-NoNES, we reach a different conclusion: In this case, the inclusion of NES has a significant impact on core deleptonization and, consequently, on the mass of the inner core at bounce. These comparisons demonstrate there is an interplay between different neutrino opacities. An improvement in one opacity may render an otherwise important second opacity relatively unimportant.}
\label{fig:fig1lentz12}
\end{figure}

Earlier in this section, we have seen the impact of adding new weak interaction channels and improving the treatment of those already included in core-collapse supernova model. Here we explore yet another dimension of this important sector of core-collapse supernova physics: the interplay of neutrino weak interaction channels (new and/or modified). \citet{LeMeMe12b} conducted an in-depth analysis focused largely on the neutrino production and interaction channels discussed above (i.e., nucleon--nucleon bremsstrahlung, non-isoenergetic scattering, and electron capture on nuclei). They demonstrated several important points: (1) While the addition of a single interaction channel may impact the dynamics of stellar core collapse and the post-bounce evolution, the addition of two interaction channels may not be additive---in fact, it may render one of the additional channels irrelevant. (2) When two or more interaction channels are included and are instead additive, the additive impact may be nonlinear. As an example, Lentz et~al.\ considered the interplay of electron capture on nuclei and neutrino--electron scattering during stellar core collapse. If we consider the nucleons as independent particles [Independent Particle Model (IPM)], electron capture on nuclei is blocked for $N>40$, where $N$ is the neutron number. In this case, the nuclear electron capture rates are given by \citet{Bruenn1985} are appropriate. In this instance, neutrino--electron scattering, which scatters neutrinos to lower energies given the core's electron degeneracy, leads to a significant increase in core deleptonization and a concommitant decrease in the inner core mass. On the other hand, if the improved nuclear electron capture rates of Langanke et~al.\ are used, which factor in nucleon interactions and correlations, nuclear electron capture is no longer blocked. In turn, low-energy neutrino states are filled, and neutrino--electron scattering is no longer able to down scatter neutrinos in energy (and contributes very little to the total neutrino opacity) and becomes rather unimportant. This is captured in Fig.~\ref{fig:fig1lentz12}. Comparing, for example, the velocity at bounce in the upper left panel of Fig.~\ref{fig:fig1lentz12} for the cases ``Base,'' which includes the full set of neutrino weak interactions with ``Base--noNES,'' which leaves out neutrino--electron scattering, it is obvious there is no difference. This is also true of all of the other quantities plotted. On the other hand, a comparison between ``Base'' and ``IPA,'' which includes nuclear electron capture in the independent particle approximation, it is evident that neutrino--electron scattering had a significant impact during collapse and on the final shock formation location.

\begin{figure}[htb]
\centering
\includegraphics[width=0.8\textwidth]{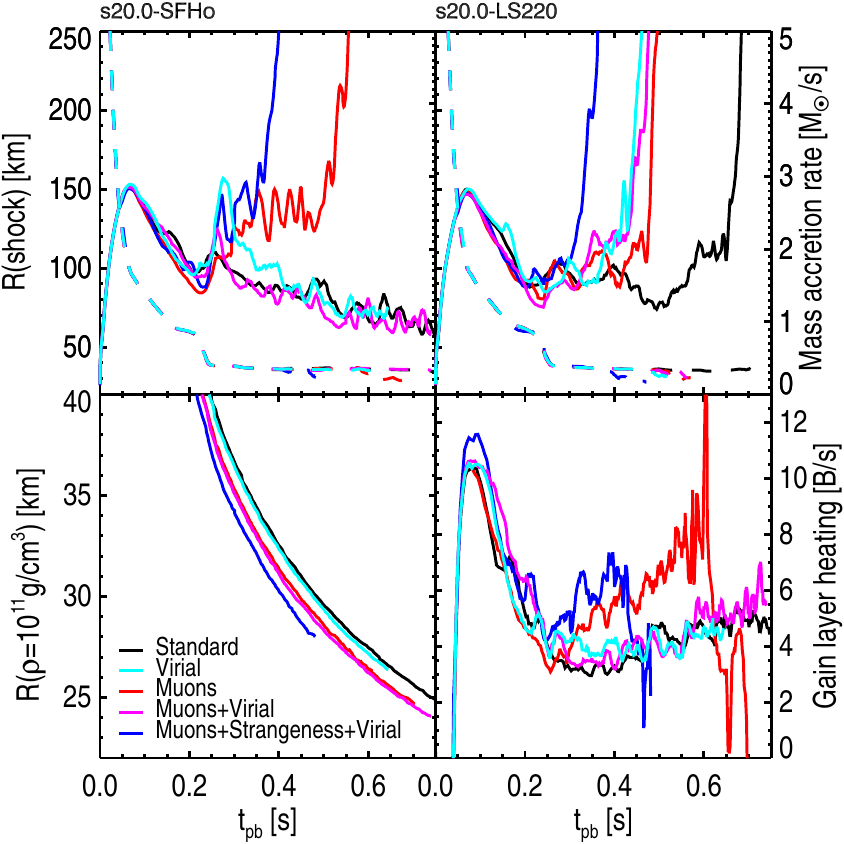}
\caption{In the upper left panel, from \citet{BoJaLo17}, the angle-averaged shock trajectories for several models, excluding and including muons, using the Steiner--Fischer--Hempel (SFHo) equation of state, are plotted. In the upper right panel, plotted are the results from the models that instead use the Lattimer--Swesty equation of state with bulk compression modulus $K=220$ MeV (LS220). Here ``Standard'' indicates models without muons.}
\label{fig:shockwithmuons}
\end{figure}

That the search for all core collapse-supernova relevant neutrino weak interactions is an ongoing activity is no better illustrated than by the very recent example provided by \citet{BoJaLo17}, whose work illuminated the importance of including muons and neutrino--muon weak interactions in core-collapse supernova models. Past models assumed that the population of muons in the stellar core during collapse, bounce, and the post-bounce neutrino shock reheating epoch would remain low given the large rest mass of the muon. Bollig et~al.\ point out that such arguments are not well motivated. The electron chemical potential in the proto-neutron star at this time exceeds the muon rest mass, and the core temperature is large, as well. In the context of two-dimensional supernova models using the \textsc{Vertex} code and initiated from a $20\,M_\odot$ progenitor, they demonstrated that significant populations of muons are in fact produced and, more importantly, that the inclusion of muons in their supernova models impacted the outcomes quantitatively in all cases and even qualitatively in some cases, depending on the nuclear equation of state used. For the SFHo equation of state, models with muons exhibited explosion whereas counterpart models without them did not. For models with the LS220 equation of state, models with muons exhibited earlier explosions, indicating that explosion was facilitated in these models. Bollig et al.'s results are encapsulated in Fig.~\ref{fig:shockwithmuons}.

We close this section with an emphasis on one final important point: Like all weak interaction cross sections, those the community has found to be important to core-collapse supernova evolution and has included in its supernova models have uncertainties associated with them, which can arise from experimental uncertainties in the few cases where the cross sections have been measured directly, or from uncertainties in the theory used to predict them, which in the end the supernova modeling community must rely on given it is impossible to measure all relevant cross sections under all relevant thermodynamic conditions and at all relevant neutrino energies found in a supernova environment. Thus, it is important to explore the potential impact of such uncertainties on the quantitative and qualitative core-collapse supernova model outcomes. 

\begin{figure}[htb]
\includegraphics[width=\textwidth]{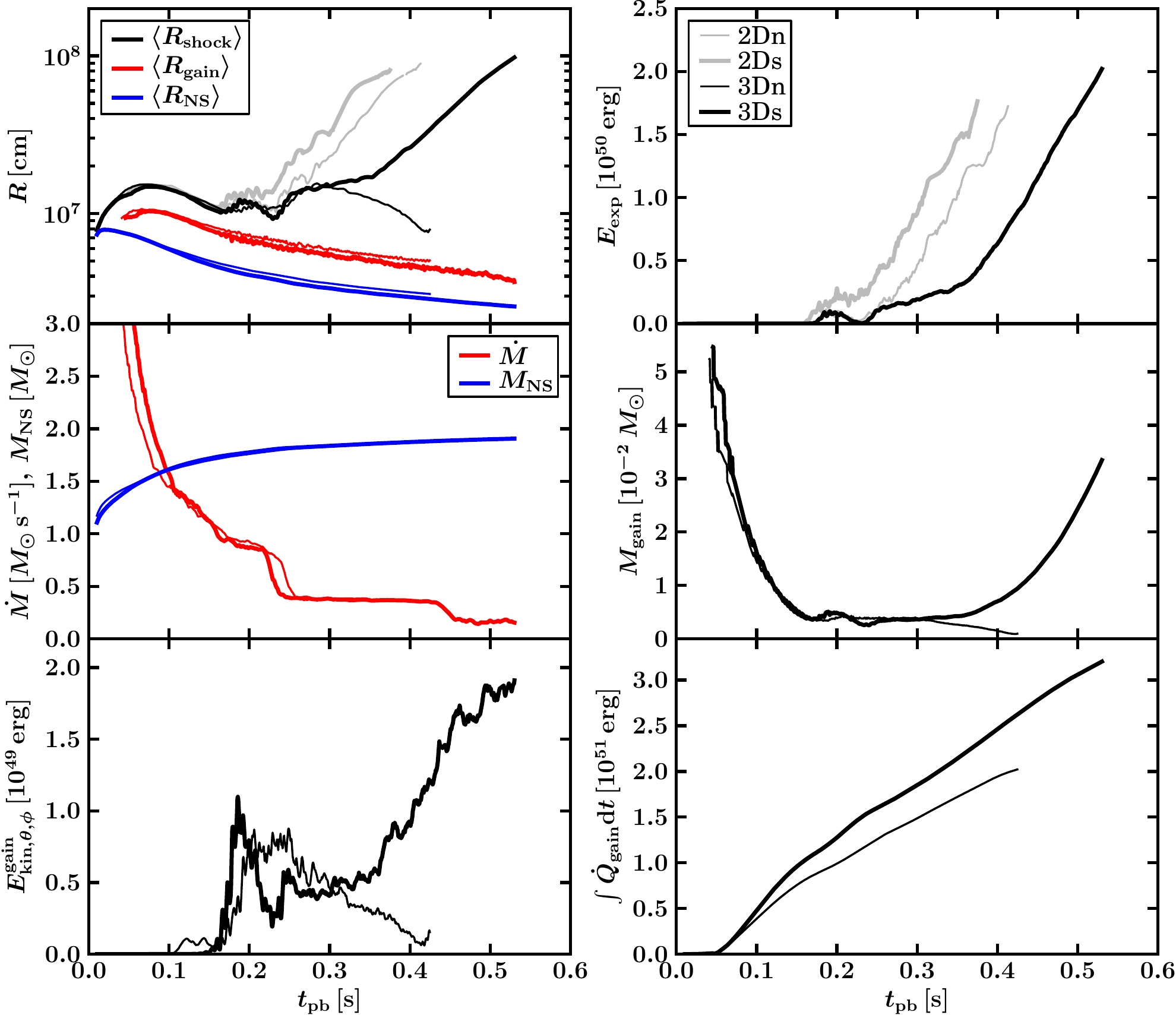}
\caption{Results are shown here from the core-collapse supernova studies of \citet{MeJaBo15}. In particular, in the uppermost left panel, the angle-averaged shock radius is plotted for two pairs of models, one for the two-dimensional case and one for the three-dimensional case. All cases are launched from a $20\,M_\odot$ progenitor and were performed with the \textsc{Vertex} code. Within each case, two simulations were performed, one using the standard weak interaction cross section for neutrino--nucleon scattering and the other including a correction to the strangeness content of the nucleon, which results in a correction to the coupling constants. In two dimensions, both models explode, with some quantitative differences observed in the shock trajectories. In the more important three-dimensional case, the outcomes with and without the correction are \emph{qualitatively} different. Specifically, explosion is not obtained in their model unless the opacity correction is included.}
\label{fig:strangeness}
\end{figure}

Case in point: The exploration of the impact of the uncertainty in the neutrino--nucleon cross section. \citet{MeJaBo15} performed two state-of-the-art three-dimensional simulations of the core-collapse supernova explosion of a $20\,M_\odot$ progenitor. In one case, they included what the modeling community 
regarded at the time as the state-of-the-art neutrino weak interaction set, with no modification to any of the cross sections. In the other, they varied one of the cross sections, albeit a critical one: the cross section for neutrino scattering on nucleons. This cross section is one of the most important for neutrino transport below the neutrinospheres, as the leading opacity source and, as we saw above, as an additional heating source for matter within the proto-neutron star. Uncertainty in the cross section for neutrino--nucleon scattering arises from, among other things, uncertainty in the strangeness content of the nucleon, which can alter the coupling constants. In particular, Melson et~al.\ varied the cross section by $\sim$10\%, consistent with the experimental uncertainties, and in so doing found they could \emph{qualitatively} alter the outcome of the model. When the standard weak interaction set was used, they did not obtain an explosion in the model. When they varied the neutrino--nucleon cross section, they did. The results are shown in Fig.~\ref{fig:strangeness}.
Of course, we have already seen that variations in a particular cross section can interact with variations in another. The only way the supernova modeling community can accurately assess the impact of variations in a single cross section is to vary all of them, in a statistically meaningful way---i.e., perform a sensitivity study. And, obviously, this should be performed, at least ultimately, in the context of three-dimensional models. Unfortunately, the last requirement cannot be met at this time. Such a study would require that many three-dimensional models be performed, which at the moment, even with the significant computing power afforded the modeling community by today's supercomputers, is prohibitive. Such studies should be conducted, but they will have to wait for future supercomputing capabilities.

\begin{landscape}
\begin{table}
\caption{Relevant modern neutrino emissivities and opacities, most or all of which have been adopted in three-dimensional core-collapse supernova models.}
\label{tab:neutrinoopacities}
\begin{tabular}{|c|c|c|}
\hline
Category                                       & Weak Ineraction & Opacity Source \\
\hline
Absorption and Emission              & $e^{-}+p\rightleftarrows n+\nu_e$ & \citet{BuSa98}, \citet{RePrLa98}, \citet{Horowitz2002} \\
                                                      & $e^{+}+n\rightleftarrows p+\bar{\nu}_e$ &  \citet{BuSa98}, \citet{RePrLa98}, \citet{Horowitz2002} \\
                                                      & $e^{-}+A(Z,N)\rightleftarrows A(Z-1,N+1)+\nu_e$ & \citet{LaMaSa03} \\
\hline
Coherent Isoenergetic Scattering  & $\nu_{e,\mu,\tau},\bar{\nu}_{e,\mu,\tau} + A \rightarrow  \nu_{e,\mu,\tau},\bar{\nu}_{e,\mu,\tau} + A$ & \citet{BrMe97}, \citet{Horowitz1997}\\
\hline
Non-Isoenergetic Scattering          & $\nu_{e,\mu,\tau},\bar{\nu}_{e,\mu,\tau} + e^{-,+} \rightarrow  \nu_{e,\mu,\tau},\bar{\nu}_{e,\mu,\tau} + e^{+,-}$ & \citet{MeBr93c} \\
                                                       & $\nu_{e,\mu,\tau},\bar{\nu}_{e,\mu,\tau} + n,p \rightarrow  \nu_{e,\mu,\tau},\bar{\nu}_{e,\mu,\tau} + n,p$ & \citet{BuSa98}, \citet{RePrLa98}, \citet{Horowitz2002} \\
\hline
Pair Creation and Annihilation        & $e^{+}+e^{-} \rightleftarrows \nu_{e,\mu,\tau} + \bar{\nu}_{e,\mu,\tau}$ & \citet{Bruenn1985} \\
                                                       & $N+N \rightleftarrows N+N+\nu_{e,\mu,\tau} + \bar{\nu}_{e,\mu,\tau}$ & \citet{HaRa98} \\
                                                       & $\nu_{e} + \bar{\nu}_{e}\rightleftarrows \nu_{\mu,\tau} + \bar{\nu}_{\mu,\tau}$ & \citet{BuJaKe03} \\
\hline
\end{tabular}
\end{table}
\end{landscape}

\subsection{The relevant neutrino interactions}
\label{sec:neutrinoInteractions}

The previous section makes clear that the effort to ascertain which neutrino weak interactions are important to core-collapse supernovae theory is an ongoing activity. To date, the list included in Table \ref{tab:neutrinoopacities} is what is deemed to be the essential list. Most, if not all, of the weak interactions in the list have been included in the state of the art simulations whose underlying numerical methods have been the focus of this review. Motivated by the recent example documented in the previous section, in Table \ref{tab:muonopacities} we also include a list of the relevant neutrino weak interactions involving muons. At present, these have been included by only one group \citep{BoJaLo17} and, as discussed, have been found to be important by this group. In light of this, adoption of these weak interactions by other groups is certainly warranted.

\subsubsection{Boltzmann collision term}

\begin{table}
\caption{Relevant neutrino--muon weak interactions.}
\begin{center}
\begin{tabular}{|c|}
\hline
$\mu^{-} + p \rightleftarrows n+ \nu_{\mu}$ \\
$\mu^{-} + \nu_{e} \rightleftarrows e^{-} + \nu_{\mu}$ \\
$\mu^{-} + \bar{\nu}_{\mu}+ \rightleftarrows e^{-} + \bar{\nu}_{e}$ \\
$\mu^{-} \rightleftarrows e^{-} + \bar{\nu}_{e} + \nu_{\mu}$ \\
\hline
$\mu^{+} + n \rightleftarrows p + \bar{\nu}_\mu$ \\
$\mu^{+} + \bar{\nu}_e \leftrightarrows e^{+} + \bar{\nu}_\mu$ \\
$\mu^{+} + \nu_\mu \leftrightarrows e^{+} + \nu_e$ \\
$\mu^{+} \leftrightarrows e^{+} + \nu_e + \bar{\nu}_\mu$ \\
\hline
$\nu_{e,\mu,\tau} + \mu^{+,-} \rightleftarrows \nu_{e,\mu,\tau} + {\mu^{+,-}}$ \\
\hline
\end{tabular}
\end{center}
\label{tab:muonopacities}
\end{table}

We write the collision term as the sum of terms corresponding to the main processes---emission and absorption, scattering, and pair creation and annihilation---listed in Table \ref{tab:neutrinoopacities}:
\begin{equation}
  \mathcal{C}[f_{s}](p) = \mathcal{C}_{\ABEM}[f_{s}](p) + \mathcal{C}_{\SCAT}[f_{s}](p) + \mathcal{C}_{\PAIR}[f_{s}](p).  
  \label{eq:collisionTermBoltzmannSum}
\end{equation}
For each of the terms, we focus on its functional form, which is closely related to the computational complexity of including a particular weak interaction in a core-collapse supernova model. Each term---hence, each added interaction---warrants tailored consideration.

The term due to neutrino emission and absorption is written as
\begin{equation}
  \f{1}{\varepsilon}\,\mathcal{C}_{\ABEM}[f_{s}](p) = [1-f_{s}(p)]\eta_{s} - \chi_{s}\,f_{s}(p),
  \label{eq:collisionTermAmEm}
\end{equation}
where $\eta_{s}$ and $\chi_{s}$ are the emissivity and absorption opacity of neutrino species $s$ and are assumed to be isotropic in the momentum-space angle (independent of $\omega$), but depend on the neutrino energy $\varepsilon$.  
The blocking factor, $1-f_{s}(p)$, is included to account for the Fermi--Dirac statistics of neutrinos, and suppresses neutrino emission when the phase-space occupancy is high (i.e., when $f_{s}\lesssim1$).  
It is common to introduce $\tilde{\chi}_{s}=(\eta_{s}+\chi_{s})$, associated in this case with ``stimulated absorption'' (as opposed to the stimulated emission of photons), and to define $f_{0,s}=\eta_{s}/\tilde{\chi}_{s}$, in which case Eq.~\eqref{eq:collisionTermAmEm} can be written in relaxation form:
\begin{equation}
  \f{1}{\varepsilon}\,\mathcal{C}_{\ABEM}[f_{s}](p) = \tilde{\chi}\,\big(\,f_{0,s}-f_{s}\,\big).  
  \label{eq:collisionTermAmEmRelaxation}
\end{equation}
In this form it is easy to see that the collision term drives the distribution function towards the equilibrium distribution, $f_{0,s}$.  
Also note, this interaction is local in momentum-space; i.e., there is no coupling across momentum-space.  

Neutrino--matter scattering (the second and third category in Table~\ref{tab:neutrinoopacities}) is described by
\begin{align}
  \f{1}{\varepsilon}\,\mathcal{C}_{\SCAT}[f_{s}](p)
  &=\big(1-f_{s}(p)\big)\int_{V_{p}}\mathcal{R}_{\SCAT}^{\IN}(p,p')\,f_{s}(p')\,d^{3}p' \nonumber \\
  &\hspace{12pt}
  -f_{s}(p)\int_{V_{p}}\mathcal{R}_{\SCAT}^{\OUT}(p,p')\,(1-f_{s}(p'))\,d^{3}p',
  \label{eq:collisionTermScat}
\end{align}
where $\mathcal{R}_{\SCAT}^{\IN}(p,p')$ is the scattering rate from momentum $p'$ into $p$, and $\mathcal{R}_{\SCAT}^{\OUT}(p,p')$ is the scattering rate out of momentum $p$ into $p'$.  
When compared with the collision term in Eq.~\eqref{eq:collisionTermAmEmRelaxation}, the coupling in momentum-space (due to the integral operators) increases the computational complexity of evaluating the collision operator.  
If momentum-space is discretized into $N_{p}$ bins, a brute force evaluation of Eq.~\eqref{eq:collisionTermScat} for all $p$ requires $\mathcal{O}(N_{p}^{2})$ operations.  
Note also the blocking factors in Eq.~\eqref{eq:collisionTermScat}, which suppress scattering to high-occupancy regions of momentum-space.  
The second category in Table~\eqref{tab:neutrinoopacities} (coherent, isoenergetic scattering) is obtained as a simplification of Eq.~\eqref{eq:collisionTermScat} by letting
\begin{equation}
  \mathcal{R}_{\SCAT}^{\IN/\OUT}(p,p') \to \mathcal{R}_{\ISO}(|p|,\cos\alpha)\,\delta(|p|-|p'|),
\end{equation}
where $\cos\alpha=p \cdot p'/(|p||p'|)$. For this type of interaction, with $d^{3}p'=|p'|^{2}\,d|p'|\,d\omega'$, the collision term is given by
\begin{align}
  \f{1}{\varepsilon}\,\mathcal{C}_{\ISO}[f_{s}](|p|,\omega)
  &=\int_{\mathbb{S}^{2}}\mathcal{R}_{\ISO}(|p|,\cos\alpha)\,|p|^{2}\,f_{s}(|p|,\omega')\,d\omega' \nonumber \\
  &\hspace{12pt}
  -f_{s}(|p|,\omega)\int_{\mathbb{S}^{2}}\mathcal{R}_{\ISO}(|p|,\cos\alpha)\,|p|^{2}\,d\omega',
  \label{eq:collisionTermScatIso}
\end{align}
which is considerably simplified relative to the scattering operator in Eq.~\eqref{eq:collisionTermScat}.  

Finally, neutrino-antineutrino pair creation and annihilation (e.g., from electron-positron pairs; the fourth category in Table~\ref{tab:neutrinoopacities}) is described by
\begin{align}
  \f{1}{\varepsilon}\,\mathcal{C}_{\PAIR}[f_{s}](p)
  &=(1-f_{s}(p))\int_{V_{p}}\mathcal{R}_{\PAIR}^{\IN}(p,p')\,(1-\bar{f}_{s}(p'))\,d^{3}p' \nonumber \\
  & \hspace{12pt}
  -f_{s}(p)\int_{V_{p}}\mathcal{R}_{\PAIR}^{\OUT}(p,p')\,\bar{f}_{s}(p')\,dp',
  \label{eq:collisionTermPair}
\end{align}
where $\mathcal{R}_{\PAIR}^{\IN}(p,p')$ and $\mathcal{R}_{\PAIR}^{\OUT}$ are the neutrino-antineutrino pair production and annihilation rates, respectively, and $\bar{f}_{s}$ is the antineutrino distribution function.  
We note that the functional form of the collision term for the last of the pair processes included in Table \ref{tab:neutrinoopacities} is not represented by the functional form for pair creation and annihilation presented here. In this particular case, both in-states and both out-states correspond to neutrinos, which, when treated without 
approximation, results in a collision term involving four distribution functions. This non-approximate treatment of the process has yet to be implemented 
in core-collapse supernova models. As a result, we do not include its functional form here.

All of the above rates $\eta_{s}$, $\chi_{s}$, $\mathcal{R}_{\SCAT}^{\IN/\OUT}$, and $\mathcal{R}_{\PAIR}^{\IN/\OUT}$ depend on the thermodynamic state of the stellar core fluid (e.g., $\rho$, $T$, and $Y_{e}$).  

Symmetries in some of the collision kernels exist \citep[e.g.,][]{Bruenn1985,Ce94}, which should be leveraged in computations.  
First, because the total number of neutrinos is conserved in neutrino--matter scattering, 
\begin{equation}
  \int_{V_{p}}\mathcal{C}_{\SCAT}[f_{s}](p)\,\f{d^{3}p}{\varepsilon}=0,
\end{equation}
and the following in--out invariance holds:
\begin{equation}
  \mathcal{R}_{\SCAT}^{\IN}(p,p') = \mathcal{R}_{\SCAT}^{\OUT}(p',p).  
  \label{eq:inOutInvarianceScattering}
\end{equation}
Second, when the neutrino distribution function equals the local Fermi--Dirac distribution, $f_{s}=f_{0,s}=1/[e^{(\varepsilon-\mu_{\nu,s})/T}+1]$, where $T$ is the matter temperature and $\mu_{\nu}$ is the equilibrium neutrino chemical potential, the net energy and momentum transfer between neutrinos and matter due to scattering must vanish.  
Thus, requiring
\begin{equation}
  \int_{V_{p}}\mathcal{C}_{\SCAT}[f_{0,s}](p)\,g(p)\,\f{d^{3}p}{\varepsilon}=0
\end{equation}
for an arbitrary function $g(p)$, gives
\begin{equation}
  \mathcal{R}_{\SCAT}^{\IN}(p,p') = \mathcal{R}_{\SCAT}^{\OUT}(p,p')\,e^{-(\varepsilon-\varepsilon')/T}=\mathcal{R}_{\SCAT}^{\IN}(p',p)\,e^{-(\varepsilon-\varepsilon')/T},
\end{equation}
where Eq.~\eqref{eq:inOutInvarianceScattering} is used in the rightmost expression.  

\subsubsection{Two-moment collision terms}
\label{sec:collisionTermsTwoMoment}

Collision terms for the two-moment model are derived by taking angular moments of the collision term in Eq.~\eqref{eq:collisionTermBoltzmannSum}.  
Such terms have been discussed in the context of multidimensional two-moment models by, e.g., \citet{ShKiSe11}.  
For completeness, we list two-moment collision terms corresponding to angular moments of Eqs.~\eqref{eq:collisionTermAmEmRelaxation}, \eqref{eq:collisionTermScat}, and \eqref{eq:collisionTermPair} here.  
Considering the two-moment models delineated in Sect.~\ref{sec:TwoMoment}, the relevant angular moments of the Boltzmann collision term are
\begin{equation}
  \f{1}{4\pi}\int_{\mathbb{S}^{2}}\mathcal{C}[f_{s}]\,\f{d\omega}{\varepsilon}
  \quad\text{and}\quad
  \f{1}{4\pi}\int_{\mathbb{S}^{2}}\mathcal{C}[f_{s}]\,\ell_{j}\,\f{d\omega}{\varepsilon}.  
\end{equation}
(The first of these terms also appears in the one-moment model discussed in Sect.~\ref{sec:oneMomentKinetics}; cf.\ Eq.~\eqref{eq:spectralLagrangianEnergyEquationFLD_3p1}.)

\paragraph{Emission/absorption}
For emission and absorption, the evaluation is straightforward since the emissivity and opacity are isotropic in momentum space angle.  
The zeroth moment gives
\begin{align}
  \f{1}{4\pi}\int_{\mathbb{S}^{2}}\mathcal{C}_{\ABEM}[f_{s}]\,\f{d\omega}{\varepsilon}
  =\big(1-\mathcal{D}_{s}\big)\,\eta_{s} - \chi_{s}\,\mathcal{D}_{s}
  =\tilde{\chi}_{s}\,\big(\,\mathcal{D}_{0,s}-\mathcal{D}_{s}\,\big),
\end{align}
where the zeroth moment of the equilibrium distribution is defined as
\begin{equation}
  \mathcal{D}_{0,s}=\f{1}{4\pi}\int_{\mathbb{S}^{2}}f_{0,s}\,d\omega.  
\end{equation}
The first moment gives
\begin{equation}
  \f{1}{4\pi}\int_{\mathbb{S}^{2}}\mathcal{C}_{\ABEM}[f_{s}]\,\ell_{j}\,\f{d\omega}{\varepsilon}
  =-\tilde{\chi}_{s}\,\mathcal{I}_{s,j},
\end{equation}
since the angular moment of $\ell_{j}$ vanishes.  

\paragraph{Angular kernel approximations}
To incorporate scattering and pair processes in the two-moment model, following \citet{Bruenn1985}, the kernels are expanded in a Legendre series up to linear order; e.g.,
\begin{equation}
  \mathcal{R}_{\SCAT}^{\IN}(p,p') 
  = \mathcal{R}_{\SCAT}^{\IN}(\varepsilon,\varepsilon',\Omega)
  \approx \Phi_{\SCAT,0}^{\IN}(\varepsilon,\varepsilon') + \Phi_{\SCAT,1}^{\IN}(\varepsilon,\varepsilon')\,\Omega(\omega,\omega'),
  \label{eq:kernelExpansion}
\end{equation}
where $\Omega=\ell_{\mu}(\omega)\ell^{\mu}(\omega')$ is the cosine of the scattering angle.  
From the orthogonality of the Legendre polynomials, the scattering coefficients are then evaluated form the kernels as
\begin{equation}
  \big\{\,\Phi_{\SCAT,0}^{\IN}(\varepsilon,\varepsilon'),\Phi_{\SCAT,1}^{\IN}(\varepsilon,\varepsilon')\,\big\}
  =\f{1}{2}\int_{-1}^{1}\mathcal{R}_{\SCAT}^{\IN}(\varepsilon,\varepsilon',\Omega)\,\big\{\,1,\,3\,\Omega\,\big\}\,d\Omega.  
\end{equation}
Terms beyond linear can be included in the expansion of the kernel in Eq.~\eqref{eq:kernelExpansion} at the expense of a more complicated collision operator for the two-moment model.  
\citet{SmCe96} investigated the effect of including the quadratic term for neutrino-electron scattering in a configuration during the infall phase of stellar core collapse.  
They found that including the quadratic term results in a better fit to the scattering kernel, but when comparing stationary state transport solutions with and without the quadratic term, they found no significant difference in relevant quantities such as the neutrino number density, flux, and transfer rates of lepton number, energy, or momentum to the stellar fluid.  
We also note that \citet{JuBoJa18}, in their Appendix~A, provide expressions for pair processes that include the quadratic term in the Legendre expansion of the kernels.  

\paragraph{Scattering}
Employing the expansion in Eq.~\eqref{eq:kernelExpansion} for the scattering operator gives
\begin{align}
  &\f{1}{4\pi}\int_{\mathbb{S}^{2}}\mathcal{C}_{\SCAT}[f_{s}](p)\,\f{d\omega}{\varepsilon} \\
  &=\big(1-\mathcal{D}(\varepsilon)\big)\int_{0}^{\infty}\Phi_{\SCAT,0}^{\IN}(\varepsilon,\varepsilon')\,\mathcal{D}(\varepsilon')\,dV_{\varepsilon'}
  -\mathcal{I}_{\mu}(\varepsilon)\int_{0}^{\infty}\Phi_{\SCAT,1}^{\IN}(\varepsilon,\varepsilon')\,\mathcal{I}^{\mu}(\varepsilon')\,dV_{\varepsilon'} \nonumber \\
  &\hspace{6pt}
  -\mathcal{D}(\varepsilon)\int_{0}^{\infty}\Phi_{\SCAT,0}^{\OUT}(\varepsilon,\varepsilon')\,\big(1-\mathcal{D}(\varepsilon')\big)\,dV_{\varepsilon'}
  +\mathcal{I}_{\mu}(\varepsilon)\int_{0}^{\infty}\Phi_{\SCAT,1}^{\OUT}(\varepsilon,\varepsilon')\,\mathcal{I}^{\mu}(\varepsilon')\,dV_{\varepsilon'} \nonumber
\end{align}
for the zeroth moment (recall that $dV_{\varepsilon}=4\pi\varepsilon^{2}d\varepsilon$), and
\begin{align}
  &\f{1}{4\pi}\int_{\mathbb{S}^{2}}\mathcal{C}_{\SCAT}[f_{s}]\,\ell_{j}\,\f{d\omega}{\varepsilon} \\
  &=-\mathcal{I}_{j}(\varepsilon)\int_{0}^{\infty}\Phi_{\SCAT,0}^{\IN}(\varepsilon,\varepsilon')\,\mathcal{D}(\varepsilon')\,dV_{\varepsilon'}
  +\big(\f{1}{3}\,g_{j\mu}-\widehat{\mathcal{K}}_{j\mu}(\varepsilon)\big)\int_{0}^{\infty}\Phi_{\SCAT,1}^{\IN}(\varepsilon,\varepsilon')\,\mathcal{I}^{\mu}(\varepsilon')\,dV_{\varepsilon'} \nonumber \\
  &\hspace{12pt}
  -\mathcal{I}_{j}(\varepsilon)\int_{0}^{\infty}\Phi_{\SCAT,0}^{\OUT}(\varepsilon,\varepsilon')\,\big(1-\mathcal{D}(\varepsilon')\big)\,dV_{\varepsilon'}
  +\widehat{\mathcal{K}}_{j\mu}\int_{0}^{\infty}\Phi_{\SCAT,1}^{\OUT}(\varepsilon,\varepsilon')\,\mathcal{I}^{\mu}(\varepsilon')\,dV_{\varepsilon'} \nonumber
\end{align}
for the first moment.  
Here we have used
\begin{equation}
  \f{1}{4\pi}\int_{\mathbb{S}^{2}}\ell_{\mu}(\omega)\,\ell_{\nu}(\omega)\,d\omega=\f{1}{3}\,g_{\mu\nu}.  
\end{equation}

\paragraph{Pair processes}
Employing the kernel expansion in Eq.~\eqref{eq:kernelExpansion} for the neutrino-antineutrino pair creation and annihilation operator in Eq.~\eqref{eq:collisionTermPair} gives
\begin{align}
  &\f{1}{4\pi}\int_{\mathbb{S}^{2}}\mathcal{C}_{\PAIR}[f_{s}](p)\,\f{d\omega}{\varepsilon} \\
  &=\big(1-\mathcal{D}(\varepsilon)\big)\int_{0}^{\infty}\Phi_{\PAIR,0}^{\IN}(\varepsilon,\varepsilon')\,\big(1-\bar{\mathcal{D}}(\varepsilon')\big)\,dV_{\varepsilon'}
  +\mathcal{I}_{\mu}(\varepsilon)\int_{0}^{\infty}\Phi_{\PAIR,1}^{\IN}(\varepsilon,\varepsilon')\,\bar{\mathcal{I}}^{\mu}(\varepsilon')\,dV_{\varepsilon'} \nonumber \\
  &\hspace{12pt}
  -\mathcal{D}(\varepsilon)\int_{0}^{\infty}\Phi_{\PAIR,0}^{\OUT}(\varepsilon,\varepsilon')\,\bar{\mathcal{D}}(\varepsilon')\,dV_{\varepsilon'}
  -\mathcal{I}_{\mu}(\varepsilon)\int_{0}^{\infty}\Phi_{\PAIR,1}^{\OUT}(\varepsilon,\varepsilon')\,\bar{\mathcal{I}}^{\mu}(\varepsilon')\,dV_{\varepsilon'} \nonumber
\end{align}
for the zeroth moment of the collision operator, and
\begin{align}
  &\f{1}{4\pi}\int_{\mathbb{S}^{2}}\mathcal{C}_{\PAIR}[f_{s}](p)\,\ell_{j}\,\f{d\omega}{\varepsilon} \\
  &=-\mathcal{I}_{j}(\varepsilon)\int_{0}^{\infty}\Phi_{\PAIR,0}^{\IN}(\varepsilon,\varepsilon')\,\big(1-\bar{\mathcal{D}}(\varepsilon')\big)\,dV_{\varepsilon'}
  -\big(\f{1}{3}\,g_{j\mu}-\widehat{\mathcal{K}}_{j\mu}(\varepsilon)\big)\int_{0}^{\infty}\Phi_{\PAIR,1}^{\IN}(\varepsilon,\varepsilon')\,\bar{\mathcal{I}}^{\mu}(\varepsilon')\,dV_{\varepsilon'} \nonumber \\
  &\hspace{12pt}
  -\mathcal{I}_{j}(\varepsilon)\int_{0}^{\infty}\Phi_{\PAIR,0}^{\OUT}(\varepsilon,\varepsilon')\,\bar{\mathcal{D}}(\varepsilon')\,dV_{\varepsilon'}
  -\widehat{\mathcal{K}}_{j\mu}(\varepsilon)\int_{0}^{\infty}\Phi_{\PAIR,1}^{\OUT}(\varepsilon,\varepsilon')\,\bar{\mathcal{I}}^{\mu}(\varepsilon')\,dV_{\varepsilon'} \nonumber
\end{align}
for the first moment.  
Here, $\bar{\mathcal{D}}$ and $\bar{\mathcal{I}}^{\mu}$ are the zeroth and first moments of the antineutrino distribution function $\bar{f}$.  

\subsection{Neutrino--matter coupling}

In coupling neutrinos and matter, we are primarily concerned with lepton and four-momentum exchange.  
The neutrino lepton current density is
\begin{equation}
  J_{\mbox{\tiny neutrino}}^{\nu} = \sum_{s=\nu_{e},\bar{\nu}_{e}}\mathsf{g}_{s}\,N_{s}^{\nu},
  \label{eq:neutrinoLeptonCurrentDensity}
\end{equation}
where $N_{s}^{\nu}$ is the neutrino four-current density for neutrino species $s$, defined as in Eq.~\eqref{eq:numberMoments} with distribution function $f_{s}$, and $\mathsf{g}_{s}$ is the lepton number of neutrino species $s$ ($\mathsf{g}_{s}=+1$ for neutrinos, and $\mathsf{g}_{s}=-1$ for antineutrinos).  
From the electron number conservation equation, Eq.~\eqref{eq:ElectronNumberConservation}, and the neutrino number conservation equation, Eq.~\eqref{eq:numberEquation} (one for each neutrino species), we obtain
\begin{equation}
  \nabla_{\nu}\big(\,J_{\mbox{\tiny neutrino}}^{\nu} + J_{e}^{\nu}/m_{\mbox{\tiny B}}\,\big) 
  = \sum_{s=\nu_{e},\bar{\nu}_{e}}\mathsf{g}_{s}\,\int_{V_{p}}\mathcal{C}[f_{s}]\,\pi_{m} - L,
  \label{eq:totalLeptonConservation}
\end{equation}
Lepton number conservation demands that the source term of the right-hand side of Eq.~\eqref{eq:ElectronNumberConservation} takes the form
\begin{equation}
\label{eq:electronfractionequationsourceterm}
  L = \sum_{s=\nu_{e},\bar{\nu}_{e}}\mathsf{g}_{s}\int_{V_{p}}\mathcal{C}[f_{s}]\,\pi_{m}.  
\end{equation}
Note that, for simplicity of this exposition, we have assumed that only electron neutrinos and antineutrinos are involved in lepton exchange with the fluid, but see \citet{BoJaLo17} for a discussion of additional lepton exchange channels when muons are included as a fluid component.  
When muons are included, an additional equation for the muon number density, similar to Eq.~\eqref{eq:ElectronNumberConservation}, must be evolved, and the definition of the neutrino lepton current density in Eq.~\eqref{eq:neutrinoLeptonCurrentDensity} must be extended to include contributions from muon neutrinos.  
(Technically, similar extensions should be done to accommodate tauons, but, because of their large rest mass, they can be neglected as an agent for lepton number exchange with the fluid \citep{BoJaLo17}.)

The total neutrino stress-energy tensor is defined as
\begin{equation}
  T_{\mbox{\tiny neutrino}}^{\mu\nu}
  =\sum_{s=1}^{\nSpecies}T_{s}^{\mu\nu},
\end{equation}
where the stress-energy tensor for neutrino species $s$, $T_{s}^{\mu\nu}$, is defined as in Eq.~\eqref{eq:stressEnergyMoments} with distribution function $f_{s}$ and $\nSpecies$ is the total number of neutrino species.  
Using Eqs.~\eqref{eq:fluidFourMomentumConservation} and \eqref{eq:fourMomentumEquation}, the divergence of the total (fluid plus neutrino) stress-energy is
\begin{equation}
  \nabla_{\nu}\big(\,T_{\mbox{\tiny neutrino}}^{\mu\nu}+T_{\mbox{\tiny fluid}}^{\mu\nu}\,\big)
  =\sum_{s=1}^{\nSpecies}\int_{V_{p}}\mathcal{C}[f_{s}]\,p^{\mu}\,\pi_{m} - G^{\mu}.  
\end{equation}
Then, four-momentum conservation in neutrino--matter interactions demands the right-hand side of Eq.~\eqref{eq:fluidFourMomentumConservation} takes the form
\begin{equation}
\label{eq:fourmomentumequationsourceterm}
  G^{\mu} = \sum_{s=1}^{\nSpecies}\int_{V_{p}}\mathcal{C}[f_{s}]\,p^{\mu}\,\pi_{m}.  
\end{equation}

To illustrate the complexity of the neutrino--matter coupling problem further, we consider the neutrino--matter coupling problem in the space-homogeneous case using the number conservative two-moment model discussed in Sect.~\ref{sec:TwoMoment}.  
The angular moments of the neutrino distribution function of species $s$ evolve according to
\begin{align}
  d_{t}\big(\,\sqrt{\gamma}\,\big[\,W\,\mathcal{D}_{s}+v^{i}\,\mathcal{I}_{s,i}\,\big]\,\big)
  &= \f{\alpha\,\sqrt{\gamma}}{4\pi}\int_{\mathbb{S}^{2}}\mathcal{C}[f_{s}]\,\f{d\omega}{\varepsilon}, \label{eq:spectralNumberEquationSpaceHomogeneous} \\
  d_{t}\big(\,\sqrt{\gamma}\,\big[\,W\,\mathcal{I}_{s,j}+v^{i}\,\widehat{\mathcal{K}}_{s,ij}\,\big]\,\big)
  &= \f{\alpha\,\sqrt{\gamma}}{4\pi}\int_{\mathbb{S}^{2}}\mathcal{C}[f_{s}]\,\ell_{j}\,\f{d\omega}{\varepsilon}, \label{eq:spectralNumberFluxEquationSpaceHomogeneous}
\end{align}
where we use the ordinary derivative $d_{t}=d/dt$ to indicate that we consider the space-homogeneous case where physical variables are considered functions of time only.  
The right-hand sides of Eqs.~\eqref{eq:spectralNumberEquationSpaceHomogeneous} and \eqref{eq:spectralNumberFluxEquationSpaceHomogeneous} will include the contributions from emission and absorption, scattering, pair processes (as discussed above), and other processes.  
This sub-problem is typically considered in numerical implementations where neutrino--matter interactions are solved for in a time-implicit fashion, e.g., as is done within an implicit-explicit framework for integrating the full neutrino-radiation hydrodynamics system forward in time, which we will discuss in more details later (see, e.g., Sect.~\ref{sec:numericalTwoMomentKinetics}).  
Coupled to the transport equations, are the fluid evolution equations, which are combined with the transport equations and formulated as constraints due to mass, four-momentum, and lepton number conservation:
\begin{align}
  d_{t}\big(\,\sqrt{\gamma}\,D\,\big)
  &=0, \label{eq:massConservationConstraint} \\
  d_{t}\big(\,\sqrt{\gamma}\,\big[\,S_{j}+S_{j,\mbox{\tiny neutrino}}\,\big]\,\big)
  &=0, \label{eq:momentumConservationConstraint} \\
  d_{t}\big(\,\sqrt{\gamma}\big[\,\tau_{\mbox{\tiny fluid}}+E_{\mbox{\tiny neutrino}}\,\big]\,\big)
  &=0, \label{eq:energyConservationConstraint} \\
  d_{t}\big(\,\sqrt{\gamma}\,\big[\,N_{e}+N_{\mbox{\tiny neutrino}}\,\big]\,\big)
  &=0, \label{eq:leptonNumberConservationConstraint}
\end{align}
where $N_{e}=D\,Y_{e}/m_{\mbox{\tiny B}}$, and
\begin{align}
  S_{j,\mbox{\tiny neutrino}}
  &=\sum_{s=1}^{\nSpecies}\int_{0}^{\infty}\mathcal{F}_{j,s}\,dV_{\varepsilon}, \\
  E_{\mbox{\tiny neutrino}}
  &=\sum_{s=1}^{\nSpecies}\int_{0}^{\infty}\mathcal{E}_{s}\,dV_{\varepsilon}, \\
  N_{\mbox{\tiny neutrino}}
  &=\sum_{s=\nu_{e},\bar{\nu}_{e}}\mathsf{g}_{s}\int_{0}^{\infty}\mathcal{N}_{s}\,dV_{\varepsilon},
\end{align}
and where the Eulerian angular moments $\mathcal{F}_{s,j}$, $\mathcal{E}_{s}$, and $\mathcal{N}_{s}$ are defined in Section~\eqref{sec:EulerianDecompositions}.  
The Eulerian neutrino number density $\mathcal{N}_{s}$ is expressed in terms of the Lagrangian moments in Eq.~\eqref{eq:eulerianNumberInTermsOfLagrangianMoments}, which is also the expression inside the time-derivative on the left-hand side of Eq.~\eqref{eq:spectralNumberEquationSpaceHomogeneous}.  
The Eulerian momentum and energy, can also be written as combinations of the quantities in the time-derivatives on the left-hand side of Eqs.~\eqref{eq:spectralNumberEquationSpaceHomogeneous} and \eqref{eq:spectralNumberFluxEquationSpaceHomogeneous}:
\begin{align}
  \mathcal{F}_{j,s}
  &=\varepsilon\,
  \big\{\,
    W\,v_{j}\,\big[\,W\,\mathcal{D}_{s}+v^{i}\,\mathcal{I}_{s,i}\,\big]
    +\big[\,W\,\mathcal{I}_{s,j}+v^{i}\,\widehat{\mathcal{K}}_{s,ij}\,\big]
  \,\big\}, \\
  \mathcal{E}_{s}
  &=\varepsilon\,
  \big\{\,
    W\,\big[\,W\,\mathcal{D}_{s}+v^{i}\,\mathcal{I}_{s,i}\,\big]
    +v^{j}\,\big[\,W\,\mathcal{I}_{s,j}+v^{i}\,\widehat{\mathcal{K}}_{s,ij}\,\big]
  \,\big\}.  
\end{align}

Thus, adopting a closure for the radiation moments, writing $\widehat{\mathcal{K}}_{s,ij}$ in terms of $\mathcal{D}_{s}$ and $\mathcal{I}_{s,j}$ as discussed in Section~\eqref{sec:closure}, and an equation of state for the fluid $p=p(\rho,e,Y_{e})$, the system given by Eqs.~\eqref{eq:spectralNumberEquationSpaceHomogeneous}-\eqref{eq:spectralNumberFluxEquationSpaceHomogeneous} and \eqref{eq:massConservationConstraint}-\eqref{eq:leptonNumberConservationConstraint} can be solved for the radiation moments $\mathcal{D}_{s}$ and $\mathcal{I}_{s,j}$, and the fluid states $\rho$, $v^{i}$, $e$, and $Y_{e}$.  
This is a nonlinear system of equations, where nonlinearities are due to the radiation moment closure, the fluid equation of state, the dependence of $D$, $S_{j}$, and $\tau_{\mbox{\tiny fluid}}$ on $\rho$, $v^{i}$, $e$, and $Y_{e}$, and the nonlinear dependence of the neutrino opacities discussed in Sect.~\ref{sec:collisionTermsTwoMoment} on the thermodynamic state $\rho$, $e$, and $Y_{e}$.  
Modeling this four-momentum and lepton exchange between neutrinos and the fluid --- with all the relevant neutrino--matter interactions included --- constitutes the major computational cost of core-collapse supernova models.  

\section{Phase-space discretizations and implementations}

\subsection{Boltzmann kinetics: spatial and energy finite differencing plus discrete ordinates}

\subsubsection{Phase-space coordinates}
\label{sec:PhaseSpaceCoordinates}

\begin{figure}[htb]
\includegraphics[width=\textwidth]{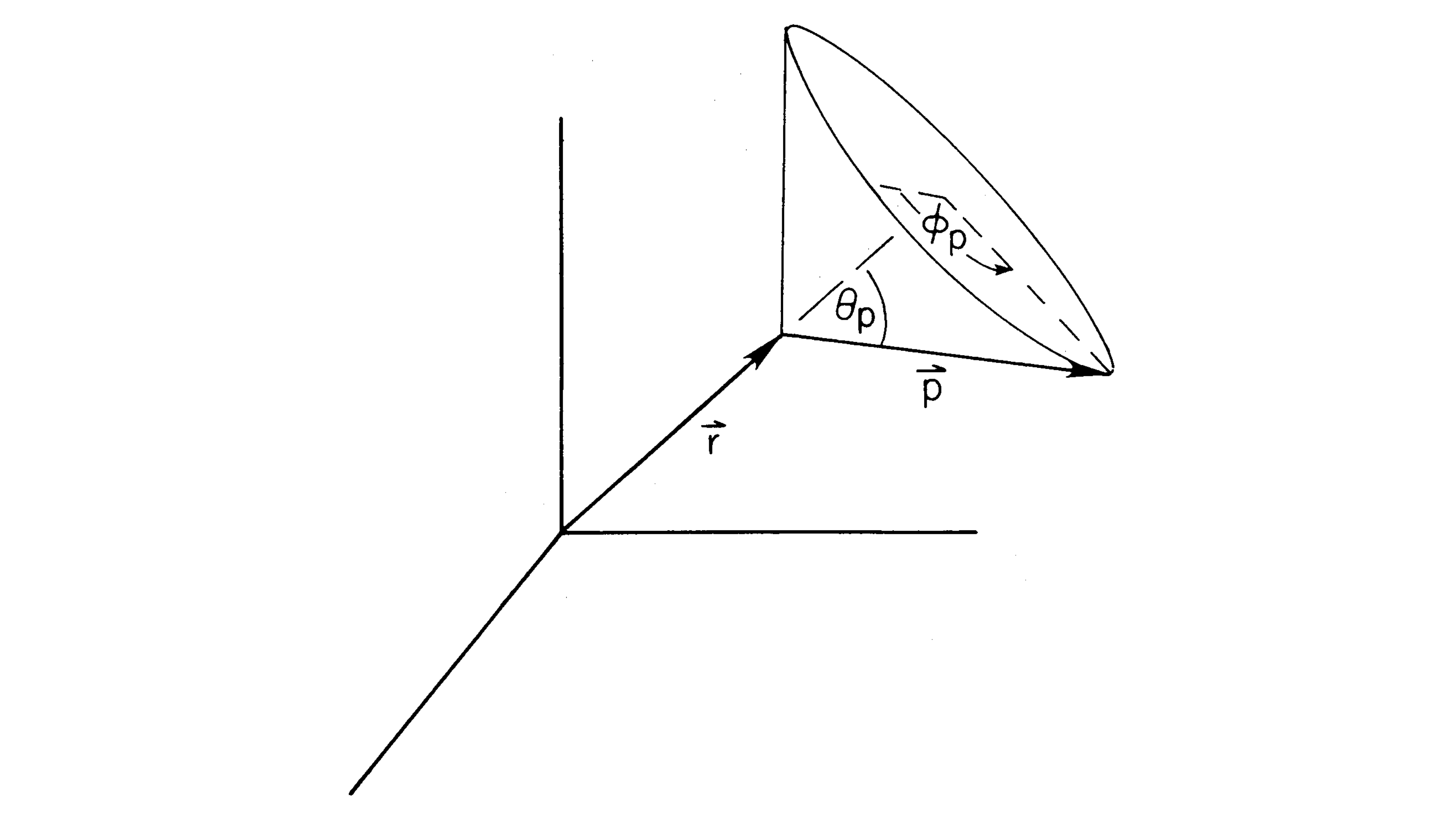}
\caption{Diagram illustrating the spherical momentum-space coordinates used in most neutrino radiation hydrodynamics implementations. The angle $\theta_p$ is the angle between the outgoing radial direction and the neutrino propagation direction, at the neutrino's location. The neutrino direction cosine, $\mu\equiv\cos\theta_p$, is defined in terms of it. $\phi_p$ is the associated momentum-space azimuthal angle. In spherical symmetry, the distribution function is only a function of $\mu$, not 
$\phi_p$.}
\label{fig:momspacevar}
\end{figure}

In a spherical spatial coordinate system, the neutrino's direction of propagation is 
specified relative to the basis vectors $\{\mathbf{e}_{r,\theta,\phi}\}$ as (see Fig.~\ref{fig:momspacevar})
\begin{equation}
\mathbf{n}=(n^{r},n^{\theta},n^{\phi}),
\label{eq:componentsofn}
\end{equation}
where
\begin{equation}
n^{r}=\cos\theta_{p},
\end{equation}
\begin{equation}
n^{\theta}= \sin\theta_{p}\cos\phi_{p},
\end{equation}
\begin{equation}
n^{\phi}= \sin\theta_{p}\sin\phi_{p}.
\end{equation}
This can be rexpressed as
\begin{equation}
n^{r}= \mu,
\end{equation}
\begin{equation}
n^{\theta}= (1 - \mu^2)^{\frac{1}{2}}\cos\phi_{p},
\end{equation}
\begin{equation}
n^{\phi}  = (1 - \mu^2)^{\frac{1}{2}}\sin\phi_{p},
\end{equation}
where $\mu\equiv\cos\theta_{p}$.
When spherical spatial and momentum-space coordinates are used, as defined above, the neutrino distribution
function has the following dependencies for no imposed symmetry, axisymmetry, and spherical symmetry,
\begin{equation}
f=f(r,\theta,\phi,\mathbf{n},E,t)=f(r,\theta,\phi,\mu,\phi_{p},E,t),
\end{equation}
\begin{equation}
f=f(r,\theta,\mathbf{n},E,t)=f(r,\theta,\mu,\phi_{p},E,t),
\end{equation}
\begin{equation}
f=f(r,\mathbf{n},E,t)=f(r,\mu,E,t),
\end{equation}
respectively, where in all three cases $E$ is the neutrino energy.

\subsubsection{Spherical symmetry}

We illustrate the approach used by \citet{MeBr93b,MeMe99,LiMeMe04,MeLiCa04,MeLiCa06} in the context of a model that assumes Newtonian gravity and is valid to $\mathcal{O}(v/c)$. The fully general relativistic case is detailed in \citet{LiMeMe04}. In the Newtonian gravity, $\mathcal{O}(v/c)$ case, the conservative neutrino Boltzmann equation reads
\begin{eqnarray}
\label{eq:boltzeq}
& & \frac{1}{c}\frac{\partial F}{\partial t} 
                +  4\pi \mu \frac{\partial (r^{2}\rho F)}{\partial m}
                +  \frac{1}{r}\frac{\partial [(1-\mu ^{2})F]}{\partial \mu } \\ \nonumber
               & + & \frac{1}{c}(\frac{\partial \rm{ln}\rho }{\partial t}+\frac{3v}{r})
                      \frac{\partial [\mu (1-\mu ^{2})F]}{\partial \mu }
                +  \frac{1}{c}[\mu ^{2}(\frac{\partial \rm{ln}\rho }{\partial t}+\frac{3v}{r})-\frac{v}{r}]
                      \frac{1}{E^{2}}\frac{\partial (E^{3}F)}{\partial E} \nonumber \\ 
                & = & \frac{j}{\rho }-\tilde{\chi }F \nonumber \\
                & + & \frac{1}{c}\frac{1}{h^{3}c^{3}}E^{2}\int d\mu 'R_{{\rm IS}}F
                 -  \frac{1}{c}\frac{1}{h^{3}c^{3}}E^{2}F\int d\mu 'R_{{\rm IS}} 
\nonumber \\
                & + & \frac{1}{h^{3}c^{4}}
                       (\frac{1}{\rho }-F)  
                      \int dE'E'^{2}d\mu '
                      \tilde{R}_{{\rm NIS}}^{{\rm in}} F
                 -  \frac{1}{h^{3}c^{4}}
                      F  
                      \int dE'E'^{2}d\mu '
                      \tilde{R}_{{\rm NIS}}^{{\rm out}} (\frac{1}{\rho }-F)  
\nonumber \\
                & + &\frac{1}{h^{3}c^{4}}
                       (\frac{1}{\rho }-F) 
                      \int dE'E'^{2}d\mu '
                      \tilde{R}_{{\rm PAIR}}^{{\rm em}} (\frac{1}{\rho }-\bar{F})
                 -  \frac{1}{h^{3}c^{4}}
                      F
                      \int dE'E'^{2}d\mu '
                      \tilde{R}_{{\rm PAIR}}^{{\rm abs}} \bar{F} \nonumber  ,
\end{eqnarray}  
where $F\equiv f/\rho$, $m$ is the Lagrangian mass coordinate, $\mu$ is the neutrino direction cosine, as defined above, and $E$ is the neutrino energy. 
In spherical symmetry, $F=F(t,m,\mu,E)$. After the time derivative term on the left-hand side of the Boltzmann equation, the remaining 
terms correspond to the transport of neutrinos in all three dimensions of phase space: $(m,\mu,E)$. The first term corresponds to 
spatial transport of neutrinos through the stellar core layers. As a neutrino propagates through the core, its direction cosine, defined 
in spherical coordinates with respect to the outward radial vector at its position, changes. This is captured by the second term. The 
third and fourth terms capture the transport of neutrinos in angle and energy due to relativistic (in this case to $\mathcal{O}(v/c)$) angular 
aberration and frequency shift, respectively. On the right-hand side, the collision term includes (1) thermal emission, with emissivity, $j$, 
(2) absorption, with absorption opacity $\tilde{\chi}\equiv j+\chi$, which accounts for stimulated absorption, (3) iso-energetic scattering, with 
scattering kernel $R_{\rm IS}$, (4) non-isoenergetic scattering, with scattering kernel, $R_{\rm NIS}$, and (5) neutrino pair creation and
annihilation, with pair-production kernel, $R_{\rm PAIR}$. The distribution function for antineutrinos are designated by $\bar{F}$. While the
left-hand side of the Boltzmann equation is linear in the distribution functions, it is important to note that the right-hand side is not. The 
nonlinearity on the right-hand side is evident due to the blocking factors corresponding to the boundedness of the neutrino distribution
functions: $f$ lies in the range $[0,1]$. There is an additional nonlinearity that is implicit in the equation. The distribution functions are updated
together with the matter internal energy and electron fraction, due to energy and lepton number exchange between the neutrinos and 
the matter as a result of the above processes. In turn, the neutrino emissivity, opacity, and scattering kernels depend on the thermodynamic
state of the matter, which depends on the matter's density, internal energy, and electron fraction. Thus, a simultaneous linearization of 
the discretized equations of neutrino radiation hydrodynamics in the neutrino distribution functions, the matter internal energy, and the 
matter electron fraction is required.

The finite differencing of the time derivative of the neutrino distribution
function in Eq.~\eqref{eq:boltzeq} is straightforward:
\begin{equation}
\label{eq:eq_ct_fd}
\frac{\partial F}{\partial t}=\frac{F_{i',j',k'}-{F}_{i',j',k'}^{n}}{dt}.
\end{equation}
For simplicity, we define the zone-center indices for each of the phase space dimensions with primed indices:
$i^{'}\equiv i+1/2$, $j^{'}\equiv j+1/2$, and $k^{'}\equiv k+1/2$.
Focusing now on the spatial advection term, the first of the $\mathcal{O}(1)$ terms:
In the free streaming limit,
the advected neutrino number in a time step (as measured by a
comoving observer) can be large relative to the neutrino number
in a zone (mass shell). Upwind differencing of the advection term is appropriate
to limit destabilizing errors in the fluxes. For discrete
direction cosines, \( \mu _{j'} \), the direction of the neutrino ``wind''
is given by the sign of \( \mu _{j'} \). 
On the other hand, in diffusive conditions,
the neutrino flux may be orders of magnitude smaller than
the nearly isotropic neutrino density in a zone.
In this situation, an asymmetric differencing 
can lead to an overestimation of the first angular
moment because of improper cancellations among the contributions of
the nearly isotropic neutrino radiation field. As a result, Mezzacappa et~al.\ interpolate between
upwind differencing in free streaming regimes 
and centered differencing in diffusive
regimes. 
Specifically, using the coefficients, \( \beta _{i,k'} \), defined as 
\begin{equation}
\label{eq:eq_transport_coefficients_fd}
\beta _{i,k'}=\left\{ \begin{array}{cc}
1/2 & {\rm if}\quad 2dr_{i}>\lambda _{i,k'},\\
\left( 2dr_{i}/\lambda _{i,k'}+1\right) ^{-1} & {\rm otherwise},
\end{array}\right. 
\end{equation}
where $\lambda_{i,k}$ is the angle-averaged neutrino mean free path, the spatial advection term is discretized as 
\begin{equation}
\label{eq:eq_fd_da}
\mu\frac{\partial r^{2}\rho F}{\partial m}=\frac{\mu _{j'}}{dm_{i'}}\left[ 4\pi r^{2}_{i+1}\rho _{i+1}F_{i+1,j',k'}
-4\pi r^{2}_{i}\rho _{i}F_{i,j',k'}\right] 
\end{equation}
with
\begin{equation}
\rho _{i}F_{i,j',k'}=\beta _{i,k'}\rho _{i'-1}F_{i'-1,j',k'}+\left( 1-\beta _{i,k'}\right)\rho _{i'}F_{i',j',k'}
\label{eq:eq_Fi_interpolation_out}
\end{equation}
for outward propagating neutrinos \( \left( \mu _{j'}>0\right)  \)
and
\begin{equation}
\label{eq:eq_Fi_interpolation_in}
\rho _{i}F_{i,j',k'}=\left( 1-\beta _{i,k'}\right)\rho _{i'-1}F_{i'-1,j',k'}+\beta _{i,k'}\rho _{i'}F_{i',j',k'}
\end{equation}
for inward propagating neutrinos \( \left( \mu _{j'}<0\right)  \).

Next, focusing on the angular advection term, Mezzacappa et~al.\ use the following discretization:
\begin{equation}
\label{eq:eq_dmu_fd}
\frac{\partial [(1-\mu^{2})F]}{r\partial\mu} =\frac{3\left[ r^{2}_{i+1}-r^{2}_{i}\right] }{2\left[ r^{3}_{i+1}-r^{3}_{i}\right]} \frac{1}{w_{j'}}\left( \zeta _{j+1}F_{i',j+1,k'}-\zeta _{j}F_{i',j,k'}\right) .
\end{equation}
 The differencing of the coefficients, \( \zeta =1-\mu ^{2} \), is
defined by
\begin{equation}
\label{eq:eq_def_angular_diff_coff}
\zeta _{j+1}-\zeta _{j}=-2\mu _{j'}w_{j'},
\end{equation}
where the $w_{j'}$ are the weights corresponding to the Gaussian quadrature values used for $\mu_{j'}$. 
The discretization of the coefficient, $1/r$, of the angular advection term is set such that in an infinite homogenous 
medium in thermal equilibrium, $\rho F= f_{\rm eq} =$ constant is a solution \citep{MeBr93b}.
The angular integration of the term $\partial 
[(1-\mu^{2})F]/r\partial\mu$ produces
the zeroth and second angular moments of the neutrino distribution
function. Its finite difference representation is therefore not as
sensitive to cancellations in the diffusive limit as the 
differencing of the spatial advection term.
Upwind differencing is justified. The angular ``wind''
always points towards \( \mu =1 \). However, for reasons of completeness
and consistency, Mezzacappa et~al.\ use centered differencing in the diffusive
regime here as well, with angular coefficients, \( \gamma 
_{i',k'}\equiv \beta _{i',k'} \), and
\begin{equation}
\label{eq:eq_Fj_interpolation}
F_{i',j,k'}=\gamma _{i',k'}F_{i',j'-1,k'}+\left( 1-\gamma _{i',k'}\right) F_{i',j',k'}.
\end{equation}

Finally, Mezzacappa et~al.\ discretize the last of the $\mathcal{O}(1)$ terms in the Boltzmann equation, the collision term, as
\begin{eqnarray} 
          & & \frac{j^{n+1}_{i^{'},k^{'}}}{\rho^{n+1}_{i^{'}}}-\tilde {\chi }^{n+1}_{i^{'},k^{'}} \, F_{i^{'},j^{'},k^{'}} 
\nonumber \\ 
                & + & \frac{1}{ch^{3}c^{3}}\, E_{k^{'}}^{2}\sum_{l=1}^{jmax}w_{l^{'}}\, 
                      (R_{\rm IS})^{n+1}_{i^{'},j^{'},l^{'},k^{'}}
                   \, F_{i^{'},l^{'},k^{'}}
                 -  \frac{1}{ch^{3}c^{3}}\, E_{k^{'}}^{2}\, F_{i^{'},j^{'},k^{'}}
                      \sum_{l=1}^{jmax}w_{l^{'}}\, (R_{\rm IS})^{n+1}_{i^{'},j^{'},l^{'},k^{'}}
\nonumber \\
                & + & \frac{1}{ch^{3}c^{3}}\,(1/\rho^{n+1}_{i^{'}}-F_{i^{'},j^{'},k^{'}})
                                             \sum_{m=1}^{kmax} \Delta E_{m^{'}} E_{m^{'}}^{2} 
                                             \sum_{l=1}^{jmax}w_{l^{'}}\, 
                 \times                    (\tilde{R}^{\rm in}_{\rm NIS})^{n+1}_{i^{'},j^{'},l^{'},k^{'},m^{'}}\, 
                                             F_{i^{'},l^{'},m^{'}} 
\nonumber \\
                & - & \frac{1}{ch^{3}c^{3}}\,F_{i^{'},j^{'},k^{'}}
                                             \sum_{m=1}^{kmax} \Delta E_{m^{'}} E_{m^{'}}^{2}
                                             \sum_{l=1}^{jmax}w_{l^{'}}\,
                 \times                    (\tilde{R}^{\rm out}_{\rm NIS})^{n+1}_{i^{'},j^{'},l^{'},k^{'},m^{'}}\, 
                                             (1/\rho^{n+1}_{i^{'}}-F_{i^{'},l^{'},m^{'}})
\nonumber \\ 
                & + & \frac{1}{ch^{3}c^{3}}\,(1/\rho^{n+1}_{i^{'}}-F_{i^{'},j^{'},k^{'}})
                                             \sum_{m=1}^{kmax} \Delta E_{m^{'}} E_{m^{'}}^{2}
                                             \sum_{l=1}^{jmax}w_{l^{'}}\,
                 \times                  (\tilde{R}^{\rm em}_{\rm PAIR})^{n+1}_{i^{'},j^{'},l^{'},k^{'},m^{'}}\,
                                             (1/\rho^{n+1}_{i^{'}}-\bar{F}_{i^{'},l^{'},m^{'}})
\nonumber \\
                & - & \frac{1}{ch^{3}c^{3}}\,F_{i^{'},j^{'},k^{'}}
		\sum_{m=1}^{kmax} \Delta E_{m^{'}} E_{m^{'}}^{2}
                                             \sum_{l=1}^{jmax}w_{l^{'}}\,
                 \times                    (\tilde{R}^{\rm abs}_{\rm PAIR})^{n+1}_{i^{'},j^{'},l^{'},k^{'},m^{'}}\,
                                             \bar{F}_{i^{'},l^{'},m^{'}}
\label{eq:collfd}
\end{eqnarray}  
It is important to note that the collision term is differenced implicitly with respect to time. All of the neutrino and 
antineutrino distribution functions in Eq.~\eqref{eq:collfd} are evaluated at the new time step. Given the implementation
of discrete ordinates in angle, the angular integrals in the collision term are evaluated with Gaussian quadrature, 
using the same quadrature set used for the angular discretizations of the distribution function and terms 
on the left-hand side of the Boltzmann equation.

\subsubsection{Challenges: relativistic effects and the simultaneous conservation of lepton number and energy}
\label{sec:relativisticEffectsAndConservationOfEnergy}

Define
\begin{eqnarray}
J^{N} & = & \int ^{1}_{-1}\int ^{\infty }_{0}FE^{2}dEd\mu 
,\\\label{eq:JN}
H^{N} & = & \int ^{1}_{-1}\int ^{\infty }_{0}FE^{2}dE\mu d\mu 
\label{eq:HN}.
\end{eqnarray}
\( J^{N} \) and \( H^{N} \) are the zeroth and first
angular \emph{number} moments of the distribution function.
Integration of Eq.~\eqref{eq:boltzeq} over $\mu$ and $E$ with 
 $E^{2}$ as the measure of integration gives the following evolution 
 equation for $J^{N}$:
\begin{equation}
\label{eq:eq_neutrino_number_conservation}
\frac{\partial J^{N}}{\partial t}+\frac{\partial }{\partial m}\left[ 4\pi r^{2}\rho H^{N}\right] -\int \frac{j}{\rho }E^{2}dEd\mu +
\int \chi FE^{2}dEd\mu =0.
\end{equation}
One more integration
over rest mass \( m \) from the center of the star to its surface
gives the evolution equation for the total neutrino (lepton) number. It is clear from Eq.~\eqref{eq:eq_neutrino_number_conservation} 
that the total neutrino (lepton) number
in the computational domain will change only as a result of an inflow or an outflow of neutrinos at the boundary of the domain and/or  as a 
result of the exchange
of lepton number between the neutrinos and the matter.
Now, in the same way, define the energy moments:
\begin{eqnarray}
J^{E} & = & \int FE^{3}dEd\mu , \\\label{eq:JE}
H^{E} & = & \int FE^{3}dE\mu d\mu , \\\label{eq:HE}
K^{E} & = & \int FE^{3}dE\mu ^{2}d\mu . \label{eq:KE}
\end{eqnarray}
By taking the zeroth and first angular moments of the energy moment 
($\int E^{3}dE\{ \partial F/\partial t = O[F]\}$) of the Boltzmann equation, 
the latter weighted by the fluid velocity, $v$,---i.e., $\int 
E^{3}dEd\mu \{ \partial F/\partial t = O[F]\}$
and $v\int E^{3}dEd\mu \mu \{ \partial F/\partial t = O[F]\}$---
one obtains two equations:
\begin{eqnarray}
\label{eq:eq_radiation_energy}
\frac{\partial J^{E}}{\partial t} & + & \frac{\partial }{\partial m}\left[ 4\pi r^{2}\rho 
H^{E}\right] -\left( \frac{\partial \rm{\ln} \rho }{\partial t}
+\frac{2v}{r}\right) K^{E}+\frac{v}{r}\left( J^{E}-K^{E}\right) \nonumber \\
 & - & \int \frac{j}{\rho }E^{3}dEd\mu +\int \chi FE^{3}dEd\mu 
 =0,
\end{eqnarray}
and
\begin{eqnarray}
\label{eq:eq_radiation_momentum} 
v\frac{\partial H^{E}}{\partial t} & + & \frac{\partial }{\partial m}\left[ 4\pi r^{2}v\rho K^{E}\right] 
-4\pi r^{2}\rho \frac{dv}{dm}K^{E}-\frac{v}{r}\left( J^{E}-K^{E}\right)
\nonumber\\
& - & v\left( \frac{\partial \rm{\ln} \rho }{\partial t}+\frac{2v}{r}\right) H^{E}
  +  v\int \chi FE^{3}dE\mu d\mu =0.
\end{eqnarray}
Eq.~\eqref{eq:eq_radiation_energy} is the evolution equation for the comoving-frame neutrino energy per gram. 
Eq.~\eqref{eq:eq_radiation_momentum} is the evolution equation for the comoving-frame neutrino momentum per gram.
Combining the two, to $\mathcal{O}(v/c)$, one obtains the laboratory-frame neutrino energy conservation equation:
\begin{eqnarray}
0 & = & \frac{\partial }{\partial t}\left( J^{E}+vH^{E}\right) +\frac{\partial }{\partial m}\left[ 4\pi r^{2}\rho \left( vK^{E}+H^{E}\right) \right] \nonumber \\
 & - & \int \frac{j}{\rho }E^{3}dEd\mu +\int \chi FE^{3}dEd\mu +v\int \chi FE^{3}dE\mu d\mu.
\label{eq:eq_ovc_radiation_energy_conservation} 
\end{eqnarray}
Note that $J^{E}+vH^{E}$ is the laboratory-frame neutrino energy per gram as 
expressed in terms of the comoving-frame moments $J^{E}$ and $H^{E}$.
Similarly, $vK^{E}+H^{E}$ is the laboratory-frame flux per gram expressed in terms of comoving-frame moments.
Integration of Eq.~\eqref{eq:eq_ovc_radiation_energy_conservation}
over enclosed mass leads to an equation for total neutrino energy conservation.
It is clear that, with the exception again of fluxes at the boundary of the computational
domain and energy exchange with the matter due to collisions (the terms involving 
$j$ and $\chi$) and neutrino stress (the term involving $v\chi$), the total neutrino energy as defined in 
the laboratory frame (where one can speak of conservation of energy) is conserved.

In arriving at Eq.~\eqref{eq:eq_ovc_radiation_energy_conservation}, the expressions $(\partial \rm{ln}\rho /\partial t +2v/r)K^{E}$
and $K^{E}4\pi r^{2}\rho \partial v/\partial m$ 
in Eqs.~\eqref{eq:eq_radiation_energy} and \eqref{eq:eq_radiation_momentum}
cancel given the continuity equation
\begin{equation}
\frac{\partial \rm{ln} \rho }{\partial t}+\frac{2v}{r}=-4\pi r^{2}\rho \frac{\partial v}{\partial m}.
\label{eq:continuity}
\end{equation}
To achieve global energy conservation in the discrete limit, one must ensure,  
these cancellations occur in the finite differencing as well. Identifying the origin
of the terms $(\partial \rm{ln}\rho /\partial t +2v/r)K^{E}$ and \( K^{E}4\pi r^{2}\rho \partial v/\partial m \),
we find that $(\partial \rm{ln}\rho /\partial t +2v/r)K^{E}$ originates from the zeroth 
moment of the energy advection term,
\begin{equation}
\left[ \mu ^{2}\left( \frac{\partial \rm{ln} \rho }{\partial t}+\frac{2v}{r}\right) -\left( 1-\mu ^{2}\right) \frac{v}{r}\right] \frac{1}{E^{2}}\frac{\partial }{\partial E}\left( E^{3}F\right) ,
\label{eq:eq_observer_frequency}
\end{equation}
in the Boltzmann equation \eqref{eq:boltzeq}, 
and \( K^{E}4\pi r^{2}\rho \partial v/\partial m \) originates from 
the first moment of the spatial advection term, 
\begin{equation}
 \mu \frac{\partial \left( 4\pi r^{2}\rho F\right)}{\partial m},
 \label{eq:spatialadvection}
 \end{equation}
in the same equation.
The terms \( \left( J^{E}-K^{E}\right) v/r \)
also stem from the zeroth moment of the energy advection term, Eq.~\eqref{eq:eq_observer_frequency},
and the first moment of the angular advection term
\begin{equation}
\frac{1}{r}\frac{\partial \left[ \left( 1-\mu ^{2}\right) F\right]}{\partial \mu}
\label{eq:angularadvection}
\end{equation}
in the Boltzmann equation \eqref{eq:boltzeq}. The requirement of global energy
conservation in the laboratory frame therefore imposes interdependencies on
the finite differencing of the $\mathcal{O}(1)$ spatial and angular advection terms, Eqs.~\eqref{eq:spatialadvection} and \eqref{eq:angularadvection},
and the $\mathcal{O}(v/c)$ energy advection term, Eq.~\eqref{eq:eq_observer_frequency} \citep{LiMeMe04}.
In particular, given a choice of finite differencing of the $\mathcal{O}(1)$ terms on the left-hand side of the Boltzmann equation 
\eqref{eq:boltzeq}, conservation of energy requires ``matched'' finite differencing for the coefficients
\begin{equation}
A \equiv 
\frac{\partial \rm{ln} \rho }{\partial 
t}+\frac{2v}{r}
\label{eq:Adef}
\end{equation}
and
\begin{equation}
B \equiv (1-\mu^{2})\frac{v}{r}
\label{eq:Bdef}
\end{equation}
of the $\mathcal{O}(v/c)$ advection terms in the same equation.

Mezzacappa et~al.\ begin by multiplying the discrete representation of the 
$\mathcal{O}(1)$ terms on the left-hand side of the Boltzmann equation \eqref{eq:boltzeq} by $1+\mu\bar{v}_{i+1}$
(in what follows, unless otherwise specified the indices are $i'$, $j'$, and $k'$):
\begin{eqnarray}
\label{eq:discrete1}
& &(1+\mu\bar{v}_{i+1})E\frac{F-\bar{F}}{cdt}
+(1+\mu\bar{v}_{i+1})E\frac{4\pi\mu}{dm}[\bar{r}^{2}_{i+1}\bar{\rho}_{i+1}F_{i+1}-\bar{r}^{2}_{i}\bar{\rho}_{i}F_{i}]
\\ \nonumber
&+&(1+\mu\bar{v}_{i+1})E\frac{3(\bar{r}^{2}_{i+1}-\bar{r}^{2}_{i})}{2(\bar{r}^{3}_{i+1}-\bar{r}^{2}_{3})}\frac{1}{w}[\zeta_{j+1}F_{j+1}-\zeta_{j}F_{j}]
\\ \nonumber
&=&\frac{(1+\mu v_{i+1})EF-(1+\mu\bar{v}_{i+1})E\bar{F}}{cdt}
-\frac{\mu v_{i+1}EF-\mu\bar{v}_{i+1}EF}{cdt}
\\ \nonumber
&+&\frac{4\pi\mu}{dm}[(1+\mu\bar{v}_{i+1})E\bar{r}^{2}_{i+1}\bar{\rho}_{i+1}F_{i+1}-(1+\mu\bar{v}_{i})E\bar{r}^{2}_{i}\bar{\rho}_{i}F_{i}] \\ \nonumber
& - & \frac{4\pi\mu^{2}}{dm}[\bar{v}_{i+1}\bar{r}^{2}_{i}\bar{\rho}_{i}EF_{i}-\bar{v}_{i}\bar{r}^{2}_{i}\bar{\rho}_{i}EF_{i}]
\\ \nonumber
&+&\frac{3(\bar{r}^{2}_{i+1}-\bar{r}^{2}_{i})}{2(\bar{r}^{3}_{i+1}-\bar{r}^{2}_{3})}\frac{1}{w}[\zeta_{j+1}EF_{j+1}-\zeta_{j}EF_{j}] \\ \nonumber
& + & \frac{3(\bar{r}^{2}_{i+1}-\bar{r}^{2}_{i})}{2(\bar{r}^{3}_{i+1}-\bar{r}^{2}_{3})}\bar{v}_{i+1}\frac{1}{w}[\mu\zeta_{j+1}EF_{j+1}-\mu\zeta_{j}EF_{j}]
\\ \nonumber
&=&\frac{(1+\mu v_{i+1})EF-(1+\mu\bar{v}_{i+1})E\bar{F}}{cdt}
- EF\frac{\mu v_{i+1}-\mu\bar{v}_{i+1}}{cdt} \\ \nonumber
&+&\frac{4\pi\mu}{dm}[(1+\mu\bar{v}_{i+1})E\bar{r}^{2}_{i+1}\bar{\rho}_{i+1}F_{i+1}-(1+\mu\bar{v}_{i})E\bar{r}^{2}_{i}\bar{\rho}_{i}F_{i}]
-\frac{4\pi\mu^{2}}{dm}\bar{r}^{2}_{i}\bar{\rho}_{i}EF_{i}[\bar{v}_{i+1}-\bar{v}_{i}]
\\ \nonumber
& + & \frac{3(\bar{r}^{2}_{i+1}-\bar{r}^{2}_{i})}{2(\bar{r}^{3}_{i+1}-\bar{r}^{2}_{3})}\frac{1}{w}[\zeta_{j+1}EF_{j+1}-\zeta_{j}EF_{j}]
\\ \nonumber
&+& \frac{3(\bar{r}^{2}_{i+1}-\bar{r}^{2}_{i})}{2(\bar{r}^{3}_{i+1}-\bar{r}^{2}_{3})}
\bar{v}_{i+1}\frac{1}{w}[\mu\zeta_{j+1}EF_{j+1}-\mu\zeta_{j}EF_{j}].
\end{eqnarray}
\noindent A bar over a variable indicates its value is to be taken at time step $t^n$. 
As noted, the total energy equation is obtained when summing Eqs.~\eqref{eq:eq_radiation_energy} 
and \eqref{eq:eq_radiation_momentum} and then integrating over $m$ (the integration in
$\mu$ and $E$ has already taken place). In this sequence of 
integrations (over $\mu$, $E$, and then $m$), the term involving $A$ in Eq.~\eqref{eq:eq_radiation_energy} 
cancels with the term $-4\pi r^{2}\rho K^{E}dv/dm$ in Eq.~\eqref{eq:eq_radiation_momentum}. 

Identifying the appropriate velocity gradient term
in Eq.~\eqref{eq:discrete1} and focusing on the appropriate
integration (in this case, over $m$), Mezzacappa et~al.\ require that 
[below, the term involving $A$ comes from the zeroth moment of 
first term in the observer correction \eqref{eq:eq_observer_frequency}
\emph{after an integration by parts in energy, $E$}; the term involving the velocity
gradient is the next to last term in Eq.~\eqref{eq:discrete1},
corresponding to the first moment of the spatial propagation term
in the Boltzmann equation \eqref{eq:boltzeq}]:
\begin{eqnarray}
& &\sum_{i=1,imax-1}\mu^2 A_{i'}F_{i'}dm_{i'}
\nonumber \\
&-&\sum_{i=1,imax-1}4\pi\mu^{2}\bar{r}_{i}^{2}\bar{\rho}_{i}F_{i}(\bar{v}_{i+1}-\bar{v}_{i})
\nonumber \\
&=&\sum_{i=1,imax-1}\mu^2 A_{i'}F_{i'}dm_{i'}
\nonumber \\
&-&\sum_{i=1,imax-1,j\leq jmax/2}4\pi\mu^{2}\bar{r}_{i}^{2}(\beta_{i}\bar{\rho}_{i'}F_{i'}+(1-\beta_{i})\bar{\rho}_{i'-1}F_{i'-1})(\bar{v}_{i+1}-\bar{v}_{i})
\nonumber \\
&-&\sum_{i=1,imax-1,j\geq 
jmax/2+1}4\pi\mu^{2}\bar{r}_{i}^{2}(\beta_{i}\bar{\rho}_{i'-1}F_{i'-1}+(1-\beta_{i})\bar{\rho}_{i'}F_{i'})(\bar{v}_{i+1}-\bar{v}_{i})
\nonumber \\
&=&\sum_{i=1,imax-1}\mu^2 A_{i'}F_{i'}dm_{i'}
\nonumber \\
&-&\sum_{i=1,imax-1,j\leq jmax/2}4\pi\mu^{2}\bar{r}_{i}^{2}(\bar{v}_{i+1}-\bar{v}_{i})\beta_{i}\bar{\rho}_{i'}F_{i'}
\nonumber \\
&-&\sum_{i=1,imax-2,j\leq jmax/2}4\pi\mu^{2}\bar{r}_{i+1}^{2}(\bar{v}_{i+2}-\bar{v}_{i+1})(1-\beta_{i+1})\bar{\rho}_{i'}F_{i'}
\nonumber \\
&-&\sum_{i=1,imax-1,j\geq jmax/2+1}4\pi\mu^{2}\bar{r}_{i}^{2}(\bar{v}_{i+1}-\bar{v}_{i})(1-\beta_{i})\bar{\rho}_{i'}F_{i'}
\nonumber \\
&-&\sum_{i=1,imax-2,j\geq 
jmax/2+1}4\pi\mu^{2}\bar{r}_{i+1}^{2}(\bar{v}_{i+2}-\bar{v}_{i+1})\beta_{i+1})\bar{\rho}_{i'}F_{i'}\nonumber \\
& = & 0,
\end{eqnarray}
\noindent which gives 
\begin{equation}
A_{i',k'}=4\pi\frac{\bar{\rho}_{i'}}{dm_{i'}}(\bar{r}_{i}^{2}(\bar{v}_{i+1}-\bar{v}_{i})\beta_{i,k'}
                                                          +\bar{r}_{i+1}^{2}(\bar{v}_{i+2}-\bar{v}_{i+1})(1-\beta_{i+1,k'}))
\label{eq:coeffdiff1}
\end{equation}
for $j\leq jmax/2$ and
\begin{equation}
A_{i',k'}=4\pi\frac{\bar{\rho}_{i'}}{dm_{i'}}(\bar{r}_{i}^{2}(\bar{v}_{i+1}-\bar{v}_{i})(1-\beta_{i,k'})
                                                           +\bar{r}_{i+1}^{2}(\bar{v}_{i+2}-\bar{v}_{i+1})\beta_{i+1,k'})							   
\label{eq:coeffdiff2}
\end{equation}
for $j\geq jmax/2 + 1$. (The case $i=imax-1$ is a boundary case, the details of which are not important for 
the present discussion.)

Similarly, defining $B^{'}$ according to
\begin{equation}
B_{i',j',k'}\equiv 
\frac{3}{2}\frac{\bar{r}_{i+1}^{2}-\bar{r}_{i}^{2}}{\bar{r}_{i+1}^{3}-\bar{r}_{i}^{3}}\bar{v}_{i+1}B^{'}_{j',k'},
\end{equation}
and again focusing on the appropriate integration (in this case, 
over $\mu$), Mezzacappa et~al.\ require that (below, the term involving $B'$ comes 
from the zeroth moment of the second term in brackets in the energy advection term
(\ref{eq:eq_observer_frequency}), \emph{after an integration by parts in angle, $\mu$}; 
the second term is the last term in Eq.~\eqref{eq:discrete1},
corresponding to the first moment of the angular advection term):
\begin{eqnarray}
\label{eq:coeffdiff3}
0 & = & \sum_{j=1,jmax}B^{'}_{j'}F_{j'}w_{j'}
+\sum_{j=1,jmax}\frac{2}{w_{j'}}[\mu_{j'}\alpha_{j+1}F_{j+1}-\mu_{j'}\alpha_{j}F_{j}]w_{j'} \\
&=&\sum_{j=1,jmax}B^{'}_{j'}F_{j'}w_{j'} \nonumber \\
& + & \sum_{j=1,jmax}2[\mu_{j'}\alpha_{j+1}(\gamma F_{j'}+(1-\gamma 
)F_{j'+1})
                 -\mu_{j'}\alpha_{j}  (\gamma F_{j'-1}+(1-\gamma )F_{j'})]
\nonumber \\
&=&\sum_{j=1,jmax}B^{'}_{j'}F_{j'}w_{j'}
\nonumber \\
&+&\sum_{j=1,jmax}2[\mu_{j'}\alpha_{j+1}\gamma -\mu_{j'}\alpha_{j}(1-\gamma )]F_{j'} \nonumber \\
& + & \sum_{j=2,jmax}2\mu_{j'-1}\alpha_{j}(1-\gamma )F_{j'}
+\sum_{j=1,jmax-1}(-2)\mu_{j'+1}\alpha_{j+1}\gamma F_{j'}
\nonumber \\
&=&\sum_{j=1,jmax}B^{'}_{j'}F_{j'}w_{j'} \nonumber \\
& + & \sum_{j=1,jmax}2\gamma     \alpha_{j+1}(\mu_{j'}-\mu_{j'+1})F_{j'}
+\sum_{j=1,jmax}2(1-\gamma )\alpha_{j}  
(\mu_{j'-1}-\mu_{j'})F_{j'}, \nonumber
\end{eqnarray}
which gives
\begin{equation}
B_{i',j',k'}
=\frac{3}{2}\frac{\bar{r}_{i+1}^{2}-\bar{r}_{i}^{2}}{\bar{r}_{i+1}^{3}-\bar{r}_{i}^{3}}\bar{v}_{i+1}
[2\gamma_{i',k'} 
\alpha_{j+1}\frac{\mu_{j'+1}-\mu_{j'}}{w_{j'}}
+2(1-\gamma_{i',k'} )\alpha_{j}
\frac{\mu_{j'}-\mu_{j'-1}}{w_{j'}}].
\label{eq:B}
\end{equation}

Given the necessary matched finite differencing for $A$ and $B$, Mezzacappa et~al.\ then consider the 
finite difference representation of the energy advection term (\ref{eq:eq_observer_frequency}). Using the definitions
(\ref{eq:Adef}) and (\ref{eq:Bdef}), they rewrite the equation corresponding to the change in the distribution function
due to relativistic energy advection as
\begin{equation}
0=E^{3}\left( \frac{\partial F}{\partial t}\right) _{E}+ \left( \mu ^{2}A-B\right) E\frac{\partial }{\partial E}\left[ E^{3}F\right] ,
\label{eq:eq_lagrangian_energy_derivative}
\end{equation}
and then solve it analytically.
To solve Eq.~\eqref{eq:eq_lagrangian_energy_derivative}, Mezzacappa et~al.\ write the prefactor of the energy derivative
as the time derivative of the quantity 
\begin{equation}
R_{f}=r^{3\mu^{2}-1}\rho^{\mu^2};
\end{equation}
i.e.,
\begin{equation}
\frac{\partial \rm{ln} R_{f}}{\partial t}=\mu ^{2}A-B.
\end{equation}
They then transform from the {}``Eulerian''
variable \( x=E \) to the {}``Lagrangian'' variable \( y=E/R_{f} \), and in so doing they transform Eq.~\eqref{eq:eq_lagrangian_energy_derivative}:
\begin{eqnarray}
0 & = & \left( \frac{\partial }{\partial t}\left[ E^{3}F\right] \right) _{E}+\frac{\partial R_{f}}{R^{2}_{f}\partial t}E\times R_{f}\frac{\partial }{\partial E}\left[ E^{3}F\right] \nonumber \\
 & = & \left( \frac{\partial }{\partial t}\left[ E^{3}F\right] \right) _{E}-\left( \frac{\partial \left[ E/R_{f}\right] }{\partial t}\right) _{E}\frac{\partial \left[ E^{3}F\right] }{\partial \left[ E/R_{f}\right] } \nonumber \\
 & = &\left( \frac{\partial }{\partial t}\left[ E^{3}F\right] \right) _{E/R_{f}}.
\end{eqnarray}
 For a small section of energy phase space \( E^{2}\Delta E=\left( E^{3}_{2}-E^{3}_{1}\right) /3 \),
this relationship leads to \begin{equation}
\left( \frac{\partial }{\partial t}\left[ E^{2}F\Delta E\right] \right) _{E/R_{f}}=0,
\label{eq:eq_bunch_enumber_evolution}
\end{equation}
which has the following interpretation:
The neutrinos in the energy interval
\( E^{2}\Delta E \) move
along constant \( E/R_{f} \) in the phase space of a comoving
observer. Given this, Mezzacappa et~al.\ are able to evolve 
any neutrino quantity in this phase-space interval---in particular, the neutrino specific energy,
\( d\epsilon =E^{3}F\Delta E \):
\begin{equation}
\left( \frac{\partial }{\partial t}\left[ E^{3}F\Delta E\right] \right)_{E/R_{f}}
=E^{2}F\Delta E\left( \frac{\partial E}{\partial t}\right) 
_{E/R_{f}}=\frac{\partial \rm{ln} R_{f}}{\partial t}d\epsilon .
\label{eq:eq_specific_energy_change}
\end{equation}

They then consider a neutrino energy group \( k' \),
with neighboring groups \( k'+dk \), \( dk=\pm 1 \). From Eq.~\eqref{eq:eq_bunch_enumber_evolution},
the number of neutrinos before energy advection, \( F_{i',j',k'}E^{2}_{k'}dE_{k'} \),
is equal to the number of neutrinos after advection. The 
distribution of these neutrinos in energy
after the advection yields a diminished number of neutrinos
\( F_{i',j',k'}E_{k'}^{2}dE_{k'}-n_{i',j',k'}^{-} \) in group \( k' \)
and an additional number of neutrinos \( n_{i',j',k'+dk}^{+} \) in
the neighboring group \( k'+dk \) such that
\begin{equation}
F_{i',j',k'}E_{k'}^{2}dE_{k'}-\left[ \left( F_{i',j',k'}E_{k'}^{2}dE_{k'}-n^{-}_{i',j',k'}\right) +n^{+}_{i',j',k'+dk}\right] =0.
\label{eq:eq_bunch_enumber_fd}
\end{equation}
 Eq.~\eqref{eq:eq_specific_energy_change} defines a similar 
 correction
for the specific neutrino energy in group $k'$:
\begin{eqnarray}
F_{i',j',k'}E_{k'}^{3}dE_{k'} & - & \left[ \left( F_{i',j',k'}E_{k'}^{3}dE_{k'}-E_{k'}n^{-}_{i',j',k'}\right) +E_{k'+dk}n^{+}_{i',j',k'+dk}\right] \nonumber \\
 & = & -\left( \mu ^{2}_{j'}A_{i',k'}-B_{i',j',k'}\right) 
 F_{i',j',k'}E^{3}_{k'}dE_{k'}dt,
\label{eq:eq_specific_energy_change_fd} 
\end{eqnarray}
where $A_{i',k'}$ and $B_{i',j',k'}$ are given by Eqs.~\eqref{eq:coeffdiff1}, \eqref{eq:coeffdiff2}, and \eqref{eq:B}.
Equations~\eqref{eq:eq_bunch_enumber_fd} and \eqref{eq:eq_specific_energy_change_fd} can be solved for $n^{-}_{i',j',k'}$ and $n^{+}_{i',j',k'}$:
\begin{eqnarray}
n^{-}_{i',j',k'} & = & \left( \mu _{j'}^{2}A_{i',k'}-B_{i',j',k'}\right) \frac{dE_{k'}}{E_{k'+dk}-E_{k'}}E^{3}_{k'}F_{i',j',k'}dt , \nonumber \\
n^{+}_{i',j',k'} & = & 
n_{i',j',k'-dk}^{-},\label{eq:eq_oe_deltaplusminus}
\end{eqnarray}
which leads, given the change in the neutrino distribution function in group $k'$ due to energy advection can be expressed as
\begin{equation}
F_{i',j',k'}=F_{i',j',k'}^{n}+\left( n^{+}_{i',j',k'}-n^{-}_{i',j',k'}\right) /\left( E_{k'}^{2}dE_{k'}\right),
\end{equation}
to the following finite difference representation of the energy advection term in the Boltzmann equation \eqref{eq:boltzeq}:
\begin{eqnarray}
 \frac{1}{E^{2}_{k'}dE_{k'}}
   \left[ \left( \mu _{j'}^{2}A_{i',k'-dk}-B_{i',j',k'}\right) \frac{dE_{k'-dk}}{E_{k'}-E_{k'-dk}}E_{k'-dk}^{3}F_{i',j',k'-dk}\right. \nonumber \\
  -  \left. \left( \mu _{j'}^{2}A_{i',k'}-B_{i',j',k'}\right) 
 \frac{dE_{k'}}{E_{k'+dk}-E_{k'}}E_{k'}^{3}F_{i',j',k'}\right] 
 .\label{eq:eq_oe_fd} 
\end{eqnarray}

Mezzacappa et~al.\ are then left with the task of finding a finite difference representation
for the angular advection term in Eq.~\eqref{eq:boltzeq}. 
Their finite differencing of the energy advection term conserved 
specific neutrino energy. Their finite differencing of the angular advection term is 
designed to conserve specific neutrino luminosity.
With \( \zeta =1-\mu ^{2} \), the angular aberration term can be 
rewritten as
\begin{equation}
(\frac{\partial F}{\partial t})_{\mu}=\left( A+B/\zeta \right) \frac{\partial }{\partial \mu }\left[ \zeta \mu F\right].
\label{eq:aberration}
\end{equation}
As before, Mezzacappa et~al.\ seek an analytic solution to Eq.~\eqref{eq:aberration}.  
To do so, they convert the prefactor of the angular derivative to a time derivative. 
For the quantity 
\( R_{a}=r^{3}\rho  \), they find
\begin{equation}
\frac{d\rm{ln} R_{a}}{dt}=A+B/\zeta .
\end{equation}
They then rewrite Eq.~\eqref{eq:aberration} in terms 
of the {}``Lagrangian'' variable \( y=\zeta ^{-1/2}\mu /R_{a} \) instead 
of the {}``Eulerian'' variable \( x=\mu  \). After multiplication by 
\( \zeta \mu  \), Eq.~\eqref{eq:aberration} becomes:
\begin{eqnarray}
0 & = & \zeta \mu \left[ \left( \frac{\partial F}{\partial t}\right) _{\mu }+\alpha \left( A+B/\zeta \right) \frac{\partial }{\partial \mu }\left[ \zeta \mu F\right] \right] \nonumber \\
 & = & \left( \frac{\partial }{\partial t}\left[ \zeta \mu F\right] \right) _{\mu }+\zeta ^{-1/2}\mu \frac{\partial R_{a}}{R^{2}_{a}\partial t}\times \zeta ^{3/2}R_{a}\frac{\partial }{\partial \mu }\left[ \zeta \mu F\right] \nonumber \\
 & = & \left( \frac{\partial }{\partial t}\left[ \zeta \mu F\right] \right) 
 _{\mu }
  - \left( \frac{\partial \left[ \zeta ^{-1/2}\mu /R_{a}\right] }{\partial t}\right) _{\mu }\frac{\partial \left[ \zeta \mu F\right] }{\partial \left[ \zeta ^{-1/2}\mu /R_{a}\right] }\nonumber \\
  & = & \left( \frac{\partial }{\partial t}\left[ \zeta \mu F\right] \right) _{\zeta ^{-1/2}\mu /R_{a}}.
\end{eqnarray}
As before, the interpretation is clear: The neutrinos initially residing in 
the interval \( \left( 1-3\mu ^{2}\right) \Delta \mu =\zeta _{2}\mu _{2}-\zeta _{1}\mu _{1} \)
are shifted by angular aberration along constant \( \mu /\left( \sqrt{\zeta }R_{a}\right)  \)
in the phase space of a comoving observer:
\begin{equation}
\left( \frac{\partial }{\partial t}\left[ \left( 1-3\mu ^{2}\right) F\Delta \mu \right] \right) _{\zeta ^{-1/2}\mu /R_{a}}=0.
\label{eq:eq_bunch_lnumber_evolution}
\end{equation}
Given Eq.~\eqref{eq:eq_bunch_lnumber_evolution}, Mezzacappa et~al.\ are in turn able to evaluate the change in other neutrino quantities---in particular, the 
specific neutrino luminosity, \( d\ell =\left( 1-3\mu ^{2}\right) \mu F\Delta \mu  \):
\begin{eqnarray}
\left( \frac{\partial }{\partial t}\left[ \left( 1-3\mu ^{2}\right) \mu 
F\Delta \mu \right] \right)_{\zeta ^{-1/2}\mu /R_{a}}
& = &\left( 1-3\mu ^{2}\right) F\Delta \mu \left( \frac{\partial \mu}{\partial t}\right)_{\zeta ^{-1/2}/R_{a}}\nonumber \\
& = &\zeta \frac{\partial \rm{ln} R_{a}}{\partial t}d\ell .
\label{eq:eq_specific_luminosity_change}
\end{eqnarray}
Identifying their bin size \( \left( 1-3\mu _{j'}^{2}\right) \Delta \mu _{j'}=w_{j'} \)
with their Gaussian quadrature weights, Eq.~\eqref{eq:eq_bunch_lnumber_evolution}
leads to their condition for neutrino number conservation,
\begin{equation}
F_{i',j',k'}w_{j'}-\left[ \left( F_{i',j',k'}w_{j'}-n^{-}_{i',j',k'}\right) +n^{+}_{i',j'+dj,k'}\right] =0,
\end{equation}
and Eq.~\eqref{eq:eq_specific_luminosity_change} leads to their 
prescription for the numerical evolution of the specific luminosity,
\begin{eqnarray}
F_{i',j',k'}\mu _{j'}w_{j'} & - & \left[ \left( F_{i',j',k'}\mu _{j'}w_{j'}-\mu _{j'}n^{-}_{i',j',k'}\right) +\mu _{j'+dj}n^{+}_{i',j'+dj,k'}\right] \nonumber \\
 & = & -\left( \zeta _{j'}A_{i',k'}+B_{i',j',k'}\right) F_{i',j',k'}\mu _{j'}w_{j'}dt,
\end{eqnarray}
where \( dj=\pm 1 \). 
The change in the neutrino distribution from angular aberration is 
then
\begin{equation}
F_{i',j',k'}=F_{i',j',k'}^{n}+\left( n_{i',j',k'}^{+}-n_{i',j',k'}^{-}\right) /w_{j'},
\label{eq:fdot_aberration}
\end{equation}
with 
\begin{eqnarray}
n^{-}_{i',j',k'} & = & \left( A_{i',k'}+B_{i',j',k'}/\zeta _{j'}\right) \frac{w_{j'}}{\mu _{j'+dj}-\mu _{j'}}\zeta _{j'}\mu _{j'}F_{i',j',k'}dt, \nonumber \\
n^{+}_{i',j',k'} & = & n^{-}_{i',j'-dj,k'}.
\end{eqnarray}
This leads to the following finite difference representation of the 
angular aberration term in the Boltzmann equation \eqref{eq:boltzeq}:
\begin{eqnarray}
 \frac{1}{w_{j'}}\left[ \left( A_{i',k'}+B_{i',j'-dj,k'}/\zeta _{j'-dj}\right) \frac{w_{j'-dj}}{\mu _{j'}-\mu _{j'-dj}}\zeta _{j'-dj}\mu _{j'-dj}F_{i',j'-dj,k'}\right. \nonumber \\
  -  \left. \left( A_{i',k'}+B_{i',j',k'}/\zeta _{j'}\right) 
 \frac{w_{j'}}{\mu _{j'+dj}-\mu _{j'}}\zeta _{j'}\mu _{j'}F_{i',j',k'}\right] ,\label{eq:eq_omu_fd} 
\end{eqnarray}
where \( dj=+1 \) for \( \mu_{j'} \leq 0 \)
and \( dj=-1 \) for \( \mu_{j'} >0 \). 

Given the finite differencing for all of the terms in the Boltzmann equation \eqref{eq:boltzeq}---i.e., Eqs.~\eqref{eq:eq_ct_fd}, \eqref{eq:eq_fd_da}, \eqref{eq:eq_dmu_fd}, 
\eqref{eq:eq_omu_fd}, \eqref{eq:eq_oe_fd}, and \eqref{eq:collfd}---Mezzacappa et~al.\ solve the discretized equation as follows. With the exception of the discretized time derivative, which is a finite difference of the values of the distribution function at time step $t^{n+1}$ and $t^n$, the distribution function in all of the remaining terms is defined at time step $t^{n+1}$---i.e., Mezzacappa et~al.\ employ a fully implicit approach, including phase-space advection and collisions. Given the presence of blocking factors in the collision term and the presence of products of the distribution functions and the neutrino opacities, which are functions of the specific internal energy and electron fraction of the matter, which are updated together with the distribution functions given lepton number and energy exchange with the matter through collisions [see Equations (\ref{eq:fluidFourMomentumConservation}), (\ref{eq:ElectronNumberConservation}), (\ref{eq:electronfractionequationsourceterm}), and (\ref{eq:fourmomentumequationsourceterm})], linearization is necessary. Specifically, Mezzacappa et~al.\ introduce the linearizations
\begin{eqnarray}
F_{i',j',k'} & = & F^{0}_{i',j',k'}+\delta F_{i',j',k'}, \\ \label{eq:linearizationF}
\epsilon_{i'} & = & \epsilon^{0}_{i'}+\delta\epsilon_{i'}, \\ \label{eq:linearizationepsilon}
(Y_e)_{i'} & = & (Y_e)^{0}_{i'}+\delta (Y_e)_{i'}, \label{eq:linearizationYe}
\end{eqnarray}
where a $0$ superscript denotes the value of the variable at the current iterate in an outer Newton iteration of the solution algorithm. Given the dependence of 
$j$, $\tilde{\chi}$, $R_{\rm IS}$, $R_{\rm NIS}$, and $R_{\rm PAIR}$ on $\rho$, $T$, and $Y_e$, the above linearizations lead to linearizations in all of these 
quantities. For example:
\begin{equation}
j_{i',k'}=j^{0}_{i',k'}+\left[\left(\frac{\partial j}{\partial T}\right)_{\rho,Y_e}\right]^{0}_{i',k'}+\left[\left(\frac{\partial j}{\partial Y_e}\right)_{\rho,T}\right]^{0}_{i',k'}.
\end{equation}
Insertion of these linearizations into the finite differenced Boltzmann equation leads to a block-tridiagonal linear system of equations for the quantities $\delta F_{i',j',k'}$, $\delta\epsilon_{i'}$, and $\delta (Y_e)_{i'}$, which is solved for each outer iteration until a prescribed tolerance is reached for all of the variables. The block tridiagonal system has the form
\begin{equation}
-\mathbf{C}_{i}\mathbf{V}_{i-1}+\mathbf{A}_{i}\mathbf{V}_{i}-\mathbf{B}_{i+1}\mathbf{V}_{i+1}=\mathbf{U}_{i},
\label{eq:blocktridiag}
\end{equation}
where $\mathbf{B}_{i}$ and $\mathbf{C}_{i}$ are diagnoal, reflecting the fact that spatial advection couples nearest neighbors only, 
and where $\mathbf{A}_{i}$ has the form
\begin{equation}
\mathbf{A}_{i}=
\left(
\begin{array}{cc}
A_{1} & A_{2} \\
A_{3} & A_{4}  
\end{array}
\right).
\label{eq:matrixA}
\end{equation}
$\mathbf{A}_{i}$ is an $M\times M$ matrix, where $M=jmax \times kmax +2$. $jmax$ corresponds to the number of angular quadratures used
in the discrete ordinates implementation for angle, and $kmax$ corresponds to the number of energy groups. The submatrices $\mathbf{A}_{2}$
and $\mathbf{A}_{3}$ are of dimension $2\times (M-2)$ and $(M-2)\times 2$, respectively. $\mathbf{A}_{4}$ is a $2\times 2$ matrix. The 2 rightmost 
columns of $\mathbf{A}_{i}$ and the 2 bottom-most rows correspond to the coupling of the Boltzmann equation to the equations for the specific 
internal energy and electron fraction of the matter, accounting for energy and lepton number exchange. The solution vector, $\mathbf{V}_{i}$, comprising the quantities
 $\delta F_{i',j',k'}$, $\delta\epsilon_{i'}$, and $\delta (Y_e)_{i'}$, has the form
 \begin{equation}
 \left(
 \begin{array}{c}
 \delta F_{i',1',1'} \\
 \delta F_{i',2',1'} \\
   . \\
   . \\
   . \\
 \delta F_{i',1',2'} \\
 \delta F_{i',2',2'} \\
    . \\
    . \\
    . \\
 \delta\epsilon_{i'} \\
 \delta (Y_e)_{i'}
\end{array}     
 \right).
 \label{eq:solnvector}
 \end{equation}
\citep{DaMeMe05} developed a physics-based preconditioner for the above system. This ``ADI-like'' preconditioner treats the diagonal dense blocks, which correspond to coupling in momentum space, and the tridiagonal bands, which correspond to coupling in space, separately, and has proven very effective.

For Mezzacappa et al., neutrino momentum exchange with the matter is handled separately, during the hydrodynamics update, and is differenced explicitly in time.

\subsubsection{Challenges: neutrino--nucleon (small-energy) scattering}

In the case of neutrino--electron scattering, for example, where the energy transfer is not small in comparison with the widths of the zones of our energy grid, Eq.~\eqref{eq:boltzeq} is differenced using zone-centered values of energy in both the neutrino distribution function and the scattering kernels. However, for small-energy transfers compared with our energy zone widths, the scattering kernel $R_{\rm NNS}^{\rm in/out}(\epsilon_{k}, \epsilon_{k'}, \cos\theta )$ will be effectively zero if $\epsilon_{k} \ne \epsilon_{k'}$, and the scattering will become essentially isoenergetic, with negligible energy transfer. As already discussed, while the transfer of energy between neutrinos and nucleons during a scattering event is small, there are many such scatterings, and the overall impact of the energy exchange between the neutrinos and nucleons in these events is nonnegligible. Thus, a numerical treatment of this scattering contribution that reflects the fact that the energy exchange between neutrinos and matter is important and, more important, captures this exchange accurately, must be developed.

Focusing on this term in the collision term, we have
\begin{eqnarray}
& {\ds \pderiv{ f(\mu, \epsilon) }{ t } 
= [ 1 - f(\mu, \epsilon) ] \frac{1}{(hc)^{3}} \int_{0}^{\infty} \epsilon'^{2} d\epsilon'
\int_{-1}^{1} d\mu' f(\mu', \epsilon') \int_{0}^{2\pi} d\beta' R_{\rm NNS}^{\rm in}(\epsilon, 
\epsilon', \cos\theta )
} & \nonumber \\ 
& {\ds - f(\mu, \epsilon) \frac{1}{(hc)^{3}} \int_{0}^{\infty} \epsilon'^{2} d\epsilon'
\int_{-1}^{1} d\mu' [1 - f(\mu', \epsilon') ] \int_{0}^{2\pi} d\beta' R_{\rm NNS}^{\rm out}(\epsilon, 
\epsilon', \cos\theta ),
} &  \label{eq:b1}
\end{eqnarray}
where, for simplicity, we have suppressed the radial and temporal dimensions. 
With the energy zone centers, $\epsilon_{k+1/2}$, defined in terms of the energy zone edges, $\epsilon_{k}$, as
\begin{equation}
\epsilon_{k+\frac{1}{2}} = \frac{1}{3} [ \epsilon_{k}^{2} + \epsilon_{k}\epsilon_{k+1} +
\epsilon_{k+1}^{2} ],
\label{eq:b2}
\end{equation}
the volume of an energy zone is given by
\begin{equation}
4\pi \epsilon_{k+\frac{1}{2}}^{2} \Delta \epsilon_{k+\frac{1}{2}} 
= \frac{4\pi}{3} [ \epsilon_{k+1}^{3} - \epsilon_{k}^{3} ],
\label{eq:b3}
\end{equation}
where
\begin{equation}
\Delta \epsilon_{k+\frac{1}{2}} = \epsilon_{k+1} - \epsilon_{k}.
\label{eq:b4}
\end{equation}
The integral over energy can now be replaced by
\begin{equation}
\int_{0}^{\epsilon_{N+1}} \epsilon^{2} d\epsilon = \sum_{k=1}^{N} \epsilon_{k+\frac{1}{2}}^{2} \Delta
\epsilon_{k+\frac{1}{2}},
\label{eq:b4}
\end{equation}
and Eq.~\eqref{eq:b1} becomes 
\begin{eqnarray}
& {\ds \left. \pderiv{ f(\mu, \epsilon) }{ t } \right|_{\rm scat} 
} 
& \nonumber \\ 
& {\ds \simeq [ 1 - f(\mu, \epsilon) ] \frac{1}{(hc)^{3}} \int_{0}^{\epsilon_{N+1}} \epsilon '^{2}
d\epsilon'
\int_{-1}^{1} d\mu' f(\mu', \epsilon') \int_{0}^{2\pi} d\beta' R_{\rm NNS}^{\rm in}(\epsilon, 
\epsilon', \cos\theta )
} & \nonumber \\ 
& {\ds - f(\mu, \epsilon) \frac{1}{(hc)^{3}} \int_{0}^{\epsilon_{N+1}} \epsilon'^{2} d\epsilon'
\int_{-1}^{1} d\mu' [1 - f(\mu', \epsilon') ] \int_{0}^{2\pi} d\beta' R_{\rm NNS}^{\rm out}(\epsilon, 
\epsilon', \cos\theta )
} & \nonumber \\ 
& {\ds = [ 1 - f(\mu, \epsilon) ] \frac{1}{(hc)^{3}} \sum_{k'=1}^{N}
\int_{\epsilon_{k'}}^{\epsilon_{k'+1}} {\epsilon'}^{2} d\epsilon'
\int_{-1}^{1} d\mu' f(\mu', \epsilon') \int_{0}^{2\pi} d\beta' R_{\rm NNS}^{\rm in}(\epsilon, 
\epsilon', \cos\theta )
} & \nonumber \\ 
& {\ds - f(\mu, \epsilon) \frac{1}{(hc)^{3}} \sum_{k'=1}^{N} \int_{\epsilon_{k'}}^{\epsilon_{k'+1}} 
\epsilon'^{2} d\epsilon'
\int_{-1}^{1} d\mu' [1 - f(\mu', \epsilon') ] \int_{0}^{2\pi} d\beta' R_{\rm NNS}^{\rm out}(\epsilon, 
\epsilon', \cos\theta )
} & \nonumber \\ 
& {\ds = [ 1 - f(\mu, \epsilon) ] \frac{1}{(hc)^{3}} \sum_{k'=1}^{N} 
\epsilon_{k'+\frac{1}{2}}^{2} \Delta \epsilon_{k'+\frac{1}{2}}
\frac{1}{ \epsilon_{k'+\frac{1}{2}}^{2} \Delta \epsilon_{k'+\frac{1}{2}} }
\int_{\epsilon_{k'}}^{\epsilon_{k'+1}} {\epsilon'}^{2} d\epsilon'
} & \nonumber \\ 
& {\ds \times \int_{-1}^{1} d\mu' f(\mu', \epsilon') \int_{0}^{2\pi} d\beta' 
R_{\rm NNS}^{\rm in}(\epsilon,  \epsilon', \cos\theta )
} & \nonumber \\ 
& {\ds - f(\mu, \epsilon) \frac{1}{(hc)^{3}} \sum_{k'=1}^{N} 
\epsilon_{k'+\frac{1}{2}}^{2} \Delta \epsilon_{k'+\frac{1}{2}}
\frac{1}{ \epsilon_{k'+\frac{1}{2}}^{2} \Delta \epsilon_{k'+\frac{1}{2}} } 
\int_{\epsilon_{k'}}^{\epsilon_{k'+1}}{\epsilon'}^{2} d\epsilon'
} & \nonumber \\ 
& {\ds \times \int_{-1}^{1} d\mu' [1 - f(\mu', \epsilon') ] \int_{0}^{2\pi} d\beta' 
R_{\rm NNS}^{\rm out}(\epsilon, \epsilon', \cos\theta ).
} &  
\label{eq:b5}
\end{eqnarray}
In Eq.~\eqref{eq:b5}, the first approximation was made by truncating the energy integral
at $\epsilon_{N+1}$. In the second equality, the integral over the entire energy domain is 
broken up into segments within the domain, corresponding to the energy zone widths. This 
is not an approximation. In the last equality, we have inserted a factor of unity 
inside the summation over energy groups, which, again, is not an approximation. Therefore, 
no approximations have been made thus far except for truncating the range of the energy 
integration.

The ultimate goal of an improved treatment of small-energy, neutrino--nucleon scattering is
to accurately compute the energy transfer between the neutrinos and the nucleons---i.e., 
to compute accurately the change in the neutrino energy within each of the groups of our 
energy grid from such scattering. The change in the neutrino energy within a group is given by
\begin{eqnarray}
& 
{\ds  
\left. \pderiv{ E_{k+\frac{1}{2}} }{ t } \right|_{\rm scat} 
= \frac{1}{(hc)^{3}} \int_{\epsilon_{k}}^{\epsilon_{k+1}} \epsilon^{3} d\epsilon
\left. \pderiv{ f(\mu, \epsilon) }{ t } \right|_{\rm scat} 
} & \nonumber \\ 
& {\ds = \frac{1}{(hc)^{3}} \int_{\epsilon_{k}}^{\epsilon_{k+1}} \epsilon^{3} d\epsilon
[ 1 - f(\mu, \epsilon) ] \frac{1}{(hc)^{3}} \sum_{k'=1}^{N} 
\epsilon_{k'+\frac{1}{2}}^{2} \Delta \epsilon_{k'+\frac{1}{2}}
\frac{1}{ \epsilon_{k'+\frac{1}{2}}^{2} \Delta \epsilon_{k'+\frac{1}{2}} }
\int_{\epsilon_{k'}}^{\epsilon_{k'+1}} {\epsilon'}^{2} d\epsilon'
} 
& \nonumber \\ 
& 
{\ds 
\times \int_{-1}^{1} d\mu' f(\mu', \epsilon') \int_{0}^{2\pi} d\beta' 
R_{\rm NNS}^{\rm in}(\epsilon,  \epsilon', \cos\theta )
} 
& \nonumber \\ 
& 
{\ds
 - \frac{1}{(hc)^{3}} \int_{\epsilon_{k}}^{\epsilon_{k+1}} \epsilon^{3} d\epsilon
f(\mu, \epsilon) \frac{1}{(hc)^{3}} \sum_{k'=1}^{N} 
\epsilon_{k'+\frac{1}{2}}^{2} \Delta \epsilon_{k'+\frac{1}{2}}
\frac{1}{ \epsilon_{k'+\frac{1}{2}}^{2} \Delta \epsilon_{k'+\frac{1}{2}}}
\int_{\epsilon_{k'}}^{\epsilon_{k'+1}} {\epsilon'}^{2} d\epsilon'
}& \nonumber \\ 
& {\ds \times \int_{-1}^{1} d\mu' [1 - f(\mu', \epsilon') ] \int_{0}^{2\pi} d\beta' 
R_{\rm NNS}^{\rm out}(\epsilon, \epsilon', \cos\theta ),
} 
&  
\label{eq:b7}
\end{eqnarray}
where we have inserted Eq.~\eqref{eq:b5} for the time derivative of the neutrino distribution
function due to neutrino--nucleon scattering. If we now choose to define the distribution function,
$f(\mu,\epsilon)$, at the energy zone centers, Eq.~\eqref{eq:b7} can be expressed as
\begin{eqnarray}
& {\ds  
\frac{1}{(hc)^{3}} \int_{\epsilon_{k}}^{\epsilon_{k+1}} \epsilon^{3} d\epsilon
\left. \pderiv{ f(\mu, \epsilon_{k+\frac{1}{2}}) }{ t } \right|_{\rm scat}
= \left. \pderiv{ f(\mu, \epsilon_{k+\frac{1}{2}}) }{ t } \right|_{\rm scat}
\frac{1}{(hc)^{3}} \int_{\epsilon_{k}}^{\epsilon_{k+1}} \epsilon^{3} d\epsilon 
} & \nonumber \\ 
& {\ds = \frac{1}{(hc)^{3}} \int_{\epsilon_{k}}^{\epsilon_{k+1}} \epsilon^{3} d\epsilon
[ 1 - f(\mu, \epsilon_{k+\frac{1}{2}}) ] \frac{1}{(hc)^{3}} \sum_{k'=1}^{N} 
\epsilon_{k'+\frac{1}{2}}^{2} \Delta \epsilon_{k'+\frac{1}{2}}
\frac{1}{ \epsilon_{k'+\frac{1}{2}}^{2} \Delta \epsilon_{k'+\frac{1}{2}} }
\int_{\epsilon_{k'}}^{\epsilon_{k'+1}} {\epsilon'}^{2} d\epsilon'
} & \nonumber \\ 
& {\ds 
\times \int_{-1}^{1} d\mu' f(\mu', \epsilon_{k'+\frac{1}{2}}) \int_{0}^{2\pi} d\beta' 
R_{\rm NNS}^{\rm in}(\epsilon,  \epsilon', \cos\theta )
} & \nonumber \\ 
& {\ds - \frac{1}{(hc)^{3}} \int_{\epsilon_{k}}^{\epsilon_{k+1}} \epsilon^{3} d\epsilon
f(\mu, \epsilon_{k+\frac{1}{2}}) \frac{1}{(hc)^{3}} \sum_{k'=1}^{N} 
\epsilon_{k'+\frac{1}{2}}^{2} \Delta \epsilon_{k'+\frac{1}{2}}
\frac{1}{ \epsilon_{k'+\frac{1}{2}}^{2} \Delta \epsilon_{k'+\frac{1}{2}} } 
\int_{\epsilon_{k'}}^{\epsilon_{k'+1}} {\epsilon'}^{2} d\epsilon'
} & \nonumber \\ 
& {\ds \times \int_{-1}^{1} d\mu' [1 - f(\mu', \epsilon_{k'+\frac{1}{2}}) ] \int_{0}^{2\pi} d\beta' 
R_{\rm NNS}^{\rm out}(\epsilon, \epsilon', \cos\theta )
} & \nonumber \\ 
& {\ds = 
[ 1 - f(\mu, \epsilon_{k+\frac{1}{2}}) ] \frac{1}{(hc)^{3}} \sum_{k'=1}^{N} 
\epsilon_{k'+\frac{1}{2}}^{2} \Delta \epsilon_{k'+\frac{1}{2}}
\int_{-1}^{1} d\mu' f(\mu', \epsilon_{k'+\frac{1}{2}}) \int_{0}^{2\pi} d\beta'
} & \nonumber \\ 
& {\ds \times 
\frac{1}{(hc)^{3}} \int_{\epsilon_{k}}^{\epsilon_{k+1}} \epsilon^{3} d\epsilon 
\frac{1}{ \epsilon_{k'+\frac{1}{2}}^{2} \Delta \epsilon_{k'+\frac{1}{2}} }
\int_{\epsilon_{k'}}^{\epsilon_{k'+1}} {\epsilon'}^{2} d\epsilon'
R_{\rm NNS}^{\rm in}(\epsilon,  \epsilon', \cos\theta )
} & \nonumber \\ 
& {\ds - 
f(\mu, \epsilon_{k+\frac{1}{2}}) \frac{1}{(hc)^{3}} \sum_{k'=1}^{N} 
\epsilon_{k'+\frac{1}{2}}^{2} \Delta \epsilon_{k'+\frac{1}{2}}
\int_{-1}^{1} d\mu' [1 - f(\mu', \epsilon_{k'+\frac{1}{2}}) ] \int_{0}^{2\pi} d\beta'
} & \nonumber \\ 
& {\ds \times 
\frac{1}{(hc)^{3}} \int_{\epsilon_{k}}^{\epsilon_{k+1}} \epsilon^{3} d\epsilon
\frac{1}{ \epsilon_{k'+\frac{1}{2}}^{2} \Delta \epsilon_{k'+\frac{1}{2}} } 
\int_{\epsilon_{k'}}^{\epsilon_{k'+1}} {\epsilon'}^{2} d\epsilon'
R_{\rm NNS}^{\rm out}(\epsilon, \epsilon', \cos\theta ).
} &  
\label{eq:b17}
\end{eqnarray}
The first equality in Eq.~\eqref{eq:b17} stems from the fact that, once the distribution function 
is evaluated at the energy zone center and, consequently, its time derivative is evaluated there, 
the time derivative becomes a constant integrand and can be taken outside
of the integral. Dividing both sides of Eq.~\eqref{eq:b17} by
\begin{equation}
\frac{1}{(hc)^3}\int_{\epsilon_{k}}^{\epsilon_{k+1}} \epsilon^{3} d\epsilon = \frac{1}{(hc)^3} \epsilon_{k+1/2}^{3} \Delta
\epsilon_{k+\frac{1}{2}},
\label{eq:b4e}
\end{equation}
we obtain
\begin{eqnarray}
& {\ds \left. \pderiv{ f(\mu, \epsilon_{k+\frac{1}{2}}) }{ t } \right|_{\rm scat} =
} & \nonumber \\ 
& {\ds = 
[ 1 - f(\mu, \epsilon_{k+\frac{1}{2}}) ] \frac{1}{(hc)^{3}} \sum_{k'=1}^{N} 
\epsilon_{k'+\frac{1}{2}}^{2} \Delta \epsilon_{k'+\frac{1}{2}}
\int_{-1}^{1} d\mu' f(\mu', \epsilon_{k'+\frac{1}{2}}) \int_{0}^{2\pi} d\beta'
} & \nonumber \\ 
& {\ds \times 
\frac{1}{ \epsilon_{k+\frac{1}{2}}^{3} \Delta \epsilon_{k+\frac{1}{2}} }
\int_{\epsilon_{k}}^{\epsilon_{k+1}} \epsilon^{3} d\epsilon 
\frac{1}{ \epsilon_{k'+\frac{1}{2}}^{2} \Delta \epsilon_{k'+\frac{1}{2}} }
\int_{\epsilon_{k'}}^{\epsilon_{k'+1}} \epsilon'^{2} d\epsilon'
R_{\rm NNS}^{\rm in}(\epsilon,  \epsilon', \cos\theta )
} & \nonumber \\ 
& {\ds - f(\mu, \epsilon_{k+\frac{1}{2}}) \frac{1}{(hc)^{3}} \sum_{k'=1}^{N} 
\epsilon_{k'+\frac{1}{2}}^{2} \Delta \epsilon_{k'+\frac{1}{2}}
\int_{-1}^{1} d\mu' [1 - f(\mu', \epsilon_{k'+\frac{1}{2}}) ] \int_{0}^{2\pi} d\beta'
} & \nonumber \\ 
& {\ds \times \frac{1}{ \epsilon_{k+\frac{1}{2}}^{3} \Delta \epsilon_{k+\frac{1}{2}} }
\int_{\epsilon_{k}}^{\epsilon_{k+1}} \epsilon^{3} d\epsilon
\frac{1}{ \epsilon_{k'+\frac{1}{2}}^{2} \Delta \epsilon_{k'+\frac{1}{2}} } 
\int_{\epsilon_{k'}}^{\epsilon_{k'+1}} \epsilon'^{2} d\epsilon'
R_{\rm NNS}^{\rm out}(\epsilon, \epsilon', \cos\theta ),
} &  
\label{eq:b21}
\end{eqnarray}
which we rewrite as
\begin{eqnarray}
& {\ds \left. \pderiv{ f(\mu, \epsilon_{k+\frac{1}{2}}) }{ t } \right|_{\rm scat} 
= [ 1 - f(\mu, \epsilon_{k+\frac{1}{2}}) ] \frac{1}{(hc)^{3}} \sum_{k'=1}^{N} 
\epsilon_{k'+\frac{1}{2}}^{2} \Delta \epsilon_{k'+\frac{1}{2}}
} & \nonumber \\ 
& {\ds \times \int_{-1}^{1} d\mu' f(\mu', \epsilon_{k'+\frac{1}{2}}) \int_{0}^{2\pi} d\beta'
\langle R_{\rm NNS}^{\rm in}(\epsilon,  \epsilon', \cos\theta ) \rangle_{E}
} & \nonumber \\ 
& {\ds - f(\mu, \epsilon_{k+\frac{1}{2}}) \frac{1}{(hc)^{3}} \sum_{k'=1}^{N} 
\epsilon_{k'+\frac{1}{2}}^{2} \Delta \epsilon_{k'+\frac{1}{2}}
} & \nonumber \\ 
& {\ds \times \int_{-1}^{1} d\mu' [1 - f(\mu', \epsilon_{k'+\frac{1}{2}}) ] \int_{0}^{2\pi} d\beta'
\langle R_{\rm NNS}^{\rm out}(\epsilon, \epsilon', \cos\theta ) \rangle_{E},
} &  
\label{eq:b22}
\end{eqnarray}
where
\begin{eqnarray}
& {\ds \langle R_{\rm NNS}^{\rm in/out}(\epsilon,  \epsilon', \cos\theta ) \rangle_{E}
} & \nonumber \\ 
& {\ds \equiv \frac{1}{ \epsilon_{k+\frac{1}{2}}^{3} \Delta \epsilon_{k+\frac{1}{2}} }
\int_{\epsilon_{k}}^{\epsilon_{k+1}} \epsilon^{3} d\epsilon
\frac{1}{ \epsilon_{k'+\frac{1}{2}}^{2} \Delta \epsilon_{k'+\frac{1}{2}} }
\int_{\epsilon_{k'}}^{\epsilon_{k'+1}} \epsilon'^{2} d\epsilon'
R_{\rm NNS}^{\rm in/out}(\epsilon, \epsilon', \cos\theta ).
} &  
\label{eq:b23}
\end{eqnarray}
With the scattering kernel defined as in Eq.~\eqref{eq:b23} in the collision term of the Boltzmann
equation, the energy transfer between the neutrinos and the nucleons resulting from the many 
neutrino--nucleon scattering events is captured accurately, despite the fact that the energy exchange 
per scattering is much less than a typical energy zone width.

\subsubsection{Axisymmetry}

The first implementation of multi-angle, multi-frequency neutrino transport in the context of spatially two-dimensional, axisymmetric core-collapse supernova models was achieved by \citet{OtBuDe08}. Their implementation was based on the neutrino transport solver developed by \citet{LiBuWa04} for the neutrino specific intensity ($I$), whose evolution is given by the following equation:

\begin{equation}
\frac{D I}{Dt}+\Omega\cdot\nabla I + \sigma I = S.
\label{eq:specificintensity}
\end{equation}

\noindent Here, $D/Dt$ is the Lagrangian time derivative, $\Omega$ is the unit vector in the direction of neutrino propagation, whose components are $(\cos\theta_{p},\sin\theta_{p}\cos\phi_{p},\sin\theta_{p}\sin\phi_{p})$, where $\theta_{p}$ and $\phi_{p}$ are spherical momentum-space coordinates defined relative to the outward radial direction, $\sigma$ is the total absorption cross section, including absorption and scattering, and $S$ is the total emissivity, including emission and scattering.

Eq.~\ref{eq:specificintensity} is temporally discretized fully implicitly. The phase space discretization is handled as follows. Space---i.e., radius and angle---is discretized using a conservative difference scheme. Momentum space---i.e., the space comprising the two dimensions corresponding to the angles of the neutrino's direction of propagation, $\theta_{p}$ and $\phi_{p}$, and the dimension corresponding to the neutrino's energy, $\epsilon_{\nu}$, is discretized as follows. The discrete ordinates method is used for the momentum-space dimensions. Further details of the discretization of Eq.~\ref{eq:specificintensity} have not yet been provided.
 
%
\subsubsection{Three spatial dimensions}

The journey down what will no doubt be a long road toward the implementation of general relativistic, three-dimensional Boltzmann neutrino transport in the context of core-collapse supernovae was begun by \citet{SuYa12}. With core-collapse supernovae in mind, they began by solving the conservative form of the Boltzmann equation for three-dimensional, static stellar core configurations: 
\begin{eqnarray}
\label{eqn:eqtransfin-spherical2a}
\frac{1}{c}\frac{\partial f}{\partial t} 
+ \frac{\mu}{r^{2}} \frac{\partial}{\partial r} (r^{2} f)
+ \frac{\sqrt{1-\mu^{2}}~{\rm cos}~\phi_{p}}{r {\rm sin}~\theta} 
  \frac{\partial}{\partial \theta} ({\rm sin}~\theta f)
+ \frac{\sqrt{1-\mu^{2}}~{\rm sin}~\phi_{p}}{r {\rm sin}~\theta} 
  \frac{\partial f}{\partial \phi} \nonumber \\
+ \frac{1}{r} 
  \frac{\partial}{\partial \mu} [(1-\mu^{2}) f]
- \frac{\sqrt{1-\mu^{2}}}{r} 
  \frac{{\rm cos}~\theta}{{\rm sin}~\theta}
  \frac{\partial}{\partial \phi_{p}} ({\rm sin}~\phi_{p} f)
= \left[ \frac{1}{c} \frac{\delta f}{\delta t} \right]_{\rm collision}.
\end{eqnarray}
In light of the use of spherical polar coordinates, there are terms that correspond to advection in momentum space
even in a static medium in flat spacetime---i.e., even in the absence of special and general relativistic effects.
For example, as a neutrino propagates, its direction cosine, $\mu\equiv\cos\theta_{p}$, which is defined relative to the outwardly
pointing radial basis vector, will necessarily change. This is described by the fourth term on the left-hand side of Eq.~\eqref{eqn:eqtransfin-spherical2a}. 
This is not a geometric effect, as spacetime is flat in this case. Rather, it is a coordinate 
effect. The last term on the left-hand side of the same equation has a similar origin and interpretation. Given the 
assumption of a static medium and flat spacetime, no other terms appear on the left-hand side, which would describe
special and general relativistic effects were they considered.

The discretization of Eq.~\eqref{eqn:eqtransfin-spherical2a} follows and extends that used in \citet{MeBr93b}---i.e.,
finite differencing in space and energy, and discrete ordinates in neutrino propagation angles.
For the second term on the left-hand side of Eq.~\eqref{eqn:eqtransfin-spherical2a}, corresponding to radial advection
of neutrinos, Sumiyoshi and Yamada use the following discretization:
\begin{equation}
\label{eqn:advection-radial}
\left[ \frac{\mu}{r^{2}} \frac{\partial}{\partial r} (r^{2} f) \right]
=
\left[       \mu         \frac{\partial}{\partial (r^3 / 3)} (r^{2} f) \right]
= {\mu}_{j} ~ \frac{3}{r_{I}^{3} - r_{I-1}^{3}} 
~ ( r_{I}^{2} ~ f_{I} - r_{I-1}^{2} ~ f_{I-1} ), 
\end{equation}
where, in their notation,  $f_{I-1}$ and $f_{I}$ are the neutrino distributions 
at the cell interfaces of the $i$-th zone.
The quantities ${\mu}_{j} f_{I}$ at the cell boundaries are defined by 
\begin{equation}
\label{eqn:fnu-radial}
{\mu}_{j} f_{I} = 
  \frac{ {\mu}_{j} - | {\mu}_{j} | }{2} 
\{ ( 1 - \beta_{I} ) f_{i} +       \beta_{I}   f_{i+1} \}
+ \frac{ {\mu}_{j} + | {\mu}_{j} | }{2} 
\{         \beta_{I} f_{i} + ( 1 - \beta_{I} ) f_{i+1} \},
\end{equation}
and $\beta_{I}$ is 
\begin{equation}
\label{eqn:beta-radial}
\beta_{I} = 1 - \frac{1}{2} 
\frac{\alpha \Delta r_{I} / \lambda_{I}}{1 + \alpha \Delta r_{I} / \lambda_{I}}. 
\end{equation}
In the diffusion (free-streaming) limit, $\beta_{I}=1/2 (1)$. 
The advection in $\mu = \cos \theta_{p}$ is discretized as  
\begin{equation}
\label{eqn:advection-munu}
\left[ \frac{1}{r} 
  \frac{\partial}{\partial \mu} [~(1-\mu^{2}) f~] \right] 
= \frac{3}{2} ~ \frac{r_{I}^{2} - r_{I-1}^{2}}{r_{I}^{3} - r_{I-1}^{3}}
~ \frac{1}{{d \mu}_{j}} 
~ \left[ (1-{\mu}^2)_{J} f_{J} - (1-{\mu}^2)_{J-1} f_{J-1} \right]. 
\end{equation}
Upwind differencing is implemented, and $f_{J} = f_{j}$.  
$\theta$-advection is first reexpressed and then discretized as 
\begin{eqnarray}
\label{eqn:advection-polar}
\left[ \frac{\sqrt{1-\mu^{2}}~{\rm cos}~\phi_{p}}{r {\rm sin}~\theta} 
  \frac{\partial}{\partial \theta} ({\rm sin}~\theta f) \right]
=
\left[ - \frac{\sqrt{1-\mu^{2}}~{\rm cos}~\phi_{p}}{r } 
  \frac{\partial}{\partial \mu} [~(1-\mu^{2})^{\frac{1}{2}} f~] \right] \nonumber \\
 =  - \frac{3}{2} ~ \frac{r_{I_r}^{2} - r_{I_r-1}^{2}}{r_{I_r}^{3} - r_{I_r-1}^{3}} 
(1-{\mu}_{j_{\theta}}^{2})^{\frac{1}{2}} {\cos \phi_{p}}_{j_{\phi}}~ \frac{1}{d \mu_{i_{\theta}}} 
~ \left[ (1-{\mu}^2)^{\frac{1}{2}}_{I_{\theta}}   f_{I_{\theta}} 
       - (1-{\mu}^2)^{\frac{1}{2}}_{I_{\theta}-1} f_{I_{\theta}-1} \right]. \nonumber  \\
\end{eqnarray}
The factor, $(1-{\mu}_{j_{\theta}}^{2})^{\frac{1}{2}} {\cos \phi_{p}}_{j_{\phi}}$, 
determines the direction of advection and 
the evaluation of $f_{I_{\theta}}$ at the cell interface.  
Given the sign of $\cos \phi_{p}$, $f_{I_{\theta}}$ is determined by
\begin{eqnarray}
\label{eqn:fnu-polar}
{\cos \phi_{p}}_{j_{\phi}} f_{I_{\theta}} = 
  \frac{ {\cos \phi_{p}}_{j_{\phi}} + | {\cos \phi_{p}}_{j_{\phi}} | }{2} 
\{ ( 1 - \beta_{I_{\theta}} ) f_{i_{\theta}} +       \beta_{I_{\theta}}   f_{i_{\theta}+1} \} \nonumber \\
+ \frac{ {\cos \phi_{p}}_{j_{\phi}} - | {\cos \phi_{p}}_{j_{\phi}} | }{2} 
\{       \beta_{I_{\theta}}   f_{i_{\theta}} + ( 1 - \beta_{I_{\theta}} ) f_{i_{\theta}+1} \}.  
\end{eqnarray}
As before, $\beta_{I_{\theta}}$ takes on values between $\frac{1}{2}$ (diffusion limit) and $1$ (free-streaming limit) and
is defined in the same way as $\beta_{I}$, using instead the angular zone widths and mean free paths.
$\phi_{p}$ advection is discretized as
\begin{eqnarray}
\label{eqn:advection-phinu}
\left[ - \frac{\sqrt{1-\mu^{2}}}{r} 
  \frac{{\rm cos}~\theta}{{\rm sin}~\theta}
  \frac{\partial}{\partial \phi_{p}} ({\rm sin}~\phi_{p} f) \right] 
=
\left[ - \frac{\sqrt{1-\mu^{2}}}{r} 
  \frac{\mu}{\sqrt{1-\mu^{2}}}
  \frac{\partial}{\partial \phi_{p}} ({\rm sin}~\phi_{p} f) \right] = \nonumber \\
 - \frac{3}{2} ~ \frac{r_{I_r}^{2} - r_{I_r-1}^{2}}{r_{I_r}^{3} - r_{I_r-1}^{3}} 
(1-{\mu}_{j_{\theta}}^{2})^{\frac{1}{2}} ~ \frac{\mu_{i_{\theta}}}{(1-{\mu_{i_{\theta}}^{2})^{\frac{1}{2}}}}
~ \frac{1}{{d \phi_{p}}_{j_{\phi}}} 
\left[ (\sin \phi_{p})_{J_{\phi}} f_{J_{\phi}} - (\sin \phi_{p})_{J_{\phi}-1} f_{J_{\phi}-1} \right]. \nonumber \\
\end{eqnarray}
In this case, the sign of $\mu_{i_{\theta}} (\sin \phi_{p})_{J_{\phi}}$ determines 
the direction of advection.  
Upwind differencing is used to determine $f_{J_{\phi}}$ at the cell interface. $f_{J_{\phi}}$ is given by 
\begin{eqnarray}
\label{eqn:fnu-phinu}
\mu_{i_{\theta}} (\sin \phi_{p})_{J_{\phi}} f_{J_{\phi}} = 
\frac{ \mu_{i_{\theta}} (\sin \phi_{p})_{J_{\phi}} + | \mu_{i_{\theta}} (\sin \phi_{p})_{J_{\phi}} | }{2} 
f_{j_{\phi}+1} \nonumber \\
+
\frac{ \mu_{i_{\theta}} (\sin \phi_{p})_{J_{\phi}} - | \mu_{i_{\theta}} (\sin \phi_{p})_{J_{\phi}} | }{2} 
f_{j_{\phi}}.  
\end{eqnarray}
Last but not least, $\phi$ advection is discretized as follows 
\begin{eqnarray}
\label{eqn:advection-azimuthal}
\left[ \frac{\sqrt{1-\mu^{2}}~{\rm sin}~\phi_{p}}{r {\rm sin}~\theta} 
  \frac{\partial f}{\partial \phi} \right] 
=
\left[ \frac{\sqrt{1-\mu^{2}}~{\rm sin}~\phi_{p}}{r \sqrt{1-\mu^{2}}} 
  \frac{\partial f}{\partial \phi} \right] \nonumber \\
= \frac{3}{2} ~ \frac{r_{I_r}^{2} - r_{I_r-1}^{2}}{r_{I_r}^{3} - r_{I_r-1}^{3}} 
(1-{\mu}_{j_{\theta}}^{2})^{\frac{1}{2}} ~ \frac{{\sin \phi_{p}}_{j_{\phi}}}{(1-{\mu_{i_{\theta}}^{2})^{\frac{1}{2}}}}
~ \frac{1}{{d \phi}_{i_{\phi}}} \left[ f_{I_{\phi}} - f_{I_{\phi}-1} \right].
\end{eqnarray}
Given the sign of ${\sin\phi_{p}}_{j_{\phi}}$ and, therefore, the advection direction, $f_{I_{\phi}}$ is given by 
\begin{eqnarray}
\label{eqn:fnu-azimuthal}
{\sin \phi_{p}}_{j_{\phi}} f_{I_{\phi}} = 
\frac{ {\sin \phi_{p}}_{j_{\phi}} + | {\sin \phi_{p}}_{j_{\phi}} | }{2} 
\{      \beta_{I_{\phi}}  f_{i_{\phi}} + (1 - \beta_{I_{\phi}}) f_{i_{\phi}+1}\} \nonumber \\
+
\frac{ {\sin \phi_{p}}_{j_{\phi}} - | {\sin \phi_{p}}_{j_{\phi}} | }{2} 
\{ (1 - \beta_{I_{\phi}}) f_{i_{\phi}} +      \beta_{I_{\phi}}  f_{i_{\phi}+1}\}. 
\end{eqnarray}
$\beta_{I_{\phi}}$ is determined in the same way as its counterparts in the radial and $\theta$ directions, 
using the appropriate angular zone widths and mean free paths.

Focusing on the temporal discretization, the phase-space discretizations spelled out in Eqs.~\eqref{eqn:advection-radial} through \eqref{eqn:fnu-azimuthal} are assembled and evaluated in a fully implicit manner, as shown schematically below (i.e., the phase-space
discretizations themselves are not inserted; each term is represented by its continuum counterpart):

\begin{eqnarray}
\label{eqn:boltzmann-implicit}
\frac{1}{c}\frac{f_{i}^{n+1} - f_{i}^{n}}{\Delta t} 
+ \left[ \frac{\mu}{r^{2}} \frac{\partial}{\partial r} (r^{2} f) \right]^{n+1}
+ \left[ \frac{\sqrt{1-\mu^{2}}~{\rm cos}~\phi_{p}}{r {\rm sin}~\theta} 
  \frac{\partial}{\partial \theta} ({\rm sin}~\theta f) \right]^{n+1} \nonumber \\
+ \left[ \frac{\sqrt{1-\mu^{2}}~{\rm sin}~\phi_{p}}{r {\rm sin}~\theta} 
  \frac{\partial f}{\partial \phi} \right]^{n+1} 
+ \left[ \frac{1}{r} 
  \frac{\partial}{\partial \mu} [(1-\mu^{2}) f] \right]^{n+1} \nonumber \\
+ \left[ - \frac{\sqrt{1-\mu^{2}}}{r} 
  \frac{{\rm cos}~\theta}{{\rm sin}~\theta}
  \frac{\partial}{\partial \phi_{p}} ({\rm sin}~\phi_{p} f) \right]^{n+1} 
= \left[ \frac{1}{c} \frac{\delta f}{\delta t} \right]_{\rm collision}^{n+1},
\end{eqnarray}

\noindent where $n$ designates the current time slice. The left-hand side of Eq.~\eqref{eqn:boltzmann-implicit} 
is linear in the distribution function, but the right-hand side is not. Consequently, as in the spherically symmetric case,
both sides of Eq.~\eqref{eqn:boltzmann-implicit} are linearized in $f$. (In this case, Sumiyoshi and Yamada are working
with a hydrostatic and thermally frozen stellar core profile. As a result, linearizations in $\epsilon$ and $Y_e$ are not 
necessary.) This gives rise to a linear system of equations for $\delta f_i$. To solve the combination of the outer nonlinear
system of equations and the corresponding inner linear system of equations, Sumiyoshi and Yamada implement a 
Newton--Krylov approach---specifically, they implement Newton--BiCGSTAB, with point-Jacobi preconditioning.

The extension of these lepton-number conservative methods to the special relativistic case was documented by \citet{NaSuYa14}. 
They deployed novel momentum-space gridding based on three considerations: 
(1) The invariant emissivity and opacity, which together define an invariant collision term on the right-hand side of the 
Boltzmann equation, can be computed in either the inertial, Eulerian frame or the inertial frame of an observer instantaneously 
comoving with the stellar core fluid. The value obtained in both cases would be the same if the neutrino angles and energies 
used in either case were related by the Lorentz transformation between the two inertial frames.
(2) The Lorentz transformations of angles and energies between the Eulerian and comoving frames decouple---i.e., one 
is free to define one's energy grid in either of the two frames independently of one's angular grids, allowing the choices 
that would simplify the numerics while respecting the physics. 
(3) The dominant opacity during stellar core collapse stems from coherent, isoenergetic scattering on nuclei---i.e., any 
novel gridding should be constructed with this opacity in mind.

In Nagakura et al.'s notation, the invariance of the collision term can be expressed as

\begin{eqnarray}
\label{eq:collisionrela}
\varepsilon^{\rm{lb}}\Bigl( \frac{\delta f}{\delta t} \Bigr)_{{\rm col}}^{{\rm lb}} = 
\varepsilon^{\rm{fr}}\Bigl( \frac{\delta f}{\delta \tilde{t}} \Bigr)_{{\rm col}}^{\rm{fr}}.
\end{eqnarray}
where $t (\tilde{t})$ is the Eulerian (comoving) frame time and where the labels ${\rm lb (fr)}$ correspond to the Eulerian 
(comoving) frames. The equality in Eq.~\eqref{eq:collisionrela} is to be understood as follows: If one evaluates the 
left-hand side at a particular neutrino angle and energy as measured by the Eulerian observer, the equality is guaranteed
provided the righ-hand side is evaluated at the corresponding Lorentz transformed neutrino angle and energy, which 
would be the angle and energy measured by the comoving observer.
The neutrino energies in the two frames, $\varepsilon^{\rm{lb}}$ and $\varepsilon^{\rm{fr}}$, are related by 
\begin{equation}
\varepsilon^{\rm{fr}} = \varepsilon^{{\rm lb}} \gamma 
(1 - \mathbf{n}^{\rm lb} \cdot \mathbf{v}) , \label{eqn:Lorentz-energy}
\end{equation}
where $\gamma$ is the Lorentz factor, $\mathbf{n}^{\rm lb}$ is the neutrino propagation direction as measured
in the Eulerian frame, and $\mathbf{v}$ is the fluid velocity in the same frame.
The unit neutrino propagation direction vectors in the two frames are related by
\begin{eqnarray}
\varepsilon^{\rm{fr}} \mathbf{n}^{\rm{fr}} = 
\varepsilon^{{\rm lb}}
\left[\mathbf{n}^{\rm lb} + \left( - \gamma + (\gamma - 1) 
\frac{\mathbf{n}^{\rm lb} \cdot \mathbf{v}}{v^2}\right)
\mathbf{v}\right], \label{eq:energytrans} 
\end{eqnarray}
where $\mathbf{n}^{\rm{fr}}$ denotes the unit neutrino propagation direction vector in the comoving frame.

\begin{figure}[htb]
\includegraphics[width=\textwidth]{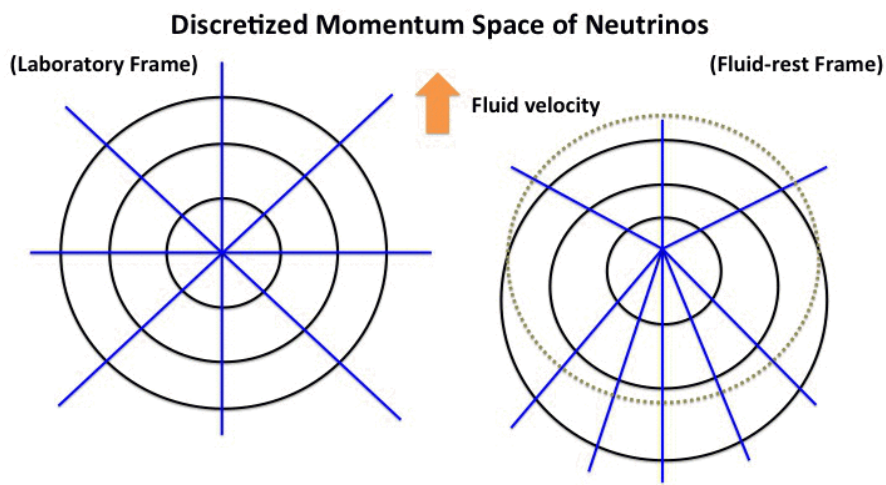}
\caption{The left panel shows a schematic of uniform momentum-space angular and energy grids in the Laboratory frame. Constant-energy grid lines are represented by concentric circles \citep{NaSuYa14}. Constant angles are indicated by radial lines. The right panel shows the corresponding contours and lines in the comoving frame. Also added (dotted line) is a constant comoving-frame neutrino energy contour.}
\label{fig:momspacegrids1}
\end{figure}

\begin{figure}[htb]
\includegraphics[width=\textwidth]{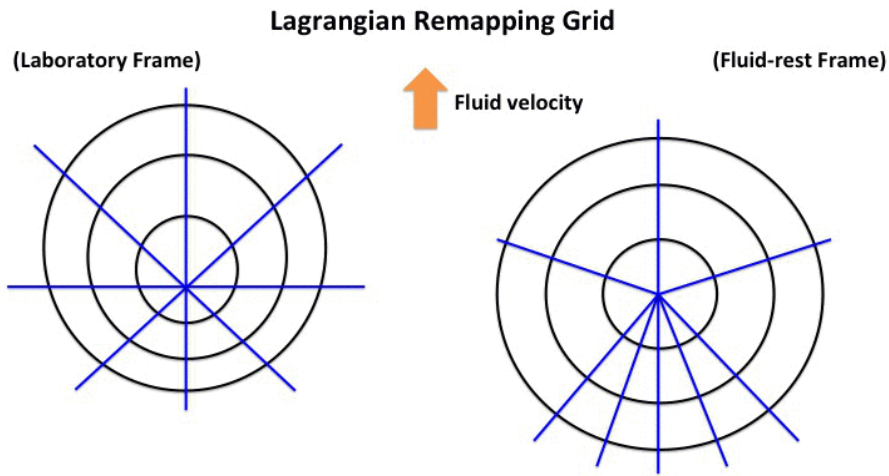}
\caption{The left panel shows the Lagrangian Remapping Grid (LRG) used by \citet{NaSuYa14} in their Boltzmann transport implementation. On the LRG, the neutrino angular grid is uniform, but the energy grid corresponds to a uniform energy grid in the comoving frame, shown in the right panel by concentric circles. The two energy grids are related by a Lorentz transformation. Given that the angular grid is uniform in the Laboratory frame, the corresponding angular grid in the comoving frame is not uniform. The angular grids, too, are related by a Lorentz transformation between the frames.}
\label{fig:momspacegrids2}
\end{figure}

Figure \ref{fig:momspacegrids1} from Nagakura et~al.\ shows two momentum-space grids associated with momentum-space spherical coordinates. The grid on the left corresponds to a choice of uniform gridding in both angle and energy in the Eulerian frame. (Uniform gridding
is typically not used for either, but for simplicity Nagakura et~al.\ consider this case to illustrate the essential features of their 
approach.) The grid on the right corresponds to the Lorentz-transformed Eulerian grid---i.e., the counterpart grid in the comoving
frame. This grid is no longer uniform in either angle or energy.  On the comoving-frame grid, an isoenergetic scattering 
event, wherein the neutrino's angle changes but its energy does not, would necessitate an interpolation in energy given the 
fact the energy grid is not uniform in angle. The number (typically 
$\sim$20) of energy ``groups'' used in most core-collapse supernova simulations is low, and to make matters worse, the groups are typically 
spaced exponentially, with coarser resolution at higher energies. Interpolation on such grids is problematic for these reasons and for the conservation of neutrino (lepton) number. To overcome these difficulties, Nagakura et~al.\ use the independence of the Lorentz transformation for neutrino angles and energy and choose a hybrid-grid approach. They introduce the Lagrangian Remap Grid (LRG) for the Eulerian observer, which is shown on the left-hand side of Fig.~\ref{fig:momspacegrids2}, which is the primary grid used in their work. On the LRG, the angular grid is uniform but the energy grid is not. The energy grid on the LRG is the Lorentz transform of the energy grid on the right-hand side of the same figure, which corresponds to the comoving-frame observer's energy grid, which is uniform. Of course, by virtue of the Lorentz transformation and the fact that the angular grid is uniform in the Eulerian frame, the angular grid in the comoving frame cannot be uniform. This presents no difficulties in their approach, so Nagakura et~al.\ opt for the simplicity of the uniform angular grid on the LRG, their primary grid. The evaluation of the collision term on the LRG, which is how the collision term is evaluated in Nagakura et al.'s approach to the discretization and solution of the Boltzmann equation, is the same as its evaluation on the comoving-frame grid, given the invariance of the collision term for such Lorentz-transform-related grids. Since the latter energy grid is uniform across angles, no interpolation in energy is required in evaluating, for example, isoenergetic scattering. The Lorentz transformation between the two grids is spatially and temporally dependent, so the LRG must be continually redefined as the evolution proceeds, but the comoving-frame grid does not change. As the LRG evolves, a conservative remapping procedure is used to remap the neutrino distributions on the previous LRG to the new LRG.

With all of the above in mind, and focusing on isoenergetic scattering, the right-hand side of the Boltzmann equation, Eq.~\eqref{eqn:eqtransfin-spherical2a}, is evaluated on the LRG as
\begin{eqnarray}
& & \left(\frac{\delta f}{\delta t}\right)_{\rm collision} \\ \nonumber
& & = \gamma \left( 1-\mathbf{n}^{\rm lb}\cdot \mathbf{v}\right)\left(\frac{\delta f}{\delta\tilde{t}}\right)_{\rm collision} \\ \nonumber
& & =  
\gamma \left( 1-\mathbf{n}^{\rm lb}\cdot \mathbf{v}\right)
\left[
\frac{-(\epsilon^{\rm fr})^2}{(2\pi)^3}
\int d{{\Omega}^{'}}^{\rm fr}
R^{\rm fr}(\Omega^{\rm fr},{\Omega^{'}}^{\rm fr})
[f^{\rm fr}(\epsilon^{\rm fr},\Omega^{\rm fr})-f^{\rm fr}(\epsilon^{\rm fr},{\Omega^{'}}^{\rm fr})]\right] \\ \nonumber
& & = 
\gamma \left( 1-\mathbf{n}^{\rm lb}\cdot \mathbf{v}\right)
[
\frac{-[\epsilon^{\rm fr}(\epsilon^{\rm lb})]^2}{(2\pi)^3}\int d{{\Omega}^{'}}^{\rm lb}\frac{d{{\Omega}^{'}}^{\rm fr}}{d{{\Omega}^{'}}^{\rm lb}}
R^{\rm lb}[\Omega^{\rm fr}(\Omega^{\rm lb}),{\Omega^{'}}^{\rm fr}({\Omega^{'}}^{\rm lb})] \\ \nonumber
& & \times
\{f^{\rm lb}[\epsilon^{\rm fr}(\epsilon^{\rm lb}),\Omega^{\rm fr}(\Omega^{\rm lb})]-f^{\rm lb}[\epsilon^{\rm fr}(\epsilon^{\rm lb}),{\Omega^{'}}^{\rm fr}({\Omega^{'}}^{\rm lb})]\}
] \\ \nonumber
& & = 
\gamma \left( 1-\mathbf{n}^{\rm lb}\cdot \mathbf{v}\right)
[
\frac{-[\epsilon^{\rm fr}(\epsilon^{\rm lb})]^2}{(2\pi)^3}\int d{{\Omega}^{'}}^{\rm lb}\frac{d{{\Omega}^{'}}^{\rm fr}}{d{{\Omega}^{'}}^{\rm lb}}
R^{\rm lb}[\Omega^{\rm fr}(\Omega^{\rm lb}),{\Omega^{'}}^{\rm fr}({\Omega^{'}}^{\rm lb})] \\ \nonumber
& & \times
\{
f^{\rm lb}(\epsilon^{\rm lb},\Omega^{\rm lb})-f^{\rm lb}(\epsilon^{\rm lb},{\Omega^{'}}^{\rm lb})
\}
] .\\ \nonumber
\end{eqnarray}
The last equality follows from the invariance of the distribution function.

\begin{figure}[htb]
\includegraphics[width=\textwidth]{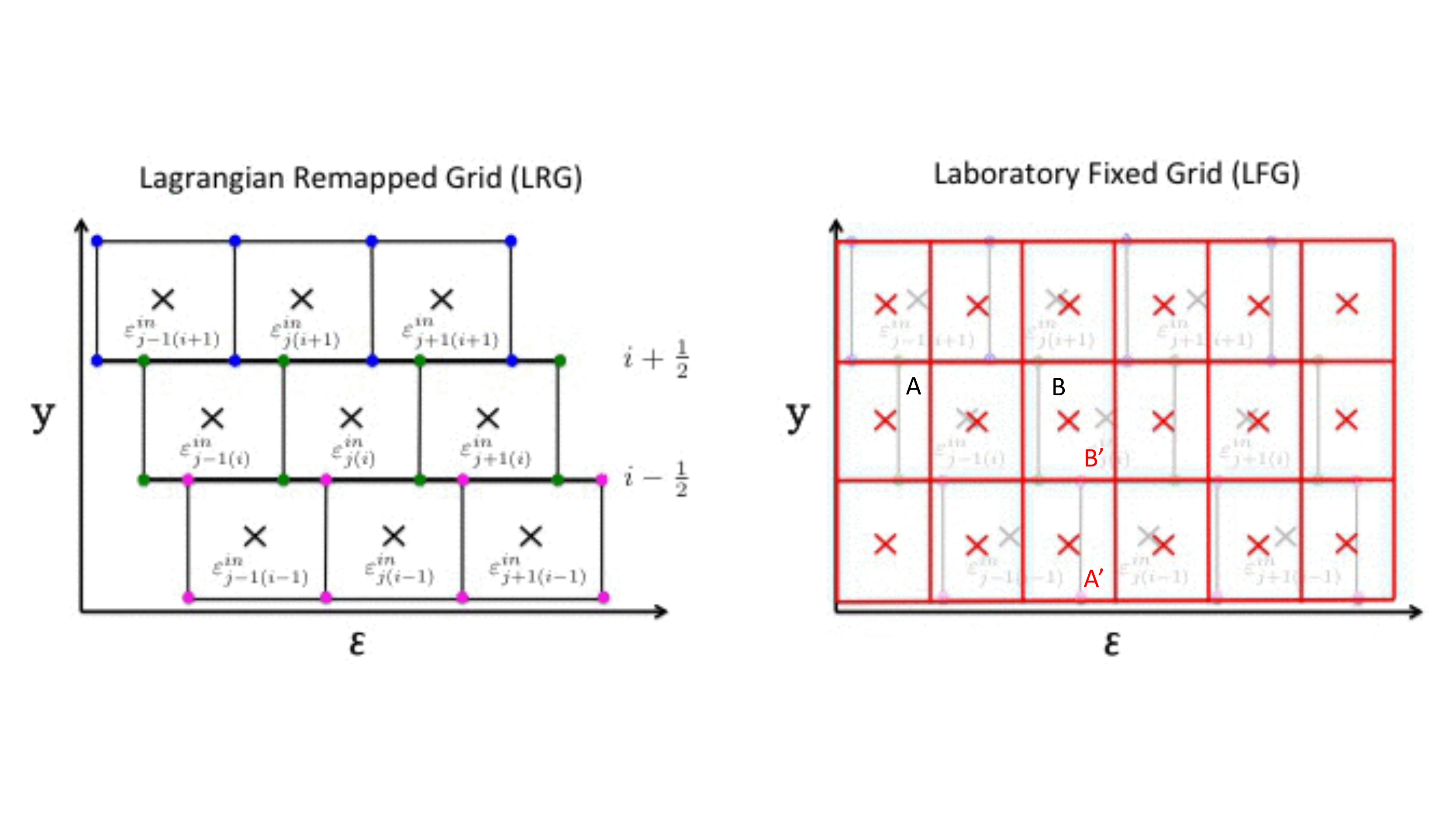}
\caption{In the left panel, energy zones on the LRG are shown for adjacent radial or angular grid points, designated here by $y$ \citep{NaSuYa14}. In the right panel, the higher-resolution Laboratory Fixed Grid (LFG) is shown, superimposed on the LRG.}
\label{fig:LRGLFG}
\end{figure}

While the use of the LRG simplifies the evaluation of the collision term and avoids the need to introduce velocity-dependent angle and 
energy advection on the left-hand side of the Boltzmann equation, there is a cost: It complicates spatial advection. To overcome this
inherited difficulty, Nagakura et~al.\ invoke yet another grid, the Laboratory Fixed Grid (LFG). 
The LFG is like the grid depicted on the left-hand side of 
Fig.~\ref{fig:momspacegrids1}. 
It is the same for all Eulerian 
observers at different spatial locations and is constant in time. And, in Nagakura et al.'s implementation, it is of higher resolution in 
energy relative to the LRG. This is evident in Fig.~\ref{fig:LRGLFG}.

Given the LFG, the treatment of spatial and angular advection occurs in the following steps: 
(1) Using the subgrid energy distribution, $f_{\rm subgrid}$, the values of the distribution function, $f$, at the zone centers of the LFG grid are determined by $f_{\rm subgrid}(\epsilon_{\rm LFG_{A^{'},B^{'},...}})$,  where $\epsilon_{\rm LFG_{A^{'},B^{'},...}}$ are the value of the energies corresponding to the zone centers on the LFG grid for zones A$^{'}$, B$^{'}$, ..., respectively. (For the example points selected here, the LFG energies are the same.)
(2) Once the values of the distribution function are defined at the zone centers of the LFG, they can be used to define the spatial and angular fluxes on the LRG as follows. Consider Fig.~\ref{fig:LRGLFG}. On the left-hand side of the figure, the LRG is shown. On the right, the LFG is overlaid on the LRG. Note, too, here we are 
considering advection in space and angle, denoted on the vertical axis by $y$ to represent both. Let us consider LFG zones A$^{'}$ and B$^{'}$. The flux at the interface between these two zones is determined from the value of the distribution function there, which is determined by interpolating between the values of the distribution function at the A$^{'}$ and B$^{'}$ zone centers, as outlined by \citet{SuYa12}. (When invoking the LFG, this interpolation involves only two zones, not three as it would in the case of the LRG.)
(3) Given the fluxes on the LFG, we are ready to define the fluxes that will be used on the LRG to update the distribution function in each of the LRG's zones due to advection. Note that advection into (for example) LFG zone B$^{'}$ from A$^{'}$ involves advection into a single zone. However, it is easy to see from Fig.~\ref{fig:LRGLFG} 
that advection from A$^{'}$ into B$^{'}$ involves advection into two zones of the LRG: A and B. To divide the contribution of the LFG flux into B$^{'}$ into LRG fluxes into zones A and B, we split the flux as follows: 
\begin{equation}
F_{A^{'}|B^{'}}=\gamma F_{A^{'}|B^{'}} + (1-\gamma)F_{A^{'}|B^{'}},
\end{equation}
\label{eq:apportioningoff}
where $F_{A^{'}|B^{'}}$ is the LFG flux at the interface between LFG zones A$^{'}$ and B$^{'}$ and where
\begin{eqnarray}
\label{eq:NLNR}
\gamma & = &\frac{N_L}{N_L+N_R}, \\
N_L & = & |\epsilon^{3}_{AB}-\epsilon^{3}_{L}|f_A, \\
N_R & = & |\epsilon^{3}_{AB}-\epsilon^{3}_{R}|f_B,
\end{eqnarray}
with $\epsilon_{AB}$ corresponding to the value of the energy at the interface of the LRG zones A and B and where $\epsilon_{L(R)}$ corresponds to the energy value associated with the left (right) boundary of the LFG zone B$^{'}$. $f_{A(B)}$ corresponds to the value of the distribution function on the LRG in zone A(B). In other words,
the LFG flux at the interface of LFG zones A$^{'}$ and B$^{'}$ is split, according to the distribution-weighted energy volume, between LRG fluxes into zones A and B. Note that zone B, for example, has multiple LFG fluxes advecting into it. The total LRG flux for zone B would therefore be the sum of all of the relevant LFG fluxes into it determined in the manner described here.
(4) Once the LRG interface fluxes are defined as in step 3, the spatial (or angular) advection on the LRG is carried out as outlined by \citet{SuYa12}.

Nagakura et al.'s novel method has been designed to conserve lepton number. A demonstration that it simultaneously conserves energy at an appropriate 
level remains to be demonstrated.

With regard to the temporal discretization with special relativistic effects included, Nagakura et~al.\ use a semi-implicit method. This is necessitated by the fact 
that the methods outlined above for the treatment of advection on the LRG cannot be made fully implicit. With the temporal descretization alone in mind, the
Boltzmann equation can be written as
\begin{eqnarray}
\frac{f^{n+1} - f^{n}}{\Delta t} = - F_{\rm{adv}}(f^{gs},f^{n+1}) + \Bigl( \frac{\delta f}{\delta t} \Bigr)_{{\rm col}}^{{\rm lb}}(f^{n+1}), \label{eq:conBoltzrewrite_fullimp}
\end{eqnarray}
where
\begin{eqnarray}
&&F_{\rm{adv}}(f^{gs},f^{n+1}) = F^{SR}_{\rm{adv}}(f^{\rm{gs}}) + 
\kappa \Bigl(    
F^{\rm{NR}}_{\rm{adv}}(f^{n+1}) - F^{\rm{NR}}_{\rm{adv}}(f^{\rm{gs}})
 \Bigr). \label{eq:stabilizationB}
\end{eqnarray}
The first term on the right-hand side of Eq.~\eqref{eq:stabilizationB} is the advection term for the special relativistic case. It is evaluated explicitly
at the value of the current iterate, $f^{gs}$. The second two terms correspond to what the advection terms would be in the non-relativistic case, evaluated both implicitly and explicitly (at the current iterate), respectively. Together they represent a ``correction'' to the first term and are introduced for numerical stability. When $f^{gs}\rightarrow f^{n+1}$, the second two terms cancel, and the right-hand side of Eq.~\eqref{eq:stabilizationB} becomes $F^{SR}_{adv}(f^{n+1})$, as desired. The parameter, $\kappa$, is a limiter and prevents the correction from becoming larger than the first term, which Nagakura et~al.\ note can happen when the fluid velocities
become several tens of percent of the speed of light.

Given the solution of the distribution function and, in particular, the numerical determination of the collision term, the update to the matter electron fraction and 
stress--energy tensor (including both energy and momentum exchange) are computed as follows [see Eqs.~\eqref{eq:fluidFourMomentumConservation}, \eqref{eq:ElectronNumberConservation}, \eqref{eq:electronfractionequationsourceterm}, and \eqref{eq:fourmomentumequationsourceterm}]:
\begin{eqnarray}
T^{\mu\nu}_{\hspace{3.5mm} ,\nu} &=& - G^{\mu}, \label{eq:TandGfinal} \\
N_{e \hspace{0.5mm} ,\nu}^{\nu} &=& - \Gamma, \label{eq:NandGammafinal}
\end{eqnarray}
where
\begin{eqnarray}
G^{\mu} &\equiv& \sum_{\rm{i}} G_{\rm{i}}^{\mu}, \label{eq:Gsumdef} \\
G_{\rm{i}}^{\mu} &\equiv& \int p_{\rm{i}}^{\mu} \Bigl( \frac{\delta f}{\delta \tau} \Bigr)_{\rm{col}(\rm{i})} dV_p, \label{eq:Gdef} \\
\Gamma &\equiv& \Gamma_{\nu_{e}} - \Gamma_{\bar{\nu_{e}}}, \label{eq:Gammasumdef} \\
\Gamma_{i} &\equiv& \int \Bigl( \frac{\delta f}{\delta \tau} \Bigr)_{\rm{col}(\rm{i})} dV_p, \label{eq:Gammadef}
\end{eqnarray}
and where, for Nagakura et al., $N_{e}^{\nu}$ (our $J_{e}^{\nu}$) is the electron density current, $dV_p$ (our $\pi_m$) is the invariant momentum-space volume element, and $i$ indicates the neutrino species.

\begin{figure}[htbp]
\centering
\includegraphics[width=0.9\textwidth]{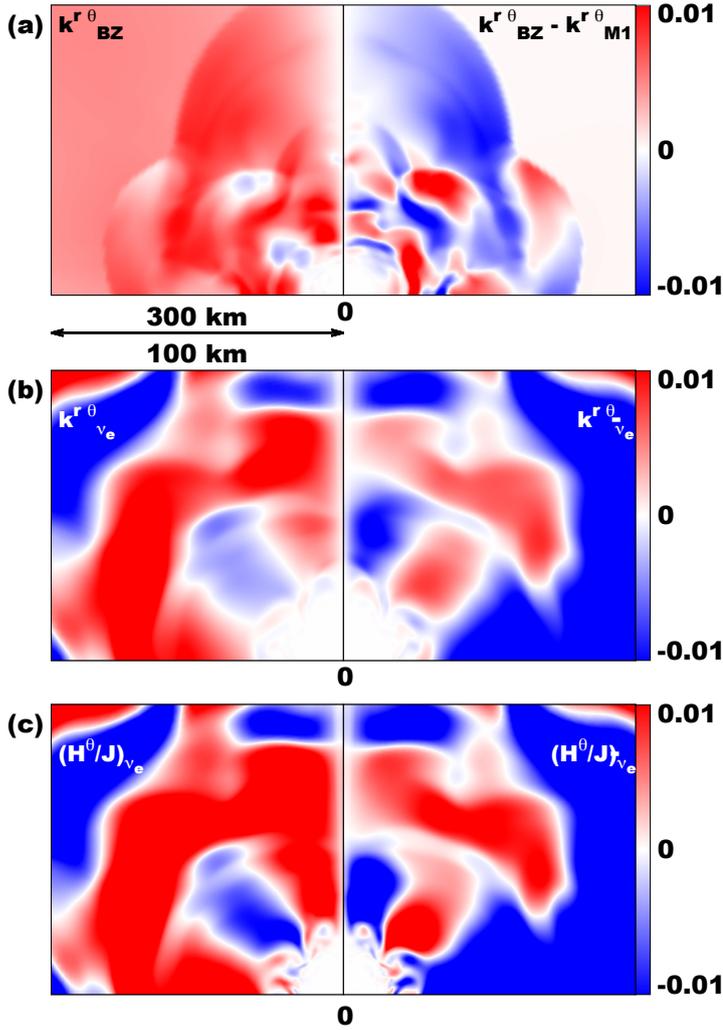}
\caption{In the top left panel, \citet{NaIwFu18} plot the $r-\theta$ component of the Eddington tensor, $k^{r\theta}$, at 190 ms after bounce in a core-collapse supernova simulation they performed with their Boltzmann neutrino transport solver, initiated from a progenitor of $11.2\,M_\odot$. In the corresponding upper right panel, they plot the (absolute) difference between $k^{r\theta}$ computed with both Boltzmann neutrino transport and two-moment neutrino transport with M1 closure. In both cases, $k^{r\theta}$ is evaluated at the mean neutrino energy at each point of the spatial grid shown here. Nagakura et~al.\ classify such absolute differences in the off-diagonal components of the Eddington tensor in their model as substantial, indicating that Boltzmann transport is needed to accurately compute the components of the neutrino Eddington tensor. In their model, $k^{r\theta}$ was demonstrated to dictate the evolution of the lateral neutrino fluxes, not $k^{\theta\theta}$, in the critical semitransparent regime.}
\end{figure}
\label{fig:krtheta}

\subsection{Boltzmann kinetics: spatial discontinuous Galerkin discretization plus spectral multigroup $P_{N}$}

A numerical treatment of Boltzmann kinetics that implements a finite-element discretization---specifically, a Discontinuous Galerkin (DG) 
discretization---for the spatial degrees of freedom together with a spectral decomposition in momentum space was developed by \citet{RaAbRe13}
for the Boltzmann equation:
\begin{equation}\label{eq:relativistic.boltzmann}
  p^\mu \frac{\partial F}{\partial x^{\mu}} = \mathbb{C}[F]\,.
\end{equation}
In this scheme, the distribution function, $F$, is first decomposed in momentum space as
\begin{equation}\label{eq:solution.ansatz}
  F(x^\alpha, \nu, \varphi, \theta) =
  \sum_{n=0}^{N_\nu} \sum_{\ell = 0}^{N} \sum_{m = -\ell}^\ell
  F^{n\ell m}(x^\alpha)\, \chi_{n}(\nu)\, Y_{\ell m}(\varphi,\theta)\,,
\end{equation}
where the orthonormal basis functions in the energy dimension are defined by
\begin{align}\label{eq:energy.basis}
  \chi_n(\nu) &=
  \begin{cases}
    {1}/{\sqrt{V_n}}, & \textrm{if } \nu \in [\nu_n, \nu_{n+1}]\,, \\ 0,
    & \textrm{otherwise}\,,
  \end{cases}\,,&
  V_n &= \int_{\nu_n}^{\nu_{n+1}} h^3 \nu^2\, \dd \nu
      = \frac{h^3}{3} (\nu_{n+1}^3 - \nu_n^3)\,.
\end{align}
Using the orthonormality of the spherical harmonics and $\chi_{n}(\nu)$, the coefficients in the momentum-space expansion of the distribution function, Eq.~\eqref{eq:solution.ansatz},
are given by
\begin{equation}
  F_{n \ell m}(x^\alpha) = \int_{0}^{\infty} h^3 \nu^2\, \dd \nu
  \int_{\mathcal{S}_1} \dd \Omega\, F(x^\alpha, \nu, \varphi, \theta)\,
  Y_{\ell m}(\varphi, \theta) \, \chi_{n}(\nu)\,.
\end{equation}
Radice et~al.\ introduce the shorthand notation:
\begin{equation}
  \Psi_A(\nu, \varphi, \theta) \equiv \chi_{n}(\nu)\, Y_{\ell
  m}(\varphi,\theta)\,,
\end{equation}
and reexpress Eq.~\eqref{eq:solution.ansatz} as
\begin{equation}\label{eq:solution.ansatz.short}
  F(x^\alpha, \epsilon, \varphi, \theta) = \sum_A F^A(x^\alpha) \Psi_A(\epsilon,
  \varphi, \theta) = F^A \Psi_A\,.
\end{equation}
Inserting the expansion (\ref{eq:solution.ansatz.short}) into the Boltzmann equation \eqref{eq:relativistic.boltzmann} leads to
a coupled system of equations for the expansion coefficients that must be solved to determine them as a function of time and space:
\begin{equation}\label{eq:scheme.derivation.step1}
  p^0 \frac{\partial F^B}{\partial t} \Psi_B + p^k \frac{\partial
  F^B}{\partial x^k} \Psi_B = \mathbb{C}[F]\,.
\end{equation}
Multiplying Eq.~\eqref{eq:scheme.derivation.step1} by $\Psi^A$ (in the notation of Radice et al., a superscript $A$ indicates a complex
conjugate), integrating over momentum space, and using the orthonormality
of the basis functions $\Psi_A$ gives
\begin{equation}\label{eq:charon.scheme}
  \frac{\partial F^A}{\partial t} + {{\mathcal{P}}^{kA}}_{B}
  \frac{\partial F^B}{\partial x^k} = \mathbb{S}^A[F]\,,
\end{equation}
where
\begin{equation}\label{eq:charon.stiff}
  {{\mathcal{P}}^{kA}}_{B} \equiv \int p^k\, \Psi^A\, \Psi_B\, \dd \Pi\,
\end{equation}
and
\begin{equation}\label{eq:charon.source}
  \mathbb{S}^A[F] \equiv \int \mathbb{C}[F]\, \Psi^A\, \dd \Pi\,.
\end{equation}
In Eqs~\eqref{eq:charon.stiff} and \eqref{eq:charon.source}, $\dd \Pi$ is the invariant momentum-space volume element.
Once the expansion coefficients are obtained by solving Eq.~\eqref{eq:charon.scheme}, the solution to the original Boltzmann equation is 
given by Eq.~\eqref{eq:solution.ansatz}.

Radice et~al.\ illustrate their approach to solving Eq.~\eqref{eq:charon.scheme} by considering the one-dimensional, collisionless case:
\begin{equation}
  \frac{\partial F^A}{\partial t} +
  {{\mathcal{P}}^{1A}}_{B} \frac{\partial F^B}{\partial x} = 0\,.
\label{eq:charon.scheme1D}
\end{equation}
In a DG discretization in $x$, the distribution function is written as an expansion in Lagrange polynomials:
\begin{equation}
F^{A}(x,t)=\sum_{A}{F^A}_{i}(t)u(x),
\label{eq:1DDGexpansion}
\end{equation}
where
\begin{equation}
  u(x) = u_{i-1/2} l_{i-1/2}(x) + u_{i+3/2} l_{i+3/2}(x)\,,
\end{equation}
and where the Lagrange polynomials are defined by
\begin{align}
  l_{i-1/2}(x) &= 1 - \frac{x - x_{i-1/2}}{x_{i+3/2} - x_{i-1/2}}\,,&
  l_{i+3/2}(x) &= \frac{x - x_{i-1/2}}{x_{i+3/2} - x_{i-1/2}}\,.
\end{align}
Insertion of the expansion \eqref{eq:1DDGexpansion} in Eq.~\eqref{eq:charon.scheme1D} yields the following set of coupled ordinary differential equations for the 
coefficients ${F^A}_{i}$:
\begin{equation}
  \Delta x \frac{\dd {F^A}_i}{\dd t} = {{\mathbb{F}}^A}_i\,,
\label{eq:ODE}
\end{equation}
where the flux factors are given by
\begin{align}
\label{eq:dg_fluxes}
  {{\mathbb{F}}^A}_i &\equiv \frac{3}{2} \mathcal{F}^- -
  \overline{\mathcal{F}} - \frac{1}{2} \mathcal{F}^+ \,,&
  {{\mathbb{F}}^A}_{i+1} &\equiv \frac{1}{2}\mathcal{F}^- +
    \overline{\mathcal{F}} - \frac{3}{2} \mathcal{F}^+\,,
\end{align}

\[
  \overline{\mathcal{F}} \equiv \frac{1}{2} \Big[
   \big({{\mathcal{P}}^{1A}}_{B}\big)_i
      {F^B}_i +
   \big({{\mathcal{P}}^{1A}}_{B}\big)_{i+1} {F^B}_{i+1}\Big]\,,
\]

\[
  \mathcal{F}^- \equiv \frac{1}{2}
  \bigg[
    {{\mathcal{P}}^{1A}}_{B} \big({F^B}_L + {F^B}_R\big) -
    {{\mathcal{R}}^{1A}}_{C}  \mathrm{max}(v, |{\Lambda^{1C}}_{D}|)
    {{\mathcal{L}}^{1D}}_{B}  \big({F^B}_R - {F^B}_L\big)
  \bigg]\,,
\]

\[
  {{\mathcal{P}}^{1A}}_{B} = {{\mathcal{R}}^{1A}}_{C}
   {\Lambda^{1C}}_{D} {{\mathcal{L}}^{1D}}_{B}\,.
\]
In Eq.~\eqref{eq:dg_fluxes}, $\overline{\mathcal{F}}$ is the average flux; $\mathcal{F}^-$ is the flux computed at the boundary $x_{i+1/2}$ of the $i^{\rm th}$ 
element through an exact solution of the Riemann problem with left and right states, ${F^B}_{L}$ and ${F^B}_{R}$, respectively; $\mathcal{F}^+$ is defined in the same way, at the boundary $x_{i+3/2}$ ; ${{\mathcal{R}}^{1A}}_{C}$ is the matrix of right eigenvectors of ${{\mathcal{P}}^{1A}}_{B}$; ${{\mathcal{L}}^{1D}}_{B}$ is the matrix of left eigenvectors of ${{\mathcal{P}}^{1A}}_{B}$; ${\Lambda^{1C}}_{D}$ is the matrix of eigenvalues of ${{\mathcal{P}}^{1A}}_{B}$; and $v$ is a parameter taken to be the first abscissa of the adopted Legendre quadrature (this parameter is introduced by Radice et~al.\ to dissipate numerically zero-speed modes).
The three-dimensional extension of the scheme is given by constructing the flux factors in each of the three dimensions in the same way, which gives

\begin{equation}\label{eq:ldg.full}
  \frac{\dd {F^A}_{i,j,k}}{\dd t} =
  \mathbb{S}^A[F] +
  \frac{1}{\Delta x} {{\mathbb{F}}^A}_{i,j,k} +
  \frac{1}{\Delta y} {{\mathbb{G}}^A}_{i,j,k} +
  \frac{1}{\Delta z} {{\mathbb{H}}^A}_{i,j,k} \,.
\end{equation}

Now, focusing on the temporal discretization of Eq.~\eqref{eq:ldg.full} and using Radice et al.'s rewrite of the equation as
\begin{equation}\label{eq:ldg.short}
  \frac{\dd F^A}{\dd t} = \mathbb{S}^A[F] + \mathcal{A}^A[F]\,,
\end{equation}
the authors evolve the coefficients of the distribution function's DG--spectral expansion, Eqs.~\eqref{eq:solution.ansatz} and \eqref{eq:1DDGexpansion}, in a 
two-step, semi-implicit, asymptotic-preserving scheme \citep{McEvLo08}, staged as a predictor step, 
\begin{equation}
\label{eq:predictorstep}
  \frac{{F^A}_{k+1/2} -
  {F^A}_{k}}{\Delta t/2} = {\mathcal{A}}^{A}[F_k] + \mathbb{S}^A[F_{k+1/2}]\,\,,
\end{equation}
followed by a corrector step,
\begin{equation}
\label{eq:correctorstep}
  \frac{{F^A}_{k+1} -
  {F^A}_{k}}{\Delta t} = {\mathcal{A}}^{A}[F_{k+1/2}] + \mathbb{S}^A[F_{k+1}]\,.
\end{equation}

Given that Radice et~al.\ choose to use a partially spectral scheme, like all others deploying such schemes they had to contend with the
Gibbs phenomenon. To do so, they were informed by the seminal work of \citet{McHa10}, who developed a method, 
using filtering, to mitigate Gibbs phenomena in $P_{N}$ schemes. Unfortunately, as pointed out by McLerran and Hauck and by Radice et al.,
the filtered $P_{N}$ scheme does not have a unique continuum limit---i.e., it cannot be shown to be a discretization of a system of partial
differential equations. In Radice et al.'s approach, the spherical harmonic expansion of the solution is filtered at each time step using 
a spherical-spline filter:
\begin{equation}
  \label{eq:filter_expansion}
  \big[\mathscr{F}(F)\big](\varphi, \theta) = \sum_{\ell=0}^N \sum_{m =
  -\ell}^\ell \bigg[\sigma\Big(\frac{\ell}{N+1}\Big)\bigg]^s F^{\ell m}
  Y_{\ell m}(\varphi, \theta)\,,
\end{equation}
where $\sigma(\eta)$ is a filter function of order $p$ such that
\begin{align}
  \sigma(0) &= 1\,, &
  \sigma^{(k)}(0) = 0\,, \ \textrm{for } k = 1,2, \ldots p-1\,,
\end{align}
and where $s$ is a strength parameter, which is chosen to be a function of the time step:
\begin{equation}
s=\beta\Delta t,
\end{equation}
where $\beta$ is a parameter. Radice et~al.\ document success using a modified, second-order Lanczos filter:
\begin{equation}
\sigma(\eta)=\frac{\sin\eta}{\eta}.
\label{eq:Lanczosfilter}
\end{equation}
With the introduction of filtering, the time stepping algorithm, Eqs.~\eqref{eq:predictorstep} and \eqref{eq:correctorstep}, is modified as follows:
\begin{align}
  \label{eq:filter.a}
  \frac{{F^A}_{*} -
  {F^A}_k}{\Delta t/2} & = {\mathcal{A}^A}[F_k] + {\mathbb{S}^A}[F_{k+1/2}], \\
  \label{eq:filter.1}
  {F^A}_{k+1/2} & = {\mathscr{F}^A}_{B} {F^B}_{*}, \\
  \label{eq:filter.b}
   \frac{{F^A}_{**} -
  {F^A}_{k}}{\Delta t} &= {\mathcal{A}^A}[F_{k+1/2}] + {\mathbb{S}^A}[F_{k+1}], \\
  \label{eq:filter.2}
   {F^A}_{k+1} &= {\mathscr{F}^A}_{B} {F^B}_{**},
\end{align}
where ${\mathscr{F}^A}_{B}$ is a diagonal matrix that instantiates the filtering operation. Moreover, Radice et~al.\ were able to show that their filtering 
method represents the first-order, operator-split discretization of a term added to the underlying system of partial differential equations, Eq.~\eqref{eq:charon.scheme}:
\begin{equation}\label{eq:filtered.pn}
  \frac{\partial F^A}{\partial t} + {{\mathcal{P}}^{kA}}_{B}
  \frac{\partial F^B}{\partial x^k} = e^A + {S^A}_{B} F^B + \beta
  {L^A}_{B}  F^B\,,
\end{equation}
where ${L^A}_{B}$ is a diagonal matrix with coefficients $\log\sigma(l/(N+1))$. That is, their filtering method is equivalent to the addition of a forward-scattering 
term [$\sigma(0)=1$] to Eq.~\eqref{eq:charon.scheme}, and their overall method is a unique discretization of an underlying system of coupled partial differential equations, Eq.~\eqref{eq:filtered.pn}.

While the filtering effectively mitigates the Gibbs phenomenon, the distribution function can still become negative, which is unphysical.  
To contend with negative distribution functions in the context of the filtered $P_{N}$ scheme, \citet{LaHa19} developed and analyzed so-called positivity limiters, which can be used to ensure positivity of distribution function in each step of a time integration scheme.  

\subsection{Boltzmann kinetics: spectral decomposition across phase space}

\citet{PePeNo14} opt for a fully spectral approach to the solution of the 3+1 general relativistic Boltzmann equation in the CFC approximation
in non-conservative form:
\begin{equation}
  \label{eq:boltzefcfc}
  \frac{1}{\alpha}\frac{\partial f}{\partial t} + \left(
    \frac{p^i}{\Psi^2 \epsilon} -
  \frac{\beta^i}{\alpha} \right) \frac{\partial f}{\partial x^i} -
\bar{\Gamma}^j\,\!_{\mu \nu} p^\mu p^\nu J^i\,\!_j  \frac{1}{\epsilon}\frac{\partial f}{\partial p^i} = \frac{1}{\epsilon}\mathcal{C}[f].
\end{equation}
In this case, the 3+1 line element is
\begin{equation}
  \label{eq:lineelem}
  ds^2 = -\alpha^2 dt^2 + \gamma_{\tilde{i}\tilde{j}}(dx^{\tilde{i}} + \beta^{\tilde{i}} dt)(dx^{\tilde{j}} + \beta^{\tilde{j}} dt),
\end{equation}
where the spatial geometry is assumed to be conformally flat---i.e., 
\begin{equation}
  \label{eq:defcfc}
  \gamma_{\tilde{i}\tilde{j}} = \Psi^4 f_{\tilde{i}\tilde{j}}.
\end{equation}
In Eq.~\eqref{eq:defcfc}, $f_{\tilde{i}\tilde{j}}$ is the flat metric and $\Psi$ is the conformal factor,
\begin{equation}
  \label{e:def_Psi}
  \Psi = \left( \frac{\det \gamma_{\tilde{i}\tilde{j}}}{\det
      f_{\tilde{i}\tilde{j}}} \right)^{1/12}.
\end{equation}
In Eq.~\ref{eq:boltzefcfc}, $p^{\mu}$ and $\epsilon$ correspond to the neutrino four-momenta and energy, respectively, 
measured by an Eulerian observer. $\bar{\Gamma}^j\,\!_{\mu \nu}$ are the Ricci rotation coefficients. Peres et al.'s 
choice of phase-space coordinates is motivated by the known challenge time derivatives present for spectral methods. Specifically, 
were comoving-frame four-momenta chosen instead, the coefficients of the advection terms on the left-hand side of Eq.~\ref{eq:boltzefcfc} 
would contain time derivatives associated with, for example, relativistic Doppler shift. Of course, the collision term is best evaluated 
in the comoving frame, using comoving-frame four momenta, so the choice of Eulerian frame four-momenta necessitates additional
work to treat collisions. Peres et~al.\ leave the detailed treatment of this term to future publication. They also acknowledge the benefits
of beginning instead with the conservative form of Eq.~\eqref{eq:boltzefcfc} and leave that to future publication, as well.

In their approach, the distribution function is written as an expansion in terms of the basis functions across all six dimensions of phase space---in this case, 
spherical coordinates in both space and momentum space:
\begin{equation}
  \label{eq:6_variables}
  f \left(t, r, \theta, \phi, \epsilon, \Theta, \Phi \right) \simeq
\sum_{i=0}^{n_r} \sum_{j=0}^{n_\theta} \sum_{k=0}^{n_\phi} \sum_{l=0}^{n_\epsilon}
\sum_{m=0}^{n_\Theta} \sum_{p=0}^{n_\Phi} C_{ijklmp}(t)\, T_i (\bar{r})\,
F_j(\theta)\, F_k(\phi)\, T_l( \bar{\epsilon} )\, T_m( \bar{\Theta}
)\, F_p(\Phi).
\end{equation}
Chebyshev basis functions are used for $r$, $\epsilon$, and $\Theta$---i.e., for the expected non-periodic nature of the distribution function in
these dimensions. Fourier basis functions are used for $\theta$, $\phi$, and $\Phi$---i.e., for the expected periodic nature of the distribution function
in these dimensions.
Barred variables in Eq.~\eqref{eq:6_variables} are in the range $[-1,1]$ and are related to the standard coordinates by affine transformations. In the case
of the radial coordinate, the affine transformation is written explicitly as
\begin{equation}
  \label{eq:map_af}
  r = \alpha_r \bar{r} + \beta_r, \qquad \bar{r} \in [-1,1],
\end{equation}
where $\alpha_r$ and $\beta_r$ are constants, with $R_{\rm min}
= \beta_r - \alpha_r$ and $R_{\rm max} = \alpha_r + \beta_r$. 
$R_{\rm min}$ and $R_{\rm max}$ are the minimum and maximum radii of the spherical shell considered in the Peres et~al.\ 
analysis, respectively. (The extension of their method to $r=0$ is left for future development.) 
Ignoring the collision term in Eq.~\eqref{eq:boltzefcfc}, it can be written in terms of the Liouville operator, $\tilde{L}[f]$, as
\begin{equation}
  \label{eq:time_pde}
  \frac{\partial f}{\partial t} = - \tilde{L}[f]. 
\end{equation}
Substituting the expansion (\ref{eq:6_variables}) into Eq.~\eqref{eq:time_pde} results in a system of coupled ordinary differential equations for the 
solution vector, $U_N(t)$, where $N=n_r\times n_\theta \times n_\phi \times n_\epsilon \times n_\Theta \times n_\Phi$. The elements 
of the solution vector are the coefficients $C_{ijklmp}(t)$. Under this substitution, the operator, $\tilde{L}[f]$, in Eq.~\eqref{eq:time_pde} 
becomes an $N\times N$ matrix. To solve this system of equations, Peres et~al.\ employ an explicit, 
third-order, Adams--Bashforth scheme,
\begin{equation}
  \label{e:ODE_integration}
  U_N^{n+1} = U_N^n - \Delta t \left( \frac{23}{12} \tilde{L}_N U_N^n
    - \frac{4}{3} \tilde{L}_N U_N^{n-1} + \frac{5}{12} \tilde{L}_N
    U_N^{n-2} \right),
\end{equation}
though they emphasize they are not restricted to explicit updates but could also deploy semi-implicit and implicit 
methods. 

\subsection{Boltzmann kinetics: Monte Carlo methods}
\label{sec:MC}

Up to now, we have focused on deterministic methods for the solution of the Boltzmann neutrino transport equations
in core-collapse supernovae. But nondeterministic---specifically Monte Carlo---methods have also been used. Until 
recently, they have been confined to ``snapshot'' studies in a particular slice of an evolving stellar core and have been
used most extensively as a gauge of the accuracy of deterministic, but approximate, methods. Although it has yet to
be used in the context of a core-collapse supernova simulation as the method of choice for treating time-dependent
neutrino transport, a lepton-number and energy conserving Monte Carlo scheme for such transport has been developed by \citet{AbBuOt12} 
for the $O(v/c)$ limit of special relativistic effects and Newtonian gravity.

In their paper, Abdikamalov et~al.\ illustrate their method assuming spherical symmetry. They begin with the equation for 
the neutrino intensity for each neutrino species, here written generically without a species label:
\begin{eqnarray}
\label{eq:te}
\frac{1}{c}\frac{\p I (r,\mu,\ve,t)}{\p t} + \frac{\p 
  I(r,\mu,\ve,t)}{\p r} + \frac{1-\mu^2}{r} \frac{\p I
  (r,\mu,\ve,t)}{\p \mu} \nonumber\\\nonumber\\ =
  \kappa_a(\ve,T) \left[ B(\ve,T) - I
  (x,\mu,\ve,t)\right] - \kappa_s(\ve,T) I
  (r,\mu,\ve,t) \nonumber\\\nonumber\\ + 2 \pi
  \int_{-1}^{+1} \int_0^\infty \vk_s(\ve',\mu' \to
  \ve,\ \mu) I(x,\mu',\ve',t) d\mu' d\ve' \, .
\end{eqnarray} 
The first term term on the right-hand side is the familiar term for emission and absorption of neutrinos. $\kappa_a(s)$ is the total
absorption (scattering) opacity. The last term describes the additional source of neutrino as a result of inscattering into the 
neutrino ``beam'' with direction $\mu$ and energy $\epsilon$. Eq.~\eqref{eq:te} is solved using the boundary conditions:
\begin{equation}
\label{eq:tebcR}
I (R,\mu,\ve,0) = I_R (\mu,\ve,t)\, , \quad
-1\le\mu\le 0 \, .
\end{equation}
In their set of evolution equations, Eq.~\eqref{eq:te} is coupled to the material energy equation and the equation for the 
evolution of the electron fraction:
\begin{eqnarray}
  \label{eq:u0}
  \rho \frac{d U_m} {d t} = 2 \pi \sum_i 
   \int_{-1}^1 \int_0^\infty \kappa_{ai} (I_i - B_i) \, d\mu
   d\ve \nonumber\\\nonumber\\ + \sum_i S_i \,  , \\\nonumber\\
  \label{eq:Ye0}
  \rho N_A \frac{d Y_e} {d t} = 2\pi \sum_i s_i \int_{-1}^1
 \int_0^\infty \frac{\kappa_{ai}}{\ve} (I_i - B_i) \, 
 d\mu d\ve \, .
\end{eqnarray}
The sum over $i$ is over neutrino species, which will be dropped in what follows. $s_i=+1, -1, 0$ for electron neutrinos, 
electron antineutrinos, and heavy-flavor neutrinos, respectively, and will be carried through the remaining presentation 
of the method. In Eq.~\eqref{eq:u0}, $S$ is the contribution to the material energy from energy-exchanging scattering
with neutrinos and is given by
\begin{eqnarray}
\label{eq:s}
S = (2\pi)^2 \int_0^\infty \!\!\!\! \int_0^\infty \!\!\! \int_{-1}^1
  \int_{-1}^1 \!\! \bigg[\frac{\ve}{\ve'} \vk_s(\ve', \mu' \!\!\to
  \ve,\mu) I (x,\mu',\ve',t) \nonumber\\\nonumber\\ - \vk_s(\ve, \mu
  \to \ve', \mu') I (x,\mu,\ve,t) \bigg] d\ve d\ve' d\mu d\mu' \, .
\end{eqnarray}
Abdikamalov et~al.\ introduce the additional quantities:
\begin{eqnarray}
  \label{eq:u_r} 
    {U_r} & = & \frac{4\pi}{c} \int_0^\infty B d \ve \, , \\ 
  \label{eq:b}
  b & = & \frac{B}{4\pi \int_0^\infty B d\ve} \, , \\ 
  \label{eq:sigma_p}
  \kappa_p & = & \frac{\int_0^\infty \kappa_a B d\ve}{\int_0^\infty B 
  d\ve} \, , \\
  \label{eq:xi_a}
  \chi_a & = & \frac{\kappa_{a}}{\ve}\, , \\
  \label{eq:xi_p}
  \chi_p & = & \frac{\int_0^\infty\chi_a B d\ve}{\int_0^\infty B d\ve} \, ,
\end{eqnarray}
where $U_r$ is the equilibrium neutrino energy density and $\kappa_p$ is the Planck-mean opacity. The evolution
equation for $U_r$ is related to the evolution equations for $U_m$ and $Y_e$ by
\begin{equation}
  \label{eq:dBidt2}
  \frac{d {U_r}}{d t} = \beta \left(\rho \frac{d U_m}{dt}
  \right) + \zeta \left( \rho N_A \frac{dY_e}{dt} \right)\, ,
\end{equation}
where 
\begin{equation}
  \label{eq:alphabeta}
  \beta = \frac{1}{\rho C_V} \left(\frac{\p {U_r}}{\p
    T}\right)_{\rho,Ye} \, 
\end{equation}  
and
\begin{equation}
\label{eq:zeta}
  \zeta = \frac{1}{\rho N_A} \left[\left(\frac{\p {U_r}}{\p
      Y_e} \right)_{\rho,T} -\frac{1}{C_V} \left(\frac{\p
      U_m}{\p Y_e} \right)_{\rho,T} \left(\frac{\p
      {U_r}}{\p T} \right)_{\rho,Ye} \right] \, .
\end{equation}
In Eqs.~\eqref{eq:dBidt2} through \eqref{eq:zeta}, $N_A$ is Avogadro's Number and $C_V$ is the material heat 
capacity.

As with deterministic methods, the first step in the solution of Eq.~\eqref{eq:te}, \eqref{eq:u0}, and \eqref{eq:Ye0} is to 
linearize them. As part of this linearization procedure, Abdikamalov also ensure that these three evolution equations
become decoupled. The first step in the linearization process involves approximating 
$\{\kappa_a, \kappa_p,\kappa_s, \vk_s, b, \chi_a, \chi_p, \beta, \zeta\}$ 
with 
$\{\tilde\kappa_a,\tilde\kappa_p, \tilde\kappa_s, \tilde\vk_s, \tilde b, \tilde\chi_a, \tilde\chi_p, \tilde\beta, \tilde\zeta\}$. Abdikamalov define the 
latter as the time-centered values of the former within the time interval $t_n\le t\le t_{n+1}$. In practice, they are chosen
at the initial time step: $t_n$. Given this linearization, Eqs.~\eqref{eq:te}, \eqref{eq:u0}, \eqref{eq:Ye0}, and \eqref{eq:dBidt2}
become:
\begin{eqnarray}
\label{eq:rt2}
\frac{1}{c} \frac{\p I(\mu,\ve)} {\p t} + \mu \frac{\p
  I(\mu,\ve)}{\p r} + \frac{1-\mu^2}{r} \frac{\p
  I(\mu,\ve)}{\p \mu} \nonumber\\\nonumber\\ = c \tilde
  \kappa_{a} \tilde b {U_r} - (\tilde \kappa_a+\tilde \kappa_s)
  I(\mu,\ve) \nonumber\\\nonumber\\ + 2 \pi \int_{-1}^{+1}
  \int_0^\infty \tilde \vk_s(\ve',\mu' \to \ve,\
  \mu) I(\mu',\ve') d\mu' d\ve' \, ,
\end{eqnarray}
\begin{eqnarray}
  \label{eq:u2}
 &&  \rho \frac{d U_m} {d t} = 2\pi \int_{-1}^1 \int_0^\infty \tilde
  \kappa_{a} I d\mu d\ve - c \tilde \kappa_{p} {U_r} + S\, ,
  \\\nonumber\\ 
  \label{eq:Ye2}
 && \rho N_A \frac{d Y_e} {d t} = 2\pi s_i \int_{-1}^1 \int_0^\infty 
  \tilde{\chi}_{a} I \, d\mu d \ve - c s_i
  \tilde{\chi}_{p} {U_r} \, ,
\end{eqnarray}
\begin{equation}
  \label{eq:dBidt5}
  \frac{d {U_r}}{d t} = 2\pi \int_{-1}^1 \int_0^\infty \tilde
  \gamma I d \, \mu d \ve - c \tilde \gamma_{p}
  {U_r} + \tilde\beta S \, ,
\end{equation}
where
\begin{eqnarray}
  \tilde \gamma &=& \tilde \beta \tilde \kappa_{a} + \tilde \zeta s_i
  \tilde{\chi}_{a} \, , \\\nonumber\\ 
  \tilde \gamma_{p} &=& \tilde \beta \tilde \kappa_{p} + \tilde \zeta
  s_i \tilde{\chi}_{p} \, .
\end{eqnarray}
It is understood in Eqs.~\eqref{eq:u2} and \eqref{eq:dBidt5} that $\vk_s$ is replaced by $\tilde\vk_s$. Abdikamalov et~al.\ 
then time average Eq.~\eqref{eq:dBidt5} and use 
\begin{equation}
  \label{eq:bbar}
  \bar{U}_r = \alpha U_{r,n+1} + (1 - \alpha) U_{r,n}^* \, ,
\end{equation}
where, as pointed out by Abdikamalov et al., $\alpha$ controls the degree to which the method is implicit, and where
\begin{equation}
U_{r,n}^* = U_{r,n}+\tilde\beta\Delta t_n\bar S 
\end{equation}
and
\begin{equation}
\label{eq:s_bar}
\bar S = \frac{1}{\Delta t_n} \int_{t_n}^{t_{n+1}} S(t) dt ,
\end{equation}
to obtain
\begin{equation}
  \label{eq:dBidt10}
  \bar{U}_r = f_{n} U_{r,n}^* + 2\pi\frac{1-f_{n}}{c \tilde
    \gamma_{p}} \int_{-1}^1 \int_0^\infty \tilde \gamma \bar{I} \,
    d\mu d\ve \, ,   
\end{equation}
where 
\begin{equation}
U_{r,n}^* = U_{r,n}+\tilde\beta\Delta t_n\bar S 
\end{equation}
and
\begin{equation}
\label{eq:ff_nu}
f_{n} = \frac{1}{1 + \alpha c \Delta t_n \tilde \gamma_{p}} .
\end{equation}
Abdikamalov et~al.\ now assume that $\bar{U}=U_r(t)$ and $\bar{I}=I(t)$ in Eq.~\eqref{eq:dBidt10} and use the resultant equation to
substitute for $U_r$ in Eq.~\eqref{eq:rt2}, to obtain their final equation for the evolution of the neutrino intensity:
\begin{eqnarray}
  \label{eq:rt5}
  \frac{1}{c} \frac{\partial I} {\partial t} + \mu \frac{\p I}{\p r} +
  \frac{1-\mu^2}{r} \frac{\p I}{\p \mu} = \tilde \kappa_{ea} c \tilde
  b U_{r,n}^*  \nonumber\\\nonumber\\
  - \tilde \kappa_{ea} I + \tilde \kappa_{es,e} I + \tilde
  \kappa_{es,l} I + \tilde \kappa_s I
  \nonumber\\\nonumber\\+ 
  2\pi \frac{\tilde \kappa_a \tilde b}{\tilde \kappa_p} \int_{-1}^1
  \int_0^\infty \tilde \kappa_{es,e} I \, d\mu d \ve
  + 2\pi \frac{\tilde \kappa_a \tilde b}{\tilde \chi_p} \int_{-1}^1
  \int_0^\infty \tilde \chi_{es,l} I \, d\mu d \ve
  \nonumber\\\nonumber\\ 
  + 2 \pi \int_{-1}^{+1} \int_0^\infty \tilde
  \vk_s(\ve',\mu' \to \ve,\ \mu)
  I(\mu',\ve') d\mu' d\ve' \, ,
\end{eqnarray}
where
\begin{eqnarray}
\label{eq:kappa_ea}
\kappa_{ea} &=& f_n \kappa_a \, , \\
\label{eq:kapppa_ese}
\kappa_{es,e} &=& (1-f_n) \frac{\tilde \beta \tilde \kappa_p}{\tilde
  \gamma_p} \kappa_a \, , \\
\label{eq:kapppa_esl}
\kappa_{es,l} &=& (1-f_n) \frac{\tilde \zeta s_i \tilde \chi_p}{\tilde
  \gamma_p} \kappa_a \,  , \\
\label{eq:chi_ese}
\chi_{es,e} &=& (1-f_n) \frac{\tilde \beta \tilde \kappa_p}{\tilde
  \gamma_p} \chi_a \, ,  \\
\label{eq:chi_esl}
\chi_{es,l} &=& (1-f_n) \frac{\tilde \zeta s_i \tilde \chi_p}{\tilde
  \gamma_p} \chi_a \, .
\end{eqnarray}
A similar procedure can be used to derive equations for the updates of $U_m$ and $Y_e$, as was performed for
$U_r$. Abdikamalov et~al.\ point out that care must be taken to use the same expression for $U_r$---specifically, 
Eq.~\eqref{eq:dBidt10} with $\bar{U}_r=U_r(t)$ and $\bar{I}=I(t)$---in the derivation of the equation for $U_m$ in order to guarantee conservation of energy, to 
arrive at
\begin{eqnarray}
  \label{eq:u5}
  U_{m,n+1} =  U_{m,n} + \frac{\Delta t_n}{\rho} \bigg\{ 2\pi\int_{-1}^1
  \int_0^\infty \tilde \kappa_{ea} \bar I \, d \mu d \ve -
  \nonumber\\\nonumber\\ c f_{n} \tilde \kappa_{p} U_{r,n} +
  2\pi \int_{-1}^1 \int_0^\infty \tilde \kappa_{es,l} \bar I
  \, d \mu  d \ve \nonumber\\\nonumber\\ -
  2\pi \frac{\tilde \kappa_p}{\tilde \chi_p} \int_{-1}^1 \int_0^\infty
  \tilde \chi_{es,l} \bar I \, d \mu  d \ve + \bar S \bigg\} \, 
\end{eqnarray}
and
\begin{eqnarray}
  \label{eq:Ye3}
  Y_{e, n+1} = Y_{e, n} + \frac{\Delta t_n}{\rho N_A} \bigg\{ 2\pi s_i
  \int_{-1}^1 \int_0^\infty \tilde \chi_{ea} \bar I \, d \mu d \ve
  - \nonumber\\\nonumber\\ c s_i f_{n} \tilde \chi_{p} U_{r,n} + 
  2\pi s_i \int_{-1}^1 \int_0^\infty \tilde \chi_{es,e} \bar I \, d
  \mu d \ve \nonumber\\\nonumber\\ - 2\pi s_i \frac{\tilde
  \chi_p}{\tilde \kappa_p} \int_{-1}^1 \int_0^\infty \tilde
  \kappa_{es,e} \bar I \, d \mu d \ve \bigg\}\, .
\end{eqnarray}

Having linearized and decoupled the equations of motion, the evolution in Abdikamalov et al.'s  Monte Carlo approach proceeds as follows:
The weight associated with each Monte Carlo paricle (MCP) is the number of particles associated with it and is assumed to be $N_0$.
The number of particles emitted by the matter in the time interval $[t_n,t_n+1]$ is
\begin{equation}
{\cal N}_T = 8\pi^2\int_{t_n}^{t_{n+1}}\int_0^R\int_0^\infty
\frac{\kappa_a(\ve, T) B(\ve, T)}{\ve} r^2 dt
dr d\ve  \, .
\end{equation}
Then, the number of MCP's emitted in this time interval is
\begin{equation}
N_T = \mathrm{RInt} \left( {\cal N}_T / N_0 \right) \, ,
\end{equation}
where $\mathrm{RInt}(x)$ returns the largest integer less than $x$. The particle energy in each MCP is chosen according to the functional form
of $\kappa B$. Since thermal emission is isotropic, the angle of propagation of each MCP emitted, $\mu$, is chosen uniformly on the interval $[-1,+1]$
using
\begin{equation}
\mu = 2\xi-1 \, ,
\label{eq:randomangledetermination}
\end{equation}
where $\xi$ is a random number that takes on values in the interval $[0,1]$. 
Similarly, the emission time is chosen uniformly on the interval $[t_n,t_{n+1}]$ using
\begin{equation}
t = t_n + (t_{n+1} - t_n) \xi \, .
\end{equation}
To choose the zone in which an MCP is emitted, Abdikamalov et~al.\ use the probability that the MCP is emitted in a particular zone, 
which is given by the total number of particles emitted in that particular zone divided by the total number of particles emitted across 
all zones. Once an MCP is emitted in a particular zone, its location (assuming spherical symmetry) within that zone is determined using
\begin{equation}
r=\left[r_{j-1/2}^3 + \left(r_{j+1/2} - r_{j-1/2} \right)^3\xi
  \right]^{1/3} \, .
\end{equation}
where $j$ is the zone index. The number of MCPs entering from the outer boundary of the domain, at radius $R$, during the interval 
$[t_n,t_{n+1}]$ is given by
\begin{equation}
  N_B = \mathrm{RInt} \left[ - \frac{8\pi^2 R^2}{N_0}
    \int_{t_n}^{t_{n+1}}\int_0^\infty \int_{-1}^0 \frac{\mu
      I_R(\mu,\ve,t)}{\ve} dt d\ve d\mu \right] \, .
\end{equation}
The number of MCPs present at the beginning of the interval is
\begin{equation}
  N_{IC} = \mathrm{RInt} \left[  \frac{8\pi^2}{cN_0} \int_0^R
    \int_{-1}^1 \int_0^\infty I_i (r, \mu, \ve) r^2 dr d\mu d\ve
    \right] \, ,
\end{equation}
where the spatial zone, propagation angle, and energy of each MCP is chosen randomly using the functional form of $I$.

During transport, an emitted MCP will either (1) travel within the zone without collision and remain in the zone, 
(2) encounter a collision within the zone, or (3) exit the zone. These three possibilities correspond to three 
different distances, given by
\begin{equation}
\label{eq:d_b}
d_b=\left\{
\begin{array}{ll}
\left|\left[r_{j+1/2}^2-r^2(1-\mu^2)\right]^{1/2} - r \mu\right|, & 
\mathrm{if} \ j=1 \ \mathrm{or} \\ & \mu>0 \, , \ \sin\theta \ge
\frac{R_{j-1/2}}{r} \,  \\ 
& \\ & \\
\left|\left[r_{j-1/2}^2-r^2(1-\mu^2)\right]^{1/2} + r \mu\right|, &
\mathrm{if} \ \mu < 0 \, , \ \sin\theta <
\frac{R_{j-1/2}}{r} \,  , \\ & \\
\end{array}
\right.
\end{equation}
\begin{equation}
  \label{eq:d_t}
  d_t = c (t_{n+1} - t) \, ,
\end{equation}
and
\begin{equation}
  \label{eq:d_c}
  d_c = - \frac{\ln \xi}{\kappa_a + \kappa_s} \, .
\end{equation}
In Eqs.~\eqref{eq:d_b}, \eqref{eq:d_t}, and \eqref{eq:d_c}, $d_b$, $d_t$, and $d_c$ are the distance to the boundary
of the zone, the distance the particle can travel in the time interval if it does not encounter a collision, and the distance 
between collisions, respectively. Once these distances are known, the MCP is moved to the location corresponding to the
smallest of the three distances, and to the associated time, according to
\begin{eqnarray}
  \label{eq:xupdate}
  r &\to& \sqrt{r^2 - 2 r d \mu + d^2} \, , \\ \nonumber\\
  \label{eq:tupdate}
  t &\to& t + d / c \, .
\end{eqnarray}
If $d=d_c$, the MCP is either absorbed or scattered. To determine which, Abdikamalov et~al.\ use the following probabilities 
corresponding to the absorption and scattering coefficients appearing in Eq.~\eqref{eq:rt5}, the equation governing the 
MCP transport:
\begin{eqnarray}
P_{ea} & = & \kappa_{ea}/(\kappa_e+\kappa_s) , \\
P_s & = & \kappa_s/(\kappa_e+\kappa_s) , \\
P_{es,e} & = & \kappa_{es,e}/(\kappa_e+\kappa_s) , \\
P_{es,l} & = & \kappa_{es,l}/(\kappa_e+\kappa_s) .
\end{eqnarray}
The sum of all of these probabilities is, of course, equal to 1. As a result, to determine which of the above interactions takes place, Abdikamalov et~al.\ sample a random number $\xi$ in the range $[0,1]$. Based on the value of $\xi$: (1) if $\xi < P_{ea}$, the MCP undergoes effective absorption, (2) if $P_{ea} < \xi < P_{ea} + P_s$, the MCP is scattered, (3) if  
$P_{ea} + P_s < \xi < P_{ea} + P_s + P_{es,e}$, the MCP undergoes effective scattering in which its total energy is conserved, and (4) if $\xi > P_{ea} + P_s + P_{es,e}$, the MCP undergoes effective scattering in which its total lepton number is conserved. Within the domain $[0,1]$, the subdomain corresponding to each of the above possibilities is proportional to the probability for each possibility to occur, which ensures that the selection procedure yields a statistically correct result. And their result does not depend on the order in which they consider the possibilities.
If the MCP is absorbed, its energy and lepton number are deposited in the zone and it is removed from the population of 
MCPs. 
If the MCP undergoes real scattering, it is moved to the location where the scattering occurs. For iso-energetic 
scattering, its angle is determined randomly using Eq.~\eqref{eq:randomangledetermination}. If its energy changes as well,
its new energy is determined by randomly sampling the functional form of the scattering kernel in energy.
If the MCP undergoes effective scattering, which is isotropic, the MCP's angle is again determined randomly using 
Eq.~\eqref{eq:randomangledetermination} and its energy is determined by randomly sampling the local emissivity spectrum since 
effective scattering mimics absorption and reemission.
If $d=d_b$ and the boundary is the zone boundary, the transport sampling process begins again, using the values
of the opacities in the new zone. If the boundary is the outer boundary, the MCP is removed from the population of MCPs.
Finally, if $d=d_t$, the MCP is stored for the next time step. The above procedure is conducted for all of the MCP's in the 
computational domain (i.e., in all zones) at the beginning of a time step.

For the case of a non-static medium, the comoving and Eulerian frames are no longer coincident and an extension of the 
Monte Carlo procedure outlined above is necessary. Abdikamalov et~al.\ extend their approach as follows: The emissivities 
and opacities are naturally computed in the comoving frame. Once calculated, the number of MCPs emitted in this frame  
in each cell is determined. Assuming spherical symmetry for simplicity, the location, $r_0$, direction of propagation, $\mu_0$, 
and energy, $\epsilon_0$ of each MCP emitted at $t_0$ is sampled based on the comoving frame emissivities. Each of these quantities is
then transformed to the Eulerian frame using the well-known transformations (reproduced here for the spherically symmetric
case):
\begin{equation}
\label{eq:e_lorentz}
\ve_0 = \gamma \ve \left(1 - \frac{V_r \mu}{c}\right) \, ,
\end{equation}
\begin{equation}
\label{eq:mu_lorentz}
\mu_0  = \frac{\mu-V_r/c}{1-\mu V_r/c} \, ,
\end{equation}
\begin{equation}
\varphi_0  = \varphi \, ,
\end{equation}
\begin{equation}
\label{eq:kappa_lorentz}
\kappa(\mu, \ve) = \frac{\ve_0}{\ve} \kappa_0 (\ve_0) \ ,
\end{equation}
\begin{equation}
r = \gamma_j \left[r_0 + V_{r,j} (t_0-t_n)\right] \, ,
\end{equation}
\begin{equation}
t = \gamma_j \left(t_0-t_n+ \frac{V_{r,j} r_0}{c^2}\right) \, .
\end{equation}
The index $j$ in the last two equations is the index of the comoving-frame cell in which the MCP is emitted. (Of course,
$V_{r,j}$ is measured in the Eulerian frame.) Once these transformations are made, the MCP is transported in the
Eulerian frame, as described in the static case. Note, however, the distance to collision must be determined 
using the Eulerian-frame values of the opacities. Most of the steps in the static case proceed in the same way, with the 
exception of scattering, which requires additional care. If the MCP scatters, Abdikamalov et~al.\ transform the angle of
propagation and the energy of the MCP into the comoving frame, determine a new comoving-frame angle and energy 
due to the scattering event, then transform this new set of momentum-space variables back into the Eulerian frame 
before the transport of the MCP proceeds. The amount of energy and momentum exchanged between the MCP and 
the matter during the scattering, determined in the comoving frame, is recorded.

One further addition to the method presented by Abdikamalov et~al.\ that should be noted is the computational 
efficiency they gain by coupling their method to a Discrete Diffusion Monte Carlo (DDMC) method, first developed 
by \citet{Densmore2007} for photon transport and extended by Abdikamalov to neutrino transport. 
The latter method is used in diffusive regimes, where the original Monte Carlo method is plagued by the short 
distances between collisions: MCP paths between collisions become very short and the number of such paths 
that have to be simulated becomes prohibitively large. However, even with the coupling to DDMC, the Monte 
Carlo approach described here remains expensive and awaits future computing architectures 
that are more capable and well-suited to such an approach in order to be used for core-collapse supernova simulations.

\subsection{Two-moment kinetics}
\label{sec:numericalTwoMomentKinetics}

Numerical methods for solving equations for two-moment kinetics in core-collapse supernovae have now been developed by multiple groups \citep{MuJaDi10,OCon15,JuObJa15,KuTaKo16,RoOtHa16,SkDoBu19}.  
There are as many variations in approach as there are groups. Here we focus on common features and highlight specific solutions.  
For example, some authors have adopted fully relativistic descriptions \citep[e.g.,][]{MuJaDi10,OCon15,KuTaKo16,RoOtHa16}, while others have resorted to approximations that seek to capture relativistic effects \citep[e.g.,][]{JuObJa15,SkDoBu19}.  
Current methods for solving the equations for neutrino-radiation hydrodynamics using the two-moment approach employ finite-volume or finite-difference type methods.  
To this end, the system of equations can be written in the compact form (cf.\ Eqs.~\eqref{eq:BaryonMassConservation3p1}-\eqref{eq:fluidMomentumEquation3p1}, and Eqs.~\eqref{eq:spectralEulerianEnergyEquation_3p1} and \eqref{eq:spectralEulerianMomentumEquation_3p1})
\begin{align}
  \pd{}{t}\mathbf{U} 
  + \pd{\mathbf{F}^{i}(\mathbf{U})}{i} 
  + \pd{}{\varepsilon}\big(\,\varepsilon\,\mathbf{F}^{\varepsilon}(\mathbf{U})\,\big)
  =\mathbf{S}(\mathbf{U}) + \mathbf{C}(\mathbf{U}) ,
  \label{eq:fluidTwoMomentSystem}
\end{align}
where the vector of evolved quantities is given by
\begin{equation}
  \mathbf{U}
  =\sqrt{\gamma}\,\big(\,D,\,S_{j},\,\tau,\,D\,Y_{e},\,\varepsilon^{2}\mathcal{E}_{1},\,\varepsilon^{2}\mathcal{F}_{1,j},\ldots,\,\varepsilon^{2}\mathcal{E}_{\nSpecies},\,\varepsilon^{2}\mathcal{F}_{\nSpecies,j}\,\big)^{T}.
  \label{eq:fluidTwoMomentState}
\end{equation}
The spatial flux vectors $\mathbf{F}^{i}$, energy-space flux vector $\mathbf{F}^{\varepsilon}$ (zero for fluid variables), ``geometry'' sources $\mathbf{S}$, and the ``collision'' source due to neutrino--matter interactions $\mathbf{C}$ can be inferred from equations given in Sections~\ref{sec:hydrodynamics3p1}, \ref{sec:MomentKineticsAndClosure}, and \ref{sec:neutrinoInteractions}.  
Here, as an example, we consider the Eulerian two-moment model described in Sect.~\ref{sec:TwoMoment} with $\nSpecies$ neutrino species.  
Note that for each neutrino species, each radiation moment is represented by $N_{\varepsilon}$ degrees of freedom to represent the energy distribution of neutrinos, giving a total of $4\times N_{\varepsilon}\times\nSpecies$ radiation degrees of freedom (compared to $6$ fluid degrees of freedom) per point in spacetime.  
In core-collapse supernova models, $N_{\varepsilon}=\mathcal{O}(20)$, while $\nSpecies=3-6$, resulting in $240-480$ degrees of freedom per spacetime point.  

Among the approaches to solve the system of equations given by Eq.~\eqref{eq:fluidTwoMomentSystem} numerically, high-resolution shock-capturing (HRSC) methods (e.g., finite-volume or finite-difference), initially developed for compressible hydrodynamics with shocks, have attracted much attention recently.  
(For simplicity of presentation, we proceed to discuss the case of one spatial dimension.)  
In the HRSC approach, the spacetime is discretized into spacelike foliations of spacetime with discrete time coordinates $\{\,t^{n}\,\}_{n=0}^{N_{t}}$, where the time step $\dt=t^{n+1}-t^{n}$ is the separation between foliations.  
On each foliation, spatial positions are assigned coordinates $\{\,x_{j-\f{1}{2}}\,\}_{j=1}^{N_{x}+1}$, separating $N_{x}$ ``cells'' with width $\dx_{j}=(x_{j+\f{1}{2}}-x_{j-\f{1}{2}})$.  
In addition, for radiation quantities, momentum (energy) space is discretized into $N_{\varepsilon}$ ``energy bins'' with edges $\{\,\varepsilon_{i-\f{1}{2}}\,\}_{i=1}^{N_{\varepsilon}+1}$ and bin widths $\de_{i}=(\varepsilon_{i+\f{1}{2}}-\varepsilon_{i-\f{1}{2}})$.  
Integration of Eq.~\eqref{eq:fluidTwoMomentSystem} over the phase-space cell $I_{ij}=I_{i}^{\varepsilon}\times I_{j}^{\varepsilon}$, where $I_{i}^{\varepsilon}=(\varepsilon_{i-\f{1}{2}},\varepsilon_{i+\f{1}{2}})$ and $I_{j}^{x}=(x_{j-\f{1}{2}},x_{j+\f{1}{2}})$, gives the semi-discretized system
\begin{equation}
  \deriv{\mathbf{U}_{ij}}{t}
  =-\f{1}{\dV_{ij}}\big(\,\mathbf{F}_{ij+\f{1}{2}}^{x}-\mathbf{F}_{ij-\f{1}{2}}^{x}\,\big)
  -\f{1}{\dV_{ij}}\big(\,\varepsilon_{i+\f{1}{2}}\mathbf{F}_{i+\f{1}{2}j}^{\varepsilon}-\varepsilon_{i-\f{1}{2}}\mathbf{F}_{i-\f{1}{2}j}^{\varepsilon}\,\big)
  +\mathbf{S}_{ij}+\mathbf{C}_{ij},
  \label{eq:fluidTwoMomentSystemSemiDiscrete}
\end{equation}
where the evolved quantities are the cell averages defined as
\begin{equation}
  \mathbf{U}_{ij}(t) = \f{1}{\dV_{ij}}\int_{I_{ij}}\mathbf{U}(\varepsilon,x,t)\,d\varepsilon\,dx
  \quad\text{and}\quad
  \dV_{ij} = \int_{I_{ij}}\sqrt{\gamma}\,\varepsilon^{2}d\varepsilon dx,
  \label{eq:cellAverage}
\end{equation}
with $\mathbf{S}_{ij}$ and $\mathbf{C}_{ij}$ defined analogously, and the fluxes defined as
\begin{align}
  \mathbf{F}_{ij\pm\f{1}{2}}^{x}(t)
  &= \int_{I_{i}^{\varepsilon}}\mathbf{F}^{x}(\varepsilon,x_{j\pm\f{1}{2}},t)\,d\varepsilon, \label{eq:fluxSpace} \\
  \mathbf{F}_{i\pm\f{1}{2},j}^{\varepsilon}(t)
  &= \int_{I_{j}^{x}}\mathbf{F}^{\varepsilon}(\varepsilon_{i\pm\f{1}{2}},x,t)\,dx. \label{eq:fluxEnergy}
\end{align}
In Eq.~\eqref{eq:fluidTwoMomentSystemSemiDiscrete}, the temporal dimension has been left continuous (semi-discrete).  
Moreover, the equation is still exact.  
Approximations enter with the specification of the fluxes in Eqs.~\eqref{eq:fluxSpace} and \eqref{eq:fluxEnergy}, and the integrals to evaluate the sources $\mathbf{S}_{ij}$ and $\mathbf{C}_{ij}$.  
These approximations ultimately result in phase-space discretization errors.  
With these specifications, the approximate system in Eq.~\eqref{eq:fluidTwoMomentSystemSemiDiscrete} can be viewed as a system of ordinary differential equations (ODEs), which can be integrated forward in time with an ODE solver, which introduces temporal discretization errors.  
This discretization approach is called the method of lines (MOL).  

\subsubsection{Spatial discretization}

The spatial fluxes in Eq.~\eqref{eq:fluxSpace} can be approximated with an appropriate numerical flux:
\begin{equation}
  \mathbf{F}_{ij+\f{1}{2}}^{x}(t)
  \approx\de_{i}\,\widehat{\mathbf{F}^{x}}\big(\mathbf{U}(\varepsilon_{i},x_{j+\f{1}{2}}^{-},t),\mathbf{U}(\varepsilon_{i},x_{j+\f{1}{2}}^{+},t)\big),
  \label{eq:numericalFluxSpace}
\end{equation}
where $\mathbf{U}(\varepsilon_{i},x_{j+\f{1}{2}}^{\pm},t)$ is an approximation of $\mathbf{U}$ to the immediate left and right of the cell interface located at $x_{j+\f{1}{2}}$ ($x_{j+\f{1}{2}}^{\pm}=\lim_{\delta\to0^{+}}x_{j+\f{1}{2}}\pm\delta$).  
(In Eq.~\eqref{eq:numericalFluxSpace}, the midpoint rule is used to approximate the integral, but a more accurate quadrature rule can be used if desired.)  
Two things must be defined when computing the interface fluxes: (1) the procedure to \emph{reconstruct} the ``left'' and ``right'' states, and (2) the numerical flux function $\widehat{\mathbf{F}^{x}}$.  
The reconstruction step for radiation variables is essentially identical to that used for hydrodynamics schemes: a polynomial of degree $k$ is reconstructed from the evolved quantities (cell averages).  
To this end, the accuracy of the numerical method depends in part on the degree of the reconstructed polynomial, and the desired polynomial degree impacts the width of the computational stencil, since values in $k+1$ cells are needed to reconstruct a polynomial of degree $k$.  
The most commonly used methods are monotonized piecewise linear \citep{vanLeer74,Lev92} and piecewise parabolic methods \citep{CoWo84}, as well as higher order monotonicity preserving (MP) \citep{SuHu97} and weighted essentially nonoscillatory (WENO) reconstruction methods \citep{LiOsCh94,Shu97}.  
Monotonicity constraints are placed on the reconstruction polynomial to ensure nonoscillatory solutions around discontinuities.  
For fluid variables, the numerical flux function can be computed with a standard Riemann solver; e.g., HLL \citep{HaLaLe83} or HLLC \citep{ToSpSp94}.  
However, when using finite-volume or finite-difference methods to solve for the radiation moments, specification of the numerical flux requires additional care.  
As elucidated by the analysis in \citet{AuChCh02} in the context of the $\mathcal{O}(v/c)$ limit of the energy integrated (gray) Lagrangian two-moment model presented in Sect.~\ref{sec:TwoMoment}, in the asymptotic diffusion limit (characterized by a short neutrino mean free path) the inherent numerical dissipation associated with the numerical flux used for hyperbolic systems overwhelms the physical radiative diffusive flux and leads to spurious evolution unless the mean free path is resolved by the spatial grid.  
We discuss this important issue further below \citep[see also][for discussions on this topic]{JiLe96,LoMo01}.  
Since it is not practical to resolve the neutrino mean free path in core-collapse supernova simulations, the numerical fluxes for the radiation moment equations are modified to better capture the evolution in diffusive regimes.  
Following \citet{AuChCh02}, \citet{OcOt13} propose the following modified HLL numerical fluxes for the two-moment model for neutrino transport \citep[see also][]{KuTaKo16}:
\begin{align}
  \widehat{F_{\mathcal{E}_{s}}^{x}}\big(\mathbf{U}_{\LEFT},\mathbf{U}_{\RIGHT}\big)
  &=\f{\lambda^{+}F_{\mathcal{E}_{s}}^{x}(\mathbf{U}_{\LEFT})+\lambda^{-}F_{\mathcal{E}_{s}}^{x}(\mathbf{U}_{\RIGHT})-\xi\lambda^{-}\lambda^{+}\big((\mathcal{E}_{s})_{\RIGHT}-(\mathcal{E}_{s})_{\LEFT}\big)}{\lambda^{-}+\lambda^{+}} \label{eq:modifiedNumericalFluxEnergy} , \\
  \widehat{F_{\mathcal{S}_{s,j}}^{x}}\big(\mathbf{U}_{\LEFT},\mathbf{U}_{\RIGHT}\big)
  &=\f{\xi^{2}\big(\lambda^{+}F_{\mathcal{S}_{s,j}}^{x}(\mathbf{U}_{\LEFT})+\lambda^{-}F_{\mathcal{S}_{s,j}}^{x}(\mathbf{U}_{\RIGHT})\big)-\xi\lambda^{-}\lambda^{+}\big((\mathcal{S}_{s,j})_{\RIGHT}-(\mathcal{S}_{s,j})_{\LEFT}\big)}{\lambda^{-}+\lambda^{+}} \nonumber \\
  &\hspace{12pt}
  +(1-\xi^{2})\,\f{1}{2}\,\big(\,F_{\mathcal{S}_{s,j}}^{x}(\mathbf{U}_{\LEFT})+F_{\mathcal{S}_{s,j}}^{x}(\mathbf{U}_{\RIGHT})\,\big),
  \label{eq:modifiedNumericalFluxMomentum}
\end{align}
where $F_{\mathcal{E}_{s}}^{x}$ and $F_{\mathcal{S}_{s,j}}^{x}$ are the radiation energy and momentum spatial fluxes, respectively, 
and $\lambda^{-}$ and $\lambda^{+}$ are estimates of the largest (absolute) eigenvalues for left-going and right-going waves, respectively \cite[see, e.g.,][for explicit expressions of estimates]{ShKiSe11}.  
In the modified numerical fluxes in Eqs.~\eqref{eq:modifiedNumericalFluxEnergy} and \eqref{eq:modifiedNumericalFluxMomentum}, $\xi$ is a local parameter depending on the ratio of the neutrino mean free path to the local grid size:
\begin{equation}
  \xi = \min\big(1,\lambda_{ij}/\dx_{j}\big),
\end{equation}
where $\lambda_{ij}$ is a local, energy-dependent neutrino mean free path (computed from the neutrino opacities).  
Thus, when the mean free path is much smaller than a grid cell ($\xi\to0$), the numerical dissipation term (proportional to the jump in the conserved variables across the interface) vanishes, and the numerical flux switches to an average of the fluxes evaluated with the left and right states (a similar approach is also taken in \citet{JuObJa15,SkDoBu19}).  
It should be noted that the average flux is appropriate for solving parabolic equations, but is in general unstable for hyperbolic equations \citep[e.g.,][]{Lev92}.  

To further illustrate the issue with the numerical flux, and to see how the modifications in Eqs.~\eqref{eq:modifiedNumericalFluxEnergy}--\eqref{eq:modifiedNumericalFluxMomentum} help, it is easiest to consider the reduced system
\begin{align}
  \pd{\mathcal{J}}{t}+\pd{\mathcal{H}}{x} &= 0, \label{eq:reducedTwoMomentEnergyEquation} \\
  \pd{\mathcal{H}}{t}+\pd{\mathcal{K}}{x} &=-\f{1}{\lambda}\,\mathcal{H} \label{eq:reducedTwoMomentMomentumEquation},
\end{align}
where
\begin{equation}
  \big\{\,\mathcal{J},\mathcal{H},\mathcal{K}\,\big\}(x,t) = \f{1}{2}\int_{-1}^{1}f(\mu,x,t)\,\mu^{\{0,1,2\}}\,d\mu,
\end{equation}
and $\lambda$ is the scattering mean free path.  
When scattering events are frequent ($\lambda\to0$), the system in Eqs.~\eqref{eq:reducedTwoMomentEnergyEquation}--\eqref{eq:reducedTwoMomentMomentumEquation} limits to parabolic behavior governed by
\begin{equation}
  \pd{\mathcal{J}}{t}+\pd{\mathcal{H}}{x}=0
  \quad\text{and}\quad
  \mathcal{H}=-\f{\lambda}{3}\pd{\mathcal{J}}{x}
  \quad\Rightarrow\quad
  \pd{\mathcal{J}}{t}-\f{\lambda}{3}\,\partial_{xx}\mathcal{J}=0,
  \label{eq:simpleDiffusionLimit}
\end{equation}
which is referred to as the diffusion limit.  
The semi-discrete form of Eqs.~\eqref{eq:reducedTwoMomentEnergyEquation}--\eqref{eq:reducedTwoMomentMomentumEquation} can be written as
\begin{align}
  d_{t}\mathcal{J}_{i}+\f{1}{\dx}\Big(\,\widehat{\mathcal{H}}_{i+\f{1}{2}}-\widehat{\mathcal{H}}_{i-\f{1}{2}}\,\Big)&=0, \label{eq:reducedTwoMomentEnergyEquationSemiDiscrete} \\
  d_{t}\mathcal{H}_{i}+\f{1}{\dx}\Big(\,\widehat{\mathcal{K}}_{i+\f{1}{2}}-\widehat{\mathcal{K}}_{i-\f{1}{2}}\,\Big)&=-\f{1}{\lambda}\,\mathcal{H}_{i}.  \label{eq:reducedTwoMomentMomentumEquationSemiDiscrete}
\end{align}
With constant reconstruction, which results in first-order spatial accuracy, the numerical fluxes in Eqs.~\eqref{eq:modifiedNumericalFluxEnergy}--\eqref{eq:modifiedNumericalFluxMomentum} at the $x_{i+\f{1}{2}}$ interface become
\begin{align}
  \widehat{\mathcal{H}}_{i+\f{1}{2}}
  &=\f{1}{2}\Big(\,\mathcal{H}_{i+1}+\mathcal{H}_{i}-\xi\,\big(\,\mathcal{J}_{i+1}-\mathcal{J}_{i}\,\big)\,\Big), \label{eq:modifiedNumericalFluxEnergySimple} \\
  \widehat{\mathcal{K}}_{i+\f{1}{2}}
  &=\f{1}{2}\Big(\,\mathcal{K}_{i+1}+\mathcal{K}_{i}-\xi\,\big(\,\mathcal{H}_{i+1}-\mathcal{H}_{i}\,\big)\,\Big), \label{eq:modifiedNumericalFluxMomentumSimple}
\end{align}
where for simplicity we set $\lambda^{+}=\lambda^{-}=1$ (i.e., the global Lax-Friedrichs flux).  
By ignoring the time derivative term in Eq.~\eqref{eq:reducedTwoMomentMomentumEquationSemiDiscrete} and using the numerical flux in Eq.~\eqref{eq:modifiedNumericalFluxMomentumSimple} with $\mathcal{K}=\mathcal{J}/3$, one can write
\begin{align}
  \mathcal{H}_{i} 
  &= - \mathrm{Kn}\,\f{1}{2}\,\Big(\,\f{1}{3}\,\big(\,\mathcal{J}_{i+1}-\mathcal{J}_{i-1}\,\big)-\xi\,\big(\,\mathcal{H}_{i-1}-2\,\mathcal{H}_{i}+\mathcal{H}_{i+1}\,\big)\,\Big), \nonumber \\
  &\approx - \mathrm{Kn}\,\f{1}{2}\,\f{1}{3}\,\big(\,\mathcal{J}_{i+1}-\mathcal{J}_{i-1}\,\big), \label{eq:momentumDensitySimpleApproximate}
\end{align}
where we have introduced the Knudsen number $\mathrm{Kn}=\lambda/\dx$, the ratio of the mean free path to the spatial grid size.  
In Eq.~\eqref{eq:momentumDensitySimpleApproximate}, we ignored the numerical dissipation term because in the diffusion limit $|\mathcal{H}|\ll\mathcal{J}$.  
Then, inserting the numerical flux, Eq.~\eqref{eq:modifiedNumericalFluxEnergySimple}, using Eq.~\eqref{eq:momentumDensitySimpleApproximate}, into Eq.~\eqref{eq:reducedTwoMomentEnergyEquationSemiDiscrete} gives the approximate semi-discrete form of Eq.~\eqref{eq:reducedTwoMomentEnergyEquation} in the diffusion limit:
\begin{align}
  &d_{t}\mathcal{J}_{i}
  -\f{1}{(2\dx)^{2}}
  \Big[\,
    \f{\lambda}{3}\Big(\mathcal{J}_{i-2}-2\,\mathcal{J}_{i}+\mathcal{J}_{i+2}\Big) \nonumber \\
  &\hspace{72pt}
    +\min(\lambda,\dx)\Big(\mathcal{J}_{i-1}-2\,\mathcal{J}_{i}+\mathcal{J}_{i+1}\Big)
  \,\Big]=0,
  \label{eq:simpleDiffusionLimitSemiDiscrete}
\end{align}
which is an approximation to the diffusion equation in Eq.~\eqref{eq:simpleDiffusionLimit}.  
Note that the last term on the left-hand side of Eq.~\eqref{eq:simpleDiffusionLimitSemiDiscrete} is due to the numerical dissipation term (proportional to $\xi$) in Eq.~\eqref{eq:modifiedNumericalFluxEnergySimple}.  
Because of the introduction of $\xi$ in Eq.~\eqref{eq:modifiedNumericalFluxEnergySimple}, Eq.~\eqref{eq:simpleDiffusionLimitSemiDiscrete} remains a reasonable approximation to a diffusion equation with the correct diffusion coefficient $\lambda/3$, even as $\lambda\ll\dx$.  
Without the modification to the numerical flux (i.e., $\xi=1$ independent of $\lambda$), we would obtain Eq.~\eqref{eq:simpleDiffusionLimitSemiDiscrete} with $\min(\lambda,\dx)\to\dx$.  
In this case the numerical diffusion term would overwhelm the physical diffusion term when $\lambda\ll\dx$, and result in spurious evolution.  
Note, in this simplified discussion, where we assumed constant spatial reconstruction, the numerical dissipation term is of the same order of magnitude as the physical dissipation term, and still contributes to the diffusive evolution.  
With higher-order accurate spatial reconstruction, the relative contribution of this term decreases.  
Also note that in arriving at Eq.~\eqref{eq:simpleDiffusionLimitSemiDiscrete}, we only relied on the modification to the numerical flux in the energy equation, as is done by \citet{SkDoBu19}.  
Finally, note that in the physical diffusion term in Eq.~\eqref{eq:simpleDiffusionLimitSemiDiscrete}, the second derivative is approximated with a wide stencil, which supports a mode with odd-even point decoupling \citep{LoMo01}.  

\subsubsection{Energy discretization}
\label{sec:EnergyDiscretization}

Next we consider the approximation of the energy fluxes in Eq.~\eqref{eq:fluxEnergy}, which contribute to shifts in the neutrino energy spectrum due to gravitational and moving fluid effects.  
\citet{MuJaDi10}, who solved the Lagrangian two-moment model in Section~\eqref{sec:TwoMoment}, developed a method to compute the energy fluxes that is inherently number conservative; i.e., with this discretization of the energy derivative, the energy equation in the Lagrangian two-moment model in Eq.~\eqref{eq:spectralLagrangianEnergyEquation_3p1} is consistent with the equation for number conservation in Eq.~\eqref{eq:spectralNumberEquation_3p1} at the discrete level.  
A key observation in achieving this is that the number conservation equation is obtained by multiplying the Lagrangian energy equation with a factor $1/\varepsilon$.  
At the continuum level, when this factor is brought inside the energy derivative, the remainder cancels with the first term on the right-hand side of Eq.~\eqref{eq:spectralLagrangianEnergyEquation_3p1}, resulting in the conservative number equation in Eq.~\eqref{eq:spectralNumberEquation_3p1}.  
The relevant equation is given by considering only the energy derivative and the (non-collisional) source term in Eq.~\eqref{eq:spectralLagrangianEnergyEquation_3p1} \citep[cf.\ Eq.~(B1) in][]{MuJaDi10}:
\begin{equation}
  \pd{J}{t}+\pd{}{\varepsilon}\big(\,\varepsilon\,F_{J}\,\big)=F_{J},
  \label{eq:energyAdvectionEquation}
\end{equation}
where we introduce the shorthand notation
\begin{align}
  J = \sqrt{\gamma}\,\varepsilon^{2}\,\big(\,W\mathcal{J}+v^{i}\mathcal{H}_{i}\,\big)
  \quad\text{and}\quad
  F_{J} = - \alpha\,\sqrt{\gamma}\,\varepsilon^{2}\,\mathcal{T}^{\mu\nu}\nabla_{\mu}u_{\nu}.  
\end{align}
Dividing Eq.~\eqref{eq:energyAdvectionEquation} by $\varepsilon$ gives the conservation equation:
\begin{equation}
  \pd{N}{t}+\pd{}{\varepsilon}\big(\,F_{J}\,\big)=0,
  \label{eq:numberAdvectionEquation}
\end{equation}
where $N=J/\varepsilon$ is the spectral Eulerian number density (cf.\ Eq.~\eqref{eq:spectralNumberEquation_3p1}).  

Similar to Eq.~\eqref{eq:fluidTwoMomentSystemSemiDiscrete}, the semi-discrete form of Eq.~\eqref{eq:energyAdvectionEquation} can be written as
\begin{equation}
  \deriv{J_{i}}{t}
  =-\f{1}{\de_{i}}\big(\,\varepsilon_{i+\f{1}{2}}{\widehat{F_{J}}}_{i+\f{1}{2}}-\varepsilon_{i-\f{1}{2}}{\widehat{F_{J}}}_{i-\f{1}{2}}\,\big) + {F_{J}}_{i},
  \label{eq:energyAdvectionEquationSemiDiscrete}
\end{equation}
where ${\widehat{F_{J}}}_{i\pm\f{1}{2}}$ are the numerical flux functions to be determined.  
(Here we drop the spatial index $j$ to simplify the notation.)
Dividing Eq.~\eqref{eq:energyAdvectionEquationSemiDiscrete} by $\varepsilon_{i}$ and defining $N_{i}=J_{i}/\varepsilon_{i}$ gives a provisionary semi-discrete form of Eq.~\eqref{eq:numberAdvectionEquation}:
\begin{align}
  \deriv{N_{i}}{t}
  &=-\f{1}{\de_{i}}
  \big(\,
    \f{\varepsilon_{i+\f{1}{2}}}{\varepsilon_{i}}{\widehat{F_{J}}}_{i+\f{1}{2}}
    -\f{\varepsilon_{i-\f{1}{2}}}{\varepsilon_{i}}{\widehat{F_{J}}}_{i-\f{1}{2}}
  \,\big) + \f{{F_{J}}_{i}}{\varepsilon_{i}} \label{eq:numberAdvectionEquationSemiDiscrete} \\
  &=-\f{1}{\de_{i}}
  \big(\,
    {\widehat{F_{J}}}_{i+\f{1}{2}} - {\widehat{F_{J}}}_{i+\f{1}{2}}
  \,\big)
  -\f{(\varepsilon_{i+\f{1}{2}}-\varepsilon_{i})}{\de_{i}}\f{{\widehat{F_{J}}}_{i+\f{1}{2}}}{\varepsilon_{i}}
  -\f{(\varepsilon_{i}-\varepsilon_{i-\f{1}{2}})}{\de_{i}}\f{{\widehat{F_{J}}}_{i-\f{1}{2}}}{\varepsilon_{i}}
  + \f{{F_{J}}_{i}}{\varepsilon_{i}}.  \nonumber
\end{align}
Without specifying the numerical fluxes $\widehat{F_{J}}_{i\pm\f{1}{2}}$, the last three terms in the second line of Eq.~\eqref{eq:numberAdvectionEquationSemiDiscrete} do in general not cancel, and the neutrino number density is not conserved in the energy advection step, which is contrary to what is suggested by Eq.~\eqref{eq:numberAdvectionEquation}.  
However, there is some freedom in choosing the numerical fluxes.  
To determine the numerical fluxes, \citet{MuJaDi10} demand total number conservation upon integration of Eq.~\eqref{eq:numberAdvectionEquationSemiDiscrete} over all energy bins; i.e.,
\begin{align}
  0=d_{t}N_{\TOT}
  &\equiv\sum_{i=1}^{N_{\varepsilon}}\deriv{N_{i}}{t}\,\de_{i}
  =-\sum_{i=1}^{N_{\varepsilon}}
  \Big\{\,\f{\varepsilon_{i+\f{1}{2}}}{\varepsilon_{i}}{\widehat{F_{J}}}_{i+\f{1}{2}}
  -\f{\varepsilon_{i-\f{1}{2}}}{\varepsilon_{i}}{\widehat{F_{J}}}_{i-\f{1}{2}}
  -\f{\de_{i}}{\varepsilon_{i}}{F_{J}}_{i}\,\Big\} \nonumber \\
  &=-\sum_{i=1}^{N_{\varepsilon}}\Big\{\,\Big(\,\f{1}{\varepsilon_{i}}-\f{1}{\varepsilon_{i+1}}\,\Big)\,\varepsilon_{i+\f{1}{2}}\,{\widehat{F_{J}}}_{i+\f{1}{2}}-\f{\de_{i}}{\varepsilon_{i}}\,{F_{J}}_{i}\,\Big\},
\end{align}
where zero flux energy space boundaries are assumed (i.e., ${\widehat{F_{J}}}_{\f{1}{2}}={\widehat{F_{J}}}_{N_{\varepsilon}+\f{1}{2}}=0$).  
Next, the numerical flux is split into ``left'' and ``right'' contributions
\begin{equation}
  {\widehat{F_{J}}}_{i+\f{1}{2}} = {F_{J}^{\LEFT}}_{i} + {F_{J}^{\RIGHT}}_{i+1},
  \label{eq:energyFluxSplit}
\end{equation}
so that the change in the total number density can be written as (assuming $\varepsilon_{\f{1}{2}}=0$ and setting ${F_{J}^{\RIGHT}}_{N_{\varepsilon}+1}=0$)
\begin{equation}
  d_{t}N_{\TOT}=-\sum_{i=1}^{N_{\varepsilon}}
  \Big\{\,
    \Big(\f{1}{\varepsilon_{i}}-\f{1}{\varepsilon_{i+1}}\Big)\,\varepsilon_{i+\f{1}{2}}\,{F_{J}^{\LEFT}}_{i}
    +\Big(\f{1}{\varepsilon_{i-1}}-\f{1}{\varepsilon_{i}}\Big)\,\varepsilon_{i-\f{1}{2}}\,{F_{J}^{\RIGHT}}_{i}
    -\f{\de_{i}}{\varepsilon_{i}}\,{F_{J}}_{i}
  \,\Big\}.  
  \label{eq:totalNumberChange}
\end{equation}
Number conservation is then obtained by demanding
\begin{equation}
  \Big(\f{1}{\varepsilon_{i}}-\f{1}{\varepsilon_{i+1}}\Big)\,\varepsilon_{i+\f{1}{2}}\,{F_{J}^{\LEFT}}_{i}
    +\Big(\f{1}{\varepsilon_{i-1}}-\f{1}{\varepsilon_{i}}\Big)\,\varepsilon_{i-\f{1}{2}}\,{F_{J}^{\RIGHT}}_{i}
    =\f{\de_{i}}{\varepsilon_{i}}\,{F_{J}}_{i}.
\end{equation}
Furthermore, \citet{MuJaDi10} introduce
\begin{align}
  \varepsilon_{i+\f{1}{2}}\,{F_{J}^{\LEFT}}_{i}
  &=\f{\de_{i}}{1-\varepsilon_{i}\varepsilon_{i+1}^{-1}}\,{F_{J}}_{i}\,\xi_{i}, \\
  \varepsilon_{i-\f{1}{2}}\,{F_{J}^{\RIGHT}}_{i}
  &=\f{\de_{i}}{\varepsilon_{i}\varepsilon_{i-1}^{-1}-1}\,{F_{J}}_{i}\,(1-\xi_{i}),
\end{align}
where $\xi_{i}$ is a local weighting factor
\begin{equation}
  \xi_{i}=\f{j_{i+\f{1}{2}}^{\sigma}}{j_{i-\f{1}{2}}^{\sigma}+j_{i+\f{1}{2}}^{\sigma}}
  \quad\text{and}\quad
  1-\xi_{i}=\f{j_{i-\f{1}{2}}^{\sigma}}{j_{i-\f{1}{2}}^{\sigma}+j_{i+\f{1}{2}}^{\sigma}} ,
\end{equation}
depending on the zeroth moment ($j$) of the distribution function at cell interfaces, $j_{i-\f{1}{2}}^{\sigma}$ and $j_{i+\f{1}{2}}^{\sigma}$, which are computed as weighted geometric means of $j$ using values from adjacent energy bins.  
In regions where $J_{i}$ varies modestly with $i$, $\xi_{i}$ is close to $1/2$, while in the high-energy tail of the neutrino spectrum, where $J_{i}$ decreases rapidly with increasing $i$, $\xi_{i}\ll1$ \citep[see Appendix~B in][for further details]{MuJaDi10}.  
Then, using the split in Eq.~\eqref{eq:energyFluxSplit}, the numerical flux, e.g., at interface $\varepsilon_{i+\f{1}{2}}$, to be used in Eq.~\eqref{eq:energyAdvectionEquationSemiDiscrete} is given by
\begin{align}
  \varepsilon_{i+\f{1}{2}}{\widehat{F_{J}}}_{i+\f{1}{2}}
  &=\varepsilon_{i+\f{1}{2}}\,{F_{J}^{\LEFT}}_{i} + \varepsilon_{i+\f{1}{2}}\,{F_{J}^{\RIGHT}}_{i+1} \nonumber \\
  &=\f{\de_{i}}{1-\varepsilon_{i}\varepsilon_{i+1}^{-1}}\,{F_{J}}_{i}\,\xi_{i} + \f{\de_{i+1}}{\varepsilon_{i+1}\varepsilon_{i}^{-1}-1}\,{F_{J}}_{i+1}\,(1-\xi_{i+1}).  
  \label{eq:numericalFluxEnergySpaceGeneral}
\end{align}
For a commonly used geometrically progressing grid where $\varepsilon_{i+\f{1}{2}}=\de_{1}\,\lambda^{i-1}$ (where $\lambda>1$ and $i=1,\ldots,N_{\varepsilon}$), it can be shown that $\de_{i}/(1-\varepsilon_{i}\varepsilon_{i+1}^{-1})=\de_{i+1}/(\varepsilon_{i+1}\varepsilon_{i}^{-1}-1)=\varepsilon_{i+\f{1}{2}}$, so that the numerical flux can be written as
\begin{equation}
  {\widehat{F_{J}}}_{i+\f{1}{2}}\big({F_{J}}_{i},{F_{J}}_{i+1}\big) = {F_{J}}_{i}\,\xi_{i} + {F_{J}}_{i+1}\,(1-\xi_{i+1}),
  \label{eq:numericalFluxEnergySpace}
\end{equation}
which is simply a weighted average with nonlinear weights $\xi_{i}$ and $(1-\xi_{i+1})$.  
If $\xi_{i},\xi_{i+1}>0$ and $\xi_{i}+\xi_{i+1}=1$, the numerical flux is a convex combination of ${F_{J}}_{i}$ and ${F_{J}}_{i+1}$, but this is not guaranteed.  
Although the numerical flux in Eq.~\eqref{eq:numericalFluxEnergySpaceGeneral} was developed by \citet{MuJaDi10} to ensure neutrino number conservation in the context of the Lagrangian two-moment model, the same approach has also been applied to the Eulerian two-moment model by \citet{OCon15,KuTaKo16}.  
(It is not at all clear that the approach developed by \citet{MuJaDi10} in the context of the Lagrangian two-moment model results in a number conservative scheme when applied to the Eulerian two-moment model. In the Lagrangian two-moment model, the spectral neutrino number and energy equations are related simply by a factor of $1/\epsilon$, whereas in an Eulerian two-moment model, the relationship is more complex, involving both the spectral neutrino energy and momentum equations \citep[cf.][]{EnCaMe12c,CaEnMe13a}.)
We also note that the numerical flux in Eq.~\eqref{eq:numericalFluxEnergySpaceGeneral} is also used by \citet{JuObJa15}, who solve the Lagrangian two-moment model in the $\mathcal{O}(v/c)$ limit.  

A few remarks should be made about the numerical flux in Eq.~\eqref{eq:numericalFluxEnergySpace}.  
First, a numerical flux is said to be consistent if, when the two arguments are set to be equal, it reduces to the common value; i.e., when ${F_{J}}_{i}={F_{J}}_{i+1}={F_{J}}$ the following holds:
\begin{equation}
  {\widehat{F_{J}}}_{i+\f{1}{2}}\big({F_{J}},{F_{J}}\big)={F_{J}}.  
  \label{eq:consistentNumericalFlux}
\end{equation}
Consistency of the numerical flux is generally required for a numerical method to be convergent \citep{CrMa80,Lev02}.  
Since it is not guaranteed that $\xi_{i}+\xi_{i+1}=1$, the numerical flux in Eq.~\eqref{eq:numericalFluxEnergySpace} is not consistent.  
Second, if one sets $\xi_{i}=1/2~\forall i$ (which makes it consistent), the numerical flux in Eq.~\eqref{eq:numericalFluxEnergySpace} reduces to a simple arithmetic average, which is known to be notoriously unstable when combined with explicit time integration \citep[e.g.,][]{Lev02}.  

\citet{SkDoBu19}, who also solve the Lagrangian two-moment model in the $\mathcal{O}(v/c)$ limit, follow a different approach adapted from \citet{VaAuDu11}.  
In this case, assuming Cartesian coordinates for simplicity, the evolved quantity and the flux in energy space in the neutrino energy equation (cf.\ Eq.~\eqref{eq:energyAdvectionEquation}) are given by
\begin{equation}
  J = \varepsilon^{2}\mathcal{J}
  \quad\text{and}\quad
  F_{J} = - \varepsilon^{2}\mathcal{K}^{i}_{\hspace{4pt}j}\,\pd{v^{j}}{i},
\end{equation}
where $\mathcal{K}^{i}_{\hspace{4pt}j}$ is the radiation stress tensor (cf.\ Eq.~\eqref{eq:radiationStressTensor}) and $v^{i}$ are components of the fluid three-velocity.  
Similarly, the evolved quantity and flux in energy space from the neutrino momentum equation are given by
\begin{align}
  H_{k} = \varepsilon^{2}\mathcal{H}_{k}
  \quad\text{and}\quad
  F_{H_{k}} = - \varepsilon^{2}\mathcal{L}^{i}_{\hspace{4pt}jk}\,\pd{v^{j}}{i},
\end{align}
where $\mathcal{L}^{i}_{\hspace{4pt}jk}$ is the heat flux tensor in Eq.~\eqref{eq:radiationHeatFluxTensor}.  
With $\mathbf{u}=\big(\,J,H_{k}\,\big)^{T}$ and $\mathbf{f}^{\varepsilon}(\mathbf{u})=\big(\,H_{k},F_{H_{k}}\,\big)^{T}$, the subsystem to be solved is then given by
\begin{equation}
  \pd{\mathbf{u}}{t}+\pd{\big(\,\varepsilon\,\mathbf{f}^{\varepsilon}(\mathbf{u})\big)}{\varepsilon} = 0, 
\end{equation}
which is a familiar advection-type equation.  
For the energy equation, the numerical flux in energy space is then given by
\begin{equation}
  {\widehat{F_{J}}}_{i+\f{1}{2}}
  = - \varepsilon_{i+\f{1}{2}}^{2}\,\widehat{\mathcal{K}}^{i}_{\hspace{4pt}j}(\varepsilon_{i+\f{1}{2}})\,\pd{v^{j}}{i},
  \label{eq:numericalFluxEnergySpaceUpwind}
\end{equation}
where an upwind approach is used to compute
\begin{equation}
  \widehat{\mathcal{K}}^{i}_{\hspace{4pt}j}(\varepsilon_{i+\f{1}{2}})
  =\left\{
  \begin{array}{rl}
  \mathcal{K}^{i}_{\hspace{4pt}j}(\varepsilon_{i+\f{1}{2}}^{-}), & \text{if } \pd{v^{j}}{i} < 0\\
  \mathcal{K}^{i}_{\hspace{4pt}j}(\varepsilon_{i+\f{1}{2}}^{+}), & \text{if } \pd{v^{j}}{i} \ge 0.
  \end{array}
  \right.
\end{equation}
A similar expression is used for the energy-space fluxes in the radiation momentum equation.  
The eigenvalues of the flux Jacobian $\partial\mathbf{f}^{\varepsilon}/\partial\mathbf{u}$ associated with the reduced system of equations governing the ``advection'' in energy space are always of the same sign \citep{VaAuDu11}.  
This is one motivation for using the upwind flux.
Although the numerical flux in Eq.~\eqref{eq:numericalFluxEnergySpaceUpwind} does not necessarily lead to exact number conservation (as is the case for the corresponding numerical flux developed by \citet{MuJaDi10}), the upwind flux has desirable properties that can improve numerical stability (e.g., the upwind flux is consistent and can be used to design monotone numerical schemes \citep[cf.][]{CrMa80,Lev92}).  

\subsubsection{Time integration approaches}

After the specification of approximations to the terms on the right-hand side of Eq.~\eqref{eq:fluidTwoMomentSystemSemiDiscrete}, the system is evolved in time with an ODE solver.  
When solving the general relativistic radiation hydrodynamics system, \citet{KuTaKo16} write the resulting ODE system in the following form:
\begin{equation}
  \deriv{\mathbf{U}}{t} + \mathbf{S}_{\mbox{\tiny adv,s}} + \mathbf{S}_{\mbox{\tiny avd,e}} + \mathbf{S}_{\mbox{\tiny grv}} + \mathbf{S}_{\nu\mbox{\tiny m}} = 0,
\end{equation}
where the spatial advection term $\mathbf{S}_{\mbox{\tiny adv,s}}$, the energy advection term $\mathbf{S}_{\mbox{\tiny avd,e}}$, the gravitational source term $\mathbf{S}_{\mbox{\tiny grv}}$, and the neutrino--matter interaction term $\mathbf{S}_{\nu\mbox{\tiny m}}$ correspond to the terms on the right-hand side of Eq.~\eqref{eq:fluidTwoMomentSystemSemiDiscrete}.  
(Here we omit phase-space indices for brevity.)  
In their time integration scheme, \citet{KuTaKo16} evaluate the spatial advection and gravitational source terms explicitly, while the energy advection and neutrino--matter interaction terms are evaluated implicitly:
\begin{align}
  \f{\mathbf{U}^{*}-\mathbf{U}^{n}}{\dt} &+ \mathbf{S}_{\mbox{\tiny adv,s}}^{n} + \mathbf{S}_{\mbox{\tiny grv}}^{n} = 0, \label{eq:kurodaExplicitUpdate} \\
  \f{\mathbf{U}^{n+1}-\mathbf{U}^{*}}{\dt} &+ \mathbf{S}_{\mbox{\tiny avd,e}}^{n+1} + \mathbf{S}_{\nu\mbox{\tiny m}}^{n+1} = 0. \label{eq:kurodaImplicitUpdate}
\end{align}
This splitting is a special case of a more general class of time integration methods referred to as implicit-explicit (IMEX) schemes \citep{AsRuSp97,PaRu05}. 
The splitting in Eqs.~\eqref{eq:kurodaExplicitUpdate}--\eqref{eq:kurodaImplicitUpdate} is first-order accurate in time, while high-order accurate methods have been developed.  
The main benefit of introducing this split is to avoid a distributed implicit solve, since the spatial advection term couples neighboring cells in space, which can reside on different processing units.  
On the downside, the time step is restricted by the speed of light, but this is acceptable for relativistic systems.  
In general, the neutrino--matter interaction term cannot be integrated efficiently in time with explicit methods because the stable time step needed to resolve the governing time scale is many orders of magnitude shorter than that governing the spatial advection term.  
There is another benefit of integrating the neutrino--matter interaction term separately with implicit methods. These terms are local in space, which makes them easier to parallelize.  
The energy advection term can be integrated with explicit or implicit methods.  
Using explicit methods for this term, an additional time step restriction is needed, but this is usually less severe than that introduced by the spatial advection term \citep[e.g.,][]{OCon15,JuObJa15}.  
On the other hand, since the neutrino--matter interaction term couple the entire momentum space, including the energy advection term in the implicit update \citep[as is also done by, e.g.,][]{MuJaDi10}, which only couples nearest neighbors in energy, does not add significantly to the computational complexity.  
One should note that in their Appendix~B, \citet{KuTaKo16} report significantly different electron fraction profiles when comparing explicit versus implicit integration of $\mathbf{S}_{\mbox{\tiny avd,e}}$, but the reason for this is not clear.  

The implicit solve in Eq.~\eqref{eq:kurodaImplicitUpdate} requires the solution of a nonlinear system of equations.  
To this end, \citet{KuTaKo16}, write the system as
\begin{equation}
  \mathbf{f}(\mathbf{P}^{n+1}) 
  \equiv \f{\mathbf{U}(\mathbf{P}^{n+1})-\mathbf{U}^{*}}{\dt} + \mathbf{S}_{\mbox{\tiny avd,e}}(\mathbf{P}^{n+1}) + \mathbf{S}_{\nu\mbox{\tiny m}}(\mathbf{P}^{n+1}) = 0,
  \label{eq:kurodaNonlinearSystem}
\end{equation}
where the unknowns are given by the vector of ``primitive'' variables:
\begin{equation}
  \mathbf{P} = \big(\,\rho,\,v_{j},\,s,\,Y_{e},\,\mathcal{E}_{1},\mathcal{F}_{1,j},\ldots,\mathcal{E}_{\nSpecies},\mathcal{F}_{\nSpecies,j}\,\big)^{T}.
  \label{eq:kurodaPrimitive}
\end{equation}
To solve the nonlinear system in Eq.~\eqref{eq:kurodaNonlinearSystem}, \citet{KuTaKo16} employ a Newton-Raphson scheme:
\begin{equation}
  \pderiv{\mathbf{f}(\mathbf{P}^{k})}{\mathbf{P}}\delta\mathbf{P}^{k} = - \mathbf{P}^{k}
  \quad\rightarrow\quad
  \mathbf{P}^{k+1} = \mathbf{P}^{k} + \delta\mathbf{P}^{k}
\end{equation}
for $k=0,1,2,\ldots$, with $\mathbf{P}^{0}=\mathbf{P}^{*}$.  
The iteration is continued until $|\delta\mathbf{P}^{k}|<\mbox{tol}\,|\mathbf{P}^{k}|$, where the tolerance is typically set to $\mbox{tol}=10^{-8}$.  
\citet{KuTaKo16} treat the problem fully implicitly, evaluating the neutrino--matter interactions at $t^{n+1}$, and thus include derivatives of opacities in $\mathbf{S}_{\nu\mbox{\tiny m}}$ with respect to $\mathbf{P}$ in the Jacobian $(\partial\mathbf{f}/\partial\mathbf{P})$.  
To help convergence in the Newton-Raphson procedure, \citet{KuTaKo16} also monitor the change in total lepton number during iterations (see their Section~3.3 for details), which improves the robustness of the method.  
Note that in the primitive vector in Eq.~\eqref{eq:kurodaPrimitive} the radiation quantities are the Eulerian moments $\big(\mathcal{E},\mathcal{F}_{j}\big)$, while the closure and the neutrino--matter interaction terms are most naturally expressed in terms of the Lagrangian moments $\big(\mathcal{J},\mathcal{H}_{j}\big)$.  
To evaluate the closure and collision terms during the Newton-Raphson iterations, the Lagrangian moments are kept consistent with the Eulerian moments through the relations:
\begin{align}
  \mathcal{J} &= u_{\mu}u_{\nu}\mathcal{T}^{\mu\nu}
  = W^{2}\,\mathcal{E} - 2\,W\,u_{i}\,\mathcal{F}^{i} + u_{i}u_{j}\mathcal{S}^{ij}, \\
  \mathcal{H}_{j} &= - u_{\nu}\,h_{j\mu}\mathcal{T}^{\mu\nu}
  =\big(\,W\,\mathcal{E}-u_{k}\,\mathcal{F}^{k}\,\big)\,h_{j\mu}\,n^{\mu}+W\,h_{jk}\,\mathcal{F}^{k}-u_{i}\,h_{jk}\,\mathcal{S}^{ik}.  
\end{align}
The number of iterations needed to reach convergence varies during a simulation.  
It is at its maximum in the center around core bounce (several tens), but settles down to $\sim10$ after the shock stalls.  

\citet{JuObJa15}, employing the $\mathcal{O}(v/c)$ limit of the Lagrangian two-moment model in Section~\eqref{sec:TwoMoment} coupled to non-relativistic hydrodynamics, also use a combination of explicit and implicit methods to integrate the coupled equations in time, but ease the computational cost by treating some interaction terms explicitly.  
They split the solution vector into radiation variables $\mathbf{X}=(\mathcal{J},\mathcal{H}_{j})$ and fluid variables $\mathbf{U}=(\rho,\rho Y_{e},\rho\mathbf{v},e_{\mathrm{t}})$, where the total fluid energy density is $e_{\mathrm{t}}=e_{\mathrm{i}}+\rho v^{2}/2$, and $e_{\mathrm{i}}$ is the internal energy density. They write the radiation hydrodynamics system as
\begin{align}
  &\pd{\mathbf{X}}{t} + \big(\delta_{t}\mathbf{X}\big)_{\mathrm{hyp}} + \big(\delta_{t}\mathbf{X}\big)_{\mathrm{vel}} = \big(\delta_{t}\mathbf{X}\big)_{\mathrm{src}}, \\
  &\pd{\mathbf{U}}{t} + \big(\delta_{t}\mathbf{U}\big)_{\mathrm{hyd}} = \big(\delta_{t}\mathbf{U}\big)_{\mathrm{src}},
\end{align}
where in the transport equations, $\big(\delta_{t}\mathbf{X}\big)_{\mathrm{hyp}}$ represents the velocity-independent hyperbolic terms, $\big(\delta_{t}\mathbf{X}\big)_{\mathrm{vel}}$ represents all the velocity-dependent terms in the transport equations, and $\big(\delta_{t}\mathbf{X}\big)_{\mathrm{src}}$ represent neutrino--matter interactions.  
The phase-space advection terms combine to $\big(\delta_{t}\mathbf{X}\big)_{\mathrm{adv}}=\big(\delta_{t}\mathbf{X}\big)_{\mathrm{hyp}}+\big(\delta_{t}\mathbf{X}\big)_{\mathrm{vel}}$.  
In the hydrodynamics equations, $\big(\delta_{t}\mathbf{U}\big)_{\mathrm{hyd}}$ represents the non-radiative physics, while $\big(\delta_{t}\mathbf{U}\big)_{\mathrm{src}}$ the radiative source terms.  
For a given time step $\dt$, when advancing the system from $t^{n}$ to $t^{n+1}=t^{n}+\dt$, a `predictor' step to $t^{n+1/2}=t^{n}+\dt/2$ is performed first:
\begin{align}
  \mathbf{X}^{n+\f{1}{2}}
  &=\mathbf{X}^{n} + \f{\dt}{2}\,\Big[\,-\big(\delta_{t}\mathbf{X}\big)_{\mathrm{hyp}}^{n}+\big(\delta_{t}\mathbf{X}\big)_{\mathrm{src}}^{n,n+\f{1}{2}}\,\Big], 
  \label{eq:justRadiationPredictor} \\
  \mathbf{U}^{n+\f{1}{2}}
  &=\mathbf{U}^{n} + \f{\dt}{2}\,\Big[\,-\big(\delta_{t}\mathbf{U}\big)_{\mathrm{hyd}}^{n}+\big(\delta_{t}\mathbf{U}\big)_{\mathrm{src}}^{n,n+\f{1}{2}}\,\Big],
  \label{eq:justHydroPredictor}
\end{align}
followed by the `corrector' step:
\begin{align}
  \mathbf{X}^{n+1}
  &=\mathbf{X}^{n} + \dt\,\Big[\,-\big(\delta_{t}\mathbf{X}\big)_{\mathrm{hyp}}^{n+\f{1}{2}}+\big(\delta_{t}\mathbf{X}\big)_{\mathrm{src}}^{n+\f{1}{2},n+1}\,\Big], 
  \label{eq:justRadiationCorrector} \\
  \mathbf{U}^{n+1}
  &=\mathbf{U}^{n} + \dt\,\Big[\,-\big(\delta_{t}\mathbf{U}\big)_{\mathrm{hyd}}^{n+\f{1}{2}}+\big(\delta_{t}\mathbf{U}\big)_{\mathrm{src}}^{n+\f{1}{2},n+1}\,\Big],
  \label{eq:justHydroCorrector}
\end{align}
where double superscripts indicate that the source terms can be evaluated using radiation and hydrodynamics variables in the old and the new state.  
(The implicit neutrino--matter solve can be simplified considerably by time-lagging some terms. See the discussion below.)  
When comparing with the scheme of \citet{KuTaKo16} in Eqs.~\eqref{eq:kurodaExplicitUpdate}-\eqref{eq:kurodaImplicitUpdate}, the scheme used by \citet{JuObJa15} uses two explicit evaluations and two implicit evaluations, instead of one of each.  
Also note that \citet{KuTaKo16} treat the velocity-dependent terms implicitly in time, while these terms are treated explicitly by \citet{JuObJa15}.  
While being formally first-order accurate in time, it can be shown that the scheme in Eqs.~\eqref{eq:justRadiationPredictor}-\eqref{eq:justHydroCorrector} is second-order accurate with respect to the explicit part.  
Except for the use of both old and new variables in the implicit part, it is equivalent to the scheme presented by \citet{McEvLo08}.  

The prospect of evaluating some variables in the implicit neutrino--matter solve in the old state is potentially rewarding, since this part of the solve usually accounts for the majority of the computational cost in simulations.  
When doing this, stability and accuracy concerns are important to consider, and this could be investigated with rigorous analysis.  
Methods with time lagging can be considered unconverged or partially converged implicit methods, and can be quite accurate, but this depends on the chosen time step and the degree of nonlinearity of the problem \cite[see, e.g.,][]{KnRiOl01,Lowr04}.  
For stability of the explicit part of the IMEX scheme in Eqs.~\eqref{eq:justRadiationPredictor}-\eqref{eq:justHydroCorrector}, an upper bound on the time step is given by the advection time scale $\tau_{\mathrm{adv}}=\dx/c\approx 3~\mu\mbox{s}\times(\dx/1~\mbox{km})$.  
On the other hand, the neutrino--matter interaction time scale can be estimated as $\tau_{\mathrm{int}}=\lambda_{\nu}/c\approx10~\mbox{ns}\times(\lambda_{\nu}/3\times10^{-3}~\mbox{km})$, where $\lambda_{\nu}$ is the neutrino mean-free path (cf.\ Figure~\ref{fig:tmfp} in Sect.~\ref{sec:needForKineticDescription}).  
In the core of a core-collapse supernova, $\lambda_{\nu}\approx3\times10^{-3}$~km, so that $\tau_{\mathrm{int}}\ll\tau_{\mathrm{adv}}$, which implies that the 
neutrino--matter interactions terms should be integrated with implicit methods in order to keep $\dt/\tau_{\mathrm{adv}}=\mathcal{O}(1)$.  
However, $\tau_{\mathrm{int}}$ should be viewed as the time scale for neutrino--matter equilibration, and neutrinos have practically equilibrated with the matter for densities above $10^{12}\mathrm{\ g\ cm}^{-3}$.  
Since in near equilibrium, the matter quantities (i.e., $\rho$, $e_{\mathrm{i}}$, and the electron density $n_{e}$) evolve on time scales that typically exceed $\tau_{\mathrm{adv}}$, it is reasonable to ask whether some neutrino opacities, which depend nonlinearly on $\rho$, $e_{\mathrm{i}}$, and $n_{e}$, can be evaluated in a lagged fashion in order to avoid costly reevaluations during an iterative implicit solve.  
Numerical experiments can give valuable insights into this question.  
To this end, \citet{JuObJa15} considered three cases for comparison
\begin{itemize}
  \item[(a)] The radiation moments $\mathbf{X}$ and the fluid variables $e_{\mathrm{i}}$ and $n_{e}$ appearing in the source terms $\big(\delta_{t}\mathbf{X}\big)_{\mathrm{src}}$ and $\big(\delta_{t}\mathbf{U}\big)_{\mathrm{src}}$ are defined at $t^{n+1}$.  
  Only the Eddington and heat flux factors ($\mathfrak{k}$ and $\mathfrak{q}$) and the coefficients of the Legendre expansion of energy-coupling interactions (e.g., scattering; cf.\ Eq.~\eqref{eq:kernelExpansion}) are evaluated at $t^{n}$.  
  \item[(b)] Like case (a), but $e_{\mathrm{i}}$ and $n_{e}$ in the source terms are evaluated at $t^{n}$ for all the opacities.  
  This alleviates the computational cost of recomputing the opacities within the iteration procedure.  
  Iterations are still performed in this case because the radiation moments appearing in the blocking factors are treated implicitly.  
  \item[(c)] Like case (b), but all the energy-coupling interactions are treated explicitly in time.  
  This renders the matrix to be inverted in the implicit solve diagonal.  
\end{itemize}
Using case (b) where $\rho>10^{11}\mathrm{\ g\ cm}^{-3}$ and case (c) for $\rho\le10^{11}\mathrm{\ g\ cm}^{-3}$, \citet{JuObJa15} performed a detailed comparison of their scheme in spherical symmetry with results from \citet{LiRaJa05} (obtained with Boltzmann-based codes) for a $13\,M_{\odot}$ star, and found good agreement.  
In addition, they computed an additional run with the same physical specifications, but where case (b) was replaced with case (a) for $\rho>10^{11}\mathrm{\ g\ cm}^{-3}$, and found the results essentially unaltered (see their Fig.~11).  
See also \citet{JuBoJa18} for an extensive comparison of the two-moment method of \citet{JuObJa15} with the \textsc{Prometheus-Vertex} code \citep{RaJa02,BuRaJa06}, and on the impact of various approximate treatments of relevant physics.  
We also note that \citet{OCon15}, who also used explicit treatment of the matter quantities in evaluation the neutrino--matter sources, reported good agreement with \citet{LiRaJa05} across many quantities.  

After obtaining expressions for the radiation moments, the changes to the fluid momentum and kinetic energy densities due to neutrino--matter interactions are computed as
\begin{align}
  \big(\delta_{t}\rho v_{j}\big)_{\mathrm{src}} 
  &= -\sum_{\nu,\xi}\big(\delta_{t}\mathcal{H}_{j,\nu,\xi}\big)_{\mathrm{src}}, \\
  \big(\delta_{t}e_{\mathrm{k}}\big)_{\mathrm{src}}
  &=-v^{j}\sum_{\nu,\xi}\big(\delta_{t}\mathcal{H}_{j,\nu,\xi}\big)_{\mathrm{src}},
\end{align}
where the sums extend over all neutrino frequencies $\nu$ and species $\xi$, and the repeated index on the fluid velocity components $v^{j}$ imply summation over all spatial dimensions.  

\citet{SkDoBu19}, employing a very similar $\mathcal{O}(v/c)$ two-moment model as \citet{JuObJa15} coupled to non-relativistic hydrodynamics, also use explicit and implicit methods to integrate the coupled equations in time.  
They only describe their time integration scheme in the context of emission, absorption, and isotropic, isoenergetic scattering.
\citet{SkDoBu19} write the radiation hydrodynamics system as
\begin{equation}
  \pd{Q}{t}+\big(\,\mathcal{F}_{Q}^{i}\,\big)_{;i} = S_{\mbox{\tiny non-stiff}} + S_{\mbox{\tiny stiff}},
  \label{eq:skinnerRadHydro}
\end{equation}
where the evolved quantities are $Q=\big(\rho,\rho v_{j},\rho e, \rho Y_{e},\mathcal{J},\mathcal{H}_{j}\big)$, where $e$ is the total specific energy of the gas, and $\mathcal{J}$ and $\mathcal{H}_{j}$ are respectively the comoving frame spectral radiation energy density and momentum density, representing all species and groups.  
Components of $\mathcal{J}$ and $\mathcal{H}_{j}$ are denoted $\mathcal{J}_{sg}$ and $\mathcal{H}_{j,sg}$, where $s$ denotes neutrino species and $g$ denotes frequency group.  
In Eq.~\eqref{eq:skinnerRadHydro}, $\big(\,\mathcal{F}_{Q}^{i}\,\big)_{;i}$ and $S_{\mbox{\tiny non-stiff}}$ represent terms from the phase-space advection operator, while $S_{\mbox{\tiny stiff}}$ represents neutrino--matter interactions.  

\citet{SkDoBu19} use operator splitting to integrate the coupled system of equations.  
The phase-space advection terms are integrated with the optimal second-order SSP-RK scheme of \citet{ShOs88}, while the update due to neutrino--matter interactions is followed by a backward Euler solve.  
This scheme applied to Eq.~\eqref{eq:skinnerRadHydro} can be written as
\begin{align}
  Q^{(1)}
  &= Q^{n} + \dt\,\Big\{\,-\big(\,\mathcal{F}_{Q}^{i}\,\big)_{;i}^{n} + S_{\mbox{\tiny non-stiff}}^{n}\,\Big\}, \label{eq:skinnerRK1} \\
  Q^{-}
  &=\f{1}{2}\,Q^{n} + \f{1}{2}\,\Big[\,Q^{(1)} + \dt\,\Big\{\,-\big(\,\mathcal{F}_{Q}^{i}\,\big)_{;i}^{(1)} + S_{\mbox{\tiny non-stiff}}^{(1)}\,\Big\}\,\Big], \label{eq:skinnerRK2} \\
  Q^{n+1}
  &=Q^{-} + \dt\,S_{\mbox{\tiny stiff}}^{n+1}, \label{eq:skinnerImplicit}
\end{align}
which requires two evaluations of $\big(\,\mathcal{F}_{Q}^{i}\,\big)_{;i}$ and $S_{\mbox{\tiny non-stiff}}$ and one implicit solve to evaluate $S_{\mbox{\tiny stiff}}$ per time step.  

After the explicit update in Eqs.~\eqref{eq:skinnerRK1}--\eqref{eq:skinnerRK2}, a nested iteration scheme is employed, where for each spatial point, the coupled system
\begin{align}
  \f{u^{n+1}-u^{-}}{\dt}
  &=-\sum_{s}\sum_{g}\big(j_{sg}^{n+1}-\kappa_{sg}^{n+1}\mathcal{J}_{sg}^{n+1}\big), \label{eq:skinnerFluidEnergyImplicit} \\
  \rho\f{\big(Y_{e}^{n+1}-Y_{e}^{-}\big)}{\dt}
  &=\sum_{s}\sum_{g}\xi_{sg}\big(j_{sg}^{n+1}-\kappa_{sg}^{n+1}\mathcal{J}_{sg}^{n+1}\big), \label{eq:skinnerElectronFractionImplicit} \\
  \f{\mathcal{J}_{sg}^{n+1}-\mathcal{J}_{sg}^{-}}{\dt}
  &=j_{sg}^{n+1}-\kappa_{sg}^{n+1}\mathcal{J}_{sg}^{n+1}, \label{eq:skinnerRadiationEnergyImplicit}
\end{align}
is solved for the material internal energy density, $u$ and electron fraction, $Y_e$---or equivalently, the temperature, $T$, and $Y_{e}$---and the spectral radiation energy density, $\mathcal{J}$.  
In Eqs.~\eqref{eq:skinnerFluidEnergyImplicit}--\eqref{eq:skinnerRadiationEnergyImplicit}, $j_{sg}$ and $\kappa_{sg}$ are the emission and absorption coefficients (depending on $\rho$, which is fixed in this step, $T$, and $Y_{e}$), and
\begin{equation}
  \xi_{sg}
  =\left\{
  \begin{array}{cc}
    -(N_{A}\,\nu)^{-1}, & s=\nu_{e} \\
    +(N_{A}\,\nu)^{-1}, & s=\bar{\nu}_{e} \\
    0, & s=\nu_{x}
  \end{array}
  \right.,
\end{equation}
where $N_{A}$ is Avogadro's number and $\nu$ is the neutrino frequency.  
In the nested iteration scheme, the updates are separated into ``inner'' and ``outer'' parts.  
In the $k$-th outer iteration, the radiation energy density is updated implicitly in the inner iteration as
\begin{equation}
  \f{\mathcal{J}_{sg}^{k}-\mathcal{J}_{sg}^{-}}{\dt} = j_{sg}^{k-1}-\kappa_{sg}^{k-1}\mathcal{J}_{sg}^{k}
  \quad\Rightarrow\quad
  \mathcal{J}_{sg}^{k} = \f{\mathcal{J}_{sg}^{-}+\dt\,j_{sg}^{k-1}}{1+\dt\,\kappa_{sg}^{k-1}},
\end{equation}
where the opacities and emissivities are evaluated using $T^{k-1}$ and $Y_{e}^{k-1}$ (as an initial guess in the first iteration $\{T^{0},Y_{e}^{0}\}=\{T^{-},Y_{e}^{-}\}$).  
The changes in energy and electron fraction are then computed as
\begin{align}
  \Delta E^{k}
  &=\sum_{s}\sum_{g}\big(\mathcal{J}_{sg}^{k}-\mathcal{J}_{sg}^{-}\big), \\
  \Delta Y_{e}^{k}
  &=\sum_{s}\sum_{g}\xi_{sg}\big(\mathcal{J}_{sg}^{k}-\mathcal{J}_{sg}^{-}\big),
\end{align}
and the residuals as
\begin{align}
  r_{E}^{k} 
  &= u^{k}-u^{-} + \Delta E^{k}, \\
  r_{Y_{e}}^{k}
  &= \rho\,\big(Y_{e}^{k}-Y_{e}^{-}\big) - \Delta Y_{e}^{k},
\end{align}
where $u^{k}=u(T^{k},Y_{e}^{k})$ (the internal energy also depends on $\rho$, which is fixed in this part of the solve).  
Then, using a Newton-Raphson technique, the temperature $T^{k}$ and electron fraction $Y_{e}^{k}$ are found such that $r_{E}^{k}=r_{Y_{e}}^{k}=0$.  
The iteration scheme is terminated when the relative change in temperature and electron fraction, $\delta T^{k}=|T^{k}-T^{k-1}|/T^{k-1}$ and $\delta Y_{e}^{k}=|Y_{e}^{k}-Y_{e}^{k-1}|/Y_{e}^{k-1}$, are below a specified tolerance (e.g., $10^{-6}$).  
In this sense, the converged solutions satisfy Eq.~\eqref{eq:skinnerFluidEnergyImplicit}--\eqref{eq:skinnerRadiationEnergyImplicit}.  
\citet{SkDoBu19} report that in practice their iteration procedure converges in a few iterations for a wide range of conditions.  
An obvious benefit of this nested approach is that nonlinear iterations are performed on a smaller system with only two unknowns ($T$ and $Y_{e}$).  
Note however that modifications to this algorithm are needed if energy coupling interactions such as scattering and pair processes are to be included in an implicit fashion as in cases (a) and (b) from \citet{JuObJa15} discussed above.  

After obtaining $u^{n+1}$, $Y_{e}^{n+1}$, $T^{n+1}$, and $\mathcal{J}_{sg}^{n+1}$ by solving Eqs.~\eqref{eq:skinnerFluidEnergyImplicit}-\eqref{eq:skinnerRadiationEnergyImplicit}, the radiation momentum density is updated implicitly as
\begin{align}
  &\f{\mathcal{H}_{j,sg}^{n+1}-\mathcal{H}_{j,sg}^{-}}{\dt} = -\big(\kappa_{sg}^{n+1}+\sigma_{sg}^{n+1}\big)\,\mathcal{H}_{j,sg}^{n+1} \nonumber \\
  &\Rightarrow
  \mathcal{H}_{j,sg}^{n+1} = \f{\mathcal{H}_{j,sg}^{-}}{1+\dt\,\big(\,\kappa_{sg}^{n+1}+\sigma_{sg}^{n+1}\,\big)},
\end{align}
where $\sigma_{sg}$ is the scattering coefficient.  
Finally, the fluid momentum and kinetic energy densities ($(\rho v_{j})$ and $(\rho e_{\mathrm{k}})$, respectively) are updated as
\begin{align}
  (\rho v_{j})^{n+1}
  &=(\rho v_{j})^{-} - \sum_{s}\sum_{g}\big(\mathcal{H}_{j,sg}^{n+1}-\mathcal{H}_{j,sg}^{-}\big), \label{eq:skinnerFluidMomentumUpdate} \\
  (\rho e_{\mathrm{k}})^{n+1}
  &=(\rho e_{\mathrm{k}})^{-} - \sum_{s}\sum_{g}(v^{j})^{-}\big(\mathcal{H}_{j,sg}^{n+1}-\mathcal{H}_{j,sg}^{-}\big), \label{eq:skinnerFluidKineticEnergyUpdate}
\end{align}
where, in the last equation, the repeated index $j$ implies summation over spatial dimensions.  
The total energy density of the gas at $t^{n+1}$ is then obtained from
\begin{align}
  (\rho e)^{n+1}
  &=u^{n+1} + (\rho e_{\mathrm{k}})^{n+1} \label{eq:skinnerTotalFluidEnergyUpdate} \\
  &=\underbrace{\big(u^{-}+(\rho e_{\mathrm{k}})^{-}\big)}_{(\rho e)^{-}}
  -\sum_{s}\sum_{g}\big(\mathcal{J}_{sg}^{n+1}-\mathcal{J}_{sg}^{-}\big)
  - \sum_{s}\sum_{g}(v^{j})^{-}\big(\mathcal{H}_{j,sg}^{n+1}-\mathcal{H}_{j,sg}^{-}\big), \nonumber
\end{align}
where Eq.~\eqref{eq:skinnerFluidEnergyImplicit}, with Eq.~\eqref{eq:skinnerRadiationEnergyImplicit} inserted, and Eq.~\eqref{eq:skinnerFluidKineticEnergyUpdate} are used.  
Note that Eq.~\eqref{eq:skinnerTotalFluidEnergyUpdate} differs from the total energy update listed in \citet{SkDoBu19}; see their Eq.~(32), which is equivalent to Eq.~\eqref{eq:skinnerFluidKineticEnergyUpdate}, but with $\rho e_{\mathrm{k}}\to \rho e$.  
We believe Eq.~\eqref{eq:skinnerTotalFluidEnergyUpdate} is correct in this context since it accounts for changes in internal \emph{and} kinetic energy due to neutrino--matter interactions.  

\subsubsection{Lepton number and energy conservation}
\label{sec:lepenergycons}

We end this section on discretization techniques for two-moment models with a discussion on the topic of lepton number and energy conservation.  
These are conservation laws inherit in the system of equations evolved, and provide a crucial consistency check on the numerical solution.  The 
challenges discussed here in the context of the two-moment model mirror the challenges discussed in Section (\ref{sec:relativisticEffectsAndConservationOfEnergy}) for Boltzmann transport.
The concept of lepton number conservation is easily understood by considering Eqs.~\eqref{eq:ElectronNumberConservation3p1} and \eqref{eq:numberEquation_3p1}, which are evolution equations for the electron density and neutrino number density, respectively.  
The Eulerian electron number is given by $N_{e}=D\,Y_{e}/m_{\mbox{\tiny B}}=W\,n_{e}$, and the Eulerian neutrino lepton number density and lepton number flux density are
\begin{equation}
  N_{\nu} = \sum_{s=1}^{\nSpecies}\mathsf{g}_{s}\,N_{s}
  \quad\text{and}\quad
  G_{\nu}^{i} = \sum_{s=1}^{\nSpecies}\mathsf{g}_{s}\,G_{s}^{i},
\end{equation}
respectively.  
Then, combining Eq.~\eqref{eq:ElectronNumberConservation3p1}, using the source term in Eq.~\eqref{eq:electronfractionequationsourceterm}, with Eq.~\eqref{eq:numberEquation_3p1} results in the conservation law for the total lepton number $N_{\mbox{\tiny Lep}}=N_{e}+N_{\nu}$
\begin{equation}
  \f{1}{\alpha\sqrt{\gamma}}
  \big[\,
    \pd{}{t}\big(\,\sqrt{\gamma}\,N_{\mbox{\tiny Lep}}\,\big)
    +\pd{}{i}\big(\,\sqrt{\gamma}\,\big[\,\alpha\,G_{\mbox{\tiny Lep}}^{i}-\beta^{i}\,N_{\mbox{\tiny Lep}}\,\big]\,\big)
  \,\big]
  =0,
  \label{eq:leptonNumberConservation_3p1}
\end{equation}
where $G_{\mbox{\tiny Lep}}^{i} = N_{e}\,v^{i}+G_{\nu}^{i}$.  
A similar conservation statement for the total energy is not available in the relativistic case because the matter and neutrino equations governing the evolution of the four-momentum---Eqs.~\eqref{eq:fluidEnergyEquation3p1}, \eqref{eq:fluidMomentumEquation3p1}, \eqref{eq:EulerianEnergyEquation_3p1}, and \eqref{eq:EulerianMomentumEquation_3p1}---are not local conservation laws.  
Instead, the so-called ADM mass, $M_{\mbox{\tiny ADM}}$, \citep{BaSh10} (defined as a global quantity) is conserved.  
(See, e.g., \citet{KuTaKo16}, their Eq.~(71), for a definition applicable to the CCSN context.)  
In this case, conservation of the ADM mass can be monitored as a consistency check.  
\citet{KuTaKo16}, see their Figure~7, report violations of ADM mass conservation, $\Delta M_{\mbox{\tiny ADM}}$, (i.e., deviations from the initial value) of order $\Delta M_{\mbox{\tiny ADM}}\approx8\times10^{50}$~erg early after core bounce.  
\citet{MuJaDi10}, see their Figure~12, report violations of ADM mass conservation of similar magnitude in a simulation extending beyond $500$~ms after core bounce.  
In their simulation, $\Delta M_{\mbox{\tiny ADM}}$ jumps by about $5\times10^{50}$~erg at bounce, and keeps increasing more gradually to $\Delta M_{\mbox{\tiny ADM}}\approx2\times10^{51}$~erg at the end of the simulation.  
This change in the ADM mass is only about $0.5$\% relative to the initial value.  

\citet{MuJaDi10} argue that the velocity-dependent terms in the transport equations are the most critical terms responsible for the violation of energy (or ADM mass) conservation.  
To see this, it is illustrative to consider the equations they solve in the special relativistic limit with Cartesian coordinates and no neutrino--matter interactions.  
Neutrino--matter interactions are entirely local, and lepton number and four-momentum conservation in this sector can be enforced by constraints as in Eqs.~\eqref{eq:momentumConservationConstraint}-\eqref{eq:leptonNumberConservationConstraint}.  
The challenge stems from the discretization of the phase-space advection operators; i.e., the left-hand side of the moment equations.  
In the special relativistic limit with Cartesian coordinates and no neutrino--matter interactions, the Lagrangian two-moment model corresponding to the one used by \citet{MuJaDi10} is given by the energy equation (cf.\ Eq.~\eqref{eq:spectralLagrangianEnergyEquation_3p1})
\begin{equation}
  \pd{}{\nu}\big(\,\hat{\mathcal{J}}u^{\nu}+\hat{\mathcal{H}}^{\nu}\,\big)
  -\pd{}{\varepsilon}\big(\,\varepsilon\,\hat{\mathcal{T}}^{\mu\nu}\,\pd{u_{\mu}}{\nu}\,\big)
  =-\hat{\mathcal{T}}^{\mu\nu}\,\pd{u_{\mu}}{\nu}
  \label{eq:spectralLagrangianEnergyEquation_SR}
\end{equation}
and the momentum equation (cf.\ Eq.~\eqref{eq:spectralLagrangianMomentumEquation_3p1})
\begin{equation}
  \pd{}{\nu}\big(\,\hat{\mathcal{H}}_{j}\,u^{\nu}+\hat{\mathcal{K}}_{j}^{\hspace{2pt}\nu}\,\big)
  -\pd{}{\varepsilon}\big(\,h_{j\rho}\,\hat{\mathcal{Q}}^{\rho\mu\nu}\,\pd{u_{\mu}}{\nu}\,\big)
  =\hat{\mathcal{T}}^{\mu\nu}\,\pd{h_{j\mu}}{\nu},
  \label{eq:spectralLagrangianMomentumEquation_SR}
\end{equation}
where the ``hat'' is used to denote that a factor $\varepsilon^{2}$ has been absorbed into the definition of the moments; i.e.,
\begin{equation}
  \big\{\,\hat{\mathcal{J}},\hat{\mathcal{H}}^{\nu},\hat{\mathcal{K}}^{\mu\nu},\ldots\,\big\}
  =\varepsilon^{2}\,\big\{\,\mathcal{J},\mathcal{H}^{\nu},\mathcal{K}^{\mu\nu},\ldots\,\big\}.  
\end{equation}
Note that neither Eq.~\eqref{eq:spectralLagrangianEnergyEquation_SR} nor Eq.~\eqref{eq:spectralLagrangianMomentumEquation_SR} are local conservation laws.  
Therefore, a numerical method based on these equations requires care in the discretization process to achieve neutrino number, energy, and momentum conservation.  
(Neutrino energy \emph{and} momentum contribute to the ADM mass.)

First, note that by dividing Eq.~\eqref{eq:spectralLagrangianEnergyEquation_SR} by $\varepsilon$ results in
\begin{equation}
  \pd{\hat{\mathcal{N}}^{\nu}}{\nu} - \pd{}{\varepsilon}\big(\,\hat{\mathcal{T}}^{\mu\nu}\,\pd{u_{\mu}}{\nu}\,\big) = 0,
  \label{eq:spectralNumberEquation_SR}
\end{equation}
which is a local phase-space conservation law for the spectral number density.  
In arriving at Eq.~\eqref{eq:spectralNumberEquation_SR}, the remainder after bringing $\varepsilon^{-1}$ inside the energy derivative in Eq.~\eqref{eq:spectralLagrangianEnergyEquation_SR} cancels with the right-hand side.  
This is exactly what the discretization of the energy derivative term developed by \citet{MuJaDi10} (discussed in Sect.~\ref{sec:EnergyDiscretization}) is designed to do in order to achieve lepton number conservation.  

On the other hand, 
\begin{equation}
  -n_{\mu}\hat{\mathcal{T}}^{\mu\nu}
  = \big(\,\hat{\mathcal{E}}n^{\nu}+\hat{\mathcal{F}}^{\nu}\,\big)
  = W\,\big(\,\hat{\mathcal{J}}u^{\nu}+\hat{\mathcal{H}}^{\nu}\,\big) + v^{j}\,\big(\,\hat{\mathcal{H}}_{j}u^{\nu}+\hat{\mathcal{K}}_{j}^{\hspace{2pt}\nu}\,\big),
\end{equation}
where both the Eulerian and Lagrangian decompositions of $\hat{\mathcal{T}}^{\mu\nu}$ are used; cf.\ Eqs.~\eqref{eq:stressEnergyEulerianDecomposition} and \eqref{eq:stressEnergyLagrangianDecomposition}, respectively.  
Thus, by adding $W$ times Eq.~\eqref{eq:spectralLagrangianEnergyEquation_SR} and the contraction of $v^{j}$ with Eq.~\eqref{eq:spectralLagrangianMomentumEquation_SR} gives
\begin{equation}
  \pd{}{\nu}\big(\,\hat{\mathcal{E}}n^{\nu}+\hat{\mathcal{F}}^{\nu}\,\big)
  -\pd{}{\varepsilon}\big(\,(-n_{\rho})\,\hat{\mathcal{Q}}^{\rho\mu\nu}\,\pd{u_{\mu}}{\nu}\,\big) = 0,
  \label{eq:spectralEnergyEquation_SR}
\end{equation}
which is a local phase-space conservation law for the spectral energy density.  
When arriving at Eq.~\eqref{eq:spectralEnergyEquation_SR}, the remainders after bringing $W$ inside the spacetime derivative in Eq.~\eqref{eq:spectralLagrangianEnergyEquation_SR} and $v^{j}$ inside the spacetime derivative of Eq.~\eqref{eq:spectralLagrangianMomentumEquation_SR} cancel with the terms due to the sources on the right-hand sides of Eqs.~\eqref{eq:spectralLagrangianEnergyEquation_SR} and \eqref{eq:spectralLagrangianMomentumEquation_SR} in a nontrivial way:
\begin{align}
  &\big(\,\hat{\mathcal{J}}u^{\nu}+\hat{\mathcal{H}}^{\nu}\,\big)\,\pd{W}{\nu}
  +\big(\,\hat{\mathcal{H}}_{j}\,u^{\nu}+\hat{\mathcal{K}}_{j}^{\hspace{2pt}\nu}\,\big)\,\pd{v^{j}}{\nu}
  -\hat{\mathcal{T}}^{\mu\nu}\,\big(\,W\pd{u_{\mu}}{\nu}-v^{j}\pd{h_{j\mu}}{\nu}\,\big) \nonumber \\
  &=-\big(\,u_{\mu}\,\pd{W}{\nu}-h_{j\mu}\,\pd{v^{j}}{\nu}+W\pd{u_{\mu}}{\nu}-v^{j}\pd{h_{j\mu}}{\nu}\,\big)\,\hat{\mathcal{T}}^{\mu\nu} \nonumber \\
  &=-\pd{}{\nu}\big(\,W\,u_{\mu}-h_{j\mu}\,v^{j}\,\big)\,\hat{\mathcal{T}}^{\mu\nu} = - \hat{\mathcal{T}}^{\mu\nu}\,\pd{n_{\mu}}{\nu} = 0,
  \label{eq:spectralEnergyConservationConstraint}
\end{align}
since, in special relativity, $n_{\mu}=(-1,0,0,0)$.  
Similarly, 
\begin{equation}
  \gamma_{j\mu}\,\hat{\mathcal{T}}^{\mu\nu}
  = \big(\,\hat{\mathcal{F}}_{j}\,n^{\nu}+\hat{\mathcal{S}}_{j}^{\hspace{2pt}\nu}\,\big)
  = Wv_{j}\,\big(\,\hat{\mathcal{J}}u^{\nu}+\hat{\mathcal{H}}^{\nu}\,\big) + \big(\,\hat{\mathcal{H}}_{j}u^{\nu}+\hat{\mathcal{K}}_{j}^{\hspace{2pt}\nu}\,\big).  
\end{equation}
Then, by adding $Wv_{j}$ times Eq.~\eqref{eq:spectralLagrangianEnergyEquation_SR} and Eq.~\eqref{eq:spectralLagrangianMomentumEquation_SR} one obtains
\begin{equation}
  \pd{}{\nu}\big(\,\hat{\mathcal{F}}_{j}\,n^{\nu}+\hat{\mathcal{S}}_{j}^{\hspace{2pt}\nu}\,\big)
  -\pd{}{\varepsilon}\big(\,\gamma_{j\rho}\,\hat{\mathcal{Q}}^{\rho\mu\nu}\,\pd{u_{\mu}}{\nu}\,\big) = 0,
  \label{eq:spectralMomentumEquation_SR}
\end{equation}
which is a local conservation law for the spectral momentum density.  
Again, in arriving at Eq.~\eqref{eq:spectralMomentumEquation_SR}, the remainder after bringing $Wv_{j}$ inside the spacetime derivative in Eq.~\eqref{eq:spectralLagrangianEnergyEquation_SR} cancels with the sources in Eqs.~\eqref{eq:spectralLagrangianEnergyEquation_SR} and \eqref{eq:spectralLagrangianMomentumEquation_SR} in a nontrivial way:
\begin{align}
  &\big(\,\hat{\mathcal{J}}u^{\nu}+\hat{\mathcal{H}}^{\nu}\,\big)\,\pd{}{\nu}\big(Wv_{j}\big)
  -Wv_{j}\,\hat{\mathcal{T}}^{\mu\nu}\,\pd{u_{\mu}}{\nu} + \hat{\mathcal{T}}^{\mu\nu}\,\pd{h_{j\mu}}{\nu} \nonumber \\
  &=-\big(\,u_{\mu}\,\pd{}{\nu}\big(Wv_{j}\big) + Wv_{j}\,\pd{u_{\mu}}{\nu} - \pd{h_{j\mu}}{\nu}\,\big)\,\hat{\mathcal{T}}^{\mu\nu} \nonumber \\
  &=-\pd{}{\nu}\big(\,Wv_{j}u_{\mu}-h_{j\mu}\,\big)\,\hat{\mathcal{T}}^{\mu\nu}
  =\hat{\mathcal{T}}^{\mu\nu}\,\pd{g_{j\mu}}{\nu}=0,
  \label{eq:spectralMomentumConservationConstraint}
\end{align}
since, in special relativity and with Cartesian coordinates, $\pd{g_{j\mu}}{\nu}=0$.  

Equations~\eqref{eq:spectralEnergyConservationConstraint} and \eqref{eq:spectralMomentumConservationConstraint} can be viewed as constraints.  
Since the discretizations of Eqs.~\eqref{eq:spectralLagrangianEnergyEquation_SR} and \eqref{eq:spectralLagrangianMomentumEquation_SR} are unlikely to satisfy these constraints, they are inconsistent with energy conservation in the sense of Eq.~\eqref{eq:spectralEnergyEquation_SR} and momentum conservation in the sense of Eq.~\eqref{eq:spectralMomentumEquation_SR}.  
In the fully relativistic case, one is faced with the same issue, namely that the discretization of the Lagrangian two-moment model (Eqs.~\eqref{eq:spectralLagrangianEnergyEquation_3p1} and \eqref{eq:spectralLagrangianMomentumEquation_3p1}) is to a certain degree inconsistent with the discretization of the Eulerian two-moment model (Eqs.~\eqref{eq:spectralEulerianEnergyEquation_3p1} and \eqref{eq:EulerianMomentumEquation_3p1}).  
Since it is the Eulerian moments that enter into the definition of the ADM mass, this inconsistency can propagate and manifest itself as violations of ADM mass conservation.  
On the other hand, by using the Eulerian two-moment model as the starting point for a numerical method---e.g., as in \citet{KuTaKo16}---it may be easier to control $\Delta M_{\mbox{\tiny ADM}}$.  
(The time evolution of the ADM mass reported by \citet{KuTaKo16} and \citet{MuJaDi10} are indeed quite different.)
However, while the use of the Eulerian two-moment model may provide an advantage with regard to controlling energy conservation, one is still left with the equally challenging task of maintaining consistency with the number equation (Eq.~\eqref{eq:spectralNumberEquation_3p1}) and controlling lepton number conservation,
as discussed in detail by \citet{CaEnMe13a}, and in this case violations of lepton number conservation in the sense of Eq.~\eqref{eq:leptonNumberConservation_3p1} may still result.  

We conclude this section by discussing number, energy, and momentum conservation in the context of the $\mathcal{O}(v/c)$ limit of the relativistic Lagrangian two-moment model discussed above, implemented by \citet{JuObJa15} and \citet{SkDoBu19}.  
(Note, we use units in which $c=1$.)  
The energy equation, Eq.~\eqref{eq:spectralLagrangianEnergyEquation_SR}, is then given by
\begin{equation}
  \pd{}{t}\big(\,\hat{\mathcal{J}}+\Theta\,v^{i}\hat{\mathcal{H}}_{i}\,\big)
  + \pd{}{i}\big(\,\hat{\mathcal{H}}^{i}+v^{i}\hat{\mathcal{J}}\,\big)
  - \pd{}{\varepsilon}\big(\,\varepsilon\,\hat{\mathcal{K}}^{i}_{\hspace{4pt}k}\,\pd{v^{k}}{i}\,\big)
  = - \hat{\mathcal{K}}^{i}_{\hspace{4pt}k}\,\pd{v^{k}}{i},
  \label{eq:spectralLagrangianEnergyEquation_VoverC}
\end{equation}
while the momentum equation, Eq.~\eqref{eq:spectralLagrangianMomentumEquation_SR}, is given by
\begin{equation}
  \pd{}{t}\big(\,\hat{\mathcal{H}}_{j}+\Theta\,v^{i}\hat{\mathcal{K}}_{ij}\,\big)
  + \pd{}{i}\big(\,\hat{\mathcal{K}}^{i}_{\hspace{4pt}j}+v^{i}\hat{\mathcal{H}}_{j}\,\big)
  - \pd{}{\varepsilon}\big(\,\varepsilon\,\hat{\mathcal{L}}^{i}_{\hspace{4pt}kj}\,\pd{v^{k}}{i}\,\big)
  = - \hat{\mathcal{H}}^{i}\,\pd{v_{j}}{i}.  
  \label{eq:spectralLagrangianMomentumEquation_VoverC}
\end{equation}
For simplicity, we ignore terms proportional to the time derivative of the fluid three-velocity, which is a reasonable approximation.  
In Eqs.~\eqref{eq:spectralLagrangianEnergyEquation_VoverC} and \eqref{eq:spectralLagrangianMomentumEquation_VoverC}, we introduced a constant parameter, $\Theta$, that is either zero or one.  
For $\Theta=0$, the two-moment model reduces to the one solved by \citet{JuObJa15} and by \citet{SkDoBu19}.  
However, when $\Theta=1$, as we will show below, the two-moment model is better aligned with number, energy, and momentum conservation.  

First, by dividing Eq.~\eqref{eq:spectralLagrangianEnergyEquation_VoverC} with the particle energy $\varepsilon$ and rearranging, one obtains  
\begin{equation}
  \pd{}{t}\big(\,\hat{\mathcal{D}}+\Theta\,v^{i}\hat{\mathcal{I}}_{i}\,\big)
  + \pd{}{i}\big(\,\hat{\mathcal{I}}^{i}+v^{i}\hat{\mathcal{D}}\,\big)
  - \pd{}{\varepsilon}\big(\,\hat{\mathcal{K}}^{i}_{\hspace{4pt}k}\,\pd{v^{k}}{i}\,\big)
  = 0, \label{eq:spectralNumberEquation_VoverC}
\end{equation}
which is a local conservation law for the spectral number density $\hat{\mathcal{D}}+\Theta\,v^{i}\hat{\mathcal{I}}_{i}$.  
Note that, when $\Theta=0$, it is the Lagrangian number density defined in Eq.~\eqref{eq:numberMomentsLagrangian} that is conserved, which is incorrect in the 
$\mathcal{O}(v/c)$ limit.  
On the other hand, when $\Theta=1$, Eq.~\eqref{eq:spectralNumberEquation_VoverC} is a conservation law for the $\mathcal{O}(v/c)$ approximation of the Eulerian number density defined in Eq.~\eqref{eq:eulerianNumberInTermsOfLagrangianMoments}, which is conserved.  

Next, we consider energy and momentum conservation.  
Following the approach in the relativistic case, by adding Eq.~\eqref{eq:spectralLagrangianEnergyEquation_VoverC} and the contraction of $v^{j}$ with Eq.~\eqref{eq:spectralLagrangianMomentumEquation_VoverC} one obtains
\begin{align}
  &\pd{}{t}\big(\hat{\mathcal{J}}+(1+\Theta)\,v^{i}\hat{\mathcal{H}}_{i}\big)
  +\pd{}{i}\big(\hat{\mathcal{H}}^{i}+v^{i}\hat{\mathcal{J}}+v^{j}\hat{\mathcal{K}}^{i}_{\hspace{4pt}j}\big) \nonumber \\
  &\hspace{12pt}
  - \pd{}{\varepsilon}\big(\,\varepsilon\,\hat{\mathcal{K}}^{i}_{\hspace{4pt}k}\,\pd{v^{k}}{i}\,\big)
  =\mathcal{O}(v^{2}),
  \label{eq:spectralEulerianEnergyEquation_VoverC}
\end{align}
which, to $\mathcal{O}(v/c)$, is a local conservation law for the Eulerian spectral energy density $\hat{\mathcal{J}}+(1+\Theta)\,v^{i}\hat{\mathcal{H}}_{i}$.  
With $\Theta=1$, this is the correct $\mathcal{O}(v/c)$ limit of the Eulerian energy density in Eq.~\eqref{eq:eulerianEnergyInTermsOfLagrangianMoments}.  
Terms of higher order in the fluid velocity have been moved to the right-hand side of Eq.~\eqref{eq:spectralEulerianEnergyEquation_VoverC}, which must remain small for the $\mathcal{O}(v/c)$ limit to be valid.  
Also note that, with $\Theta=0$, energy conservation breaks down to leading order in the fluid three-velocity (a factor of 2 should appear in the coefficient of the 
second term inside the parentheses of the time derivative).  
Similarly, by adding $v_{j}$ times Eq.~\eqref{eq:spectralLagrangianEnergyEquation_VoverC} and Eq.~\eqref{eq:spectralLagrangianMomentumEquation_VoverC} one obtains
\begin{align}
  &\pd{}{t}\big(\hat{\mathcal{H}}_{j}+v_{j}\hat{\mathcal{J}}+\Theta\,v^{i}\hat{\mathcal{K}}_{ij}\big)
  + \pd{}{i}\big(\,\hat{\mathcal{K}}^{i}_{\hspace{4pt}j}+\hat{\mathcal{H}}^{i}v_{j}+v^{i}\hat{\mathcal{H}}_{j}\,\big) \\
  &\hspace{12pt}
  - \pd{}{\varepsilon}\big(\varepsilon\,\hat{\mathcal{L}}^{i}_{\hspace{4pt}kj}\,\pd{v^{k}}{i}\big)
  =\mathcal{O}(v^{2}),
  \label{eq:spectralEulerianMomentumEquation_VoverC}
\end{align}
which, to $\mathcal{O}(v/c)$, is a local conservation law for the Eulerian spectral momentum density $\hat{\mathcal{H}}_{j}+v_{j}\hat{\mathcal{J}}+\Theta\,v^{i}\hat{\mathcal{K}}_{ij}$.  
Again, with $\Theta=1$, this is the correct $\mathcal{O}(v/c)$ limit of the Eulerian momentum density equation, Eq.~\eqref{eq:eulerianMomentumInTermsOfLagrangianMoments}.  

Thus, at the expense of some additional computational complexity, by letting $\Theta=1$ in Eqs.~\eqref{eq:spectralLagrangianEnergyEquation_VoverC} and \eqref{eq:spectralLagrangianMomentumEquation_VoverC}, the two-moment model becomes consistent with number, energy, and momentum conservation in the $\mathcal{O}(v/c)$ limit.  

\subsection{One-moment kinetics}

\subsubsection{Newtonian-gravity, $O(v/c)$, finite-difference implementation}

One moment kinetics is typically deployed in the context of neutrino transport in core-collapse supernovae
using the multigroup (i.e., multi-frequency) flux-limited diffusion approximation (MGFLD). Such MGFLD 
approaches solve the neutrino and antineutrino moment equations for the zeroth moment of the distribution
function, the multigroup neutrino/antineutrino energy density, with closure provided at the level of the first 
moment, the neutrino/antineutrino energy flux, via a diffusion-like equation, modified in such a way that 
the flux cannot be come superluminal (flux limiting). \citet{SwMy09} were the first to 
implement such an approach in axisymmetric simulations of core-collapse supernovae. The equations
for the neutrino/antineutrino multigroup energy densities used by Swesty and Myra are expressed as
\begin{equation}\label{eq:bte0}
\frac{\partial E_{\epsilon}}{\partial t} +
{\nabla} \cdot \left( E_{\epsilon} {\bf v}
\right) + {\nabla} \cdot {\bf F}_{\epsilon} -
\epsilon \frac{\partial}{\partial \epsilon} 
\left( {\mathsf P}_{\epsilon}:
{\nabla} {\bf v} \right) = {\mathbb S}_{\epsilon},
\end{equation}
\begin{equation}\label{eq:bte0bar}
\frac{\partial \bar{E}_{\epsilon}}{\partial t} +
{\nabla} \cdot \left( \bar{E}_{\epsilon} {\bf v} \right) +
{\nabla} \cdot \bar{{\bf F}}_{\epsilon} -
\epsilon \frac{\partial}{\partial \epsilon} 
\left( \bar{{\mathsf P}}_{\epsilon}:
{\nabla} {\bf v} \right) = \bar{{\mathbb S}}_{\epsilon},
\end{equation}
where $E_\epsilon$ and $\bar{E}_\epsilon$ are the neutrino and antineutrino energy densities per group,
$P_\epsilon$ and $\bar{P}_\epsilon$ are the neutrino and antineutrino stress tensors, 
and ${\mathbb S}_\epsilon$ and $\bar{{\mathbb S}}_\epsilon$ are the neutrino and antineutrino matter 
couplings, respectively. The energy flux in both equations is given by a Fick's-like relation of the form
\begin{equation}\label{eq:fick}
{\bf F}_\epsilon \equiv -D_\epsilon {\nabla} E_\epsilon,
\end{equation}
where
\begin{equation} \label{eq:diff_simple}
D_\epsilon = \frac{c}{3\kappa^T_\epsilon}
\end{equation}
is the diffusion coefficient, and $\kappa_\epsilon^T$ is the total opacity. In flux-limited diffusion schemes,
the diffusion coefficient $D_\epsilon$ is modified. A general form for such a modified diffusion coefficient
is given by
\begin{equation} \label{eq:lpd}
D_\epsilon \equiv \frac{c \lambda_\epsilon(\kn)}
{\kappa^T_\epsilon}.
\end{equation}
In particular, the so-called Levermore--Pomraning flux limiter \citep{LePo81} is given by
\begin{equation}  \label{eq:lpfl}
\lambda_\epsilon(\kn)
\equiv
\frac{2 + \kn}
{6 + 3\kn + \kn^2},
\end{equation}
where $R_\epsilon$ is the radiation Knudsen number, which is the ratio of the mean free path to some
characteristic length scale in the problem. The Knudsen number is written as
\begin{equation}
\kn
\equiv
\frac{\left|{\nabla} E_\epsilon \right|}{\kappa^T_\epsilon E_\epsilon}.
\label{eq:knudsen}
\end{equation}
Note that the Knudsen number is different for different energy groups given that the opacities are typically 
(and for neutrinos in core-collapse supernovae, definitely) energy dependent. The radiation stress tensor
takes the typical form
\begin{equation} \label{eq:edddef}
{\mathsf{P}}_\epsilon \equiv {\mathsf X}_\epsilon E_\epsilon,
\end{equation}
where
\begin{equation} \label{eq:chidef}
{\mathsf X}_\epsilon \equiv \frac{1}{2} \left(
1-\chi_\epsilon \right) 
{\mathsf{I}}
+
\frac{1}{2} \left(
  3 \chi_\epsilon - 1
\right)
{\bf n}{\bf n},
\end{equation}
where $\chi_\epsilon$ is the scalar Eddington factor, which in the case of the Levermore--Pomraning 
flux-limiting scheme becomes
\begin{equation} \label{eq:chismdef}
\chi_\epsilon = \lambda_\epsilon (\kn) + 
\left\{\lambda_\epsilon(\kn)\right\}^2 \: \kn^2.
\end{equation}
Given the choice of Levermore-Pomraning flux limiting, the evolution equations \eqref{eq:bte0} and \eqref{eq:bte0bar} 
become
\begin{equation}\label{eq:bte0f}
\frac{\partial E_{\epsilon}}{\partial t} +
{\nabla} \cdot \left( E_{\epsilon} {\bf v} \right) -
{\nabla} \cdot (D_\epsilon {\nabla} E_{\epsilon}) -
\epsilon \frac{\partial}{\partial \epsilon} 
\left\{
({\mathsf X}_{\epsilon} E_\epsilon):
{\nabla} {\bf v} \right\} = {\mathbb S}_{\epsilon},
\end{equation}
\begin{equation}\label{eq:bte0barf}
\frac{\partial \bar{E}_{\epsilon}}{\partial t} +
{\nabla} \cdot \left( \bar{E}_{\epsilon} {\bf v} \right) -
{\nabla} \cdot (\bar{D}_\epsilon {\nabla} \bar{E}_{\epsilon}) -
\epsilon \frac{\partial}{\partial \epsilon} 
\left\{
(\bar{{\mathsf X}}_{\epsilon} \bar{E}_\epsilon):
{\nabla} {\bf v} \right\} = \bar{{\mathbb S}}_{\epsilon}.
\end{equation}
Swesty and Myra note, these equations are not in conservative form. They opt to monitor conservation of 
lepton number and energy after the fact. The degree to which they achieve either was not documented.
Their equations are operator split as follows (written here for just the neutrinos, not the antineutrinos):
\begin{equation} 
\left\ldbrack \frac{ \partial E_\epsilon }{\partial t}\right\rdbrack_{\rm total}  =
\left\ldbrack \frac{ \partial E_\epsilon }{\partial t}\right\rdbrack_{\rm advection}  +
\left\ldbrack \frac{ \partial E_\epsilon }{\partial t}\right\rdbrack_\text{diff-coll},
\end{equation}
where
\begin{equation} 
\left\ldbrack \frac{ \partial E_\epsilon }{\partial t}\right\rdbrack_{\rm advection} =
-{\nabla} \cdot (E_\epsilon {\bf v})
\label{eq:nu-advect},
\end{equation}
\begin{equation} 
\left\ldbrack \frac{ \partial E_\epsilon }{\partial t}\right\rdbrack_\text{diff-coll} =
{\nabla} \cdot (D_\epsilon {\nabla} E_{\epsilon}) +
\epsilon \frac{\partial}{\partial \epsilon}
\left\{
({\mathsf X}_{\epsilon} E_\epsilon):
{\nabla} {\bf v} \right\} + {\mathbb S}_{\epsilon}.
\label{eq:nu-diff}
\end{equation}

For the purpose of describing their numerical method used to treat each of the operator split
equations shown above, Swesty and Myra note, first, that the advection equations take the 
general form
\begin{equation} \label{eq:diff_ad_sc}
\left\ldbrack \frac{ \partial \psi}{\partial t}
\right\rdbrack_{\rm advection} + 
{\nabla} \cdot \left( \psi {\bf v} \right) = 0,
\end{equation}
where $\psi$ is the scalar field ($E_\epsilon$ and $\bar{E}_\epsilon$) being advected. They 
then deploy the ZEUS consistent advection scheme of \citet{StNo92} in a directionally-split 
manner to each dimension (in their case, $x_1$ and $x_2$) of the problem. For the $x_1$ update, 
Eq.~\eqref{eq:diff_ad_sc} is discretized as follows:
\begin{eqnarray} \label{eq:sflux_comb_x1}
\lefteqn{ \frac{ \left[ \Delta V \right]_{i+\lhalf,j+\lhalf}}{\Delta t} 
\left( \left[ \psi \right]^{n+\beta}_{i+\lhalf,j+\lhalf} - 
\left[ \psi \right]^{n+\alpha}_{i+\lhalf,j+\lhalf} \right) 
= }
\nonumber \\ & & 
- \left( \left[ F_1(\psi) \right]_{i+1,j+\lhalf}^{n+\alpha}
\left[ \Delta A_1 \right]_{i+1,j+\lhalf}  -
\left[ F_1(\psi) \right]_{i,j+\lhalf}^{n+\alpha}
\left[ \Delta A_1 \right]_{i,j+\lhalf} \right).
\end{eqnarray}
The fluxes in Eq.~\eqref{eq:sflux_comb_x1} 
are given by
\begin{equation}
\left[ F_1(\psi) \right]_{i,j+\lhalf} = 
\left[ {\cal I}_1\left(\frac{\psi}{\rho}\right) \right]_{i,j+\lhalf} 
\left[ F_1(\rho) \right]_{i,j+\lhalf},
\label{eq:nca1}
\end{equation}
where
\begin{equation}\label{eq:flux1}
\left[ F_{1}(\rho) \right]_{i,j+\lhalf} = 
\left[ {\cal I}_1(\rho) \right]_{i,j+\lhalf} 
\left[ \varv_{1} \right]_{i,j+\lhalf},
\end{equation}
and where 
\begin{equation} \label{eq:i1}
\left[ {\cal I}_1(\psi) \right]_{i,j+\lhalf} = 
\begin{cases} \displaystyle{
\left[ \psi \right]_{i-\lhalf,j+\lhalf} + 
\left[ \delta_1(\psi) \right]_{i-\lhalf,j+\lhalf}
\left( 1 - \frac{\left[ \varv_{1}\right]_{i,j+\lhalf} \Delta t }
{\left[x_1\right]_{i} - \left[x_1\right]_{i-1} } \right)
}
 & \text{if} \; \left[ \varv_1 \right]_{i,j+\lhalf} > 0, \vspace{0.1in} \\
\displaystyle{
\left[ \psi \right]_{i+\lhalf,j+\lhalf} - 
\left[ \delta_1(\psi) \right]_{i+\lhalf,j+\lhalf}
\left( 1 + \frac{ \left[ \varv_{1} \right]_{i,j+\lhalf} \Delta t}
{\left[x_1\right]_{i+1} - \left[x_1\right]_{i} }
\right) }
 & \text{if} \; \left[ \varv_1 \right]_{i,j+\lhalf} < 0. \\
\end{cases}
\end{equation}
In Eq.~\eqref{eq:flux1},
$\rho$ is the fluid mass density.
${\cal I}_1(\psi)$ is the van Leer monotonic upwind advection function \citep{vanLeer1977},
given by
\begin{equation} \label{eq:i1p1}
\left[ \delta_1(\psi) \right]_{i+\lhalf,j+\lhalf} = 
\begin{cases}
\displaystyle{ 
\frac{
\left[ \Delta \psi \right]_{i,j+\lhalf} 
\left[ \Delta \psi \right]_{i+1,j+\lhalf} }
{\left[ \psi \right]_{i+\lthreehalf,j+\lhalf} -
\left[ \psi \right]_{i-\lhalf,j+\lhalf}}
} \vspace{0.1in} \\
\hspace{1.5in}
\text{if} \;
\left[ \Delta \psi \right]_{i,j+\lhalf} 
\left[ \Delta \psi \right]_{i+1,j+\lhalf} 
> 0, \vspace{0.25in} \\
0 \hspace{1.4in} \text{otherwise},
\end{cases}
\end{equation}
where
\begin{equation} \label{eq:delsig1}
\left[ \Delta \psi \right]_{i,j+\lhalf} = 
\left[ \psi \right]_{i+\lhalf,j+\lhalf} -
\left[ \psi \right]_{i-\lhalf,j+\lhalf}.
\end{equation}
The $x_2$ update is computed in the same way, with the obvious substitutions. 

The remaining term, due to neutrino diffusion, relativistic effects, and collisions:
\begin{equation}
\label{eq:dc-e}
\left\ldbrack \frac{\partial (^eE_{\epsilon})}{\partial t} 
\right\rdbrack_{\rm diff-coll} -
{\nabla} \cdot \left(^eD_\epsilon ^eE_{\epsilon} \right) -
\epsilon \frac{\partial}{\partial \epsilon} 
\left( ^e{\mathsf P}_{\epsilon}:
{\nabla} {\bf v} \right) - ^e{\mathbb S}_{\epsilon} = 0
\end{equation}
is differenced implicitly in time and as follows in phase space:
\begin{eqnarray} \label{eq:nu_tr_fd}
\frac{ \left[ E_\epsilon \right]^\npo_{k+\lhalf,i+\lhalf,j+\lhalf} - 
\left[E_\epsilon\right]^n_{k+\lhalf,i+\lhalf,j+\lhalf}}{\Delta t}-
\left[{\nabla} \cdot D_\epsilon \nabla E_{\epsilon}
\right]^\npo_{k+\lhalf,i+\lhalf,j+\lhalf} \nonumber \\
-\left[ \epsilon \frac{\partial \left({\mathsf P}_{\epsilon}:
{\nabla} {\bf v}\right)}{\partial \epsilon} 
\right]^\npo_{k+\lhalf,i+\lhalf,j+\lhalf}
-\left[{\mathbb S}_{\epsilon} \right]^\npo_{k+(1/2),i+(1/2),j+)1/2)}
= 0 , \nonumber \\
\end{eqnarray}
where
\begin{eqnarray} \label{eq:divfdiff}
\lefteqn{
\left[{\nabla} \cdot D_\epsilon \nabla E_{\epsilon}
\right]^\npo_{k+\lhalf,i+\lhalf,j+\lhalf} \equiv }
\nonumber \\ & & 
\frac{1}{ \left[ g_2 \right]_{i+\lhalf} 
\left[ g_{31} \right]_{i+\lhalf} 
\left[ g_{32} \right]_{j+\lhalf}}
\Biggl\{
\frac{1}{ \left[ x_1 \right]_{i+\lthreehalf} - 
\left[ x_1 \right]_{i+\lhalf}} 
\Biggr.
\nonumber \\ & & 
\quad \quad
\Biggl.
\times 
\left( \left[ g_2 \right]_{i+1}
\left[ g_{31} \right]_{i+1} 
\left[ g_{32} \right]_{j+\lhalf} 
\left [D_\epsilon(x_1) \right]^{n+t}_{k+\lhalf,i+1,j+\lhalf}
\Biggr. \right.
\nonumber \\ & & 
\quad \quad \; \; \;
\Biggl. \left.
\times
\frac{ \left[ E_\epsilon \right]^\npo_{k+\lhalf,i+\lthreehalf,j+\lhalf} - 
\left[ E_\epsilon \right]^\npo_{k+\lhalf,i+\lhalf,j+\lhalf}}
{ \left[ x_1 \right]_{i+\lthreehalf} - 
\left[ x_1 \right]_{i+\lhalf}}
\Biggr. \right.
\nonumber \\ & & 
\quad \quad \; \; \;
- \Biggl. \left. \left[ g_2 \right]_{i}
\left[ g_{31} \right]_{i} 
\left[ g_{32} \right]_{j+\lhalf}
\left[ D_\epsilon(x_1) \right]^{n+t}_{k+\lhalf,i,j+\lhalf} 
\Biggr. \right.
\nonumber \\ & & 
\quad \quad \; \; \;
\Biggl. \left.
\times
\frac{ \left[ E_\epsilon \right]^\npo_{k+\lhalf,i+\lhalf,j+\lhalf} - 
\left[ E_\epsilon \right]^\npo_{k+\lhalf,i-\lhalf,j+\lhalf}}
{ \left[ x_1 \right]_{i+\lhalf} - 
\left[ x_1 \right]_{i-\lhalf}}  \right) 
\Biggr.
\nonumber \\ & & 
\quad \quad \quad \quad \quad \quad \quad \quad \quad \quad \quad \quad \quad \quad 
\Biggl.
+ \frac{1}{ \left[ x_2 \right]_{j+\lthreehalf} - 
\left[ x_2 \right]_{j+\lhalf}}
\Biggr.
\nonumber \\ & & 
\quad \quad
\Biggl.
\times
\left( \frac{ \left[ g_{31} \right]_{i+\lhalf} 
\left[ g_{32} \right]_{j+1}}
{ \left[ g_2 \right]_{i+\lhalf}} 
\left[ D_\epsilon(x_2) \right]^{n+t}_{k+\lhalf,i+\lhalf,j+1}
\Biggr. \right.
\nonumber \\ & & 
\quad \quad \; \; \;
\Biggl. \left.
\times
\frac{ \left[ E_\epsilon \right]^\npo_{k+\lhalf,i+\lhalf,j+\lthreehalf} -
\left[ E_\epsilon \right]^\npo_{k+\lhalf,i+\lhalf,j+\lhalf}}
{\left[ x_2 \right]_{j+\lthreehalf} - 
\left[ x_2 \right]_{j+\lhalf}} 
\Biggr. \right.
\nonumber \\ & & 
\quad \quad \; \; \;
\Biggl. \left.
- \frac{ \left[ g_{31} \right]_{i+\lhalf} 
\left[ g_{32} \right]_{j}}
{ \left[ g_2 \right]_{i+\lhalf}} 
\left[ D_\epsilon(x_2) \right]^{n+t}_{k+\lhalf,i+\lhalf,j}
\Biggr. \right.
\nonumber \\ & & 
\quad \quad \; \; \;
\Biggl. \left.
\times
\frac{ \left[ E_\epsilon \right]^\npo_{k+\lhalf,i+\lhalf,j+\lhalf} -
\left[ E_\epsilon \right]^\npo_{k+\lhalf,i+\lhalf,j-\lhalf}}
{ \left[ x_2 \right]_{j+\lhalf} - 
\left[ x_2 \right]_{j-\lhalf}}
\right) 
\Biggr\}
\end{eqnarray}
and
\begin{eqnarray} \label{eq:pv_diff}
\lefteqn{
\left[ \epsilon \frac{\partial \left({\mathsf P}_{\epsilon}:
{\nabla} {\bf v}\right)}{\partial \epsilon} 
\right]^\npo_{k+\lhalf,i+\lhalf,j+\lhalf}
\equiv
\frac{\left[ \epsilon \right]_{k+\lhalf}}
{\left[ \epsilon \right]_{k+1} - \left[ \epsilon \right]_k} 
 }
\nonumber \\
& &
\times \biggl( \left[ \mathsf X_{\epsilon}:{\nabla} {\bf v} 
\right]^{n+t}_{k+(3/2),i+(1/2),j+(1/2)} 
\left[ E_\epsilon \right]^\npo_{k+(3/2),i+(1/2),j+(1/2)} 
\nonumber \\
& &  - \left[ \mathsf X_{\epsilon}:{\nabla} {\bf v} 
\right]^{n+t}_{k-(1/2),i+(1/2),j+(1/2)} 
\left[ E_\epsilon \right]^\npo_{k-(1/2),i+(1/2),j+(1/2)} 
\biggl)
\end{eqnarray}
and
\begin{eqnarray} \label{eq:nu_tr_src_trm}
\lefteqn{
\left[{\mathbb S}_{\epsilon} \right]^\npo_{k+(1/2),i+(1/2),j+(1/2)}
 \equiv 
}
 \nonumber \\
& &                    
 -\left[ S_\epsilon \right]^{n+t}_{k+(1/2),i+(1/2),j+(1/2)}
\left( 1 + \frac{\eta\alpha}{\left( \left[\epsilon \right]_{k+(1/2)}\right)^3}
\left[E_\epsilon\right]^\npo_{k+(1/2),i+(1/2),j+(1/2)} \right)
\nonumber \\
& &                    
+ c \left[\kappa^a_\epsilon\right]^{n+t}_{k+(1/2),i+(1/2),j+(1/2)}
\left[E_\epsilon\right]^\npo_{k+(1/2),i+(1/2),j+(1/2)}
\nonumber \\
& &                    
-
\left( 1 + \frac{\eta\alpha}{\left( \left[\epsilon \right]_{k+(1/2)}\right)^3}
\left[E_\epsilon\right]^\npo_{k+(1/2),i+(1/2),j+(1/2)} \right)
\left[ \epsilon \right]_{k+(1/2)}
\nonumber \\
& &
\times
\sum_{\ell=0}^{N_g-1} \left[\Delta\epsilon\right]_{\ell+(1/2)}
\left[G\right]^{n+t}_{k+(1/2),\ell+(1/2),i+(1/2),j+(1/2)}
\left( 1 + \frac{\eta\alpha}{\left( \left[\epsilon \right]_{\ell+(1/2)}\right)^3}
\left[\bar{E}_\epsilon\right]^\npo_{\ell+(1/2),i+(1/2),j+(1/2)} \right)
\nonumber \\
& &                    
-
c\left( 1 + \frac{\eta\alpha}{\left( \left[\epsilon \right]_{k+(1/2)}\right)^3}
\left[E_\epsilon\right]^\npo_{k+(1/2),i+(1/2),j+(1/2)} \right)
\nonumber \\
& & 
\times
\sum_{\ell=0}^{N_g-1} \left[\Delta\epsilon\right]_{\ell+(1/2)}
\left[\kappa^s\right]^{n+t}_{k+(1/2),\ell+(1/2),i+(1/2),j+(1/2)}
\left[E_\epsilon\right]^\npo_{\ell+(1/2),i+(1/2),j+(1/2)}
\nonumber \\
& &                    
+ 
c\left[E_\epsilon\right]^\npo_{k+(1/2),i+(1/2),j+(1/2)}
\nonumber \\
& & 
\times
\sum_{\ell=0}^{N_g-1} \left[\Delta\epsilon\right]_{\ell+(1/2)}
\left[\kappa^s\right]^{n+t}_{\ell+(1/2),k+(1/2),i+(1/2),j+(1/2)}
\left( 1 + \frac{\eta\alpha}{\left( \left[\epsilon \right]_{\ell+(1/2)}\right)^3}
\left[E_\epsilon\right]^\npo_{\ell+(1/2),i+(1/2),j+(1/2)} \right)
.
\nonumber \\
\end{eqnarray}
In Eq.~\eqref{eq:divfdiff}, the factors $g_2$, $g_{31}$, and $g_{32}$ derive from the 3-covariant 
form of the spatial metric used by Swesty and Myra, which is given by
\begin{equation}\label{eq:3metric}
ds^2 = 
(g_1)^2 dx_1^2 +
(g_2)^2 dx_2^2 +
(g_{31} g_{32})^2 dx_3^2
\end{equation}
and is written to accommodate Cartesian, cylindrical, and spherical coordinates. In Eq.~\eqref{eq:nu_tr_src_trm},
$\kappa^a$ and $\kappa^s$ are the absorption and scattering opacities, respectively, and $G(\epsilon ,\epsilon^{'})$ is
the pair annihilation kernel. The factors $\alpha$ and $\eta$ are constants. $N_g$ is the number of energy groups,
and the superscript $n+t$, with $t$ taking on different values, designates the update stages for the electron, muon, 
and tau neutrino distributions in the overall update scheme used by Swesty and Myra, shown in their Figure 3. To solve 
Eq.~\eqref{eq:nu_tr_fd} and its counterpart for antineutrinos, simultaneously, given their coupling, Swesty and Myra implement the Newton--BiCGSTAB
subclass of Newton--Krylov iterative methods. Eq.~\eqref{eq:nu_tr_fd} and its antineutrino counterpart are first linearized. 
BiCGSTAB is used for a solution to the resultant ``inner'' linear system of equations for the updates to the iterates of the 
outer Newton iteration.

Once the quantities $^l\mathbb S_\epsilon$, where $\ell$ denotes neutrino flavor, are known from the solution of 
Eq.~\eqref{eq:nu_tr_fd} and its counterparts for heavy-flavor neutrinos, Swesty and Myra 
then update the fluid electron fraction and energy density using the following operator split equations:
\begin{equation} \label{eq:ye-source}
\left\ldbrack \frac{\partial n_e}{\partial t}\right\rdbrack_{\rm collision}
= - \int \frac{1}{\epsilon} 
\left( ^e{\mathbb S}_\epsilon - ^e\bar{{\mathbb S}}_\epsilon \right) d\epsilon,
\end{equation}
\begin{equation}
\left\ldbrack \left(\frac{\partial E}{\partial t} \right)\right\rdbrack_{\rm collision-\ell} = 
- \int 
\left({ ^\ell{\mathbb S}_\epsilon +
^\ell\bar{{\mathbb S}}_\epsilon}\right) d\epsilon,
\label{eq:e-coll-F}
\end{equation}
where $n_e$ is the electron number density and $E$ is the matter energy density. 
Equation~\eqref{eq:e-coll-F} is solved in operator split fashion for each flavor.
The discretizations for Eqs.~\eqref{eq:ye-source} and \eqref{eq:e-coll-F} for 
electron-neutrino flavor neutrinos (where both lepton number and energy are exchanged) are:
\begin{equation} \label{eq:n_xch_fd}
\left[n_e \right]^{n+1}_{i+(1/2),j+(1/2)} = 
\left[n_e \right]^{n+b}_{i+(1/2),j+(1/2)}
- \Delta t \sum_{\ell=0}^{N_g-1} \left[\Delta\epsilon\right]_{\ell+(1/2)} 
\left( 
\frac{ \left[^e{\mathbb S}_\epsilon\right]^{n+b}_{i+(1/2),j+(1/2)} 
-
\left[^e\bar{{\mathbb S}}_\epsilon\right]^{n+b}_{i+(1/2),j+(1/2)}}
{\left[\epsilon\right]_{\ell+(1/2)}}
\right),
\end{equation}
\begin{equation} \label{eq:e_xch_fde}
\left[E \right]^{n+d}_{i+(1/2),j+(1/2)} = 
\left[E \right]^{n+b}_{i+(1/2),j+(1/2)}
- \Delta t \sum_{\ell=0}^{N_g-1} \left[\Delta\epsilon\right]_{\ell+(1/2)} 
\left( 
{ \left[^e{\mathbb S}_\epsilon\right]^{n+c}_{i+(1/2),j+(1/2)} 
-
\left[^e\bar{{\mathbb S}}_\epsilon\right]^{n+c}_{i+(1/2),j+(1/2)}}
\right).
\end{equation}

In a similar manner, the neutrino--matter momentum exchanged is computed.

\subsubsection{General-relativistic, finite-difference implementation}

A general relativistic implementation of MGFLD was developed by \citet{RaJuJa19}. 
They begin with the 3+1 metric:
\begin{eqnarray}
	\mathrm{d}s^2 & \equiv & g_{ab} \mathrm{d}x^a \mathrm{d}x^b \nonumber \\ 
	& = & - \alpha^2 \mathrm{d}t^2 + \gamma_{ij} (\mathrm{d}x^i + \beta^i \mathrm{d}t)(\mathrm{d}x^j + \beta^j \mathrm{d}t)~,
	\label{eq:gr_linele}
\end{eqnarray}
and the following definitions of the comoving-frame spectral neutrino energy density, momentum density, and stress tensor:
\begin{eqnarray}
	\mathcal{J}(x^{\mu},\epsilon) & \equiv & \epsilon^3 \int f(x^{\mu},p^{\hat \mu})~\mathrm{d}\Omega~, \nonumber \\
	\mathcal{H}^{\hat i}(x^{\mu},\epsilon) & \equiv & \epsilon^3 \int l^{\hat i} f(x^{\mu},p^{\hat \mu})~\mathrm{d}\Omega~, \nonumber \\
	\mathcal{K}^{\hat i \hat j}(x^{\mu},\epsilon) & \equiv & \epsilon^3 \int l^{\hat i} l^{\hat j} f(x^{\mu},p^{\hat \mu})~\mathrm{d}\Omega~,
	\label{eq:tr_moment}
\end{eqnarray}
respectively. 
$p^{\hat \mu}\equiv \epsilon (1,l^{\hat i})$ denotes the comoving-frame, momentum-space coordinates. $l^{\hat i}$
is a unit comoving-frame, momentum-space three-vector. 
With these definitions and choice of phase-space coordinates, Rahman et~al.\ express the evolution equation for the comoving-frame neutrino energy density as given by Eq.~\eqref{eq:spectralLagrangianEnergyEquationFLD_3p1} in Sect.~\ref{sec:oneMomentKinetics}.  
Given the approximations discussed there, the neutrino energy density equation solved by Rahman et~al.\ becomes
\begin{eqnarray}
	&&\frac{1}{\alpha} \frac{\partial}{\partial t} (W \mathcal{\hat J})
    + \frac{1}{\alpha} \frac{\partial}{\partial x^j} [\alpha W (v^j-\beta^j/\alpha) \mathcal{\hat J}] \nonumber \\
    &&- \frac{1}{\alpha} \frac{\partial}{\partial x^j} \Big[\alpha^{-2} \sqrt{\gamma} \Big\{ \gamma^{i k}
    + W \Big(\frac{W}{W+1}v^j-\beta^j/\alpha \Big) v^k \Big\} D \partial_k (\alpha^3 \mathcal{J}) \Big] \nonumber \\
    && - \frac{e^{k \hat i}}{\alpha^4} \frac{\partial}{\partial t}
    (W \sqrt{\gamma} \bar v_{\hat i}) D \partial_k(\alpha^3 \mathcal{J})
    +\hat{R}_\epsilon - \frac{\partial}{\partial \epsilon} (\epsilon \hat{R}_\epsilon) \nonumber \\
    &&= \kappa_\mathrm{a} (\mathcal{\hat J}^{eq}-\mathcal{\hat J})~,
    \label{eq:tr_fld_energy_eqn}
\end{eqnarray}
where the relation $e^{\hat{j}}_{\hspace{2pt}\hat{i}}e^{k\hat{i}}=\gamma^{jk}$ was used.  

Rahman et~al.\ divide the numerical update into three steps, operator splitting Eq.~\eqref{eq:tr_fld_energy_eqn} into the
source term, the radial and spectral shift terms, and the nonradial terms. In step 1, the focus is on the source term, and the 
corresponding terms in the matter specific internal energy and electron fraction equations. The set of equations to be solved 
is given by
\begin{eqnarray}
	\frac{W}{\alpha} \partial_t \mathcal{J}_{\nu,\xi} &=&
    \bigg[ \kappa_\mathrm{a} (\mathcal{J}^{\mathrm{eq}} - \mathcal{J}) \bigg]_{\nu,\xi}, \nonumber \\
    \frac{W}{\alpha} \rho \partial_t e (T,Y_\mathrm{e}) &=&
    - \sum_{\nu,\xi} \bigg[ \kappa_\mathrm{a} (\mathcal{J}^{\mathrm{eq}} - \mathcal{J} ) \Delta \epsilon_\xi \bigg]_{\nu,\xi}, \nonumber \\
    \frac{W}{\alpha} \rho \partial_t Y_\mathrm{e} &=&
    - m_u \sum_{\xi} \bigg[ \big[\kappa_\mathrm{a} (\mathcal{J}^{\mathrm{eq}} - \mathcal{J}) \Delta \epsilon_\xi \big]_{\nu_\mathrm{e}} \nonumber \\
    &&- \big[\kappa_\mathrm{a} (\mathcal{J}^{\mathrm{eq}} - \mathcal{J}) \Delta \epsilon_\xi \big]_{\bar \nu_\mathrm{e}} \bigg]_{\xi},
	\label{eq:ns_source_term}
\end{eqnarray}
where $\nu$ and $\xi$ indicate the neutrino species and energy bin, respectively, and $\Delta \epsilon_\xi$ the energy bin width. $m_{u}$ is the atomic mass unit.
These equations are differenced fully implicitly in time and solved using Newton--Raphson iteration. Linearization of the equations in $\mathcal{J}_{\nu,\xi}$, $e$, and $Y_e$ leads to a system of linear equations that must be solved for each iteration. To do so, Rahman et~al.\ use a direct (LAPACK) solver. The quantities
$\rho$, $\alpha$, $W$, and $\kappa_a$ are all held constant during the Newton--Raphson procedure.

In step 2, the following equation is solved:
\begin{eqnarray}
	&&W\partial_t \mathcal{\hat J} + \mathcal{R}_r = 0~,
  \label{eq:2nd_step_equation}
\end{eqnarray}
where
\begin{eqnarray}
	&& \mathcal{R}_r \equiv \partial_t (W) \mathcal{\hat J} +
    \partial_r [\alpha W(v^r-\beta^r \alpha^{-1})\mathcal{\hat J}] \nonumber \\
    &&- \partial_r \Big[ \alpha^{-2} \sqrt{\gamma} \Big\{ \gamma^{rr} 
    + W \Big(\frac{W}{W+1}v^r-\beta^r \alpha^{-1}\Big) v^r \Big\} 
    D_1 \partial_r(\alpha^3 \mathcal{J}) \Big] \nonumber \\
    &&- \alpha^{-3} e^{r \hat i} \partial_t
    (W \sqrt{\gamma} \bar v_{\hat i})D_1 \partial_r (\alpha^3 \mathcal{J})
    + \alpha \Big[\hat R_\epsilon - \frac{\partial}{\partial \epsilon} (\epsilon \hat R_\epsilon) \Big]~
    \label{eq:tr_rad_red_term}
\end{eqnarray}
includes radial advection, diffusion, and acceleration, as well as spectral shifts. $D_1$ denotes the radial diffusion coefficient.
Equation~(\ref{eq:2nd_step_equation}) is solved using the Crank--Nicolson scheme:
\begin{eqnarray}
	&& (W \sqrt{\gamma}) \frac{\mathcal{J}^{n+1}_i-\mathcal{J}^{n}_i}{\Delta t}
	= -\frac{1}{2} (\mathcal{R}^{n+1}_{r,i}+\mathcal{R}^{n}_{r,i})~.    
    \label{eq:tr_rad_red_term_discrete}
\end{eqnarray}
All gravitation and hydrodynamics variables are kept fixed during transport updates. $\mathcal{R}^{n}_{r,i}$ on the right-hand side 
of equation (\ref{eq:tr_rad_red_term_discrete}) is evaluated at both $t^n$ and $t^{n+1}$. For $t^{n+1}$, Rahman et~al.\ provide the
following discretizations.
The diffusion term is discretized as
\begin{eqnarray}
	&&\Big[ \partial_r \{A^r D_1 \partial_r(\alpha^3 \mathcal{J})\} \Big]^{n+1}_{i} = \nonumber \\
    &&\frac{1}{\Delta r} \Big[A^{r}_{i+1/2}D^{n}_{1,i+1/2} 
    \frac{\alpha^3_{i+1} \mathcal{J}^{n+1}_{i+1}-\alpha^3_{i} \mathcal{J}^{n+1}_{i}}{\Delta r} \nonumber \\
    &&- A^{r}_{i-1/2}D^{n}_{1,i-1/2}
    \frac{\alpha^3_{i} \mathcal{J}^{n+1}_i-\alpha^3_{i-1} \mathcal{J}^{n+1}_{i-1}}{\Delta r} \Big]~,
    \label{eq:tr_rad_red_diff_discrete0}
\end{eqnarray}
where
\begin{eqnarray}
    A^r &\equiv& \alpha^{-2} \sqrt{\gamma} \Big\{ \gamma^{rr} 
    + W \Big(\frac{W}{W+1}v^r-\beta^r \alpha^{-1}\Big) v^r \Big\}~. 
    \label{eq:tr_rad_red_diff_discrete}
\end{eqnarray}
$i-1/2$ and $i+1/2$ denote the left and right zone edges for zone $i$, respectively. Values of the gravity and hydrodynamics
variables at zone edges are determined by linear interpolation of their zone-center counterparts.
The fluid acceleration term is discretized as
\begin{eqnarray}
	&&\Big[B^r D_{1} \partial_r(\alpha^3 \mathcal{J})\Big]^{n+1}_{i} = \nonumber \\
    &&\frac{B^{r}_i}{2} \Big[ D^{n}_{1,i+1/2} 
    \frac{\alpha^3_{i+1} \mathcal{J}^{n+1}_{i+1}-\alpha^3_{i}\mathcal{J}^{n+1}_{i}}{\Delta r} \nonumber \\
    &&+ D^{n}_{1,i-1/2} \frac{\alpha^3_{i}\mathcal{J}^{n+1}_i-\alpha^3_{i-1}\mathcal{J}^{n+1}_{i-1}}{\Delta r} \Big]~, 
    \label{eq:tr_rad_red_aber_discrete}
\end{eqnarray}
where
\begin{eqnarray}
	B^r &\equiv&  \alpha^{-3} e^{r \hat i} \partial_t(W \sqrt{\gamma} \bar v_{\hat i})~.
    \label{eq:tr_rad_red_aber_discrete_B}
\end{eqnarray}
The metric and hydrodynamics variables before and after the metric and hydrodynamics updates 
are used to evaluate the time derivative in equation (\ref{eq:tr_rad_red_aber_discrete_B}).
The advection term is discretized in an upwind fashion as
\begin{align}
	\Big[ \partial_r (C^r {\mathcal{J}}) \Big]^{n+1}_{i} &=&
    \frac{1}{\Delta r} \Big[C^{r}_{i+1/2} 
    \mathcal{J}^{n+1}_{\iota(i+1/2)}
    - C^{r}_{i-1/2}
    \mathcal{J}^{n+1}_{\iota(i-1/2)} \Big]~,
    \label{eq:tr_rad_red_adv_discrete}
\end{align} 
where
\begin{eqnarray}
    C^r &\equiv& \alpha \sqrt{\gamma} W(v^r-\beta^r \alpha^{-1})
    \label{eq:tr_rad_red_adv_discrete_C}
\end{eqnarray}
and
\begin{eqnarray}
    \iota(i+1/2) &\equiv&  
	\begin{cases}
    i, & \text{if } v^r_{i+1/2} > 0  ~ ,\\
    i+1, & \text{otherwise}~.
	\end{cases}
    \label{eq:tr_rad_red_adv_discrete_jota}
\end{eqnarray}
Spectral shifts---the last term in equation (\ref{eq:tr_rad_red_term})---are discretized using the number-conservative scheme
of \citet{MuJaDi10} discussed in Sect.~\ref{sec:EnergyDiscretization}. The flux factor, $f^{\hat{i}}$, and the Eddington tensor, $\chi^{\hat{i}\hat{j}}$,
are used to replace $\mathcal{H}^{\hat{i}}$ and $\mathcal{K}^{\hat{i}\hat{j}}$ by $f^{\hat{i}}\mathcal{I}$ and $\chi^{\hat{i}\hat{j}}\mathcal{I}$,
respectively. In evaluating the spectral shift terms, both the flux factor and the Eddington tensor are evaluated at $t^n$, whereas 
the energy density, $\mathcal{I}$, is evaluated at $t^{n+1}$.
The remaining advection and diffusion terms are included in the last transport step, 
encapsulated in the equation
\begin{align}
    &&W \sqrt{\gamma} \partial_t (\mathcal{J})
	= \mathcal{R}(\mathcal{J})~,
	\label{eq:3rd_step_equation}
\end{align}
where
\begin{eqnarray}
    &&\mathcal{R}(\mathcal{J})
	\equiv -\partial_\theta [\alpha W (v^\theta-\beta^\theta \alpha^{-1}) \mathcal{\hat J}] 
    - \partial_\phi [\alpha W (v^\phi-\beta^\phi \alpha^{-1}) \mathcal{\hat J}] \nonumber \\
    &&+ \partial_\theta \Big[ \alpha^{-2} \sqrt{\gamma} \Big\{ \gamma^{\theta \theta} 
    + W \Big(\frac{W}{W+1}v^\theta-\beta^\theta \alpha^{-1}\Big) v^\theta \Big\} 
    D_{2} \partial_\theta(\alpha^3 \mathcal{J}) \Big] \nonumber \\
    &&+ \partial_\phi \Big[ \alpha^{-2} \sqrt{\gamma} \Big\{ \gamma^{\phi \phi} 
    + W \Big(\frac{W}{W+1}v^\phi-\beta^\phi \alpha^{-1}\Big) v^\phi \Big\} 
    D_{3} \partial_\phi(\alpha^3 \mathcal{J}) \Big] \nonumber \\
    &&+ \alpha^{-3} e^{\theta \hat i} \partial_t(W \sqrt{\gamma} \bar v_{\hat i}) D_2 \partial_\theta(\alpha^3 \mathcal{J}) \nonumber \\
    &&+ \alpha^{-3} e^{\phi \hat i} \partial_t(W \sqrt{\gamma} \bar v_{\hat i}) D_3 \partial_\phi(\alpha^3 \mathcal{J}).
	\label{eq:ns_lat_term}
\end{eqnarray}
$D_2$ and $D_3$ are the diffusions coefficients in the $\theta$ and $\phi$ directions, respectively. Equation 
(\ref{eq:ns_lat_term}) is evolved using one of two explicit methods: Allen--Cheng \citep{AlCh70} 
and Runge--Kutta--Legendre (RKL2) \citet{MeBaAs12}. The latter method is a conditionally stable method 
expressly designed for the diffusion equation. In the Allen--Cheng method, a predictor step provides the 
following partial update:
\begin{eqnarray}
    &&\frac{(W\sqrt{\gamma})}{\Delta t}(\mathcal{J}^{*}_k - \mathcal{J}^{n}_k)
	= - \frac{1}{2 \Delta y} (F_{k+1} \mathcal{J}^n_{k+1} - F_{k-1} \mathcal{J}^n_{k-1}) \nonumber \\
    &&+ \frac{1}{\Delta y^2} 
    [E_{k+1/2} (\alpha^3_{k+1} \mathcal{J}^n_{k+1}-\alpha^3_{k} \mathcal{J}^{*}_k) \nonumber \\
    &&- E_{k-1/2} (\alpha^3_{k} \mathcal{J}^{*}_k-\alpha^3_{k-1} \mathcal{J}^n_{k-1})]~ \nonumber \\
    && + \frac{G_{k}}{2\Delta y} \big[ D_{k+1/2} (\alpha^3_{k+1} \mathcal{J}^{n}_{k+1}-\alpha^3_{k} \mathcal{J}^{*}_{k}) \nonumber \\
    &&+ D_{k-1/2} (\alpha^3_{k} \mathcal{J}^{*}_{k}-\alpha^3_{k-1} \mathcal{J}^{n}_{k-1}) \big]~,
    \label{eq:ns_allen_cheng_pred}
\end{eqnarray}
which, in turn, is followed by a corrector step that provides the complete update:
\begin{eqnarray}
    &&\frac{(W\sqrt{\gamma})}{\Delta t}(\mathcal{J}^{n+1}_k - \mathcal{J}^{n}_k)
	= - \frac{1}{2 \Delta y} (F_{k+1} \mathcal{J}^{*}_{k+1} - F_{k-1} \mathcal{J}^{*}_{k-1}) \nonumber \\
    &&+ \frac{1}{\Delta y^2} 
    [E_{k+1/2} (\alpha^3_{k+1} \mathcal{J}^{*}_{k+1}-\alpha^3_{k} \mathcal{J}^{n+1}_k) \nonumber \\
    &&- E_{k-1/2} (\alpha^3_{k} \mathcal{J}^{n+1}_k-\alpha^3_{k-1} \mathcal{J}^{*}_{k-1})] \nonumber \\
    &&+ \frac{G_{k}}{2\Delta y} \big[ D_{k+1/2} (\alpha^3_{k+1} \mathcal{J}^{*}_{k+1}-\alpha^3_{k} \mathcal{J}^{n+1}_{k}) \nonumber \\
    &&+ D_{k-1/2} (\alpha^3_{k} \mathcal{J}^{n+1}_{k}-\alpha^3_{k-1} \mathcal{J}^{*}_{k-1}) \big]~,
    \label{eq:ns_allen_cheng_cor}
\end{eqnarray}
where
\begin{eqnarray}
    &&E \equiv \alpha^{-2} \sqrt{\gamma} \Big\{ \gamma^{jj} 
    + W \Big(\frac{W}{W+1}v^j-\beta^j \alpha^{-1}\Big) v^j \Big\} D~, \nonumber \\
    &&F \equiv \alpha \sqrt{\gamma} W (v^j-\beta^j \alpha^{-1})~, \nonumber \\
    &&G \equiv \alpha^{-3} e^{j \hat i} \partial_t(W \sqrt{\gamma} \bar v_{\hat i}).
	\label{eq:ns_allen_cheng_FE}
\end{eqnarray}
In equations (\ref{eq:ns_allen_cheng_pred}) and (\ref{eq:ns_allen_cheng_cor}), only one spatial index, $k$, is explicitly shown and represents
a zone index in either the $\theta$ or the $\phi$ direction. Moreover, in the discretizations shown, the gridding in the single dimension 
is assumed to be uniform, with zone width $\Delta y$.
In the (s-stage) RKL2 method, which Rahman et~al.\ deploy as a 4-stage method, the update in each
of the four stages is given by
\begin{eqnarray}
    \mathcal{J}_{0} &=& \mathcal{J}^{n}~, \nonumber \\
    \mathcal{J}_{1} &=& \mathcal{J}_{0}
	+ \frac{2}{27} \frac{\Delta t}{W\sqrt{\gamma}} \mathcal{R}(\mathcal{J}_{0})~, \nonumber \\
    \mathcal{J}_{2} &=& \frac{3}{2} \mathcal{J}_{1} - \frac{1}{2} \mathcal{J}_{0}
	+ \frac{\Delta t}{W\sqrt{\gamma}} \Bigg( \frac{1}{3} \mathcal{R}(\mathcal{J}_{1})
	- \frac{2}{9} \mathcal{R}(\mathcal{J}_{0}) \Bigg)~, \nonumber \\
    \mathcal{J}_{3} &=& \frac{25}{12} \mathcal{J}_{2} - \frac{5}{6} \mathcal{J}_{1} - \frac{1}{4} \mathcal{J}_{0}
	+ \frac{\Delta t}{W\sqrt{\gamma}} \Bigg( \frac{25}{54} \mathcal{R}(\mathcal{J}_{2})
	- \frac{25}{81} \mathcal{R}(\mathcal{J}_{0}) \Bigg)~, \nonumber \\
    \mathcal{J}_{4} &=& \frac{189}{100} \mathcal{J}_{3} - \frac{81}{80} \mathcal{J}_{2} + \frac{49}{400} \mathcal{J}_{0} \nonumber \\
	&& + \frac{\Delta t}{W\sqrt{\gamma}} \Bigg( \frac{21}{50} \mathcal{R}(\mathcal{J}_{3})
	- \frac{49}{200} \mathcal{R}(\mathcal{J}_{0}) \Bigg)~, \nonumber \\
	\mathcal{J}^{n+1} &=& \mathcal{J}_{4}~.
    \label{eq:ns_runge_kutta_legendre}
\end{eqnarray}
For the $s$-th stage and zone $k$, $\mathcal{R}(\mathcal{J})$ is discretized as
\begin{eqnarray}
    &&\mathcal{R}_k(\mathcal{J}_s)
	= - \frac{1}{2 \Delta y} (F_{k+1} \mathcal{J}_{s,k+1} - F_{k-1} \mathcal{J}_{s,k-1}) \nonumber \\
	&& + \frac{1}{\Delta y^2} 
    (E_{k+1/2} (\alpha^3_{k+1} \mathcal{J}_{s,k+1}-\alpha^3_{k} \mathcal{J}_{s,k}) \nonumber \\
    &&- E_{k-1/2} (\alpha^3_{k} \mathcal{J}_{s,k}-\alpha^3_{k-1} \mathcal{J}_{s,k-1})) \nonumber \\
    && + \frac{G_{k}}{2\Delta y} \big[ D_{k+1/2} (\alpha^3_{k+1} \mathcal{J}_{s,k+1}-\alpha^3_{k} \mathcal{J}_{s,k}) \nonumber \\
    &&+ D_{k-1/2} (\alpha^3_{k} \mathcal{J}_{s,k}-\alpha^3_{k-1} \mathcal{J}_{s,k-1}) \big].
    \label{eq:ns_runge_kutta_legendre_R}
\end{eqnarray}

Finally, it is important to note that Rahman et~al.\ go to great lengths to ensure that their definitions of the diffusion coefficients
preserve causality for both the individual and the total radiative fluxes. To accomplish this, they compute the gradient of the energy 
density as
\begin{eqnarray}\label{eq:C.1}
& &    |\nabla \mathcal{J}|_{i,j,k} \\ 
&    = & \sqrt{\Bigg(\frac{\mathcal{J}_{i+1,j,k}-\mathcal{J}_{i-1,j,k}}{r_{i+1}-r_{i-1}}\Bigg)^2
    +\Bigg(\frac{\mathcal{J}_{i,j+1,k}-\mathcal{J}_{i,j-1,k}}{r_i(\theta_{j+1}-\theta_{j-1})}\Bigg)^2
    +\Bigg(\frac{\mathcal{J}_{i,j,k+1}-\mathcal{J}_{i,j,k-1}}{r_i\sin{\theta_j}(\phi_{k+1}-\phi_{k-1})}\Bigg)^2} \nonumber
\end{eqnarray}
and the Knudsen number as
\begin{eqnarray}\label{eq:C.2}
    R_{i,j,k} = \frac{|\nabla \mathcal{J}|_{i,j,k}}{(\kappa_\mathrm{t})_{i,j,k} \mathcal{J}_{i,j,k}},
\end{eqnarray}
where $({\kappa_\mathrm{t}})_{i,j,k}$ is the transport opacity at the cell center $(i,j,k)$. Equation~\eqref{eq:limiterLPW} is then 
used to compute the flux limiter, and the causality-preserving diffusion coefficients are given by
\begin{eqnarray}\label{eq:C.3}
    D_{i,j,k} = \frac{\lambda_{i,j,k}}{(\kappa_\mathrm{t})_{i,j,k}}.
\end{eqnarray}

Rahman et~al.\ do not report on the conservation of lepton number in their code, but given their use of the method 
developed by \citet{MuJaDi10}, which is specifically designed to conserve lepton number, it should be quite good. 
They do report on their conservation of energy. They report a change in total energy of $1.85\times10^{51}$ erg 
at 60 ms after bounce, most of which results at bounce, and a much more gradual increase between 60 and 525 ms 
after bounce to their final value of $\Delta E$ of $2.0\times10^{51}$ erg. As discussed in Sect.~\ref{sec:lepenergycons}, 
their use of the Lagrangian two-moment model as the starting point for their MGFLD implementation does not lend 
itself to conserving energy, nor does their use of flux-limited diffusion, as discussed in \citet{JuObJa15} and in references
cited therein.

\subsubsection{Newtonian-gravity, $O(v/c)$, finite-volume implementation}

As part of the development of the CASTRO code, \citet{ZhHoAl13} developed a
MGFLD solver using finite-volume methods. They express the equations of multigroup 
radiation hydrodynamics as
\begin{align}
  \frac{\partial \rho}{\partial t} + \nabla \cdot (\rho \vec{u})
  = { } & 0, \label{eq:mgrhd-rho} \\
    \frac{\partial (\rho \vec{u})}{\partial t} + \nabla \cdot (\rho \vec{u}
  \vec{u}) + \nabla p + \sum_{g}
  \lambda_g \nabla E_g = { } &  \vec{F}_G, \label{eq:mgrhd-rhou} \\
  \frac{\partial (\rho E)}{\partial t} + \nabla \cdot (\rho E \vec{u} + p
  \vec{u}) + \vec{u} \cdot \sum_{g}\lambda_g
  \nabla E_g  = { } & \sum_{g} c
  (\kappa_gE_{g}-j_g) + \vec{u}\cdot\vec{F}_G, \label{eq:mgrhd-rhoE} \\
  \frac{\partial (\rho Y_e)}{\partial t} + \nabla \cdot (\rho Y_e \vec{u})
  = { } & \sum_{g} c \xi_g (\kappa_gE_{g}-j_g)
  , \label{eq:mgrhd-Ye} \\
\frac{\partial E_g}{\partial t} + \nabla \cdot 
  \left(\frac{3-f_g}{2} E_g \vec{u}\right) - \vec{u} \cdot \nabla
  \left(\frac{1-f_g}{2} E_g\right)  = { } & 
    - c (\kappa_gE_{g}-j_g)
  + \nabla \cdot \left(\frac{c\lambda_g}{\chi_g} \nabla E_g \right)
   \label{eq:mgrhd-Eg} \\
  + \int_g \frac{\partial}{\partial \nu} \Bigg{[}\left(\frac{1-f}{2}
    \nabla \cdot \vec{u} + \frac{3f-1}{2} \hat{\vec{n}}\hat{\vec{n}}
    : \nabla \vec{u} \right) & \nu E_\nu \Bigg{]} \mathrm{d}\nu -
    \frac{3f_g-1}{2} E_g \hat{\vec{n}}\hat{\vec{n}} : \nabla \vec{u}, \nonumber
\end{align}
where the group quantities are defined as
\begin{equation}
  E_g = \int_{\nu_{g-1/2}}^{\nu_{g+1/2}} E_\nu \mathrm{d}\nu, \label{eq:Egdef}
\end{equation}
\begin{equation}
\label{eq:emissivity-g}
  j_g = \frac{4\pi}{c}\eta(\nu_g) \Delta \nu_g,
\end{equation}
and
\begin{equation}
\label{eq:xi}
  \xi_g = s \frac{m_{\mathrm{B}}}{h \nu_g}.
\end{equation}
In Eq.~\eqref{eq:Egdef}, the neutrino energy density per frequency, $E_\nu$, is integrated over the frequency group defined by the interval 
$[\nu_{g}-1/2,\nu_{g}+1/2]$ to yield the energy density per group. Eq.~\eqref{eq:emissivity-g} defines the group emissivity in terms of the 
emissivity, $\eta$, and the group width $\Delta\nu_g=\nu_g+1/2-\nu_g-1/2$. In order of appearance in the equations, the remaining quantities 
are, $\lambda_g$, $\kappa_g$, and $f_g$, and are defined by evaluating the flux limiter, $\lambda$, the absorption coefficient, $\kappa$, and 
the Eddington factor, $f$, at a representative group frequency, $\nu_g$---i.e., they are all group-mean values. Finally, for neutrinos, $\xi_g$ is 
given by Eq.~\eqref{eq:xi}, with $s=+1$ for electron neutrinos and $s=-1$ for electron antineutrinos.
Zhang et~al.\ split these equations into three subsets, based on their mathematical characteristics and in an effort to minimize issues arising from 
operator splitting. There is a hyperbolic subsystem that includes the evolution of the electron fraction (it also includes pieces of the evolution equation 
for the neutrino energy density, but the neutrino energy density is not evolved using this subsystem, as will be discussed):
\begin{align}
  \frac{\partial \rho}{\partial t} + \nabla \cdot (\rho \vec{u})
  = { } & 0, \label{eq:hyper-Eg1} \\
    \frac{\partial (\rho \vec{u})}{\partial t} + \nabla \cdot (\rho \vec{u}
  \vec{u}) + \nabla p + \sum_{g}
  \lambda_g \nabla E_g = { } & \vec{F}_G , \label{eq:hyper-rhou} \\
  \frac{\partial (\rho E)}{\partial t} + \nabla \cdot (\rho E \vec{u} + p
  \vec{u}) +\vec{u} \cdot \sum_{g} \lambda_g
   \nabla E_g = { } & \vec{u} \cdot \vec{F}_G , \label{eq:hyper-rhoE} \\
  \frac{\partial (\rho Y_e)}{\partial t} + \nabla \cdot (\rho Y_e \vec{u})
  = { } & 0, \label{eq:hyper-Ye} \\
    \frac{\partial E_g}{\partial t} + \nabla \cdot 
  \left(\frac{3-f_g}{2} E_g \vec{u}\right) - \vec{u} \cdot
  \nabla \left( \frac{1-f_g}{2} E_g\right)  = { } & 0 . \label{eq:hyper-Eg5}
\end{align}
There is a second set of hyperbolic equations that governs the evolution of the neutrino energy density \emph{sans} the diffusion term and the term 
that describes the coupling of neutrinos to the matter:
\begin{align}
  \frac{\partial E_g}{\partial t}  = { } &
   -\nabla \cdot (E_g \vec{u}), \label{eq:Eg2} \\
\frac{\partial E_\nu}{\partial t} = { } &
  \frac{\partial}{\partial
  \ln{\nu}} \left[ \left(\frac{1-f}{2} \nabla \cdot \vec{u} +
  \frac{3f-1}{2} \hat{\vec{n}}\hat{\vec{n}} : 
  \nabla \vec{u}\right) E_\nu \right]. \label{eq:fspace2}
\end{align}
This second set of equations results from a splitting of their equation for the neutrino energy density per frequency, $E_\nu$, prior 
to integration over group frequencies:
\begin{align}
\frac{\partial E_{\nu}}{\partial t} 
+ \nabla \cdot (E_{\nu} \vec{u})
= { } & \nabla \cdot \left(\frac{c\lambda}{\chi} \nabla E_{\nu}
\right) - (c\kappa E_{\nu} - 4\pi\eta) \nonumber \\
{ } & + \frac{\partial}{\partial
  \ln{\nu}} \left(\frac{1-f}{2} E_\nu \nabla \cdot \vec{u} +
  \frac{3f-1}{2} E_\nu \hat{\vec{n}}\hat{\vec{n}} : 
  \nabla \vec{u}\right) . \label{eq:fdrhd-Enu} 
\end{align}
Finally, there is a parabolic system of equations that describes the evolution of the neutrino energy density due to the diffusion of neutrinos in the stellar core, 
as well as the evolution of the matter internal energy and electron fraction as a result of neutrino--matter interactions:
\begin{align}
\frac{\partial (\rho e)}{\partial t} = { } & \sum_{g} c
  (\kappa_gE_{g}-j_g), \label{eq:dEg-i1} \\
\frac{\partial (\rho Y_e)}{\partial t} = { } & \sum_{g} c \xi_g
  (\kappa_gE_{g}-j_g), \label{eq:dEg-i2} \\
\frac{\partial E_g}{\partial t} = { } & -c (\kappa_gE_{g}-j_g)
  + \nabla \cdot \left(\frac{c\lambda_g}{\chi_g} \nabla E_g \right).
   \label{eq:dEg-i3}
\end{align}

The equations in the first hyperbolic subsystem, Eqs.~\eqref{eq:hyper-Eg1} through \eqref{eq:hyper-Eg5}, are solved using an explicit, unsplit, PPM method, 
with characteristic limiting, full corner coupling, and the approximate Riemann solver of \citet{BeCoTr89}. Given the Godunov states computed, the radiation 
field energy density is in turn updated via Eq.~\eqref{eq:Eg2}. Finally, Eq.~\eqref{eq:fspace2}, which takes the form of an advection equation in neutrino-energy space, is solved using a second, explicit Godunov method, based on the method of lines. In this explicit part of the update scheme, a third-order, TVD, Runge--Kutta scheme developed by \citet{ShOs88} is used.

The parabolic system, Eqs.~\eqref{eq:dEg-i1} through \eqref{eq:dEg-i3}, is instead solved implicitly. Zhang et~al.\ reformulate the equations as
\begin{align}
  F_e = { } & \rho e - \rho e^{-} - \Delta t \sum_g c (\kappa_g E_g -
  j_g) = 0 , \label{eq:Fe} \\
  F_Y = { } & \rho Y_e - \rho Y_e^{-} - \Delta t \sum_g c \xi_g (\kappa_g
  E_g - j_g) = 0 , \label{eq:FY} \\
  F_g = { } & E_g - E_g^{-} - \Delta t\, \nabla \cdot
  \left(\frac{c\lambda_g}{\chi_g} \nabla E_g \right)  
  + \Delta t\, c (\kappa_g E_g - j_g) = 0 ,
\end{align}
and linearize in $T$, $Y_e$, and $E_g$ to obtain the (outer) linear system
\begin{equation}
  \left[\begin{array}{ccc}
      ({\partial F_e}/{\partial T})^{(k)}
      & ({\partial F_e}/{\partial Y_e})^{(k)}
      & ({\partial F_e}/{\partial E_g})^{(k)} \\[3pt]
      ({\partial F_Y}/{\partial T})^{(k)}
      & ({\partial F_Y}/{\partial Y_e})^{(k)}
      & ({\partial F_Y}/{\partial E_g})^{(k)} \\[3pt]
      ({\partial F_g}/{\partial T})^{(k)}
      & ({\partial F_g}/{\partial Y_e})^{(k)}
      & ({\partial F_g}/{\partial E_g})^{(k)}
      \end{array} \right]
  \left[\begin{array}{c}
      \delta T^{(k+1)}\\
      \delta Y_e^{(k+1)}\\
      \delta E_g^{(k+1)}
      \end{array}\right]
= \left[\begin{array}{c}
      - F_e^{(k)}\\
      - F_Y^{(k)}\\
      - F_g^{(k)}
      \end{array}\right]. \label{eq:newton}
\end{equation}
They point out that if the derivatives of the diffusion coefficient, $c\lambda_g/\chi_g$, with respect to $T$, $Y_e$, and $E_g$ are ignored, 
the linear system of equations collapses to an equation for the $(k+1)$$^{\rm st}$ iterate, $E^{(k+1)}_{g}$:
\begin{equation} \begin{split}
  \left(c\kappa_g + \frac{1}{\Delta t}\right) E_g^{(k+1)} & - \nabla \cdot \left(
      \frac{c\lambda_g}{\chi_g} \nabla E_g^{(k+1)} \right) = c j_g +
      \frac{E_g^{-}}{\Delta t} \\ 
   & + H_g
  \left[c\sum_{g^{\prime}}\left(\kappa_{g^{\prime}} E_{g^{\prime}}^{(k+1)} -
    j_{g^{\prime}} \right) - \frac{1}{\Delta t} (\rho e^{(k)} - \rho e^{-}) \right] \\ 
  & +  \Theta_g
  \left[c \sum_{g^{\prime}}\xi_{g^{\prime}}\left(\kappa_{g^{\prime}}
      E_{g^{\prime}}^{(k+1)} - j_{g^{\prime}} \right) - \frac{1}{\Delta t}
    (\rho Y_e^{(k)} - \rho Y_e^{-}) \right] , \label{eq:MGdiff0} 
\end{split}\end{equation}
where $\lambda_g$, $\kappa_g$, $\chi_g$, and $j_g$ are evaluated at the $k^{\rm th}$ iterate, and where
\begin{align}
H_g = { } & \left(\frac{\partial j_g}{\partial T} - \frac{\partial
    \kappa_g}{\partial T} E_g^{(k)} \right) \eta_T -
    \left(\frac{\partial j_g}{\partial Y_e} - \frac{\partial
    \kappa_g}{\partial Y_e} E_g^{(k)}\right)  \eta_Y ,\\  
\Theta_g = { } & \left(\frac{\partial j_g}{\partial Y_e} - \frac{\partial
    \kappa_g}{\partial Y_e} E_g^{(k)}\right) \theta_Y - 
    \left(\frac{\partial j_g}{\partial T} - \frac{\partial
    \kappa_g}{\partial T} E_g^{(k)} \right) \theta_T , 
\end{align}
and
\begin{align}
  \eta_T = { } & \frac{c\Delta t}{\Omega} \left[\rho + c \Delta t
    \sum_g \xi_g \left(\frac{\partial j_g}{\partial Y_e} - \frac{\partial
    \kappa_g}{\partial Y_e} E_g^{(k)}\right) \right]  , \\
  \eta_Y = { } & \frac{c\Delta t}{\Omega} \left[c \Delta t \sum_g
    \xi_g \left(\frac{\partial j_g}{\partial T} - \frac{\partial
    \kappa_g}{\partial T} E_g^{(k)} \right) \right] , \\
  \theta_T = { } & \frac{c\Delta t}{\Omega} \left[ \rho \frac{\partial
      e}{\partial Y_e} + c \Delta t \sum_g 
      \left(\frac{\partial j_g}{\partial Y_e} - \frac{\partial
      \kappa_g}{\partial Y_e} E_g^{(k)}\right) \right] , \\ 
  \theta_Y = { } & \frac{c\Delta t}{\Omega} \left[\rho \frac{\partial
      e}{\partial T} + c\Delta t \sum_g 
      \left(\frac{\partial j_g}{\partial T} - \frac{\partial
    \kappa_g}{\partial T} E_g^{(k)} \right) \right] , \\
  \Omega = { } & \left[\rho \frac{\partial
      e}{\partial T} + c\Delta t \sum_g 
      \left(\frac{\partial j_g}{\partial T} - \frac{\partial
    \kappa_g}{\partial T} E_g^{(k)} \right) \right]
      \left[\rho + c \Delta t
    \sum_g \xi_g \left(\frac{\partial j_g}{\partial Y_e} - \frac{\partial
    \kappa_g}{\partial Y_e} E_g^{(k)}\right) \right]
  \nonumber \\
   & - \left[ \rho \frac{\partial
      e}{\partial Y_e} + c \Delta t \sum_g 
      \left(\frac{\partial j_g}{\partial Y_e} - \frac{\partial
      \kappa_g}{\partial Y_e} E_g^{(k)}\right) \right]
     \left[c \Delta t \sum_g
    \xi_g \left(\frac{\partial j_g}{\partial T} - \frac{\partial
    \kappa_g}{\partial T} E_g^{(k)} \right) \right],
\end{align}
all of which are evaluated at the $k^{\rm th}$ iterate.
Eq.~\eqref{eq:MGdiff0} couples $E_g$ to its values across all energy groups. To decouple the groups, Zhang et~al.\ choose 
to use an (inner) iterative procedure by evaluating the right-hand-side at the $k^{\rm th}$ iterate of $E_g$ and iterating the solution 
of Eq.~\eqref{eq:MGdiff0} to convergence. Once $E^{k+1}_g$ is known, the updates for $\rho e$ and $Y_e$ are 
determined by
\begin{align}
  \rho e^{(k+1)} = { } &  H \rho e^{(k)} + (1-H) \rho e^- + \Theta (\rho Y_e^{(k)} -
  \rho Y_e^-) \nonumber \\
  & + c \Delta t \sum_g\left[(\kappa_gE_g^{\ell+1} - j_g) - (H
    + \Theta \xi_g) (\kappa_gE_g^{\ell} - j_g)\right] , \label{eq:uprhoe}\\ 
 \rho Y_e^{(k+1)} = { } & \bar{\Theta} \rho Y_e^{(k)} +
  (1-\bar{\Theta}) \rho Y_e^- + \bar{H} (\rho e^{(k)} - \rho
  e^-) \nonumber \\
  & + c\Delta t \sum_g\left[\xi_g(\kappa_gE_g^{\ell+1} - j_g) -
    (\bar{H} + \bar{\Theta} \xi_g) (\kappa_gE_g^{\ell} -
    j_g)\right] , \label{eq:uprhoYe}
\end{align}
which stem from Eqs.~\eqref{eq:Fe} and \eqref{eq:FY} upon linearization and are conservative for energy and lepton number. 
In Eqs.~\eqref{eq:uprhoe} and \eqref{eq:uprhoYe}, $H$, $\Theta$, $\bar{H}$, and $\bar{\Theta}$
are defined by
\begin{eqnarray}
H & = & \sum_g H_g, \\
\Theta & = & \sum_g \Theta_g, \\
\bar{H} & = & \sum_g \xi_g H_g, \\
\bar{\Theta} & = & \sum_g \xi_g \Theta_g.
\end{eqnarray}
In turn, $T$ is updated, and the next outer iteration is initiated. Zhang et~al.\ 
deploy the synthetic acceleration scheme of \citet{MoLaMa85,MoYaWa07}, extended in this case by them to neutrino transport,
to accelerate convergence of their outer iteration.
Note that the system given by Eqs.~\eqref{eq:dEg-i1}-\eqref{eq:dEg-i3} does not include energy coupling interactions (e.g., inelastic scattering).  
Inclusion of these interactions in a fully implicit solve requires modifications to the solution procedure.

The degree to which the approach outlined here conserves lepton number and energy was not documented.

\subsection{Structure-preserving methods}
\label{sec:structurePreservingMethods}

Structure-preserving methods are advanced numerical methods that aim to capture key properties of the underlying, continuous PDEs, and include methods that preserve physical bounds on solutions (e.g., positive distribution functions), achieve asymptotic limits of a multi-scale model (e.g., diffusion limit in radiation transport and steady states), preserve constraints (e.g., the divergence-free condition in magnetohydrodynamics), or conserve secondary quantities (e.g., simultaneous conservation of neutrino number and energy).  
As such, structure-preserving methods are more faithful to the physics, and often improves accuracy and robustness.  
The energy conserving discretization of the spherically symmetric Boltzmann equation by \citet{LiMeMe04} discussed in Sect.~\ref{sec:relativisticEffectsAndConservationOfEnergy}, and the number conserving discretization of the energy equation in the Lagrangian two-moment model by \citet{MuJaDi10} discussed in \ref{sec:numericalTwoMomentKinetics} are examples of structure-preserving discretizations already in use in simulations.  
These aim to preserve secondary quantities that are not evolved directly by the numerical method.  
Below we discuss discretizations that aim to preserve physical bounds on evolved quantities.  

\subsubsection{Preamble: discontinuous Galerkin methods}
\label{sec:dgPreamble}

Since the following subsections employ the discontinuous Galerkin (DG) method, which has yet to be adapted to modeling CCSN, we include a short description of key elements here by considering the scalar conservation law,
\begin{equation}
  \pd{u}{t}+\pd{f(u)}{x} = 0,
  \label{eq:dg_scalarConservationLaw}
\end{equation}
with a linear flux $f(u)=a\,u$, where $a$ is a constant in space and time.  
We refer to \citet{CoSh89,CoLiSh89,CoHoSh90,CoSh91,CoSh98} for pioneering, in-depth expositions on the early development of DG methods.  
See also \citet{CoSh01,Shu16} for reviews.  

To solve Eq.~\eqref{eq:dg_scalarConservationLaw}, the computational domain $D$ is divided into a triangulation $\mathcal{T}$ of non-overlapping elements $K=(x_{\lo},x_{\hi})$, so that $D = \cup_{K \in \mathcal{T}}$.  
On each element, the solution will then be approximated by functions in the approximation space
\begin{equation}
  \mathbb{V}_{h}^{k}=\{\varphi_{h} : \varphi_{h}\big|_{K} \in \mathbb{P}^{k}(K), \, \, \forall\ K\in \mathcal{T} \},
  \label{eq:dg_approximationSpace}
\end{equation}
where $\mathbb{P}^{k}(K)$ denotes the space of one-dimensional polynomials of maximal degree $k$ (e.g., Legendre polynomials).  
Functions in $\mathbb{V}_{h}^k$ can be discontinuous across element interfaces (hence discontinuous Galerkin).  
One then writes the approximate solution to Eq.~\eqref{eq:dg_scalarConservationLaw} on element $K$ as the expansion
\begin{equation}
  u_{h}^{K}(x,t) = \sum_{i=1}^{k+1}u_{i}^{K}(t)\,b_{i}^{K}(x),
  \label{eq:dg_approximation}
\end{equation}
where the expansion coefficients $u_{i}^{K}$ are the unknowns for which we solve the equations, and $b_{i}^{K}\in\mathbb{V}_{h}^{k}$ are the basis functions.  
Next, one defines in what sense $u_{h}^{K}$ will approximate $u$, the solution to Eq.~\eqref{eq:dg_scalarConservationLaw}.  
To this end, the residual
\begin{equation}
  R(u_{h}^{K}) = \pd{u_{h}^{K}}{t}+\pd{f(u_{h}^{K})}{x}
  \label{eq:dg_residual}
\end{equation}
is defined, which is required to be orthogonal to all test functions $\varphi_{h}\in\mathbb{V}_{h}^{k}$; i.e.,
\begin{equation}
  \int_{K}R(u_{h}^{K})\,\varphi_{h}^{K}\,dx = 0, \quad\forall\varphi_{h}^{K}\in\mathbb{V}_{h}^{k}.
  \label{eq:dg_ansatz}
\end{equation}
Inserting Eq.~\eqref{eq:dg_residual} into Eq.~\eqref{eq:dg_ansatz}, and performing an integration by parts on the flux term gives
\begin{equation}
  \int_{K}(\pd{u_{h}^{K}}{t})\,\varphi_{h}^{K}\,dx + \big[\,f(u_{h}^{K})(x_{\hi}^{-})\,\varphi_{h}^{K}(x_{\hi}^{-})-f(u_{h}^{K})(x_{\lo}^{+})\,\varphi_{h}^{K}(x_{\lo}^{+})\,\big] - \int_{K}f(u_{h}^{K})\,\pd{\varphi_{h}^{K}}{x}\,dx = 0,
  \label{eq:dg_ansatz_weak}
\end{equation}
where $x_{\lo/\hi}^{\pm}=\lim_{\delta^{+}\to0}x_{\lo/\hi}\pm\delta$.  
However, the entirely local formulation in Eq.~\eqref{eq:dg_ansatz_weak} is problematic because it does not specify how solutions in adjacent elements interact.  
In addition, a unique flux must be defined on the element interfaces at $x_{\lo/\hi}$ to recover the conservation statement inherent in Eq.~\eqref{eq:dg_scalarConservationLaw}.  
To resolve this, the fluxes on the element interfaces are replaced by a unique value, which then gives the semi-discrete DG method in weak form:
\textit{Find $u_{h}^{K} \in \mathbb{V}_{h}^{k}$ such that}
\begin{equation}
  \int_{K}(\pd{u_{h}^{K}}{t})\,\varphi_{h}^{K}\,dx + \big[\,\widehat{f(u_{h}^{K})}(x_{\hi})\,\varphi_{h}^{K}(x_{\hi}^{-})-\widehat{f(u_{h}^{K})}(x_{\lo})\,\varphi_{h}^{K}(x_{\lo}^{+})\,\big] - \int_{K}f(u_{h}^{K})\,\pd{\varphi_{h}^{K}}{x}\,dx = 0
  \label{eq:dg_semiDiscrete_weak}
\end{equation}
holds for all $\varphi_{h}\in\mathbb{V}_{h}^{k}$ and all $K\in\mathcal{T}$.  
In Eq.~\eqref{eq:dg_semiDiscrete_weak}, $\widehat{f(u_{h}^{K})}(x)$ is a unique numerical flux defined on the interface.  
For the scalar problem considered here, the familiar upwind flux can be used:
\begin{equation}
  \widehat{f(u_{h}^{K})}(x)
  =\f{1}{2}\,\big(\,f(u_{h}^{K}(x^{-}))+f(u_{h}^{K}(x^{+}))-|a|\,(u_{h}^{K}(x^{+})-u_{h}^{K}(x^{-}))\,\big),
  \label{eq:dg_upwindFlux}
\end{equation}
which is defined in terms of approximations to the immediate left and right of $x$, which can be different.  

Undoing the integration by parts that resulted in Eq.~\eqref{eq:dg_semiDiscrete_weak} gives the semi-discrete DG method in strong form:
\textit{Find $u_{h}^{K} \in \mathbb{V}_{h}^{k}$ such that}
\begin{align}
  &\int_{K}R(u_{h}^{K})\,\varphi_{h}^{K}\,dx \label{eq:dg_semiDiscrete_strong} \\
  &= \big[\,\big(f(u_{h}^{K}(x_{\hi}^{-}))-\widehat{f(u_{h}^{K})}(x_{\hi})\big)\,\varphi_{h}^{K}(x_{\hi}^{-})-\big(f(u_{h}^{K}(x_{\lo}^{+}))-\widehat{f(u_{h}^{K})}(x_{\lo})\big)\,\varphi_{h}^{K}(x_{\lo}^{+})\,\big], \nonumber
\end{align}
for all $\varphi_{h}^{K}\in\mathbb{V}_{h}^{k}$ and all $K\in\mathcal{T}$.  
Here, the weak and the strong formulations (Eqs.~\eqref{eq:dg_semiDiscrete_weak} and \eqref{eq:dg_semiDiscrete_strong}, respectively) are equivalent statements.  
By comparing the strong formulation with Eq.~\eqref{eq:dg_ansatz}, one sees that the residual in the DG solution is orthogonal to $\varphi_{h}$ only in the convergent limit when $f(u_{h}^{K}(x^{\pm}))\to\widehat{f(u_{h}^{K})}(x)$.  
In Sections \ref{sec:boundPreserving} and \ref{sec:realizabilityPreserving}, we will only refer to the weak formulation in Eq.~\eqref{eq:dg_semiDiscrete_weak}.  

To further illustrate how the weak formulation in Eq.~\eqref{eq:dg_semiDiscrete_weak} is used in practice, let
\begin{equation}
  \mathbf{u}^{K}(t)
  =\big(\,u_{1}^{K}(t),\ldots,u_{k+1}^{K}(t)\,\big)^{T}
  \quad\text{and}\quad
  \mathbf{b}^{K}(x)
  =\big(\,b_{1}^{K}(x),\ldots,b_{k+1}^{K}(x)\,\big)^{T}.
\end{equation}
Then, by inserting Eq.~\eqref{eq:dg_approximation} into Eq.~\eqref{eq:dg_semiDiscrete_weak}, and letting $\varphi_{h}=b_{j}~(j=1,\ldots,k+1)$, one obtains an equation for the expansion coefficients:
\begin{equation}
  \deriv{\mathbf{u}^{K}}{t}
  =-(M^{K})^{-1}\,\Big\{\,\big[\,\widehat{f(u_{h}^{K})}(x_{\hi})\,\mathbf{b}^{K}(x_{\hi}^{-})-\widehat{f(u_{h}^{K})}(x_{\lo})\,\mathbf{b}^{K}(x_{\lo}^{+})\,\big] - S^{K}\,\mathbf{u}^{K}\,\Big\},
  \label{eq:dg_weak_ode}
\end{equation}
where components of the \emph{mass matrix} and \emph{stiffness matrix} are defined as
\begin{equation}
  M_{ij}^{K} = \int_{K}b_{i}^{K}\,b_{j}^{K}\,dx
  \quad\text{and}\quad
  S_{ij}^{K} = a\,\int_{K}(\pd{b_{i}^{K}}{x})\,b_{j}^{K}\,dx,
  \label{eq:dg_matrices}
\end{equation}
respectively.  
Since the basis functions are polynomials, the integrals in Eq.~\eqref{eq:dg_matrices} can be computed exactly with, e.g., Gaussian quadratures.  
Eq.~\eqref{eq:dg_weak_ode} is now a system of ODEs, which can be integrated in time with an ODE solver.  
For non-stiff problems, explicit Runge--Kutta methods can be used.  

The DG method has been used to develop structure-preserving methods in a range of applications; see for example \citet{ZhSh10b} and \citet{WuTa16} for physical-constraint-preserving methods for the non-relativistic and relativistic Euler equations, respectively, \citet{LiXi18} for a steady-state preserving method for the Euler equations with gravitation, and \citet{JuHaTe18} for an energy-conserving DG method for kinetic plasma simulations.  
We also mention the work of \cite{HeHa20}, where DG and finite-volume methods are combined to a hybrid transport scheme that captures the diffusion limit and is more efficient in terms of memory usage and computational time than the corresponding DG-only scheme.

\subsubsection{Bound-preserving methods}
\label{sec:boundPreserving}

\citet{ZhSh10a} developed a general framework for ``maximum-principle-preserving'', high-order methods for scalar conservation laws \citep[see also][]{ZhSh11}.  
Inspired by this work, \citet{EnHaXi15} developed bound-preserving methods in the context of neutrino transport, aiming to maintain a distribution function satisfying $f\in[0,1]$, as dictated by Pauli's exclusion principle.  
They considered the (collisionless) phase-space advection problem in curvilinear coordinates, and included a general relativistic example in spherical symmetry with a time-independent spacetime metric given by
\begin{equation}
  ds^{2} = -\alpha^{2}\,dt^{2} + \gamma_{ij}\,dx^{i}\,dx^{j}, \quad\text{with}\quad\gamma_{ij}=\psi^{4}\mbox{diag}\big[\,1,r^{2},r^{2}\sin^{2}\theta\,\big],
\end{equation}
where $\alpha$ is the lapse function and $\psi$ the conformal factor.  
Under these assumptions, the Boltzmann takes the form
\begin{align}
  &
  \f{1}{\alpha}\pderiv{f}{t}
  +\f{1}{\alpha\,\psi^{6}\,r^{2}}\pderiv{}{r}\Big(\,\alpha\,\psi^{4}\,r^{2}\,\mu\,f\,\Big)
  -\f{1}{\varepsilon^{2}}\pderiv{}{\varepsilon}
  \Big(\,\varepsilon^{3}\,\f{1}{\psi^{2}\,\alpha}\pderiv{\alpha}{r}\,\mu\,f\,\Big) \nonumber \\
  & \hspace{12pt}
  +\pderiv{}{\mu}
  \Big(\,\big(1-\mu^{2}\big)\,\psi^{-2}\,
    \Big\{\,
      \f{1}{r}
      +\f{1}{\psi^{2}}\pderiv{\psi^{2}}{r}
      -\f{1}{\alpha}\pderiv{\alpha}{r}
    \,\Big\}\,f
  \,\Big)
  =0, 
  \label{eq:ConservativeBoltzmannEquationSphericalSymmetryGR}
\end{align}
where $r\ge0$ is the radius, $\mu\in[-1,1]$ the momentum space angle cosine, and $\varepsilon\ge0$ is the neutrino energy.  
By defining phase-space coordinates $z^{1}=r$, $z^{2}=\mu$, and $z^{3}=\varepsilon$, the phase space volume Jacobian $\tau=\psi^{6}\,r^{2}\,\varepsilon^{2}$, and
\begin{equation}
  H^{1} = H^{(r)}= \f{\alpha}{\psi^{2}}\mu,\quad
  H^{2} = H^{(\mu)} = \f{\alpha\big(1-\mu^{2}\big)}{\psi^{2}r}\,\Psi,\quad\text{and}\quad
  H^{3} = H^{(\varepsilon)} = - \f{\varepsilon}{\psi^{2}}\pderiv{\alpha}{r}\mu,
  \label{eq:phaseSpaceFluxCoefficients}
\end{equation}
where
\begin{equation}
  \Psi = 1+r\,\pd{\ln\psi^{2}}{r}-r\,\pd{\ln\alpha}{r},
\end{equation}
Eq.~\eqref{eq:ConservativeBoltzmannEquationSphericalSymmetryGR} can be written in the compact form
\begin{equation}
  \pderiv{f}{t}+\f{1}{\tau}\sum_{i=1}^{3}\pderiv{}{z^{i}}\Big(\,\tau\,H^{i}f\,\Big) = 0.  
  \label{eq:ConservativeBoltzmannCompact}
\end{equation}
It is straightforward to show that
\begin{equation}
  \f{1}{\tau}\sum_{i=1}^{3}\pderiv{}{z^{i}}\Big(\,\tau\,H^{i}\,\Big) = 0
  \label{eq:DivergenceFreeCondition}
\end{equation}
holds.  
The divergence-free condition on the phase-space flow in Eq.~\eqref{eq:DivergenceFreeCondition} plays an important role in maintaining $f\le1$.  

\citet{EnHaXi15} employed the discontinuous Galerkin (DG) method \citep[see, e.g.,][and references therin]{CoSh01,Shu16} to solve Eq.~\eqref{eq:ConservativeBoltzmannEquationSphericalSymmetryGR}.  
To this end, the phase space domain $D$ is divided into a triangulation $\mathcal{T}$ of elements $\mathbf{K}$, so that $D = \cup_{\mathbf{K} \in \mathcal{T}}$.  
Each element is a logically Cartesian box
\begin{equation}
  \bK=\{(r,\mu,\varepsilon)\in\mathbb{R}^{3} : r\in K^{(r)}:=(r_{\lo},r_{\hi}),\, \mu\in K^{(\mu)}:=(\mu_{\lo},\mu_{\hi}),\, \varepsilon\in K^{(\varepsilon)}:=(\varepsilon_{\lo},\varepsilon_{\hi})\},
\end{equation}
where $z_{\lo}^{i}$ and $z_{\hi}^{i}$ are, respectively, the coordinates of the lower and higher boundaries of $\mathbf{K}$ in the $i$th dimension.  
On each element, the approximation space for the DG method, $\mathbb{V}_{h}^k$, is
\begin{equation}\label{ldg:vhk}
  \mathbb{V}_{h}^{k}=\{\varphi_{h} : \varphi_{h}\big|_{\bK} \in \mathbb{Q}^{k}(\bK), \, \, \forall\ \bK\in \mathcal{T} \},
\end{equation}
where $\mathbb{Q}^{k}$ is the space of tensor products of one-dimensional polynomials of maximal degree $k$.  
The approximation to the distribution function, $f_{h}$, is then expressed as
\begin{equation}
  f_{h}(\mathbf{z},t)=\sum_{i=1}^{(k+1)^{3}}C_{i}(t)\,P_{i}(\mathbf{z}),
\end{equation}
where each $P_{i}\in\mathbb{V}_{h}^{k}$.  
Note that functions in $\mathbb{V}_{h}^k$ can be discontinuous across element interfaces.  
Then, for any $(r,\mu,\varepsilon) \in D$ and any $\varphi_{h} \in \mathbb{V}_{h}^{k}$, the DG method is as follows:
\textit{Find $f_{h} \in \mathbb{V}_{h}^{k}$ such that}
\begin{align}
  &
  \int_{\bK}\pd{}{t}f_{h}\,\varphi_{h}\,dV
  -\int_{\bK}H^{(r)}f_{h}\pd{\varphi_{h}}{r}\,dV
  -\int_{\bK}H^{(\mu)}f_{h}\pd{\varphi_{h}}{\mu}\,dV
  -\int_{\bK}H^{(\varepsilon)}f_{h}\,\pd{\varphi_{h}}{\varepsilon}\,dV \nonumber \\
  & \hspace{12pt}
  + \int_{\tK^{(r)}}\widehat{H^{(r)}f_{h}}(r_{\hi},\mu,\varepsilon)\,\varphi_{h}(r_{\hi}^{-},\mu,\varepsilon)\,\tau(r_{\hi},\varepsilon)\,d\tV^{(r)} \nonumber \\
  & \hspace{48pt}
  - \int_{\tK^{(r)}}\widehat{H^{(r)}f_{h}}(r_{\lo},\mu,\varepsilon)\,\varphi_{h}(r_{\lo}^{+},\mu,\varepsilon)\,\tau(r_{\lo},\varepsilon)\,d\tV^{(r)} \nonumber \\
   & \hspace{12pt}
  + \int_{\tK^{(\mu)}}\widehat{H^{(\mu)}f_{h}}(r,\mu_{\hi},\varepsilon)\,\varphi_{h}(r,\mu_{\hi}^{-},\varepsilon)\,\tau(r,\varepsilon)\,d\tV^{(\mu)} \nonumber \\
  & \hspace{48pt}
  - \int_{\tK^{(\mu)}}\widehat{H^{(\mu)}f_{h}}(r,\mu_{\lo},\varepsilon)\,\varphi_{h}(r,\mu_{\lo}^{+},\varepsilon)\,\tau(r,\varepsilon)\,d\tV^{(\mu)} \nonumber \\
  & \hspace{12pt}
  + \int_{\tK^{(\varepsilon)}}\widehat{H^{(\varepsilon)}f_{h}}(r,\mu,\varepsilon_{\hi})\,\varphi_{h}(r,\mu,\varepsilon_{\hi}^{-})\,\tau(r,\varepsilon_{\hi})\,d\tV^{(\varepsilon)} \nonumber \\
  & \hspace{48pt}
  - \int_{\tK^{(\varepsilon)}}\widehat{H^{(\varepsilon)}f_{h}}(r,\mu,\varepsilon_{\lo})\,\varphi_{h}(r,\mu,\varepsilon_{\lo}^{+})\,\tau(r,\varepsilon_{\lo})\,d\tV^{(\varepsilon)}=0, 
  \label{eq:ConservativeBoltzmannSphericalSymmetryGRDG}
\end{align}
where the infinitesimal phase-space volume and ``area'' elements are
\begin{equation}
  dV=\tau\,dr\,d\mu\,d\varepsilon,\quad
  d\tV^{(r)}=d\mu\,d\varepsilon,\quad
  d\tV^{(\mu)}=dr\,d\varepsilon,\quad
  d\tV^{(\varepsilon)}=dr\,d\mu,
\end{equation}
and the subelements are
\begin{equation}
  \tK^{(r)}=K^{(\mu)}\times K^{(\varepsilon)},\quad
  \tK^{(\mu)}=K^{(r)}\times K^{(\varepsilon)},\quad
  \tK^{(\varepsilon)}=K^{(r)}\times K^{(\mu)}.  
\end{equation}
In Eq.~\eqref{eq:ConservativeBoltzmannSphericalSymmetryGRDG}, upwind fluxes are used for the numerical fluxes on element interfaces:
\begin{align}
  \widehat{H^{(r)}f_{h}}(r_{\hi/\lo},\mu,\varepsilon)
  &=\mathcal{H}^{(r)}\big(f_{h}(r_{\hi/\lo}^{-},\mu,\varepsilon),f_{h}(r_{\hi/\lo}^{+},\mu,\varepsilon); r_{\hi/\lo},\mu,\varepsilon\big) \label{eq:numericalFluxFunction_R} \\
  &=\f{\alpha_{\hi/\lo}}{\psi_{\hi/\lo}^{2}}
  \Big\{\,
    \f{1}{2}\big(\mu+|\mu|\big)\,f_{h}(r_{\hi/\lo}^{-},\mu,\varepsilon)+\f{1}{2}\big(\mu-|\mu|\big)\,f_{h}(r_{\hi/\lo}^{+},\mu,\varepsilon)
  \,\Big\}, \nonumber \\
  \widehat{H^{(\mu)}f_{h}}(r,\mu_{\hi/\lo},\varepsilon)
  &=\mathcal{H}^{(\mu)}\big(f_{h}(r,\mu_{\hi/\lo}^{-},\varepsilon),f_{h}(r,\mu_{\hi/\lo}^{+},\varepsilon); r,\mu_{\hi/\lo},\varepsilon\big) \label{eq:numericalFluxFunction_Mu} \\
  &=\f{\alpha}{\psi^{2}\,r}\,(1-\mu_{\hi/\lo}^{2})\,
  \Big\{\,
    \f{1}{2}\big(\Psi+|\Psi|\big)\,f_{h}(r,\mu_{\hi/\lo}^{-},\varepsilon) \nonumber \\
    &\hspace{108pt}
    +\f{1}{2}\big(\Psi-|\Psi|\big)\,f_{h}(r,\mu_{\hi/\lo}^{+},\varepsilon)
  \,\Big\}, \nonumber \\
  \widehat{H^{(\varepsilon)}f_{h}}(r,\mu,\varepsilon_{\hi/\lo})
  &=\mathcal{H}^{(\varepsilon)}\big(f_{h}(r,\mu,\varepsilon_{\hi/\lo}^{-}),f_{h}(r,\mu,\varepsilon_{\hi/\lo}^{+}); r,\mu,\varepsilon_{\hi/\lo}\big) \label{eq:numericalFluxFunction_E} \\
  &=-\f{\varepsilon_{\hi/\lo}}{\psi^{2}}
  \Big\{\,
    \f{1}{2}\big(\pd{}{r}\alpha\,\mu-|\pd{}{r}\alpha\,\mu|\big)\,f_{h}(r,\mu,\varepsilon_{\hi/\lo}^{-}) \nonumber \\
    &\hspace{66pt}
    +\f{1}{2}\big(\pd{}{r}\alpha\,\mu+|\pd{}{r}\alpha\,\mu|\big)\,f_{h}(r,\mu,\varepsilon_{\hi/\lo}^{+})
  \,\Big\}. \nonumber
\end{align}

Key to the design of a time-explicit, bound-preserving method for Eq.~\eqref{eq:ConservativeBoltzmannEquationSphericalSymmetryGR} is to find conditions such that, after the update from $f_{h}^{n}$ to $f_{h}^{n+1}$ with time step $\dt=t^{n+1}-t^{n}$, the cell-averaged distribution function, defined as
\begin{equation}
  f_{\bK}=\f{1}{V_{\bK}}\int_{\bK}f_{h}\,dV, \quad\text{where}\quad V_{\bK} =\int_{\bK}dV,
  \label{eq:boundPreservingCellAverage}
\end{equation}
satisfies the bounds; i.e., $f_{\bK}^{n+1}\in[0,1]$.  
The standard approach is to find sufficient conditions such that these bounds hold with the first-order forward Euler method, while the extension to higher-order accuracy in time relies on the use of a strong stability-preserving (SSP) time stepping method, which can be expressed as convex combinations of forward Euler operators \citep{GoShTa01}.  
The conditions that are sought include a time step restriction.  
Then, if the bounds on the cell average at $t^{n+1}$ hold with the forward Euler method provided $\dt\le\dt_{\mathrm{FE}}$ (where $\dt_{\mathrm{FE}}$ is to be determined), the bounds will also hold when an SSP method is used, provided $\dt\le C_{\mathrm{SSP}}\times\dt_{\mathrm{FE}}$, where $0<C_{\mathrm{SSP}}\le1$.  
For the optimal second- and third-order SSP Runge--Kutta (SSP-RK) methods from \citet{ShOs88}, $C_{\mathrm{SSP}}=1$.  

The equation for the cell-average is obtained from Eq.~\eqref{eq:ConservativeBoltzmannSphericalSymmetryGRDG} with $\varphi_{h}=1$ (the lowest possible degree polynomial in the approximation space $\mathbb{V}_{h}^{k}$).  
With forward Euler time stepping, we then have
\begin{align}
  f_{\bK}^{n+1}
  &=f_{\bK}^{n}
  -\f{\Delta t}{V_{\bK}}
  \Big\{\,
    \psi^{6}(r_{\hi})\,r_{\hi}^{2}\int_{\tK^{(r)}}\widehat{H^{(r)}f_{h}^{n}}(r_{\hi},\mu,\varepsilon)\,\varepsilon^{2}\,d\tV^{(r)} \nonumber \\
    &\hspace{96pt}
    -\psi^{6}(r_{\lo})\,r_{\lo}^{2}\int_{\tK^{(r)}}\widehat{H^{(r)}f_{h}^{n}}(r_{\lo},\mu,\varepsilon)\,\varepsilon^{2}\,d\tV^{(r)} \nonumber \\
    &\hspace{72pt}
    +\int_{\tK^{(\mu)}}\widehat{H^{(\mu)}f_{h}^{n}}(r,\mu_{\hi},\varepsilon)\,\psi^{6}(r)\,r^{2}\,\varepsilon^{2}\,d\tV^{(\mu)} \nonumber \\
    &\hspace{96pt}
    -\int_{\tK^{(\mu)}}\widehat{H^{(\mu)}f_{h}^{n}}(r,\mu_{\lo},\varepsilon)\,\psi^{6}(r)\,r^{2}\,\varepsilon^{2}\,d\tV^{(\mu)} \nonumber \\
    &\hspace{72pt}
    +\varepsilon_{\hi}^{2}\int_{\tK^{(\varepsilon)}}\widehat{H^{(\varepsilon)}f_{h}^{n}}(r,\mu,\varepsilon_{\hi})\,\psi^{6}(r)\,r^{2}\,d\tV^{(\varepsilon)} \nonumber \\
    &\hspace{96pt}
    -\varepsilon_{\lo}^{2}\int_{\tK^{(\varepsilon)}}\widehat{H^{(\varepsilon)}f_{h}^{n}}(r,\mu,\varepsilon_{\lo})\,\psi^{6}(r)\,r^{2}\,d\tV^{(\varepsilon)}
  \,\Big\}.
  \label{eq:averageUpdateSphericalSymmetryGR}
\end{align}
Assuming that $f_{\bK}^{n}\in[0,1]$, the flux terms (which can be positive or negative) can bring $f_{\bK}^{n+1}$ outside the bounds.  
The contributions from these terms vanish as $\dt\to0$, and this is where restrictions on the time step comes in.  
To find these restriction, $f_{\bK}^{n}$ is split into three parts and combined with the flux terms arising from the three phase-space dimensions in the current setting.  
To this end, we define positive constants $s_{1},s_{2},s_{3}>0$, satisfying $s_{1}+s_{2}+s_{3}=1$, and write the cell-average as
\begin{align}
  f_{\bK}^{n} &= \f{s_{1}}{V_{\bK}}\int_{\tilde{K}^{(r)}}\int_{K^{(r)}}f_{h}^{n}\,\tau\,dr\,d\tV^{(r)} 
  + \f{s_{2}}{V_{\bK}}\int_{\tilde{K}^{(\mu)}}\int_{K^{(\mu)}}f_{h}^{n}\,\tau\,d\mu\,d\tV^{(\mu)} \nonumber \\
  &\hspace{12pt}
  + \f{s_{3}}{V_{\bK}}\int_{\tilde{K}^{(\varepsilon)}}\int_{K^{(\varepsilon)}}f_{h}^{n}\,\tau\,d\varepsilon\,d\tV^{(\varepsilon)}.
\end{align}
Inserting this into Eq.~\eqref{eq:averageUpdateSphericalSymmetryGR} gives
\begin{equation}
  f_{\bK}^{n+1} 
  = \f{s_{1}}{V_{\bK}}\int_{\tilde{K}^{(r)}}\Gamma^{(r)}[f_{h}^{n}]d\tV^{(r)}
  + \f{s_{2}}{V_{\bK}}\int_{\tilde{K}^{(\mu)}}\Gamma^{(\mu)}[f_{h}^{n}]d\tV^{(\mu)}
  + \f{s_{3}}{V_{\bK}}\int_{\tilde{K}^{(\varepsilon)}}\Gamma^{(\varepsilon)}[f_{h}^{n}]d\tV^{(\varepsilon)},
  \label{eq:averageUpdateInTermsOfGammaSphericalSymmetryGR}
\end{equation}
where
\begin{align}
  &\Gamma^{(r)}[f_{h}^{n}](\mu,\varepsilon) \label{eq:Gamma1} \\
  &=\int_{K^{(r)}}f_{h}^{n}\,\tau\,dr 
  - \f{\dt\,\varepsilon^{2}}{s_{1}}\Big\{\,\psi^{6}(r_{\hi})\,r_{\hi}^{2}\,\widehat{H^{(r)}f_{h}^{n}}(r_{\hi},\mu,\varepsilon)-\psi^{6}(r_{\lo})\,r_{\lo}^{2}\,\widehat{H^{(r)}f_{h}^{n}}(r_{\lo},\mu,\varepsilon)\,\Big\}, \nonumber \\
  &\Gamma^{(\mu)}[f_{h}^{n}](r,\varepsilon) \label{eq:Gamma2} \\
  &=\int_{K^{(\mu)}}f_{h}^{n}\,\tau\,d\mu
  -\f{\dt\tau}{s_{2}}\Big\{\,\widehat{H^{(\mu)}f_{h}^{n}}(r,\mu_{\hi},\varepsilon) - \widehat{H^{(\mu)}f_{h}^{n}}(r,\mu_{\lo},\varepsilon) \,\Big\}, \nonumber \\
  &\Gamma^{(\varepsilon)}[f_{h}^{n}](r,\mu) \label{eq:Gamma3} \\
  &=\int_{K^{(\varepsilon)}}f_{h}^{n}\,\tau\,d\varepsilon
  -\f{\dt\psi^{6}(r)r^{2}}{s_{3}}\Big\{\,\varepsilon_{\hi}^{2}\,\widehat{H^{(\varepsilon)}f_{h}}(r,\mu,\varepsilon_{\hi})-\varepsilon_{\lo}^{2}\,\widehat{H^{(\varepsilon)}f_{h}}(r,\mu,\varepsilon_{\lo})\,\Big\}.  \nonumber
\end{align}
With the cell-average expressed as in Eq.~\eqref{eq:averageUpdateInTermsOfGammaSphericalSymmetryGR}, in order to ensure $f_{\bK}^{n+1}\ge0$, it is sufficient to find conditions for which each of the right-hand sides in Eqs.~\eqref{eq:Gamma1}, \eqref{eq:Gamma2}, \eqref{eq:Gamma3} are nonnegative.  
We illustrate the details of this for Eq.~\eqref{eq:Gamma1} \citep[see][for full details]{EnHaXi15}.  
In the DG method, the integrals over the faces $\tK^{(r)}$, $\tK^{(\mu)}$, and $\tK^{(\varepsilon)}$ in Eq.~\eqref{eq:averageUpdateInTermsOfGammaSphericalSymmetryGR} are typically evaluated with a quadrature rule.  
In this case, it is sufficient that $\Gamma^{(r)},\Gamma^{(\mu)},\Gamma^{(\varepsilon)}\ge0$ holds in the respective quadrature points.  
As an example, we let $\tilde{\mathbf{S}}^{(r)}(\in\tK^{(r)})$ denote the set of quadrature points used to integrate over $\tK^{(r)}$ in Eq.~\eqref{eq:averageUpdateInTermsOfGammaSphericalSymmetryGR}.  

To evaluate the integral on the right-hand side of Eq.~\eqref{eq:Gamma1}, an $N^{(r)}$-point Gauss-Lobatto quadrature rule is used on the interval $K^{(r)}$, with points
\begin{align}
  \hat{S}^{(r)} = \big\{\,r_{\lo}=\hat{r}_{1},\ldots,\hat{r}_{N^{(r)}}=r_{\hi}\,\big\},
\end{align}
and weights $\hat{w}_{q}\in(0,1]$, normalized such that $\sum_{q=1}^{N^{(r)}}\hat{w}_{q}=1$.  
This quadrature integrates polynomials in $r$ of degree $2N^{(r)}-3$ exactly.  
We can then write
\begin{equation}
  \int_{K^{(r)}}f_{h}^{n}\,\tau\,dr = \dr\sum_{q=1}^{N^{(r)}}\hat{w}_{q}\,f_{h}^{n}(\hat{r}_{q},\mu,\varepsilon)\,\tau(\hat{r}_{q},\mu,\varepsilon).
  \label{eq:radialLobattoQuadrature}
\end{equation}
If the distribution function is approximated with a polynomial of degree $k$ in $r$, and $\psi^{6}$ is approximated by a polynomial of degree $k_{\psi}$, the quadrature is exact if $N^{(r)}\ge(k+k_{\psi}+5)/2$.  
The reason for using the Gauss-Lobatto quadrature for the integral over $K^{(r)}$ is because it includes the end points of the interval ($r_{\lo},r_{\hi}$).  
These end points are used to balance the flux terms in the radial dimension.  
Inserting Eq.~\eqref{eq:radialLobattoQuadrature} into Eq.~\eqref{eq:Gamma1} gives
\begin{align}
  \f{1}{\dr}\Gamma^{(r)}[f_{h}^{n}]
  &=\sum_{q=1}^{N^{(r)}}\hat{w}_{q}\,f_{h}^{n}(\hat{r}_{q})\,\tau(\hat{r}_{q}) \nonumber \\
  &\hspace{12pt}
  - \f{\dt\,\varepsilon^{2}}{s_{1}}\Big\{\,\psi^{6}(r_{\hi})\,r_{\hi}^{2}\,\mathcal{H}^{(r)}\big(f_{h}(r_{\hi}^{-}),f_{h}(r_{\hi}^{+}); r_{\hi}\big) \nonumber \\
  &\hspace{56pt}
  -\psi^{6}(r_{\lo})\,r_{\lo}^{2}\,\mathcal{H}^{(r)}\big(f_{h}(r_{\lo}^{-}),f_{h}(r_{\lo}^{+}); r_{\lo}\big)\,\Big\} \nonumber \\
  &=\sum_{q=2}^{N^{(r)}-1}\hat{w}_{q}\,f_{h}^{n}(\hat{r}_{q})\,\tau(\hat{r}_{q}) + \hat{w}_{1}\,\Phi_{1}^{(r)}\big[f_{h}^{n}(r_{\lo}^{-}),f_{h}^{n}(r_{\lo}^{+})\big]\,\tau(r_{\lo}) \nonumber \\
  &\hspace{12pt}
  +\hat{w}_{N^{(r)}}\,\Phi_{N^{(r)}}^{(r)}\big[f_{h}^{n}(r_{\hi}^{-}),f_{h}^{n}(r_{\hi}^{+})\big]\,\tau(r_{\hi}),
\end{align}
where
\begin{align}
  \Phi_{1}^{(r)}\big[f_{h}^{n}(r_{\lo}^{-}),f_{h}^{n}(r_{\lo}^{+})\big]
  &=f_{h}^{n}(r_{\lo}^{+})+\f{\dt}{s_{1}\hat{w}_{1}\dr}\,\mathcal{H}^{(r)}\big(f_{h}^{n}(r_{\lo}^{-}),f_{h}^{n}(r_{\lo}^{+}); r_{\lo}\big), \\
  \Phi_{N^{(r)}}^{(r)}\big[f_{h}^{n}(r_{\hi}^{-}),f_{h}^{n}(r_{\hi}^{+})\big]
  &=f_{h}^{n}(r_{\hi}^{-})-\f{\dt}{s_{1}\hat{w}_{N^{(r)}}\dr}\,\mathcal{H}^{(r)}\big(f_{h}^{n}(r_{\hi}^{-}),f_{h}^{n}(r_{\hi}^{+}); r_{\hi}\big).
\end{align}
(For notational brevity, we have suppressed the $(\mu,\varepsilon)$-dependence.)
Using the numerical flux function in Eq.~\eqref{eq:numericalFluxFunction_R}, one can write
\begin{align}
  &\Phi_{1}^{(r)}\big[f_{h}^{n}(r_{\lo}^{-}),f_{h}^{n}(r_{\lo}^{+})\big] \label{eq:PhiOne_R} \\
  &=
  f_{h}^{n}(r_{\lo}^{+})
  +\f{\dt}{s_{1}\hat{w}_{1}\dr}\,\f{\alpha(r_{\lo})}{\psi^{2}(r_{\lo})}
  \Big\{\,
    \f{1}{2}\big(\mu+|\mu|\big)\,f_{h}^{n}(r_{\lo}^{-})+\f{1}{2}\big(\mu-|\mu|\big)\,f_{h}^{n}(r_{\lo}^{+})
  \,\Big\} \nonumber \\
  &=\f{\dt}{s_{1}\hat{w}_{1}\dr}\,\f{\alpha(r_{\lo})}{\psi^{2}(r_{\lo})}\,\f{1}{2}\big(\mu+|\mu|\big)\,f_{h}^{n}(r_{\lo}^{-})
  +\Big\{\,1+\f{\dt}{s_{1}\hat{w}_{1}\dr}\,\f{\alpha(r_{\lo})}{\psi^{2}(r_{\lo})}\,\f{1}{2}\big(\mu-|\mu|\big)\,\Big\}\,f_{h}^{n}(r_{\lo}^{+}). \nonumber
\end{align}
On the right-hand side of Eq.~\eqref{eq:PhiOne_R} (last line), the coefficient in front of $f_{h}^{n}(r_{\lo}^{-})$ is nonnegative since $\alpha(r_{\lo}),\psi^{2}(r_{\lo})>0$ and $\big(\mu+|\mu|\big)\ge0$.  
Only the coefficient in front of $f_{h}^{n}(r_{\lo}^{+})$ can become negative since $\big(\mu-|\mu|\big)\le0$.  
Assuming $f_{h}^{n}(r_{\lo}^{-}),f_{h}^{n}(r_{\lo}^{+})\ge0$, it is easy to show that $\Phi_{1}^{(r)}\big[f_{h}^{n}(r_{\lo}^{-}),f_{h}^{n}(r_{\lo}^{+})\big]\ge0$, if
\begin{equation}
  \dt\le\f{s_{1}\hat{w}_{1}\dr}{|\mu|}\,\f{\psi^{2}(r_{\lo})}{\alpha(r_{\lo})}.  
\end{equation}
Similarly, for $f_{h}^{n}(r_{\hi}^{-}),f_{h}^{n}(r_{\hi}^{+})\ge0$, one finds that $\Phi_{N^{(r)}}^{(r)}\big[f_{h}^{n}(r_{\hi}^{-}),f_{h}^{n}(r_{\hi}^{+})\big]\ge0$, if
\begin{equation}
  \dt\le\f{s_{1}\hat{w}_{N^{(r)}}\dr}{|\mu|}\,\f{\psi^{2}(r_{\hi})}{\alpha(r_{\hi})}.  
\end{equation}
Therefore, assuming $f_{h}^{n}\ge0$ in the combined quadrature set $\mathbf{S}^{(r)}=\hat{S}^{(r)}\otimes\tilde{\mathbf{S}}^{(r)}$, where the points in $\hat{S}^{(r)}$ are used to evaluate the integral over $K^{(r)}$ in Eq.~\eqref{eq:Gamma1} and the points in $\tilde{\mathbf{S}}^{(r)}$ are used to evaluate the integral over $\tK^{(r)}$ in Eq.~\eqref{eq:averageUpdateInTermsOfGammaSphericalSymmetryGR}, a sufficient condition on the time step to guarantee $\int_{\tK^{(r)}}\Gamma^{(r)}[f_{h}^{n}]d\tV^{(r)}\ge0$ is given by
\begin{align}
  \dt
  &\le\min\big(\psi^{2}(r_{\lo})/\alpha(r_{\lo}),\psi^{2}(r_{\hi})/\alpha(r_{\hi})\big)\,\hat{w}_{N^{(r)}}\,s_{1}\,\Delta r. 
  \label{eq:CFLSphericalSymmetryRadius}
\end{align}
(Here, $\hat{w}_{1}=\hat{w}_{N^{(r)}}$ is used.)  
Sufficient conditions on $\dt$ for $\int_{\tK^{(\mu)}}\Gamma^{(\mu)}[f_{h}^{n}]d\tV^{(\mu)}\ge0$ and $\int_{\tK^{(\varepsilon)}}\Gamma^{(\varepsilon)}[f_{h}^{n}]d\tV^{(\varepsilon)}\ge0$ can be derived in a similar way (we refer the interested reader to \citet{EnHaXi15} for details).  
Together, these restrictions on the time step ensures $f_{\bK}^{n+1}\ge0$.  
It should be noted that the time step restrictions derived here are sufficient, not necessary, conditions.  
They are typically more restrictive than the time step required for numerical stability.  
Thus, in practical calculations, larger time steps may be taken.  
If violations of the physical bounds are detected after a time step, $\dt$ can be reduced to the sufficient conditions before the time step is redone.  

The proof for $f_{\bK}^{n+1}\le1$ relies on the divergence-free condition in Eq.~\eqref{eq:DivergenceFreeCondition}, which can be written as
\begin{align}
  \f{1}{V_{\bK}}
  \Big\{\,
    &\psi^{6}(r_{\hi})\,r_{\hi}^{2}\int_{\tK^{(r)}}H^{(r)}(r_{\hi},\mu,\varepsilon)\,\varepsilon^{2}\,d\tV^{(r)} \nonumber \\
    &\hspace{36pt}
    -\psi^{6}(r_{\lo})\,r_{\lo}^{2}\int_{\tK^{(r)}}H^{(r)}(r_{\lo},\mu,\varepsilon)\,\varepsilon^{2}\,d\tV^{(r)} \nonumber \\
    &+\int_{\tK^{(\mu)}}H^{(\mu)}(r,\mu_{\hi},\varepsilon)\,\psi^{6}(r)\,r^{2}\,\varepsilon^{2}\,d\tV^{(\mu)} \nonumber \\
    &\hspace{36pt}
    -\int_{\tK^{(\mu)}}H^{(\mu)}(r,\mu_{\lo},\varepsilon)\,\psi^{6}(r)\,r^{2}\,\varepsilon^{2}\,d\tV^{(\mu)} \nonumber \\
    &+\varepsilon_{\hi}^{2}\int_{\tK^{(\varepsilon)}}H^{(\varepsilon)}(r,\mu,\varepsilon_{\hi})\,\psi^{6}(r)\,r^{2}\,d\tV^{(\varepsilon)} \nonumber \\
    &\hspace{36pt}
    -\varepsilon_{\lo}^{2}\int_{\tK^{(\varepsilon)}}H^{(\varepsilon)}(r,\mu,\varepsilon_{\lo})\,\psi^{6}(r)\,r^{2}\,d\tV^{(\varepsilon)}
  \,\Big\} = 0.
  \label{eq:divergenceFreeSphericalSymmetryGR}
\end{align}
In Eq.~\eqref{eq:ConservativeBoltzmannSphericalSymmetryGRDG}, we approximate the derivatives $\p_{r}\alpha$ and $\p_{r}\psi^{4}$ in $K^{(r)}$ (appearing in $H^{(\mu)}$ and $H^{(\varepsilon)}$; cf.\ Eq.~\eqref{eq:phaseSpaceFluxCoefficients}) with polynomials and compute $\alpha$ and $\psi^{4}$ from
\begin{equation}
  \alpha(r)=\alpha(r_{\lo})+\int_{r_{\lo}}^{r}\pd{}{r}\alpha(r')\,dr'\quad\mbox{and}\quad
  \psi^{4}(r)=\psi^{4}(r_{\lo})+\int_{r_{\lo}}^{r}\pd{}{r}\psi^{4}(r')\,dr',
\end{equation}
where the Gaussian quadrature rule is used to evaluate the integrals exactly.  
Two-dimensional Gaussian quadrature rules are also used to evaluate the integrals over $\tK^{(r)}$, $\tK^{(\mu)}$, and $\tK^{(\varepsilon)}$, using $L^{(r)}$, $L^{(\mu)}$, and $L^{(\varepsilon)}$ points in the $r$, $\mu$, and $\varepsilon$ dimensions, respectively.  
With this choice, it is straightforward to show that the discretization satisfies the divergence-free condition \eqref{eq:divergenceFreeSphericalSymmetryGR}, provided $L^{(\mu)}\ge1$, $L^{(\varepsilon)}\ge2$, while $L^{(r)}$ depends on the degree of the polynomials approximating $\pd{}{r}\alpha$ and $\pd{}{r}\psi^{4}$.  

Using the definitions in Eqs.~\eqref{eq:Gamma1}--\eqref{eq:Gamma3}, a direct calculation shows that
\begin{align}
  &\f{s_{1}}{V_{\bK}}\int_{\tilde{K}^{(r)}}\Gamma^{(r)}[1]d\tV^{(r)}
  + \f{s_{2}}{V_{\bK}}\int_{\tilde{K}^{(\mu)}}\Gamma^{(\mu)}[1]d\tV^{(\mu)}
  + \f{s_{3}}{V_{\bK}}\int_{\tilde{K}^{(\varepsilon)}}\Gamma^{(\varepsilon)}[1]d\tV^{(\varepsilon)} \nonumber \\
  &=\f{s_{1}}{V_{\bK}}\int_{\tilde{K}^{(r)}}\int_{K^{(r)}}\tau\,dr\,d\tV^{(r)} 
  + \f{s_{2}}{V_{\bK}}\int_{\tilde{K}^{(\mu)}}\int_{K^{(\mu)}}\tau\,d\mu\,d\tV^{(\mu)}
  + \f{s_{3}}{V_{\bK}}\int_{\tilde{K}^{(\varepsilon)}}\int_{K^{(\varepsilon)}}\tau\,d\varepsilon\,d\tV^{(\varepsilon)} \nonumber \\
  &\hspace{12pt}
  -\f{\dt}{V_{\bK}}
  \Big\{\,
    \psi^{6}(r_{\hi})\,r_{\hi}^{2}\int_{\tilde{K}^{(r)}}H^{(r)}(r_{\hi},\mu,\varepsilon)\,\varepsilon^{2}\,d\tV^{(r)}
    -\psi^{6}(r_{\lo})\,r_{\lo}^{2}\int_{\tilde{K}^{(r)}}H^{(r)}(r_{\lo},\mu,\varepsilon)\,\varepsilon^{2}\,d\tV^{(r)} \nonumber \\
  &\hspace{36pt}
  +\int_{\tilde{K}^{(\mu)}}H^{(\mu)}(r,\mu_{\hi},\varepsilon)\,\psi^{6}(r)\,r^{2}\,\varepsilon^{2}\,d\tV^{(\mu)} - \int_{\tilde{K}^{(\mu)}}H^{(\mu)}(r,\mu_{\lo},\varepsilon)\,\psi^{6}(r)\,r^{2}\,\varepsilon^{2}\,d\tV^{(\mu)} \nonumber \\
  &\hspace{36pt}
  +\varepsilon_{\hi}^{2}\int_{\tilde{K}^{(\varepsilon)}}H^{(\varepsilon)}(r,\mu,\varepsilon_{\hi})\,\psi^{6}(r)\,r^{2}\,d\tV^{(\varepsilon)}-\varepsilon_{\lo}^{2}\int_{\tilde{K}^{(\varepsilon)}}H^{(\varepsilon)}(r,\mu,\varepsilon_{\lo})\,\psi^{6}(r)\,r^{2}\,d\tV^{(\varepsilon)}\,\Big\} \nonumber \\
  &=s_{1}+s_{2}+s_{3}=1,
  \label{eq:gammasWithOneSphericalSymmetryGR}
\end{align}
where the divergence-free condition in Eq.~\eqref{eq:divergenceFreeSphericalSymmetryGR} is used.  
Since the divergence-free condition holds, it is then straightforward to show that the cell-average of $g_{h}=1-f_{h}$ satisfies (cf.\ Eq.~\eqref{eq:averageUpdateInTermsOfGammaSphericalSymmetryGR})
\begin{align}
  g_{\bK}^{n+1}
  &=1 - f_{\bK}^{n+1} \\
  &= \f{s_{1}}{V_{\bK}}\int_{\tilde{K}^{(r)}}\big(\Gamma^{(r)}[1]-\Gamma^{(r)}[f_{h}^{n}]\big)d\tV^{(r)} 
  + \f{s_{2}}{V_{\bK}}\int_{\tilde{K}^{(\mu)}}\big(\Gamma^{(\mu)}[1]-\Gamma^{(\mu)}[f_{h}^{n}]\big)d\tV^{(\mu)} \nonumber \\
  &\hspace{24pt}
  +\f{s_{3}}{V_{\bK}}\int_{\tilde{K}^{(\varepsilon)}}\big(\Gamma^{(\varepsilon)}[1]-\Gamma^{(\varepsilon)}[f_{h}^{n}]\big)d\tV^{(\varepsilon)} \nonumber \\
  &= \f{s_{1}}{V_{\bK}}\int_{\tilde{K}^{(r)}}\Gamma^{(r)}[g_{h}^{n}]d\tV^{(r)}
  + \f{s_{2}}{V_{\bK}}\int_{\tilde{K}^{(\mu)}}\Gamma^{(\mu)}[g_{h}^{n}]d\tV^{(\mu)}
  + \f{s_{3}}{V_{\bK}}\int_{\tilde{K}^{(\varepsilon)}}\Gamma^{(\varepsilon)}[g_{h}^{n}]d\tV^{(\varepsilon)}, \nonumber
\end{align}
where the linearity property of the operators in Eq.~\eqref{eq:Gamma1}--\eqref{eq:Gamma3} is used; e.g., $\Gamma^{(r)}[1]-\Gamma^{(r)}[f_{h}^{n}]=\Gamma^{(r)}[1-f_{h}^{n}]=\Gamma^{(r)}[g_{h}^{n}]$.  
Thus, provided Eq.~\eqref{eq:divergenceFreeSphericalSymmetryGR} and the restrictions on $\dt$ hold, and the conditions on $f_{h}^{n}$ also hold for $g_{h}^{n}$, it follows that $g_{\bK}^{n+1}\ge0$ (or $f_{\bK}^{n+1}\le1$).  

The numerical method developed by \citet{EnHaXi15}, and outlined above, is designed to preserve the physical bounds of the cell averaged distribution function (i.e., $0\le f_{\bK}\le1$), provided sufficiently accurate quadratures are used, specific time step restrictions are satisfied, \emph{and} that the polynomial approximating the distribution function inside each phase space element $\bK$ at time $t^{n}$ is bounded in a set of quadrature points, which we denote $S$.  
After one time step, it is possible that $f_{h}^{n+1}$ violates the bounds for some points in the set $S$.  
In the DG method, the limiter proposed by \citet{ZhSh10a} is used to reenforce the bounds.  
That is, the polynomial obtained after a time step $\dt$, $f_{h}^{n+1}(\mathbf{z})$, is replaced by with the ``limited'' polynomial
\begin{equation}
  \tilde{f}_{h}^{n+1}(\mathbf{z})=\vartheta\,f_{h}^{n+1}(\mathbf{z})+(\,1-\vartheta\,)\,f_{\bK}^{n+1}, 
  \label{eq:limitedPolynomial}
\end{equation}
where the limiter parameter $\vartheta\in[0,1]$ is given by
\begin{equation}
  \vartheta=\min\Big\{\Big|\f{M-f_{\bK}^{n+1}}{M_{S}-f_{\bK}^{n+1}}\Big|,\Big|\f{m-f_{\bK}^{n+1}}{m_{S}-f_{\bK}^{n+1}}\Big|,1\Big\}, 
  \label{eq:limiter}
\end{equation}
with $m=0$ and $M=1$, and
\begin{equation}
  M_{S}=\max_{\mathbf{z} \in S}f_{h}^{n+1}(\mathbf{z}), \qquad m_{S}=\min_{\mathbf{z} \in S}f_{h}^{n+1}(\mathbf{z}), 
\end{equation}
and $S$ represents the finite set of quadrature points in $\bK$ where the bounds must hold.  
For $\vartheta=0$, the entire solution is limited to the cell-average, while for $\vartheta=1$ $\tilde{f}_{h}^{n+1}=f_{h}^{n+1}$.  
It is thus absolutely necessary to maintain the bounds on the cell-average, otherwise the limiting procedure will be futile.  
In practice, $\vartheta$ remains close to unity, and the limiting is a small correction.  
It has been shown \citep{ZhSh10a} that this ``linear scaling limiter'' maintains high order of accuracy.  
Also, note that the limiting procedure is conservative for particle number since it preserves the cell averaged distribution function; i.e., by inserting Eq.~\eqref{eq:limitedPolynomial} into the definition of the cell average in Eq.~\eqref{eq:boundPreservingCellAverage}:
\begin{equation}
  \f{1}{V_{\bK}}\int_{\bK}\tilde{f}_{h}^{n+1}\,dV
  =\f{1}{V_{\bK}}\int_{\bK}\big(\,\vartheta\,f_{h}^{n+1}+(\,1-\vartheta\,)\,f_{\bK}^{n+1}\,\big)\,dV
  =f_{\bK}^{n+1}.  
\end{equation}

In the discussion above, forward Euler time stepping is used, which is only first-order accurate.  
For explicit time integration, the bound-preserving scheme can easily be extended to higher-order accuracy in time by using high-order SSP time stepping methods \citep{ShOs88,GoShTa01}, which are multi-stage methods that can be formulated as convex combinations of forward Euler operators.  
Provided limiting is applied at each stage, the bound-preserving property follows from convexity arguments.  
For neutrino transport problems where neutrino--matter interactions are treated with implicit methods, it is difficult to achieve both high-order accuracy and bounded solutions, and this topic remains open for further research.  
We will discuss this issue further below in the context of a two-moment model.  
Another open issue is the challenge of simultaneous number and energy conservation in the phase space advection problem discussed here: The limiter in Eq.~\eqref{eq:limitedPolynomial} preserves the particle number, but not higher moments of the distribution function.  
In the present model, the space time is stationary, which implies that the so-called Komar mass ($\alpha\,\varepsilon\,f$) is conserved.
Thus, if bounded solutions and exact conservation of the Komar mass is desired, modifications to the limiter is needed.  

\subsubsection{Realizability-preserving moment methods}
\label{sec:realizabilityPreserving}

\citet{ChEnHa19} developed a numerical method for a two-moment model based on DG spatial discretization and IMEX time stepping.  
The method is specifically designed to preserve bounds on the moments as dictated by Pauli's exclusion principle.  
As such, it is an extension of the bound-preserving method discussed above, but for a nonlinear system of hyperbolic balance laws with stiff sources.  
As is reasonable for an initial investigation, the model adopted by \citet{ChEnHa19} is rather simple, when compared to the two-moment models used to model neutrino transport in contemporary core-collapse supernova simulations.  
However, the work highlighted the role of the moment closure in the design of robust two-moment methods for neutrino transport, and developed an IMEX scheme with a reasonable time step restriction that is compatible with bounded solutions.  
As such, the work put down the foundations for a framework that may help future developments of robust methods for models with improved physical fidelity.  
To simplify the discussion, we consider the model in \citet{ChEnHa19} for one spatial dimension and define moments of the distribution function as
\begin{equation}
  \big\{\,\mathcal{J},\mathcal{H},\mathcal{K}\,\big\}(x,t)=\f{1}{2}\int_{-1}^{1}f(\mu,x,t)\,\mu^{\{0,1,2\}}\,d\mu.
\end{equation}
The two-moment model can be written as a system of hyperbolic balance laws as
\begin{equation}
  \pd{\mathbf{u}}{t} + \pd{\mathbf{f}(\mathbf{u})}{x} = \mathbf{\eta} - R\,\mathbf{u} \equiv \mathbf{c}(\mathbf{u}),
  \label{eq:twoMomentModelRealizability}
\end{equation}
where the evolved moment vector is $\mathbf{u}=(\mathcal{J},\mathcal{H})^{T}$, the flux vector is $\mathbf{f}=(\mathcal{H},\mathcal{K})^{T}$, the emissivity is $\mathbf{\eta}=(\sigma_{A}\,\mathcal{J}_{0},0)^{T}$, and $R=\mbox{diag}(\sigma_{A},\sigma_{T})$.  
Here, $\mathcal{J}_{0}$ is the zeroth moment of an equilibrium distribution function, $f_{0}$, satisfying $f_{0}\in[0,1]$ (i.e., Fermi--Dirac statistics), $\sigma_{A}\ge0$ is the absorption opacity, and $\sigma_{T}=\sigma_{A}+\sigma_{S}$, where $\sigma_{S}\ge0$ is the scattering opacity (assuming isotropic and isoenergetic scattering).  
In Eq.~\eqref{eq:twoMomentModelRealizability}, a closure is assumed so that $\mathcal{K}=\mathcal{K}(\mathbf{u})$.  

\begin{figure}
  \includegraphics[width=\textwidth]{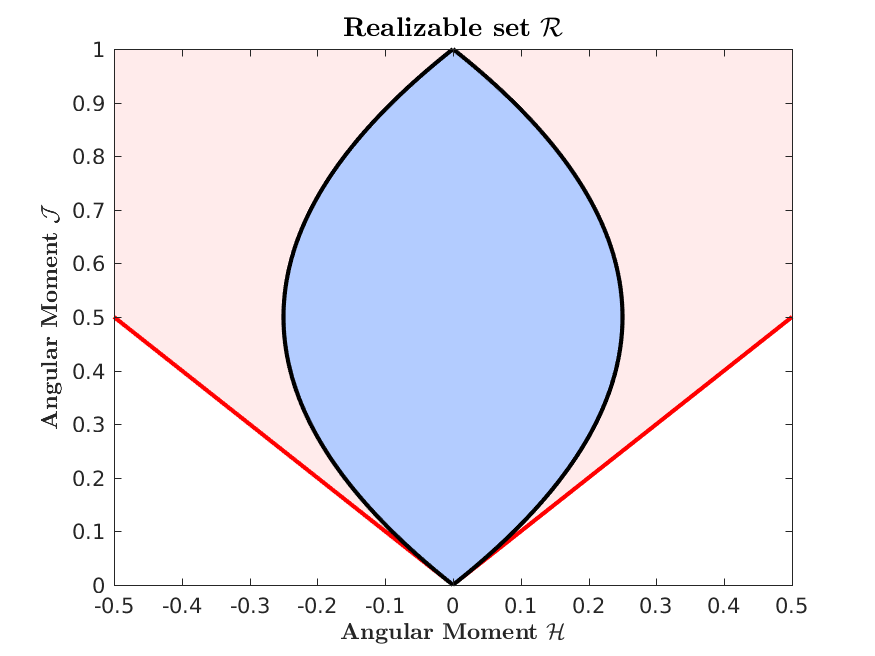}
  \caption{Illustration of the realizable set $\mathcal{R}$ (light blue region) defined in Eq.~\eqref{eq:realizableSet}.  
  The black lines define the boundary $\partial\mathcal{R}$.  
  The red lines indicate the boundary of the realizable set of positive distributions $\mathcal{R}^{+}$ (light red region).}
  \label{fig:ChEnHa19_Fig1}
\end{figure}

For fermions, the Pauli exclusion principle requires the distribution function to satisfy the condition $0 \le f \le 1$.  
This puts corresponding restrictions on realizable values for the moments of $f$.  
It is then interesting to study the design of a numerical method for solving the system of moment equations given by Eq.~\eqref{eq:twoMomentModelRealizability} that preserves realizability of the moments; i.e., the moments evolve within the set of admissible values as dictated by Pauli's exclusion principle.  
If we let
\begin{equation}
  \mathfrak{R} := \left\{\,f~|~0\le f \le 1 ~\text{and}~0<\f{1}{2}\int_{-1}^{1}f\,d\mu<1\,\right\},
\end{equation}
the moments $\mathbf{u}=(\mathcal{J},\mathcal{H})^{T}$ are realizable if they can be obtained from a distribution function $f(\mu)\in\mathfrak{R}$.
The set of all realizable moments $\mathcal{R}$ is \cite[e.g.,][]{LaBa11}
\begin{equation}
  \mathcal{R}:=\big\{\,\mathbf{u}=\big(\mathcal{J},\mathcal{H}\big)^{T}~|~\mathcal{J}\in(0,1)~\text{and}~(1-\mathcal{J})\,\mathcal{J}-|\mathcal{H}| > 0\,\big\}.
  \label{eq:realizableSet}
\end{equation}
The geometry of the set $\mathcal{R}$ in the $(\mathcal{H},\mathcal{J})$-plane is illustrated in Figure~\ref{fig:ChEnHa19_Fig1} (light blue region).  
For comparison, the realizable domain $\mathcal{R}^{+}$ of positive distribution functions (no upper bound on $f$), which is a cone defined by $\mathcal{J}>0$ and $\mathcal{J}-|\mathcal{H}|>0$ (light red region), is also shown.  
The realizable set $\mathcal{R}$ is a bounded subset of $\mathcal{R}^{+}$.  
Importantly, the set $\mathcal{R}$ is convex.  
This means that for two arbitrary elements $\mathbf{u}_{a},\mathbf{u}_{b}\in\mathcal{R}$, the convex combination $\mathbf{u}_{c} = \vartheta\,\mathbf{u}_{a} + (1-\vartheta)\,\mathbf{u}_{b}\in\mathcal{R}$, where $0\leq\vartheta\leq1$.  
This property is used repeatedly (sometimes in a nested fashion) to design the numerical method.  

The DG method for the two-moment model is in many ways very similar to that discussed in Sect.~\ref{sec:boundPreserving}.  
The computational domain $D$ is divided into elements $K=(x_{\lo},x_{\hi})$.  
One each element, the approximation space is
\begin{equation}\label{mdg:vhk}
  \mathbb{V}_{h}^{k}=\{\varphi_{h} : \varphi_{h}\big|_{K} \in \mathbb{P}^{k}(K), \, \, \forall\ K\in D \},
\end{equation}
where $\mathbb{P}^{k}$ is the space of polynomials in $x$ of maximal degree $k$.  
The approximation to the moments, $\mathbf{u}_{h}$, is then expressed as
\begin{equation}
  \mathbf{u}_{h}(x,t)=\sum_{i=1}^{k+1}\mathbf{u}_{i}(t)\,P_{i}(x),
\end{equation}
where each $P_{i}\in\mathbb{V}_{h}^{k}$ and each $\mathbf{u}_{i}$ is a two-component vector representing the unknowns per element in the DG method.  
Then, for any $x \in D$ and any $\varphi_{h} \in \mathbb{V}_{h}^{k}$, the semi-discrete DG method is as follows:
\textit{Find $\mathbf{u}_{h} \in \mathbb{V}_{h}^{k}$ such that}
\begin{align}
  \int_{K}\pd{\mathbf{u}_{h}}{t}\,\varphi_{h}\,dx
  &+\big[\,\widehat{\mathbf{f}(\mathbf{u}_{h})}(x_{\hi})\,\varphi_{h}(x_{\hi}^{-})-\widehat{\mathbf{f}(\mathbf{u}_{h})}(x_{\lo})\,\varphi_{h}(x_{\lo}^{+})\,\big] \nonumber \\
  &-\int_{K}\mathbf{f}(\mathbf{u}_{h})\,\pd{\varphi_{h}}{x}\,dx
  =\int_{K}\mathbf{c}(\mathbf{u}_{h})\,\varphi_{h}\,dx
  \label{eq:twoMomentDG}
\end{align}
holds for all $\varphi_{h}\in\mathbb{V}_{h}^{k}$ and all $K\in D$.  
In Eq.~\eqref{eq:twoMomentDG},
\begin{equation}
  \widehat{\mathbf{f}(\mathbf{u}_{h})}(x_{\hi/\lo})
  =\mathbf{h}\big(\mathbf{u}_{h}(x_{\hi/\lo}^{-}),\mathbf{u}_{h}(x_{\hi/\lo}^{+})\big)
  \label{eq:twoMomentNumericalFluxFunction}
\end{equation}
is a numerical flux, where $\mathbf{h}$ is a numerical flux function.  
In the DG method, any standard numerical flux designed for hyperbolic conservation laws can be used.  
However, \citet{ChEnHa19} used the global Lax-Friedrichs flux, where
\begin{equation}
  \mathbf{h}\big(\mathbf{u}_{h}(x_{\hi/\lo}^{-}),\mathbf{u}_{h}(x_{\hi/\lo}^{+})\big)
  =\f{1}{2}
  \Big[
    \mathbf{f}\big(\mathbf{u}_{h}(x_{\hi/\lo}^{-})\big)+\mathbf{f}\big(\mathbf{u}_{h}(x_{\hi/\lo}^{+})\big)
    -\big(\mathbf{u}_{h}(x_{\hi/\lo}^{+})-\mathbf{u}_{h}(x_{\hi/\lo}^{-})\big)
  \Big].  
  \label{eq:realizableTwoMomentNumericalFlux}
\end{equation}
It should be noted that when using the DG method for radiation transport, as long as the approximation space includes at least linear elements, it is not necessary to switch between centered and upwind-type fluxes (e.g., as is done in Eqs.~\eqref{eq:modifiedNumericalFluxEnergy}--\eqref{eq:modifiedNumericalFluxMomentum} for finite-volume and finite-difference methods to capture both the streaming and diffusive regimes).  
As such, the DG spatial discretization is naturally structure-preserving with respect to the diffusion limit, and well-suited for radiation transport \citep[e.g.,][]{LaMo89,Adams01}.  
In fact, the dissipation term in the numerical flux in Eq.~\eqref{eq:realizableTwoMomentNumericalFlux}, which is not present in the diffusive regime when employing switching between centered and upwind fluxes, plays an important role in the proof of the realizability-preserving property of the two-moment method presented here.  
It may therefore be difficult, if not impossible, to design realizability-preserving methods for the two-moment model without this term.  
Note that in the diffusion limit, $|\mathcal{H}|\ll\mathcal{J}$, the moment vector $\mathbf{u}$ is close to the line connecting $(0,0)$ and $(0,1)$ in Figure~\ref{fig:ChEnHa19_Fig1}.  
Then, if the particle density is low ($\mathcal{J}\ll1$) the moment vector is safely inside $\mathcal{R}$.  
On the other hand, if the particle density is high ($\mathcal{J}\lesssim1$), which, e.g., is the case for electron neutrinos in the supernova core, the moment vector is dangerously close to the boundary of $\mathcal{R}$, and care is needed in order to maintain $\mathbf{u}\in\mathcal{R}$.  
Further away from the supernova core, where neutrinos transition to streaming conditions, $|\mathcal{H}|\lesssim(1-\mathcal{J})\,\mathcal{J}$ ($\approx\mathcal{J}$ when $\mathcal{J}\ll1$), the moment vector is again close to the boundary of $\mathcal{R}$, and care in the numerics is again warranted.  
Maintaining $\mathbf{u}\in\mathcal{R}$ is necessary to ensure the well-posedness of the moment closure procedure \citep{Le96,Ju98,HaLeTi08}.  
Realizability-preserving methods maintain $\mathbf{u}\in\mathcal{R}$ and thus improve robustness.  

The semi-discretization of the two-moment model in Eq.~\eqref{eq:twoMomentDG} results in a system of ODEs of the form
\begin{equation}
  \deriv{\mathbf{U}}{t} = \mathbf{T}(\mathbf{U}) + \mathbf{C}(\mathbf{U}),
  \label{eq:realizableTwoMomentODE}
\end{equation}
where $\mathbf{U}$ represents all the degrees of freedom evolved with the DG method, 
\begin{equation}
  \mathbf{U} =\Big\{\, \int_{K}\pd{\mathbf{u}_{h}}{t}\,\varphi_{h}\,dx \,\Big\}_{K\in D, \varphi_{h}\in\mathbb{V}_{h}^{k}},
\end{equation}
which includes the cell-average of $\mathbf{u}_{h}$ in each element:
\begin{equation}
  \mathbf{u}_{K} = \f{1}{\dx}\int_{K}\mathbf{u}_{h}\,dx.  
  \label{eq:twoMomentBasicCellAverage}
\end{equation}
In Eq.~\eqref{eq:realizableTwoMomentODE}, the transport operator $\mathbf{T}(\mathbf{U})$ corresponds to the second (surface) and third (volume) terms on the left-hand side of Eq.~\eqref{eq:twoMomentDG}, while the collision operator $\mathbf{C}(\mathbf{U})$ corresponds to the right-hand side of Eq.~\eqref{eq:twoMomentDG}.  
To evolve Eq.~\eqref{eq:realizableTwoMomentODE} forward in time, \citet{ChEnHa19} developed IMEX schemes, where the transport operator is treated explicitly and the collision operator is treated implicitly.  
As discussed in Sect.~\ref{sec:boundPreserving}, the extension of the bound-preserving property to high-order methods relies on the strong-stability-preserving (SSP) property of the ODE solver.  
Explicit SSP Runge--Kutta (RK) methods of moderate order ($\le3$) are relatively easy to construct.  
Unfortunately, high-order (second- or higher-order temporal accuracy) SSP-IMEX methods \emph{with time step restrictions solely due to the explicit transport operator} do not exist (see for example Proposition~6.2 in \citet{GoShTa01}, which rules out the existence of implicit SSP-RK methods of order higher than one).  
Because of this, \citet{ChEnHa19} resorted to develop formally first-order accurate IMEX schemes with the following properties: (i) second-order accurate in the streaming limit, (ii) SSP (called convex-invariant in \citet{ChEnHa19}), with a time step restriction solely due to the explicit part, and (iii) well-behaved in the diffusion limit in the sense that the flux density remains proportional to the gradient of the number density with the correct constant of proportionality.  
The optimal scheme, in the sense that it is SSP with the same timestep as the forward Euler scheme applied to the explicit part, is given by
\begin{align}
  \mathbf{U}^{(1)}
  &= \Lambda_{\mathcal{R}}\Big\{\mathbf{U}^{n} + \dt\,\mathbf{T}(\mathbf{U}^{n})\Big\}, \label{eq:PDARS1} \\
  \widetilde{\mathbf{U}}^{(2)}
  &=\mathbf{U}^{(1)} + \dt\,\mathbf{C}(\widetilde{\mathbf{U}}^{(2)});
  \quad \mathbf{U}^{(2)}=\Lambda_{\mathcal{R}}\Big\{\widetilde{\mathbf{U}}^{(2)}\Big\}, \label{eq:PDARS2} \\
  \mathbf{U}^{(3)}
  &= \Lambda_{\mathcal{R}}\Big\{\mathbf{U}^{(2)} + \dt\,\mathbf{T}(\mathbf{U}^{(2)})\Big\}, \label{eq:PDARS3} \\
  \widetilde{\mathbf{U}}^{n+1}
  &= \f{1}{2}\big(\,\mathbf{U}^{n} + \mathbf{U}^{(3)}\,\big) + \f{1}{2}\dt\,\mathbf{C}(\widetilde{\mathbf{U}}^{n+1});
  \quad \mathbf{U}^{n+1}=\Lambda_{\mathcal{R}}\Big\{\widetilde{\mathbf{U}}^{n+1}\Big\}. \label{eq:PDARS4}
\end{align}
This IMEX scheme involves two explicit evaluations of the transport operator and two implicit solves to evaluate the collision operator.  
The explicit stages, Eqs.~\eqref{eq:PDARS1} and \eqref{eq:PDARS3}, are forward Euler steps, while the implicit stages, Eqs.~\eqref{eq:PDARS2} and \eqref{eq:PDARS4}, can be viewed as backward Euler steps.  
Without collisions ($\mathbf{C}=0$), the scheme reduces to the optimal second-order accurate SSP-RK scheme of \citet{ShOs88} (also referred to as Heun's method).  
Although the scheme is formally only first-order accurate in time when collisions are frequent, quantities evolve on a diffusive time scale in this case, which is much longer than the time step restriction required for stability of the explicit part.  
Therefore, temporal discretization errors remain small.  
On the other hand, second-order accuracy in the streaming limit is essential in maintaining non-oscillatory radiation solutions with the DG method in the streaming regime.  
In Eqs.~\eqref{eq:PDARS1}--\eqref{eq:PDARS4}, $\Lambda_{\mathcal{R}}$ is a realizability-enforcing limiter used to enforce point-wise realizability within each element.  
The limiter, which we describe in more detail below, assumes that the cell-average is realizable after each step.  
We begin by finding sufficient conditions for realizability-preservation of the cell-average in each step.  
For this purpose, since the remaining steps are equivalent, we consider only the explicit step in Eq.~\eqref{eq:PDARS1} and the implicit step in Eq.~\eqref{eq:PDARS2}.  

For an explicit forward Euler update, as in Eq.~\eqref{eq:PDARS1}, the equation for the cell-averaged moments (obtained from Eq.~\eqref{eq:twoMomentDG} with $\varphi_{h}=1$) is given by
\begin{equation}
  \mathbf{u}_{\bK}^{(1)} = \mathbf{u}_{\bK}^{n} 
  - \f{\dt}{\dx}\big[\,\widehat{\mathbf{f}(\mathbf{u}_{h}^{n})}(x_{\hi})-\widehat{\mathbf{f}(\mathbf{u}_{h}^{n})}(x_{\lo})\,\big].
  \label{eq:twoMomentCellAverage}
\end{equation}
To construct a realizability-preserving explicit update for the two-moment model, one seeks to find sufficient conditions such that $\mathbf{u}_{\bK}^{(1)}\in\mathcal{R}$.  
The strategy is very similar to that taken for the bound-preserving scheme discussed in Sect.~\ref{sec:boundPreserving}.  
To evaluate the integral on the right-hand side of Eq.~\eqref{eq:twoMomentCellAverage} (cf.\ Eq.~\eqref{eq:twoMomentBasicCellAverage}), an $N$-point Gauss-Lobatto quadrature rule is used on the interval $K$, with points
\begin{equation}
  \hat{S} = \big\{\,x_{\lo}=\hat{x}_{1},\ldots,\hat{x}_{N}=x_{\hi}\,\big\},
  \label{eq:twoMomentLobattoPoints}
\end{equation}
and weights $\hat{w}_{q}\in(0,1]$, normalized such that $\sum_{q=1}^{N}\hat{w}_{q}=1$.  
Using this quadrature and the numerical flux function in Eq.~\eqref{eq:twoMomentNumericalFluxFunction}, one can write Eq.~\eqref{eq:twoMomentCellAverage} as
\begin{align}
  \mathbf{u}_{\bK}^{(1)}
  &= \sum_{q=1}^{N}\hat{w}_{q}\,\mathbf{u}_{h}^{n}(\hat{x}_{q})
  -\f{\dt}{\dx}\big[\,\mathbf{h}\big(\mathbf{u}_{h}^{n}(x_{\hi}^{-}),\mathbf{u}_{h}^{n}(x_{\hi}^{+})\big)-\mathbf{h}\big(\mathbf{u}_{h}^{n}(x_{\lo}^{-}),\mathbf{u}_{h}^{n}(x_{\lo}^{+})\big)\,\big] \nonumber \\
  &= \sum_{q=2}^{N-1}\hat{w}_{q}\,\mathbf{u}_{h}^{n}(\hat{x}_{q}) 
  + (\hat{w}_{1}+\hat{w}_{N})\,\Phi\big(\mathbf{u}_{h}^{n}(x_{\lo}^{-}),\mathbf{u}_{h}^{n}(x_{\lo}^{+}),\mathbf{u}_{h}^{n}(x_{\hi}^{-}),\mathbf{u}_{h}^{n}(x_{\hi}^{+})\big),
  \label{eq:twoMomentCellAveragePhi}
\end{align}
which is a convex combination of $\{\mathbf{u}_{h}^{n}(\hat{x}_{q})\}_{q=2}^{N-1}$ and $\Phi$.  
(Note that $\hat{w}_{1}=\hat{w}_{N}$, so that $2\,\hat{w}_{1}=2\,\hat{w}_{N}=\hat{w}_{1}+\hat{w}_{N}$.)
Thus, if, for each element $K$, $\mathbf{u}_{h}^{n}(\hat{x}_{q})\in\mathcal{R},\forall q=2,\ldots,N-1$ and $\Phi\in\mathcal{R}$, since the set $\mathcal{R}$ is convex it follows that $\mathbf{u}_{\bK}^{(1)}\in\mathcal{R}$.  
In Eq.~\eqref{eq:twoMomentCellAveragePhi},
\begin{align}
  &\Phi\big(\mathbf{u}_{h}^{n}(x_{\lo}^{-}),\mathbf{u}_{h}^{n}(x_{\lo}^{+}),\mathbf{u}_{h}^{n}(x_{\hi}^{-}),\mathbf{u}_{h}^{n}(x_{\hi}^{+})\big) \nonumber \\
  &= \f{1}{2}\big[\,\mathbf{u}_{h}^{n}(x_{\lo}^{+})+\lambda\,\mathbf{h}\big(\mathbf{u}_{h}^{n}(x_{\lo}^{-}),\mathbf{u}_{h}^{n}(x_{\lo}^{+})\big)\,\big] 
  + \f{1}{2}\big[\,\mathbf{u}_{h}^{n}(x_{\hi}^{-})-\lambda\,\mathbf{h}\big(\mathbf{u}_{h}^{n}(x_{\hi}^{-}),\mathbf{u}_{h}^{n}(x_{\hi}^{+})\big)\,\big] \nonumber \\
  &=(1-\lambda)\,\Phi_{0} + \f{1}{2}\,\lambda\,\Phi_{1} + \f{1}{2}\,\lambda\,\Phi_{2},
  \label{eq:twoMomentPhi}
\end{align}
where $\lambda=\dt/(\dx\,\hat{w}_{1})=\dt/(\dx\,\hat{w}_{N})$ and
\begin{align}
  \Phi_{0}
  &=\f{1}{2}\,\Big[\,\mathbf{u}_{h}^{n}(x_{\lo}^{+})+\mathbf{u}_{h}^{n}(x_{\hi}^{-})\,\Big], \\
  \Phi_{1}
  &=\f{1}{2}\,\Big[\,\mathbf{u}_{h}^{n}(x_{\lo}^{-})+\mathbf{f}\big(\mathbf{u}_{h}^{n}(x_{\lo}^{-})\big)\,\Big]
  + \f{1}{2}\,\Big[\,\mathbf{u}_{h}^{n}(x_{\hi}^{-})-\mathbf{f}\big(\mathbf{u}_{h}^{n}(x_{\hi}^{-})\big)\,\Big] , \\
  \Phi_{2}
  &=\f{1}{2}\,\Big[\,\mathbf{u}_{h}^{n}(x_{\lo}^{+})+\mathbf{f}\big(\mathbf{u}_{h}^{n}(x_{\lo}^{+})\big)\,\Big]
  +\f{1}{2}\,\Big[\,\mathbf{u}_{h}^{n}(x_{\hi}^{+})-\mathbf{f}\big(\mathbf{u}_{h}^{n}(x_{\hi}^{+})\big)\,\Big].  
\end{align}
In the last line in Eq.~\eqref{eq:twoMomentPhi}, if $\lambda\le1$, $\Phi$ is expressed as a convex combination of $\Phi_{0}$, $\Phi_{1}$, and $\Phi_{2}$.  
Thus, if $\Phi_{0},\Phi_{1},\Phi_{2}\in\mathcal{R}$, the time step restriction
\begin{equation}
  \dt\le\hat{w}_{N}\,\dx
\end{equation}
is sufficient to guarantee $\mathbf{u}_{\bK}^{(1)}\in\mathcal{R}$.  
The condition $\Phi_{0}\in\mathcal{R}$ follows from the assumption $\mathbf{u}_{h}^{n}(x_{\lo}^{+}),\mathbf{u}_{h}^{n}(x_{\hi}^{-})\in\mathcal{R}$, while the conditions $\Phi_{1},\Phi_{2}\in\mathcal{R}$ follow from the additional assumptions $\mathbf{u}_{h}^{n}(x_{\lo}^{-}),\mathbf{u}_{h}^{n}(x_{\hi}^{+})\in\mathcal{R}$ and Lemma~2 in \citet{ChEnHa19}, which proves $\Phi_{1},\Phi_{2}\in\mathcal{R}$ provided these expressions can be generated from distributions $f\in\mathfrak{R}$.  
We note that for Lemma~2 in \citet{ChEnHa19} to hold in the current setting, the moments must be consistent with a distribution function satisfying $0\le f\le1$, which demands a two-moment closure based on Fermi--Dirac statistics (the second component of $\Phi_{1}$ and $\Phi_{2}$ involves the Eddington factor).  
The maximum entropy closures of \citet{CeBl94,LaBa11} and the Kershaw-type closure of \citet{BaLa17} are suitable.  
On the other hand, the Minerbo, M1, and Kershaw closures discussed in Sect.~\ref{sec:closure} are based on positive distribution functions (with no upper bound), and are therefore not suitable if $\mathbf{u}\in\mathcal{R}$ is desired.  
These closures are only compatible with the relaxed condition $\mathbf{u}\in\mathcal{R}^{+}$.  
(In this case the approach discussed here, with minor modifications, is still applicable; e.g., see \citet{OlHaFr12} for a method with explicit time stepping.)

For the implicit solve in Eq.~\eqref{eq:PDARS2}, the cell-average with backward Euler gives
\begin{equation}
  \mathbf{u}_{\bK}^{(2)} = \big(\,I+\dt\,R\,\big)^{-1}\big(\,\mathbf{u}_{\bK}^{(1)}+\dt\,\mathbf{\eta}\,\big).  
  \label{eq:twoMomentCellAverageImplicit}
\end{equation}
Here it is assumed that the opacity is constant within each element.  
The first component of Eq.~\eqref{eq:twoMomentCellAverageImplicit} is then
\begin{equation}
  \mathcal{J}_{\bK}^{(2)} = \f{\mathcal{J}_{\bK}^{(1)} + \dt\,\sigma_{A}\,\mathcal{J}_{0,\bK}}{1+\dt\,\sigma_{A}}.  
  \label{eq:chuCollisionMomentComponentOne}
\end{equation}
Since $\mathcal{J}_{\bK}^{(1)},\mathcal{J}_{0,\bK}\in(0,1)$, it follows that $\mathcal{J}_{\bK}^{(2)}\in(0,1)$.  
The second component of Eq.~\eqref{eq:twoMomentCellAverageImplicit} is
\begin{equation}
  \mathcal{H}_{\bK}^{(2)} = \f{\mathcal{H}_{\bK}^{(1)}}{1+\dt\,\sigma_{T}}.
  \label{eq:chuCollisionMomentComponentTwo}
\end{equation}
Then, Lemma~3 in \citet{ChEnHa19}, which considers the moments in Eqs.~\eqref{eq:chuCollisionMomentComponentOne} and \eqref{eq:chuCollisionMomentComponentTwo}, shows that $|\mathcal{H}_{\bK}^{(2)}|<(1-\mathcal{J}_{\bK}^{(2)})\,\mathcal{J}_{\bK}^{(2)}$, so that $\mathbf{u}_{\bK}^{(2)}\in\mathcal{R}$.  
Note that this assumes a very simple form of the collision operator (i.e., emission, absorption, and isotropic and isoenergetic scattering).  
For more complicated collision operators with anisotropic kernels, energy coupling interactions, and Pauli blocking factors it can become very difficult to prove that realizability of the cell-average is preserved in the implicit solve, and this must be investigated separately for each neutrino--matter interaction type.  
Moreover, the ability to prove results rigorously may then depend on the implicit solver used.  

The update in Eq.~\eqref{eq:twoMomentCellAverage} requires that for each element the polynomial approximation is realizable in each point in the quadrature set $\hat{S}$ in Eq.~\eqref{eq:twoMomentLobattoPoints}.  
Thus, after each stage in the time stepping algorithm in Eqs.~\eqref{eq:PDARS1}-\eqref{eq:PDARS4}, a limiter is applied in preparation for the next.  
Let the unlimited solution after any of the stages be $\widetilde{\mathbf{u}}_{h}=\big(\widetilde{\mathcal{J}}_{h},\widetilde{\mathcal{H}}_{h}\big)^{T}$.  
Following \citet{ZhSh10a}, a limiter from \citet{LiOs96} is first used to enforce the bounds on the zeroth moment $\widetilde{\mathcal{J}}_{h}$.  
We replace the polynomial $\widetilde{\mathcal{J}}_{h}(x)$, the first component of $\widetilde{\mathbf{u}}_{h}$, with the limited polynomial
\begin{equation}
  \widehat{\mathcal{J}}_{h}(x)
  =\vartheta_{1}\,\widetilde{\mathcal{J}}_{h}(x)+(1-\vartheta_{1})\,\mathcal{J}_{\bK},
  \label{eq:limitDensity}
\end{equation}
where the limiter parameter $\vartheta_{1}$ is given by
\begin{equation}
  \vartheta_{1}
  =\min\Big\{\,\Big|\f{M-\mathcal{J}_{\bK}}{M_{\hat{S}}-\mathcal{J}_{\bK}}\Big|,\Big|\f{m-\mathcal{J}_{\bK}}{m_{\hat{S}}-\mathcal{J}_{\bK}}\Big|,1\,\Big\},
\end{equation}
with $m=\delta$ and $M=1-\delta$, where $\delta$ is some small number (e.g., $10^{-16}$), and
\begin{equation}
  M_{\hat{S}}=\max_{x\in\hat{S}}\mathcal{J}_{h}(x)
  \quad\text{and}\quad
  m_{\hat{S}}=\min_{x\in\hat{S}}\mathcal{J}_{h}(x).  
\end{equation}
This step, which ensures $\widehat{\mathcal{J}}_{h}\in(0,1)$, corresponds to the bound-enforcing limiter described in Sect.~\ref{sec:boundPreserving}.  
After this step, we denote $\widehat{\mathbf{u}}_{h}=\big(\widehat{\mathcal{J}}_{h},\widetilde{\mathcal{H}}_{h}\big)^{T}$.  

The next step is to enforce $\gamma(\widehat{\mathbf{u}}_{h})\equiv(1-\widehat{\mathcal{J}}_{h})\,\widehat{\mathcal{J}}_{h}-|\widetilde{\mathcal{H}}_{h}|>0$ for all $x\in\hat{S}$, which follows a procedure similar to that developed by \citet{ZhSh10b} to ensure positivity of the pressure when solving the Euler equations of gas dynamics.  
If $\widehat{\mathbf{u}}_{h}$ is outside $\mathcal{R}$ for any quadrature point $x\in\hat{S}$, i.e., $\gamma(\widehat{\mathbf{u}}_{h})<0$, since $\mathbf{u}_{\bK}\in\mathcal{R}$, there exists an intersection point of the straight line $\mathbf{s}_{q}(\psi)$, connecting $\mathbf{u}_{\bK}$ and $\widehat{\mathbf{u}}_{h}$ evaluated in the troubled quadrature point $x_{q}$ (denoted $\widehat{\mathbf{u}}_{q}$), and the boundary of $\mathcal{R}$.  
This line is parameterized by
\begin{equation}
  \mathbf{s}_{q}(\psi)=\psi\,\widehat{\mathbf{u}}_{q}+(1-\psi)\,\mathbf{u}_{\bK},
\end{equation}
where $\psi\in[0,1]$.  
The intersection point $\psi_{q}$ is obtained by solving $\gamma(\mathbf{s}_{q}(\psi))=0$ for $\psi$.  
(In practice, $\psi$ needs not be accurate to many significant digits, and a bisection algorithm terminated after a few iterations is sufficient.)  
This completes the description of major steps in the scheme presented in \citet{ChEnHa19}.  

\subsection{Hybrid Methods}

From the preceding sections, it is clear that the landscape of approaches to neutrino transport, and the associated numerical methods, is growing rapidly.  
One- and two-moment models have reached a level of maturity where general relativistic core-collapse supernova modeling is feasible \citep[e.g.,][]{KuTaKo16,RaJuJa19}.  
Multidimensional models with Boltzmann neutrino transport -- e.g., using discrete ordinate or Monte Carlo methods -- are also under development and results in axial symmetry have already been published \citep{NaIwFu17}, but more work is needed to reach the same level of maturity as found in moments-based models.  
One primary reason is, of course, the computational cost associated with transport models that provide better resolution of the angular dimensions of momentum space, such as Boltzmann models.  
In particular, the computational cost of the neutrino--matter coupling problem increases dramatically with increased fidelity in this sector.  
However, the multiscale nature of the neutrino transport problem implies that Boltzmann neutrino transport is probably not necessary everywhere in a simulation.  
On the one hand, the radiation field is well captured by the low-order moment models in the collision dominated region below the neutrinospheres.  
On the other hand, higher-fidelity models may be warranted in the gain region since heating rates are sensitive to the angular shape of the neutrino distributions.  
(There is already some evidence that two-moment closures are unable to capture certain details in the radiation field; e.g., \cite{HaNaIw19}.)  
This motivates the use of hybrid methods, which, for example, aim to combine low- and high-fidelity approaches in order to provide sufficient resolution where needed, but at a reduced computational cost.  
Hybrid approaches are used in many areas of computational physics, but are not widely adopted to model neutrino transport in core-collapse supernovae.  
We note that the variable Eddington factor (VEF) method of \citet{RaJa02}, which has been shown to compare well with Boltzmann neutrino transport in spherical symmetry \citep{LiRaJa05}, can be regarded as a hybrid method, where a simplified (and less computationally expensive) Boltzmann solver is used in the context of a two-moment model to provide the moment closure.  
Adopting hybrid methods to model neutrino transport in multidimensional models is a potentially rewarding direction for near-future research, and some approaches may even be able to leverage investments in capabilities that have already been developed.  
Since these methods have not fully found their way into the core-collapse supernova modeling community, we will not go into details, but rather briefly mention some existing work, which in most cases will require further development to account for relativity and domain-specific microphysics details.  
We hope to report more on this interesting field in the future.  

So-called high-order--low-order (HOLO) approaches \citep[see, e.g., review by][]{ChChKn17} are one type of hybrid method gaining popularity for use in radiation transport (and related) applications, and combine, as the name suggests, high-fidelity solvers for the (Boltzmann) transport equation with lower-fidelity solvers (typically based on one- or two-moment models, and commonly in a gray formulation) to accelerate the process of solving the high-fidelity model --- in particular, the nonlinear coupling between radiation and a material background.  
In these applications, the radiation field is governed by a kinetic model, while the material is governed by a fluid-like model (as in the core-collapse supernova problem).  
The basic idea is that, in the collisional regime, the interaction between the kinetic and fluid components occurs in a low-dimensional subspace where only a few moments of the particle distribution function are needed to accurately capture the coupling.  
Thus, HOLO methods are effective primarily in regions where the particle mean free path is small and the problem is stiff, and one challenge is to ensure consistency between the two model components.  
Recent work on HOLO methods applied to the problem of thermal radiative transfer include applications where the high-order model is solved with continuum methods such as discrete ordinates \citep[e.g.,][]{PaKnRa12,PaKnRa13,LoMoGe19} or Monte Carlo methods \citep[e.g.,][]{PaKnRa14,BoClMo17}.  
We also point out related work on solving the linear transport equation (i.e., without nonlinear coupling to the material) with HOLO (or hybrid) methods by \cite{HaMc13,WiKeKn13,WiPaTa15,CrChGa17,CrChGa19,CrChHa20}.  

\section{Solution methods}

When ultimately expressed in computer code, all of the previously discussed deterministic methods require the use of implicit numerical methods. 
When discretized, the transport equations produce a set of nonlinear algebraic equations. When linearized, these equations in turn lead to linear 
systems of equations that relate the values of the change in the distribution functions (or moments of the distribution functions) to the neutrino--matter 
and neutrino--neutrino interactions encoded in the terms on the right-hand side of the equations: the collision term. These source terms depend 
on the changes in the neutrino radiation field, as well, giving rise to the need for implicit methods.

The solution of these linear systems is associated with the dominant computational cost for any deterministic method for neutrino transport. 
The remainder of the panoply of physics that complete a core-collapse supernova model---hydrodynamics, 
nuclear kinetics, and even the global solution of the gravitational field---are typically associated with much less computational intensity and
often require significantly less memory capacity and bandwidth. Because the 
solution techniques for the transport linear system solve depend on almost every important dimension of modern computer 
platforms---floating point performance, memory bandwidth, communication bandwidth
and latency---the particulars of individual platforms become an important consideration when a practitioner looks to instantiate a 
real implementation in the form of a production code. Therefore, the structural components of modern computers and the 
quantitative requirements for realistic modeling of transport are inextricably linked together when one looks to build a 
neutrino radiation hydrodynamics code. 

\begin{figure}[htb]
\includegraphics[width=\textwidth]{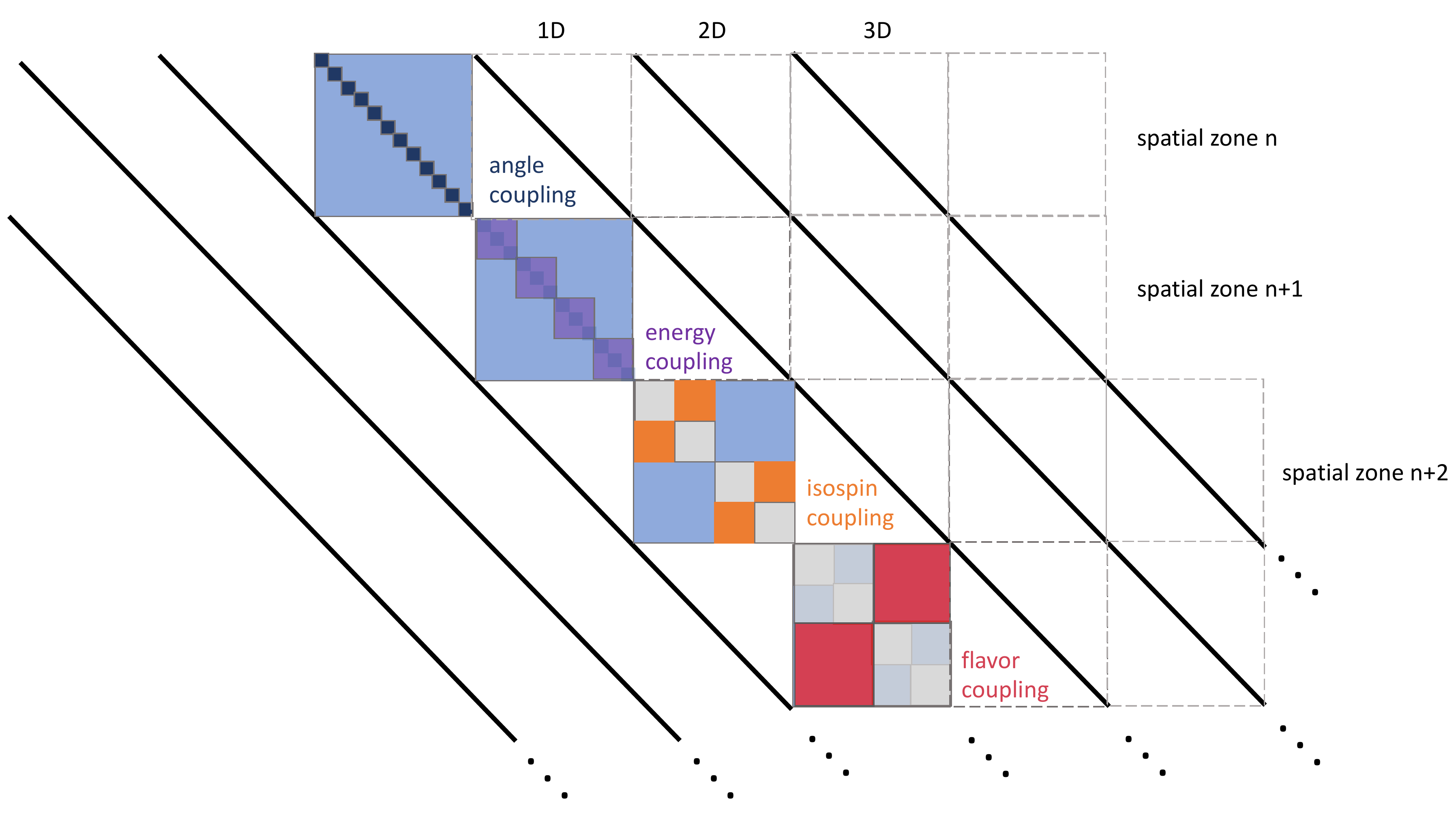}
\caption{A schematic of the structure of a typical neutrino transport linear system that must be solved at each time step. The diagonal, dense blocks are generally non-symmetric and have
characteristic substructure arising from the coupling in angle, energy, isospin (i.e. between neutrinos and antineutrinos), and neutrino flavor, though the particulars of that structure are dependent on the lexical ordering of the solution vector. 
Fully implicit methods also couple individual spatial zones to one another, producing a linear system that contains a series of outlying bands in addition to the diagonally dominant dense block structure. 
This global linear system typically requires considerable communication on parallel platforms, where domain decomposition is often used to spread the spatial extent of the problem 
across the distributed memory space. IMEX methods do not require solution of this global system, but the inversion of a similarly structured set of dense blocks is required at each spatial index. However, this reduction of the implicit problem to a purely local operation can result in considerable performance advantages. }
\label{fig:linearSystemFig}
\end{figure}

\subsection{Simulation requirements}

Regardless of the particulars of the architecture enlisted to solve the requisite equations, the computational demands of neutrino 
radiation hydrodynamics are prodigious. Some of these demands are imposed directly by the high dimensionality of the transport
equation itself. The need to discretize the neutrino phase space with adequate resolution to capture the particulars of 
the neutrino--matter interactions (cf.\ Sect.~\ref{sec:interactions}) results in energy resolutions that are typically on the order of dozens of 
groups. This requirement is amplified by the need to spatially resolve matter features in the flow that are of roughly the size of the 
neutrino mean free path at various points in the computational domain. Adaptive mesh refinement (AMR) can help ameliorate the
need to refine the grid everywhere to resolve the shortest mean free paths, but this reduction is typically only partially effective. 
Indeed, the time-dependent nature of the core-collapse supernova problem often leads to much of the grid having to be refined 
as the reheating and explosion epochs evolve. These resolution requirements directly impact the size of the linear systems that must 
be solved via deterministic methods, typically resulting in quadratic growth in the size of the system for increases in any given 
phase space dimension. 

Therefore, the product of required energy resolution, spatial resolution, number of neutrino flavors and their distribution functions or their angular moments directly
translates into a need for \emph{scalable} implementations of the solution algorithms. Any implementation needs to be able to effectively take advantage
of any future platform. This type of scalability is typically termed weak scaling. The figure of merit for weak scaling 
is how close to a constant runtime can be achieved as the computational load is increased commensurately with the amount of resources. For example, 
as problem size is increased along with the number of MPI ranks used in a simulation, good weak scalability is achieved if the runtime remains constant.  
Weak scalability is often highly dependent on effective distributed-memory parallelism, including possibly overlapping slow inter-node
communication with on-node computation. 

However, this is a necessary, but not sufficient, condition for effective investigation. The
resultant simulations must also be capable of execution in reasonable amounts of wall-clock time. Runtimes of several months 
are untenable if one wishes to explore a more-or-less complete set of supernova progenitors. Therefore, reducing the wall-clock time for 
transport computations is equally important.  This so-called strong scalability is achievable if node-level execution is made faster. On modern platforms, this
has very much become a question of the effective use of hybrid-node architectures. 

\subsection{Implementation on heterogeneous architectures}

Currently, the most widely available and performant microarchitectures are based on graphical processing units (GPUs).  As suggested by their name, GPUs were originally designed to handle computer graphics-intensive tasks in applications ranging from scientific visualization to video games. However, the very high intensity with which they compute and their relatively low power-consumption traits (as compared to modern CPUs) led to their adoption as engines for a variety of scientific computing tasks. Indeed, at this writing, GPU-based architectures dominate much of the highest-end HPC platforms, and \emph{all} planned near-future exascale platforms will employ GPUs as the primary source of compute power. 

The primary characteristic that provides the compute power of modern GPUs is the large number of compute cores, as compared to traditional CPUs. Modern GPUs (e.g. the NVIDIA V100) contain more than 5000 cores, compared to the few dozen that are present on contemporary CPUs. Each core may have a relatively low clock speed compared to a CPU, but the sheer number of processors available on a GPU leads to a much higher intensity of computation.

The architecture of the GPUs is wholly shaped by the single-instruction, multiple-data (SIMD) execution model. In this execution model, each execution unit takes as input two vectors, performs identical operations on both sets of operands (one operand from each vector), and produces a resultant vector. Modern CPUs also typically contain SIMD units: MMX, SSE, and AVX instructions are available on Intel architectures, and POWER and ARM architectures have similar extensions to execution sets to take advantage of similar units. In the case of GPUs, however, these instructions are essentially the only ones available, restricting the amount of branching and conditional execution that can be effectively carried out by the device. 

All modern GPU architectures make use of a similar set of hardware components and associated software abstractions. Here, we will primarily make use of the nomenclature used by NVIDIA to describe their GPU devices, but other vendors make use of virtually identical concepts and constructions, albeit with slightly different naming. In all cases, \emph{kernels} are launched on the device as a set of \emph{threads}. Each of these threads executes a single SIMD pipeline.  Within the kernel launch of threads, threads are grouped into a number of \emph{blocks}. These thread blocks are mapped to individual \emph{streaming multiprocessors (SMs)}. Each SM executes threads in groups of parallel threads termed a warp (the number of threads in a warp, or wave, is typically some multiple of 32). Inside each warp, a single, common instruction is executed during a clock cycle. This lockstep execution can be broken by conditionals (e.g. if-then-else instructions). When this occurs, the effect of this \emph{thread divergence} within a warp breaks the parallelization of the warp. The execution on the conditional thread continues in a serial fashion, and all the other threads are stalled. 

This execution model is further complicated by the hierarchical memory on GPUs. 
Global memory is accessible by all cores. This global memory is typically several GBs on each device. The bandwidth of this memory is often termed high-bandwidth, as it typically has bandwidths several times that for DRAM that might be attached to the CPU host. Closer to each multiprocessor there is a \emph{shared memory} that offers a space accessible to all cores inside the multiprocessor. It is typically used as a user-managed cache of the global memory. The bandwidth to this cache is typically much faster than fetching addresses from the global memory for each core. Ultimately, each core has a certain number of registers that provide the greatest memory bandwidth, but, concomitantly, have the smallest capacities.

Programming GPUs relies on providing as many operands as possible at the maximum possible rate to all of the SMs on a device. The complexity of the memory hierarchy, the execution model, and the possibility of thread divergences can make this a formidable programming task. 

Several programming models have been introduced to program GPUs. These include:

\begin{enumerate}
\item CUDA: a minor extension of C/C++ for GPU thread programming. CUDA is a proprietary programming model created and supported by NVIDIA.
\item ROCm: an extension of C/C++, much like CUDA in purpose and syntax. ROCm was created by AMD and is Open Source.
\item OpenCL: a multi-vendor standard. OpenCL is designed to work on a wide variety of platforms, not just GPUs. This makes the model very powerful, but also introduces a measure of irreducible complexity to accommodate this power. 
\item OpenACC: A directive-based approach to GPU programming, OpenACC uses code decoration much like OpenMP or other directives-based models. OpenACC provides a straightforward path for GPU programming in Fortran.
\item OpenMP (with offload): Modern OpenMP standards include a set of extensions to provide facilities for thread-level programming on GPU devices. 
\end{enumerate}

The choice for any programmer between these options depends on the code to be produced and the relative agility of the development team. For neutrino radiation transport, the Oak Ridge group, for example, has chosen to work primarily in Fortran, with OpenMP directives to marshal the GPUs. This approach allows them to extend legacy code (in Fortran) in a straightforward and performant manner. Using OpenMP provides them with a measure of platform independence, as it is the only programming model currently envisaged to be supported on all major GPU hardware (i.e. NVIDIA, AMD, and Intel devices). The partial loss of thread-level control ceded by not using a more low-level model like CUDA or ROCm is not so important for radiation transport, as the vectorized computational kernels produced in evaluation of both the left and right-hand sides of the transport equation provide plenty of floating-point operations to saturate any modern GPU streaming multiprocessor. Therefore, decorating the multi-level loop nests that contain these vectorized operations at their deepest levels with directives is an effective model. In addition, this programming model can be effectively and easily extended with GPU-enabled scientific libraries (e.g., the GPU-accelerated version of BLAS), regardless of the model used by those libraries internally. 

Many computational radiation transport practitioners have moved to Monte Carlo (MC) approaches in recent years, driven to this choice by the relative abundance of compute power available on GPUs. However, these approaches are not without complexities on GPUs, as the widely disparate sizes of the memory spaces described above (i.e., GBs to kBs to bytes as one moves from global memory to shared memory to registers) mean that MC histories are not so simply preserved. These complications mean that the relative expense of Monte Carlo methods (cf.\ Sect.~\ref{sec:MC}) cannot be fully ameliorated by porting to GPUs. Because the dense linear algebra underpinning their implementations do make effective use of GPU compute architectures, IMEX and discrete ordinates approaches have the potential to compete with MC approaches with reduced memory footprint. But, this strong reliance on a single class of numeric operations means that the success of these approaches is almost wholly dependent on the performance of linear algebra subprograms on GPUs. This is especially true for so-called batched execution of the solution of linear systems of equations, wherein several matrices and right-hand sides are computed by a single kernel invocation and the solver effectively divides the work among SMs.

\section{Summary and outlook}

The last decade has seen considerable, and accelerated, progress made on multiple fronts: (1) Ascertaining the explosion mechanism of core-collapse supernovae. (2) The development of the theory of general relativistic neutrino radiation hydrodynamics. (3) The development of robust numerical methods for the solution of the neutrino radiation hydrodynamics equations in core-collapse supernova environments. (4) And the application of these methods in increasingly sophisticated three-dimensional core-collapse supernova models. At this point, it is fair to say that we are theory and methods rich and that the frontier lies more in the application of these methods in three-dimensional core-collapse supernova models, although further method development is certainly needed. Three-dimensional, fully general relativistic models with all of the relevant neutrino physics in multi-frequency one- or two-moment approaches are on the horizon, the leading examples of which are documented in the work of \citet{KuTaKo16,RoOtHa16,RaJuJa19}. But counterpart models in three dimensions using Boltzmann neutrino transport are farther off, though here too there is a leading example in the work of \citet{NaIwFu17}. Adding a new dimension to the discussion, three-dimensional Boltzmann-based models are limited right now more by supercomputing capabilities than anything else. We have documented both moments and Boltzmann approaches here that have been developed and used by multiple research groups. Boltzmann approaches have been used in core-collapse supernova models with reduced spatial dimensionality and have served to gauge moments approaches in multidimensional models for some time. Recent developments emphasize even more the need for Boltzmann-based models. The history of core-collapse supernova theory has seen quantum leaps on a number of occasions over the past more than fifty years, often associated with an increased glimpse of the rich physics that drive such supernovae. In the past five years, evidence has mounted that neutrino quantum effects---specifically, due to neutrino--neutrino coupling in the proto-neutron star surface region---may impact the electron-flavor neutrino luminosities and spectra responsible for neutrino shock reheating and, consequently, may play a role in the supernova mechanism itself. These early conclusions will require the same extensive development to supplant them as has been documented here for the classical neutrino transport problem. We are far from the equivalent three-dimensional, general relativistic, full-physics models that deploy neutrino quantum kinetics. Early serious work on the implementation of neutrino quantum kinetics in supernova-like environments (e.g., see \citealt{RiMcKn19}) has illuminated yet new numerical challenges that will in turn require augmented methods, to handle both the classical and the quantum mechanical evolution of the three-flavor neutrino radiation field. In this context, then, it is very clear that a Boltzmann kinetic approach, which is a component of a complete quantum kinetics approach, must be a major step toward instantiating full neutrino quantum kinetics. We look forward to watching progress on this front and reporting on these developments as well, as they mature. The core-collapse supernova problem continues to manifest itself as a generational problem, one that will continue to serve as a fertile testbed for the development of transport and radiation hydrodynamics methods.

\begin{acknowledgements}
The authors would like to acknowledge extensive and fruitful discussions with Ernazar Abdikamalov, Thomas Janka, Oliver Just, Takami Kuroda, Hiroki Nagakura, Martin Obergaulinger, Nimoy Rahman, and Doug Swesty regarding their methods, as well as discussions with Cory Hauck. The authors would also like to acknowledge Robert Bollig, Marc Herant, Thomas Janka, Tobias Melson, Bernhard M\"uller, and Hiroki Nagakura for their willingness to include figures from their manuscripts in this review. AM and EE would like to acknowledge support from the National Science Foundation Gravitational Physics Theory program, through grant PHY 1806692. EE and OEBM are supported by the U.S.\ Department of Energy (DOE) Nuclear Physics and/or Advanced Scientific Computing Research programs at the Oak Ridge National Laboratory, which is supported by the Office of Science of the DOE under Contract DE-AC05-00OR22725. AM, EE, and OEBM acknowledge support from the Exascale Computing Project (17-SC-20-SC), a collaborative effort of the U.S.\ Department of Energy Office of Science and the National Nuclear Security Administration. AM acknowledges support from the U.S. Department of Energy, Office of Science, Office of Nuclear Physics and Office of Advanced Scientific Computing Research, Scientific Discovery through Advanced Computing (SciDAC) program under Award Number DE-SC0018232.
\end{acknowledgements}



\end{document}